\documentclass{SciPost_better_arXiv}

\hypersetup{
    colorlinks,
    linkcolor={red!50!black},
    citecolor={blue!50!black},
    urlcolor={blue!80!black}
}

\usepackage[bitstream-charter]{mathdesign}
\DeclareSymbolFont{usualmathcal}{OMS}{cmsy}{m}{n}
\DeclareSymbolFontAlphabet{\mathcal}{usualmathcal}

\usepackage{hyperref}
\usepackage{cleveref}
\usepackage{booktabs}
\usepackage{xcolor}
\usepackage{multirow}
\usepackage{pdflscape}
\usepackage{placeins}
\usepackage{graphicx}

\renewcommand*{\thefootnote}{\fnsymbol{footnote}}
\newcommand{\MH}{M_\mathrm{H}}
\newcommand{\VBF}{\mathrm{VBF}}
\newcommand{\DIS}{\mathrm{DIS}}

\newcommand{\ELWK}{\mathrm{EW}}

\newcommand{\NNNLO}{\mathrm{N3LO}}
\newcommand{\alphas}{\alpha_{\mathrm{s}}}
\newcommand{\GeV}{~\mathrm{GeV}}
\newcommand{\OS}{\mathrm{OS}}

\fancypagestyle{SPstyle}{
\fancyhf{}
\lhead{\colorbox{scipostblue}{\bf \color{white} ~SciPost Physics Community Reports }}
\rhead{{\bf \color{scipostdeepblue} ~Submission }}

\fancyfoot[C]{\textbf{\thepage}}
}

\begin{document}

\pagestyle{SPstyle}
\begin{flushright}
LHCHWG-2026-005, FR-PHENO-2026-012, KA-TP-25-2026, MPP-2026-150, OUTP-26-05P, P3H-26-083, TIF-UNIMI-2025-29, TTK-26-35, TTP26-035
\end{flushright}
\begin{center}{\Large \textbf{\color{scipostdeepblue}{
Predictions for Inclusive Production Cross Sections of the Higgs~Boson at the LHC and HL-LHC\\
}}}\end{center}

\begin{center}\textbf{
    {\it Editors$^\star$:} Robin~Hayes$^{1\dagger}$, Alexander~Karlberg$^{2\ddagger}$, Bernhard~Mistlberger$^{3\circ}$,
    Dermot~Moran$^{4\S}$\\[0.3cm]
    {\it Contributors:}
    Hannah~Arnold$^{5}$,
    Gaetano~Barone$^{6}$,
    Christian~Biello$^{7}$,
    Johannes~Bl\"umlein$^{8,9}$,
    Valeria~Botta$^{10}$,
    Christian~Br{\o}nnum-Hansen$^{11}$,
    Alessandro~Calandri$^{12}$,
    Benjamin~Campillo$^{13}$,
    Alessandra~Cappati$^{14}$,
    Andrea~Cardini$^{15}$,
    Suman~Chatterjee$^{16}$,
    Jiayi~Chen$^{17}$,
    Michal~Czakon$^{18}$,
    Valerio~Dao$^{5}$,
    Goutam~Das$^{19}$,
    Felix~Eschment$^{18}$,
    Silvia~Ferrario~Ravasio$^{20}$,
    Giancarlo~Ferrera$^{21}$,
    Alessandro~Gavardi$^{20}$,
    Yanping~Huang$^{22}$,
    Alexander~Huss$^{23}$,
    Stephen~Jones$^{24}$,
    Judith~Katzy$^{16}$,
    Sven~Yannick~Klein$^{18}$,
    Matthew~Lim$^{25,26}$,
    Stefano~Manzoni$^{27}$,
    Josh~McFayden$^{25}$,
    Sven-Olaf~Moch$^{28}$,
    Alexander~M\"uck$^{18}$,
    Marco~Niggetiedt$^{29}$,
    Mathieu~Pellen$^{30}$,
    Christian~T.~Preuss$^{18}$,
    Tom~Schellenberger$^{18}$,
    Lukas~Simon$^{20,31}$,
    Michael~Spira$^{32}$,
    Lourdes~Urda~G\'omez$^{33}$,
    Davide~Valsecchi$^{34}$,
    Marco~Vitti$^{11,35}$,
    Marius~Wiesemann$^{36}$,
    Malgorzata~Worek$^{18}$,
    Sebastian~Wuchterl$^{27}$,
    Silvia~Zanoli$^{37}$,
    Marco~Zaro$^{21}$ and
    Licheng~Zhang$^{38}$
}\end{center}

\begin{center}
{\normalsize
  {\bf 1} Nikhef, Science Park 105, 1098 XG Amsterdam, The Netherlands\\
  {\bf 2} Max Planck Institute for Physics, Boltzmannstr. 8, 85748 Garching, Germany\\
  {\bf 3} SLAC National Accelerator Laboratory, Stanford University, Stanford, CA 94039, USA\\
  {\bf 4} CIEMAT, Av. Complutense, 40, 28040 Madrid, Spain\\
  {\bf 5} Departments of Physics and Astronomy, Stony Brook University, Stony Brook, NY, USA\\
  {\bf 6} Department of Physics, Brown University, 182 Hope Street, Providence, RI 02912, USA\\
  {\bf 7} Institute for Theoretical Physics, ETH Zurich, 8093 Z\"urich, Switzerland\\
  {\bf 8} Deutsches Elektronen-Synchrotron DESY, Platanenallee 6, 15738 Zeuthen, Germany\\
  {\bf 9} Institut f\"ur Theoretische Physik III, IV, TU Dortmund, Otto-Hahn Stra\ss{}e 4, 44227 Dortmund, Germany\\
  {\bf 10} I. Physikalisches Institut, RWTH Aachen University, Aachen, Germany\\
  {\bf 11} Institute for Theoretical Particle Physics, Karlsruhe Institute of Technology (KIT), Wolfgang-Gaede-Stra\ss{}e 1, 76131 Karlsruhe, Germany\\
  {\bf 12} Universit\`{a} degli Studi di Firenze \& INFN, Sezione di Firenze, Via Sansone 1, 50019 Sesto Fiorentino, Italy\\
  {\bf 13} Institute for Theoretical Physics, Karlsruhe Institute of Technology (KIT), Wolfgang-Gaede-Stra\ss{}e 1, 76131 Karlsruhe, Germany\\
  {\bf 14} Universit\'{e} Catholique de Louvain, Louvain-la-Neuve, Belgium\\
  {\bf 15} Universidad de Oviedo, Instituto Universitario de Ciencias y Tecnolog\'{\i}as Espaciales de Asturias (ICTEA), Oviedo, Spain\\
  {\bf 16} Deutsches Elektronen-Synchrotron DESY, Notkestrasse 85, 22607 Hamburg, Germany\\
  {\bf 17} Department of Physics, 8888 University Dr W, Burnaby, BC V5A1S6, Canada\\
  {\bf 18} Institut f\"ur Theoretische Teilchenphysik und Kosmologie, RWTH Aachen University, 52056 Aachen, Germany\\
  {\bf 19} Department of Physics, Indian Institute of Technology Kanpur, 208016 Kanpur, India\\
  {\bf 20} Universit\`{a} degli Studi di Torino \& INFN, Sezione di Torino, Via Giuria 1, 10125 Torino, Italy\\
  {\bf 21} Dipartimento di Fisica, Universit\`{a} di Milano \& INFN, Sezione di Milano, I-20133 Milan, Italy\\
  {\bf 22} Institute of High Energy Physics, Chinese Academy of Sciences, Beijing, China\\
  {\bf 23} Theoretical Physics Department, CERN, 1211 Geneva 23, Switzerland\\
  {\bf 24} Institute for Particle Physics Phenomenology, Durham University, South Road, Durham DH1 3LE, UK\\
  {\bf 25} Department of Physics and Astronomy, School of Mathematical and Physical Sciences, University of Sussex, Brighton, BN1 9RH, UK\\
  {\bf 26} Universit\`{a} degli Studi di Milano-Bicocca \& INFN Sezione di Milano-Bicocca, Piazza della Scienza 3, Milano 20126, Italy\\
  {\bf 27} Experimental Physics Department, CERN, 1211 Geneva 23, Switzerland\\
  {\bf 28} II. Institute for Theoretical Physics, Hamburg University, Luruper Chaussee 149, 22761 Hamburg, Germany\\
  {\bf 29} Physik-Institut, Universit\"at Z\"urich, Winterthurerstrasse 190, CH-8057 Z\"urich, Switzerland\\
  {\bf 30} Physikalisches Institut, Albert-Ludwigs-Universit\"at Freiburg, D-79104 Freiburg, Germany\\
  {\bf 31} Laboratoire de Physique Th\'eorique et Hautes Energies (LPTHE), UMR 7589, Sorbonne Universit\'e et CNRS, 4 place Jussieu, 75252 Paris Cedex 05, France\\
  {\bf 32} PSI Center for Neutron and Muon Sciences, CH-5232 Villigen PSI, Switzerland\\
  {\bf 33} The Institute of Modern Physics (IMP) of the Chinese Academy of Sciences, Fudan University, Shanghai (China)\\
  {\bf 34} Institute for Particle Physics and Astrophysics, D-PHYS Department, ETH Zurich, Zurich, Switzerland\\
  {\bf 35} Institut f\"ur Astroteilchenphysik, Karlsruhe Institute of Technology (KIT), Hermann-von-Helmholtz-Platz 1, 76344 Eggenstein-Leopoldshafen, Germany\\
  {\bf 36} Department of Physics, TU Dortmund University, 44227 Dortmund, Germany\\
  {\bf 37} Rudolf Peierls Centre for Theoretical Physics, Clarendon Laboratory, Parks Road, Oxford OX1 3PU, UK\\
  {\bf 38} Department of Physics, University of Maryland, College Park (UMD), 20742 College Park, USA
 }
\\[\baselineskip]
$\star$~\href{mailto:lhc-higgs-xsbr-convener@cern.ch}{\small lhc-higgs-xsbr-convener@cern.ch}\,,\quad
$\dagger$~\href{mailto:robin.hayes@cern.ch}{\small robin.hayes@cern.ch}\,,\quad
$\ddagger$~\href{mailto:alexander.karlberg@cern.ch}{\small alexander.karlberg@cern.ch}\,,\quad
$\circ$~\href{mailto:bernhard@slac.stanford.edu}{\small bernhard@slac.stanford.edu}\,,\quad
$\S$~\href{mailto:dermot.anthony.moran@cern.ch}{\small dermot.anthony.moran@cern.ch}
\end{center}
\clearpage
\section*{\color{scipostdeepblue}{Abstract}}
\textbf{\boldmath{%
We present updated recommendations of the LHC Higgs Working Group (LHCHWG) for
the inclusive production cross sections of a Standard Model Higgs boson in
proton--proton collisions at centre-of-mass energies between $7$ and $14$~TeV,
including the Run-3 energy of $13.6$~TeV, and for Higgs-boson masses between
$120$ and $130$~GeV. All the major production modes are covered.
The central values remain consistent with those of Yellow Report~4
(YR4)~\cite{LHCHiggsCrossSectionWorkingGroup:2016ypw} within the
uncertainties quoted there, while the theoretical uncertainties are
reduced for most production modes. The improvements rest on specific
calculations rather than on a uniform advance: the complete N$^3$LO
result in the heavy-top limit together with exact NNLO top-quark mass
effects for gluon fusion; the N$^3$LO structure-function calculation
supplemented by non-factorisable corrections for vector-boson fusion;
the completion of the NLO QCD corrections to the loop-induced $gg\to
ZH$ channel for $VH$; NNLO QCD matched to NNLL soft-gluon resummation
and combined with NLO electroweak corrections for $t\bar{t}H$, and
updated predictions for $tH$; and the matched four- and
five-flavour-scheme prediction, with its explicit resummation-scale
uncertainty, for $b\bar{b}H$. Since no PDF4LHC recommendation exists
beyond NNLO, the mismatch between an N$^3$LO matrix element and NNLO
distributions has to be carried as an uncertainty, estimated process
by process: it is well determined for vector-boson fusion, whereas for
gluon fusion it is doubled relative to YR4. PDFs are now the leading
source of uncertainty for the most precisely predicted modes, while
residual perturbative uncertainties dominate for the remainder. These
results supersede the inclusive cross sections of YR4 and the ad
interim recommendations of Ref.~\cite{Karlberg:2024zxx}, and form part
of Report~5 of the LHCHWG.
}}
\vspace{\baselineskip}

\noindent\textcolor{white!90!black}{%
\fbox{\parbox{0.975\linewidth}{%
\textcolor{white!40!black}{\begin{tabular}{lr}%
  \begin{minipage}{0.6\textwidth}%
    {\small Copyright attribution to authors. \newline
    This work is a submission to SciPost Phys. Comm. Rep. \newline
    License information to appear upon publication. \newline
    Publication information to appear upon publication.}
  \end{minipage} & \begin{minipage}{0.4\textwidth}
    {\small Received Date \newline Accepted Date \newline Published Date}%
  \end{minipage}
\end{tabular}}
}}
}

\vspace{10pt}
\noindent\rule{\textwidth}{1pt}
\tableofcontents
\noindent\rule{\textwidth}{1pt}
\vspace{10pt}


\section{Introduction}
\label{sec:intro}
\renewcommand*{\thefootnote}{\arabic{footnote}}
\setcounter{footnote}{0} \setcounter{page}{1}
The discovery of a scalar particle compatible with the Standard Model
(SM) Higgs boson by the ATLAS and CMS collaborations in
2012~\cite{ATLAS:2012yve,CMS:2012qbp} completed the particle content
of the SM and inaugurated a systematic and ever more precise
experimental study of the Higgs sector. In the years since, the LHC
experiments have measured the mass, spin, parity, and couplings of the
Higgs boson in ever finer detail. Combined analyses of the full LHC
Run-2 dataset by ATLAS and CMS~\cite{ATLAS:2022vkf,CMS:2022dwd},
together with more recent updates based on the complete Run-2
sample~\cite{ATLAS:2026comb,CMS:2026nce}, have established the
couplings of the Higgs boson to third-generation fermions and to the
electroweak gauge bosons at the level of a few to ten percent,
found evidence for its decay to muons, and set increasingly stringent limits on
its couplings to light quarks, its total width, possible CP-violating
admixtures, and on Higgs boson pair production, which probes the
trilinear self-coupling directly.  The LHC has recently concluded its Run 3
at a centre-of-mass energy of $\sqrt{s}=13.6$~TeV, and the
High-Luminosity LHC (HL-LHC) is expected to ultimately deliver up to
$3$--$4\,\mathrm{ab}^{-1}$ per experiment, which will reduce
the statistical uncertainties on many Higgs-boson observables to the
percent level or below~\cite{Cepeda:2019klc}.

This remarkable experimental progress, and the even higher precision
anticipated at the HL-LHC, places correspondingly stringent demands on
the accuracy of the theoretical predictions used to interpret the
data. For several production and decay channels, the precision of
current measurements is already comparable to, or in some cases better
than, the theoretical uncertainties quoted in the most recent
comprehensive reference for Higgs production cross sections, the CERN
Yellow Report~4 (YR4)~\cite{LHCHiggsCrossSectionWorkingGroup:2016ypw},
compiled by the LHC Higgs Cross Section Working Group (LHCHWG) in
2016. Moreover, YR4 does not provide predictions at
$\sqrt{s}=13.6$~TeV, the current LHC operating energy; an interim
interpolation was subsequently provided by the working
group~\cite{Karlberg:2024zxx} as a stop-gap measure pending a full
update. Both of these facts make clear that a systematic update of the
SM predictions for all Higgs-boson production modes, incorporating the
theoretical progress of the past decade, is both timely and necessary
if the LHC and HL-LHC precision programmes are to be fully exploited.

Since the publication of YR4, the theoretical description of every
production mode considered here has advanced. For gluon fusion, the
dominant mode, the inclusive cross section in the heavy-top limit is
now known exactly at N$^3$LO in
QCD~\cite{Anastasiou:2015vya,Anastasiou:2016cez,Mistlberger:2018etf},
rather than through a threshold expansion. It has been supplemented by
exact finite-quark-mass corrections at NNLO, both for the top-only
contribution~\cite{Czakon:2020vql,Czakon:2021yub,Czakon:2024ywb} and
for the top--bottom interference~\cite{Czakon:2023kqm,Czakon:2024ywb},
by mixed QCD--electroweak corrections to the gluon-induced channel
retaining the full dependence on the electroweak boson
masses~\cite{Bonetti:2016brm,Bonetti:2017ovy,Bonetti:2018ukf,Becchetti:2020wof},
and by an approximate N$^4$LO result obtained in the soft-virtual
approximation~\cite{Das:2020adl}. Vector-boson fusion is available at
N$^3$LO in the structure-function approximation~\cite{Dreyer:2016oyx},
and the non-factorisable corrections neglected in that approximation
have since been computed~\cite{Liu:2019tuy,Dreyer:2020urf,Asteriadis:2023nyl}.
Associated production with a $W$ or $Z$ boson has reached N$^3$LO QCD
accuracy for the Drell--Yan-like
contribution~\cite{Baglio:2022wzu}, and the NLO QCD corrections to the
loop-induced $gg\to ZH$ channel have been completed with full
top-quark mass
dependence~\cite{Chen:2020gae,Wang:2021rxu,Chen:2022rua,Degrassi:2022mro,Davies:2025out},
with first NNLO results in the large-$m_t$ limit appearing
recently~\cite{Davies:2025otz}. For $t\bar{t}H$, the NNLO QCD
corrections~\cite{Catani:2022mfv,Devoto:2024nhl} have been matched to
NNLL soft-gluon resummation and combined with the complete NLO
electroweak corrections~\cite{Balsach:2025jcw}. Single-top associated
production remains known at NLO QCD~\cite{Demartin:2015uha,Demartin:2016axk};
NLO electroweak corrections are also available, but only for the three
production channels taken together, since the separation into channels
is no longer well defined once they are
included~\cite{Pagani:2020mov}. Production in association with bottom
quarks has reached N$^3$LO accuracy in the five-flavour
scheme~\cite{Duhr:2019kwi} and has been matched to the four-flavour
scheme at that order using the FONLL method~\cite{Duhr:2020kzd}; the
recommendation presented below is instead based on the
NLO+NNLL\textsubscript{part} matching of
Refs.~\cite{Bonvini:2015pxa,Bonvini:2016fgf}, which allows an explicit
resummation-scale uncertainty to be assigned.

The parton distribution functions (PDFs) entering all of these
predictions have been refined in parallel. The PDF4LHC21
combination~\cite{PDF4LHCWorkingGroup:2022cjn} is the recommended NNLO
baseline and is used throughout this report. Beyond NNLO, approximate
N$^3$LO sets have been produced by the MSHT and NNPDF
collaborations~\cite{McGowan:2022nag,NNPDF:2024nan}, benchmarked
against one another~\cite{Cooper-Sarkar:2024crx} and subsequently
combined~\cite{Cridge:2024icl}. Separately, the effect of QED
corrections to the DGLAP evolution has been quantified for Higgs
production for the first time~\cite{Manohar:2016nzj,Cooper-Sarkar:2025sqw}.
Both developments bear on gluon fusion more than on any other mode,
since the hard-scattering cross section is known one order beyond the
PDFs and since sub-percent changes in the gluon luminosity are no
longer negligible at the precision now reached. They are introduced in
Section~\ref{sec:setup} and their consequences for the recommendation
are worked out in Section~\ref{sec:ggF}.

The LHCHWG is currently preparing a set of updates which will
collectively replace YR4, published as a series of reports in SciPost
Physics Community Reports and referred to in the following as Report~5
(R5). Where numerical predictions are compared, R5 denotes the recommendations
of the present report and YR4 those of its predecessor. Several of these have already appeared, among them dedicated
studies of vector-boson fusion~\cite{Barone:2025jey}, of
$t\bar{t}H$~\cite{Balsach:2025jcw}, of the $gg\to ZH$
channel~\cite{CampilloAveleira:2025rbh}, and of the modelling of
$b\bar{b}H$ production~\cite{10.21468/SciPostPhysCommRep.18}. The
present report is the contribution to R5 concerned with inclusive
production cross sections. It collects the developments listed above
into an updated set of LHCHWG recommendations at
$\sqrt{s}=7,\,8,\,13,\,13.6$, and $14$~TeV and for Higgs-boson masses
between $120$ and $130$~GeV, covering gluon fusion (ggF), vector-boson
fusion (VBF), associated production with a $W$ or $Z$ boson (VH),
associated production with a top-quark pair ($t\bar{t}H$) or a single
top quark ($tH$), and associated production with bottom quarks
($b\bar{b}H$). For $t\bar{t}H$ and $tH$, predictions are provided at
$13$, $13.6$ and $14$~TeV only. For each production mode we document
the ingredients entering the recommendation, define the procedure by
which they are combined, and state the residual theoretical and
parametric uncertainties. We further place the resulting precision in
the context of the most precise current LHC measurements and of
projected HL-LHC
sensitivities~\cite{Cepeda:2019klc,ATLAS:2022hsp}, in order to assess
to what extent the present theoretical uncertainties will remain
sufficient, or become a limiting factor, over the coming decade of LHC
running. Within this scope the report supersedes
YR4~\cite{LHCHiggsCrossSectionWorkingGroup:2016ypw} and the ad interim
recommendations of Ref.~\cite{Karlberg:2024zxx}; branching ratios,
differential distributions and off-shell effects are not addressed
here.

The chapters that follow differ considerably in length, for two
reasons. For most production modes, the calculations entering the
recommendation and the reasoning behind the choices made in combining
them are discussed at length elsewhere in R5 and in the literature it
builds on. Those chapters therefore concentrate on the ingredients
that define the present prediction, on the combination formula and on
the uncertainty prescription, and point the reader onwards for the
supporting material. Gluon fusion is the exception. No such account
exists elsewhere, and the number of separate decisions that had to be
taken is in any case larger. The renormalisation scheme of the
top-quark mass, the finite quark-mass corrections now available at
NNLO, the mismatch between an N$^3$LO matrix element and NNLO PDFs,
and the inclusion of QED evolution each shift either the central value
or the uncertainty at the percent level, and they do not all act in
the same direction. Section~\ref{sec:ggF} consequently treats each
ingredient and its uncertainty in turn, in the manner of YR4, and is
correspondingly longer than the others. This unevenness reflects where
the work was needed rather than the relative importance of the
production modes.

The remainder of this report is organised as
follows. Section~\ref{sec:setup} defines the common input parameters
and conventions used throughout. Sections~\ref{sec:ggF}--\ref{sec:bbH}
present the state of the art and the resulting recommendations for
ggF, VBF, VH, $t\bar t H$/$tH$, and $b\bar b H$ production,
respectively. Section~\ref{sec:Outlook} discusses the resulting
uncertainties in comparison with YR4 and with current and projected
experimental precision, and Section~\ref{sec:conclusions}
concludes. Numerical tables for all production modes, collision
energies and Higgs-boson masses considered are collected in
Appendix~\ref{app:tables}; for gluon fusion these also contain the
information required to convert the recommendation to approximate
N$^3$LO PDFs. The numbers are additionally available at:
\begin{center}\url{https://gitlab.cern.ch/LHCHIGGSXS/LHCHXSWG1/crosssections}.\end{center}

\section{Common setup}
\label{sec:setup}
Here we specify the common input parameters and baseline conventions used for
the predictions presented in this report. Unless stated otherwise in the
individual production-mode sections, the numerical inputs given below are used
throughout. Where applicable, these inputs follow the Review of Particle
Physics (PDG)~\cite{ParticleDataGroup:2020ssz}.\footnote{The 2020 edition of the
PDG was the most recent one when this work was started.} Departures from this
common setup are stated explicitly in the corresponding sections, and the most
significant of them are collected in Section~\ref{sec:pdf}.

\subsection{Fermion masses}
We use the following quark-mass inputs:
\begin{align}
\label{eq:qmasses}
&M_t^\mathrm{OS}=172.5\,\mathrm{GeV},&
&M_b^{\overline{\mathrm{MS}}}(M_b^{\overline{\mathrm{MS}}})=4.18\,\mathrm{GeV},&
&M_c^{\overline{\mathrm{MS}}}(3.0\,\mathrm{GeV})=0.986\,\mathrm{GeV}.&
\end{align}
Quark widths are neglected. The mass-renormalisation scheme used in each
calculation is specified in the corresponding production-mode section. Where an
$\overline{\mathrm{MS}}$ top-quark mass is required, we use
\begin{align}
M_t^{\overline{\mathrm{MS}}}(M_t^{\overline{\mathrm{MS}}})
=162.5\,\mathrm{GeV}.
\end{align}

\subsection{Gauge boson masses}

The $W$- and $Z$-boson masses and widths are taken on-shell,
\begin{align}
M^\OS_W &= 80.379\GeV\,, \qquad &\Gamma^\OS_W &= 2.085\GeV\,, \notag\\
M^\OS_Z &= 91.1876\GeV\,, \qquad &\Gamma^\OS_Z &= 2.4952\GeV\,,
\end{align}
and are converted, where an electroweak calculation requires it, to the
corresponding pole masses and widths~\cite{BARDIN1988539} according to
\begin{align}
M_V &=
\frac{M^\OS_V}{\sqrt{1+(\Gamma^\OS_V/M^\OS_V)^2}}\,,
\qquad
\Gamma_V =
\frac{\Gamma^\OS_V}{\sqrt{1+(\Gamma^\OS_V/M^\OS_V)^2}}\,.
\end{align}

The electromagnetic coupling is defined in the $G_\mu$
scheme~\cite{Denner:2000bj}. In terms of the Fermi constant and the on-shell
gauge-boson masses,
\begin{align}
\alpha_{G_\mu}
&= \frac{\sqrt{2}}{\pi}G_F(M^\OS_W)^2\sin^2\theta_W\,,
&
\sin^2\theta_W
&=1-\frac{(M^\OS_W)^2}{(M^\OS_Z)^2}\,,
&
G_F&=1.16638\cdot10^{-5}\GeV^{-2}\,,
\end{align}
which gives
\begin{equation}
\alpha_{G_\mu}=0.007565210\,.
\end{equation}

\subsection{Higgs masses and collider energies}
Predictions are provided on a grid of Higgs-boson masses in the range
$120\leq \MH\leq130\GeV$. The grid is chosen to resolve the region around the
measured value, and includes $\MH=125.09\GeV$, the ATLAS+CMS Run-1
combination~\cite{ATLAS:2015yey}, and $\MH=125.38\GeV$, as measured by
CMS~\cite{CMS:2020xrn}:
\begin{align}
\MH = \{&120, 122, 124, \notag\\
        &124.6, 124.8, 125, 125.09, 125.2, 125.3, 125.38, 125.6, \notag\\
        &126, 128, 130\}\,\mathrm{GeV}.
\end{align}

The centre-of-mass energies cover the legacy LHC analyses, the current
programme, and the HL-LHC:
\begin{align}
\sqrt{s}=\{7,8,13,13.6,14\}\,\mathrm{TeV}.
\end{align}
The only exception is $t\bar tH$ and $tH$ production, for which predictions are
given at $13$, $13.6$ and $14$~TeV; for the two lower energies we continue to
recommend the numbers of Ref.~\cite{Karlberg:2024zxx}.

\subsection{Parton distributions and \texorpdfstring{$\alphas$}{alpha\_s}}
\label{sec:pdf}

The baseline set for this report is PDF4LHC21\_40 at NNLO, and the strong
coupling is
\begin{equation}
    \alphas(M_Z)=0.118\pm0.001\,,
\end{equation}
with PDF and $\alphas$ uncertainties evaluated following the PDF4LHC
recommendation~\cite{PDF4LHCWorkingGroup:2022cjn}. The PDF4LHC21 set is a Monte Carlo
combination of the CT18, MSHT20 and NNPDF3.1
sets~\cite{Hou:2019efy,Bailey:2020ooq,NNPDF:2017mvq}, reduced here to a
$40$-member Hessian representation. The working group takes the PDF4LHC
recommendation as its reference prescription, and PDF4LHC21 is the most recent
such recommendation; this, rather than a process-by-process optimisation, is
what fixes the baseline. It also explains why the use of parton distributions
beyond NNLO is a more delicate matter, since no PDF4LHC recommendation exists
at that accuracy; we return to this point in Section~\ref{sec:an3lo-pdfs}.
Whenever a calculation is performed in the four-flavour scheme, the
corresponding four-flavour set is used.

Two departures from this baseline should be stated plainly.

The first concerns $b\bar bH$ production, which does not use PDF4LHC21. A
consistent matching of the four- and five-flavour schemes requires the bottom
distribution to be evolved across the flavour threshold in a uniform way and
for every value of $m_b$ and of the matching scale $\mu_b$ entering the
matched prediction. The sets entering the PDF4LHC21 combination adopt
different values of $m_b$, so the combination itself cannot serve this purpose.
Dedicated sets are used instead, obtained by re-evolving the
\texttt{PDF4LHC15\_nnlo\_mc} input from an initial scale of $2\GeV$, as
described in Ref.~\cite{Bonvini:2016fgf} and in Section~\ref{sec:bbH}.

The second concerns the photon. PDF4LHC21\_40 carries no photon distribution,
so wherever photon-initiated channels are included in the electroweak
corrections we take the photon from
LUXqed17\_plus\_PDF4LHC15\_nnlo\_100~\cite{Manohar:2017eqh} while retaining
PDF4LHC21\_40 for the quarks and gluons. This hybrid is not internally
consistent: a photon distribution is inseparable from the QED corrections to
the DGLAP evolution, which also modify the quark and gluon distributions. It is
useful to keep the two effects apart. Photon-initiated partonic channels
contribute explicitly to the hard-scattering cross section, whereas QED
evolution changes the parton luminosities entering all remaining channels. The
prescription above accounts for the first and neglects the second.

It should be said that this prescription was adopted at an early stage of the
present work, before dedicated studies of QED effects in Higgs production were
available, and that it has been retained for the production modes whose
predictions were completed under it. The explicit photon-initiated contribution
is process dependent but numerically stable: for VBF at $\sqrt{s}=14$~TeV and
$\MH=125.09\GeV$ it amounts to $40.6\,\mathrm{fb}$, or $0.94\%$ of the total
cross section, and to $0.92\%$ and $0.93\%$ at $13$ and $13.6$~TeV
respectively. The neglected QED-evolution effect has since been quantified in
Ref.~\cite{Cooper-Sarkar:2025sqw}, which constructs a PDF4LHC21 QED correction
from the corresponding CT18, MSHT20 and NNPDF3.1 QED-enhanced sets and finds
shifts at $14$~TeV of $-0.7\%$ for VBF, $-0.6\%$ for $ZH$ and $W^+H$ and
$-0.55\%$ for $W^-H$. For VBF we have evaluated the same correction directly in
the setup of this report, obtaining $-0.66\%$ at each of $\sqrt{s}=13$, $13.6$
and $14$~TeV, in agreement with the value quoted above. These are to be compared with a PDF$+\alphas$
uncertainty of $2.2\%$ and a total uncertainty of $2.4\%$ for VBF, and with
uncertainties of around $2\%$ for $W^\pm H$. The omitted effect is therefore
smaller than the uncertainties already assigned, and we do not introduce a
further uncertainty to cover it. This is not to say that the effect is
negligible, only that accounting for it would not change the central
predictions or their uncertainty budgets appreciably. We stress that the
photon-initiated contribution and the QED-evolution shift are distinct and
should not be treated as correlated; their partial cancellation in VBF is
incidental.

Gluon fusion is treated differently, and the report therefore does not apply a
single QED prescription throughout. There the cross section is known at
N$^3$LO, sub-percent changes in the gluon luminosity are not negligible against
the other uncertainties, and the QED-evolution shift is instead included
explicitly following Ref.~\cite{Cooper-Sarkar:2025sqw}; photon-initiated Higgs
production is not included in that case. The implementation and its numerical
impact are described in Section~\ref{sec:ggf_qed_pdf}. A uniform treatment of
photon-initiated contributions and QED evolution across all production modes
would be desirable in a future update.

\subsection{Parton distributions beyond NNLO}
\label{sec:an3lo-pdfs}

Inclusive predictions for gluon fusion, for vector-boson fusion and for the
Drell--Yan-like part of $VH$ production are available at N$^3$LO in
QCD~\cite{Mistlberger:2018etf,Dreyer:2016oyx,Baglio:2022wzu}, while the
baseline parton distributions defined above are determined and evolved at NNLO.
Of these, the gluon-fusion and vector-boson-fusion recommendations of this
report are built on the N$^3$LO results; the $VH$ recommendation is based on
the NNLO calculation, as described in Section~\ref{sec:VH}, and is therefore
not affected.
A formally consistent prediction would require both at the same accuracy. No
PDF4LHC recommendation exists beyond NNLO, so this mismatch cannot be resolved
within the prescription the working group follows, and we retain NNLO parton
distributions as the baseline. The mismatch is instead carried as an
uncertainty, which is numerically most important for gluon fusion and is
discussed, together with the alternative of using approximate N$^3$LO sets, in
Section~\ref{sec:ggf_an3lo_pdf}.

The ingredients needed to go beyond NNLO have nevertheless advanced
considerably in recent years, in the massless evolution kernels and in the
treatment of heavy quarks alike. For the four-loop splitting functions, even-$N$
moments of the quark-to-gluon and gluon-gluon kernels have been computed and,
combined with the known endpoint behaviour, turned into controlled $x$-space
approximations~\cite{Falcioni:2024xyt,Falcioni:2024qpd}; the resulting N$^3$LO
contribution to the scale derivative of the gluon distribution amounts to $1\%$
or less for $x\gtrsim10^{-4}$. The non-singlet sector is now known exactly: all
four-loop non-singlet splitting functions have been obtained in fully analytic
form~\cite{Gehrmann:2026qbl}. Independent implementations of approximate
N$^3$LO evolution in \texttt{APFEL++}, \texttt{EKO}, \texttt{HOPPET} and
\texttt{Candia}~\cite{Bertone:2017gds,Candido:2022tld,Karlberg:2025hxk,Hampson:2025pvi}
allow these ingredients to be cross-checked, and dedicated benchmarking of the
MSHT and NNPDF evolution codes against the FHMRUVV approximation finds
excellent agreement once common ingredients are
used~\cite{Cooper-Sarkar:2024crx,Bertone:2024dpm}.

Heavy quarks require separate treatment. Their contributions to the structure
functions exhibit scaling violations that differ from the massless
case~\cite{Blumlein:2016xcy}, and must be accounted for through the
heavy-flavour Wilson coefficients in deep-inelastic scattering and in other
hard-scattering processes. The same corrections fix the parton densities in
variable-flavour-number schemes (VFNS), in the single-mass~\cite{Buza:1996wv}
and two-mass~\cite{Ablinger:2017err} cases, through the associated massive
operator matrix elements (OMEs). For $F_2(x,Q^2)$ the asymptotic description is
accurate at the $O(1\%)$ level for $Q^2/m_Q^2\gtrsim10$~\cite{Buza:1995ie}, a
region matching the cuts usually imposed in PDF fits to suppress higher-twist
contributions. Until recently these corrections were known only to two
loops~\cite{Laenen:1992zk,Laenen:1992xs,Buza:1995ie,Buza:1996wv,
Bierenbaum:2007qe,Bierenbaum:2008yu,Bierenbaum:2009zt,Bierenbaum:2022biv}.
Starting from the three-loop Mellin moments of
Ref.~\cite{Bierenbaum:2009mv}, the three-loop contributions to the massive OMEs
were obtained over the period 2009--2025 in the unpolarised and polarised
single- and two-mass cases, together with the corresponding asymptotic
heavy-flavour Wilson coefficients~\cite{Ablinger:2010ty,Ablinger:2014vwa,
Ablinger:2014nga,Ablinger:2014lka,Behring:2014eya,Ablinger:2017err,
Ablinger:2018brx,Ablinger:2017xml,Ablinger:2022wbb,Ablinger:2023ahe,
Ablinger:2024xtt}; the two-mass contributions to
$\widetilde{A}_{Qg}^{(3)}$ of Ref.~\cite{Ablinger:2025nnq} complete the
calculation of all unpolarised and polarised massive three-loop OMEs. A survey
of the mathematical methods developed along the way is given in
Ref.~\cite{Blumlein:2023aso}. In the singlet sector the heavy-quark corrections
to the structure functions are large and their scaling violations differ
markedly from the massless case. These results make consistent NNLO non-singlet
analyses~\cite{Blumlein:2021lmf} and singlet analyses with a correlated
treatment of $\alphas$, $m_c$, $m_b$ and the parton densities possible for the
first time. They also lift the VFNS matching relations from two-loop to
three-loop order in the single-mass case~\cite{Ablinger:2025joi}, giving a
refined description of a massive quark becoming light at large virtualities,
and allowing heavy-quark parton distributions to be defined in that framework.

Two approximate N$^3$LO (aN$^3$LO) global determinations exist, from
MSHT~\cite{McGowan:2022nag} and from NNPDF~\cite{NNPDF:2024nan}, which differ
in how they treat the ingredients that were unknown at the time of the
respective fits, and they have been combined into a single
set~\cite{Cridge:2024icl}. An update of the MSHT analysis incorporating the
newer splitting-function and transition-matrix-element results finds changes to
the PDFs at most comparable to, and typically well below, the quoted
uncertainties, with still smaller effects on the parton
luminosities~\cite{Cridge:2025oel}. The differences these sets induce in the
Higgs cross sections are nevertheless significant, in several cases exceeding
the PDF uncertainty itself~\cite{Cridge:2024icl}.

How that mismatch is best estimated depends on the process, and in a way worth
stating here since the two production modes affected resolve it differently.
The most direct estimate is the difference between the NNLO and aN$^3$LO
predictions themselves, but that difference is only as meaningful as the
agreement between the two independent aN$^3$LO determinations. That agreement
is strongly process dependent: for vector-boson fusion the MSHT20 and NNPDF4.0
aN$^3$LO sets give cross sections agreeing to the quoted precision, while for
gluon fusion they differ by around $3\%$, on an effect of comparable size in
the two cases~\cite{Cridge:2024icl}. Where the two determinations agree the
direct difference is a meaningful measure of the effect in its own right; this
is how it is quantified for VBF in Section~\ref{sec:vbf_an3lo}, where it is
found to be comparable to the uncertainty already assigned and is therefore not
applied to the central prediction. Where they do not agree, a more conservative
prescription is required, and the treatment adopted for gluon fusion is set out
in Section~\ref{sec:ggf_an3lo_pdf}.

\section{ggF}
\label{sec:ggF}
\begin{table}[h!]
\centering
\begin{tabular}{c|c}
\hline
\textbf{Calculations} & \textbf{References} \\ 
\hline
Approximate N$^4$LO HTL QCD & \cite{Das:2020adl} \\
N$^3$LO HTL QCD   & \cite{Anastasiou:2015vya,Anastasiou:2016cez,Mistlberger:2018etf} \\
NNLO HTL QCD   & \cite{Anastasiou:2002yz,Harlander:2002wh,Ravindran:2003um} \\
NLO HTL QCD   & \cite{Dawson:1990zj,Djouadi:1991tka} \\
NNLO QCD & \cite{Czakon:2020vql,Czakon:2021yub,Czakon:2023kqm,Czakon:2024ywb}\\
NLO QCD & \cite{Graudenz:1992pv,Spira:1995rr} \\
EW \& Mixed QCD-EW Corrections & \cite{Actis:2008ts,Actis:2008ug,Aglietti:2004nj,Becchetti:2020wof} \\
N$^3$LL Threshold Resummation & \cite{Anastasiou:2016cez,Bonvini:2014joa,Catani:2014uta} \\
\hline
\textbf{PDFs} & \textbf{References} \\
\hline
Approximate N$^3$LO PDFs & \cite{Cridge:2024icl,McGowan:2022nag,NNPDF:2024nan,Cooper-Sarkar:2024crx} \\
QED Evolution PDFs & \cite{Manohar:2016nzj,Manohar:2017eqh,Bertone:2017bme,Harland-Lang:2019pla,Xie:2021equ,Cridge:2021pxm,Cridge:2023ryv,Barontini:2024dyb,NNPDF:2024djq} \\
NNLO PDFs & \cite{PDF4LHCWorkingGroup:2022cjn,Hou:2019efy,Bailey:2020ooq,Alekhin:2017kpj,NNPDF:2017mvq}\\
\hline
\textbf{Codes} & \textbf{References} \\ 
\hline
N$^3$LO HTL QCD & \cite{Dulat:2018rbf}
\end{tabular}
\caption{Summary of theory references for Higgs boson production in gluon fusion at the LHC upon which this recommendation is based.
}
\label{tab:ggf_cite}
\end{table}

In this section, we briefly recall the definition and state of the art for each ingredient entering the cross section for Higgs boson production in gluon fusion.
We summarise which ingredients enter the current recommendation and document the details of their combination.

The partonic cross section may be decomposed up to order N$^3$LO as~\cite{Dulat:2018rbf},
\begin{align}
\sigma_{ij} = & 
\mathrm{R}_\mathrm{LO} C^2 \left( 
\sigma_{ij}^{\mathrm{LO}, \mathrm{HTL}} 
+ \sigma_{ij}^{\mathrm{NLO}, \mathrm{HTL}} 
+ \sigma_{ij}^{\mathrm{NNLO}, \mathrm{HTL}} 
+\sigma_{ij}^{\mathrm{N}^3\mathrm{LO}, \mathrm{HTL}} \right) \nonumber \\ 
&+ \delta \sigma_{ij}^{\mathrm{LO}, (t,b,c)}
+ \delta \sigma_{ij}^{\mathrm{NLO}, (t,b,c)}
+ \delta \sigma_{ij}^{\mathrm{NNLO}, (t,b,c)}\,, 
\label{eq:ggf_master}
\end{align}
where \begin{align}
  \label{eq:RLO}
  \mathrm{R}_\mathrm{LO} = \sigma_{ij}^{\mathrm{LO},(t)}/\sigma_{ij}^{\mathrm{LO}, \mathrm{HTL}}\,,\end{align}
and $\sigma_{ij}^{\mathrm{LO},(t)}$~\cite{Georgi:1977gs,Wilczek:1977zn} is the full theory result including the contribution from top-quark loops only. We discuss the definition, calculation and recommendations for each of these contributions in the following subsections.
In Section~\ref{sec:ggf_htl}, we summarise the results known in  the Heavy Top Limit (HTL), $\sigma_{ij}^{\mathrm{N}^n\mathrm{LO},\mathrm{HTL}}$.
In Section~\ref{sec:ggf_ew}, we discuss the electroweak and mixed QCD--EW corrections, which enter via the combined Wilson coefficient, $C$.
In Section~\ref{sec:ggf_mq}, we discuss corrections related to the finite quark masses $\delta \sigma_{ij}^{\mathrm{N}^n\mathrm{LO},(t,b,c)}$.
In Section~\ref{sec:ggf_an3lo_pdf}, the impact of approximate N$^3$LO PDFs is discussed.
Section~\ref{sec:ggf_qed_pdf} presents the impact of including QED evolution within the PDFs on the Higgs boson total cross section.
The impact of threshold resummation effects is small beyond N$^3$LO~\cite{Anastasiou:2016cez,Bonvini:2014joa,Catani:2014uta}, and we do not include these effects in our central recommendation.
Results are also known to N$^4$LO in the HTL using a soft-virtual approximation~\cite{Das:2020adl}, although these results do not directly enter the recommendation, we perform a comparison to this work in Section~\ref{sec:ggf_n4lo_sv}.
In Table~\ref{tab:ggf_cite} we collect a list of references for theory work either entering directly or indirectly informing decisions and/or uncertainties entering the recommendation, we encourage the reader to cite all of these works along with the current report when using the gluon-fusion numbers presented in Appendix~\ref{app:tables}.



In addition to the parameters specified in Section~\ref{sec:setup}, we further choose a central scale of $\mu_R=\mu_F=\MH/2$ and use a 3-point scale variation to assess scale uncertainties.
For our central recommendation we use the quark masses as specified in Eq.~\eqref{eq:qmasses}.

The uncertainty associated with each contribution is computed as described in the following sections. 
In the current recommendation, we group the various sources of uncertainty into three categories, 
\begin{align}
\delta(\mathrm{theory})\,, \qquad 
\delta(\mathrm{PDF}+\alpha_s)\,, \qquad
\delta(\mathrm{PDF\text{--}TH})\,,
\end{align}
with,
\begin{align}
&\delta(\mathrm{theory)} = \delta(\mathrm{scale}) + \delta(\mathrm{EWK}) + \delta(t,b,c),&
&\delta(\mathrm{PDF}+\alpha_s) = \sqrt{\delta(\mathrm{PDF})^2 + \delta(\alpha_s)^2}.&
\end{align}
The theory uncertainty, $\delta(\mathrm{theory})$, collects uncertainties related to missing or approximated fixed-order contributions.
The $\delta(t,b,c)$ uncertainty collects uncertainties related to finite quark mass effects and missing higher order uncertainties (MHOU), it is given by,%
\footnote{
At NLO the scale variation of the top-only, $\sigma^{(t)}$, and top-bottom interference, $\sigma^{(t\times b)}$, contributions are found to be \emph{anti-correlated}, we anticipate that taking a linear combination of the uncertainties yields a conservative estimate.
}
\begin{align}
& \delta(t,b,c) = \delta^\mathrm{scheme}(t) + \delta^\mathrm{MHOU}(t) + \delta^{\mathrm{scheme}}(t \times b) + \delta^\mathrm{MHOU}(t \times b) + \delta^\mathrm{MHOU}(b, c, t \times c, b \times c),
\end{align}
where each of the scheme and higher order uncertainties are assessed in Section~\ref{sec:ggf_mq}.
The PDF and $\alpha_s$ uncertainty, $\delta(\mathrm{PDF}+\alpha_s)$, collects the usual PDF uncertainty, as estimated from the replicas of the PDF4LHC21\_40 PDF set and the uncertainty due to $\alpha_s$.
Finally, the $\delta(\mathrm{PDF\text{--}TH})$ uncertainty collects the uncertainty related to the mismatch between the N$^3$LO matrix element and the NNLO PDFs.
Alternatively, this mismatch can be avoided using the approximate N$^3$LO PDFs, we discuss results obtained with the aN$^3$LO PDFs in Section~\ref{sec:ggf_an3lo_pdf}.

In the recommendation reference tables presented in Appendix~\ref{app:tables} we also report the following combined uncertainties, which are defined for consistency with YR4~\cite{LHCHiggsCrossSectionWorkingGroup:2016ypw}. Relevant differences between the calculations in this report and in YR4 are explained further in Section~\ref{sec:ggf_yr4}.
\begin{align}
\delta(\mathrm{Total}) &= \sqrt{ (\delta(\mathrm{theory}) + \delta(\mathrm{PDF-TH}))^2 + \delta(\mathrm{PDF}+\alpha_s)^2 }, \\
\delta(\mathrm{Gauss}) &= (\max(+\delta(\mathrm{theory}),-\delta(\mathrm{theory})) + \delta(\mathrm{PDF-TH}))/\sqrt{3}, \\
\delta(\mathrm{Gauss\ Total}) &= \sqrt{ \delta(\mathrm{Gauss})^2 + \delta(\mathrm{PDF}+\alpha_s)^2}.
\end{align}

\subsection{N$^3$LO Heavy-Top Limit Results}
\label{sec:ggf_htl}

The terms $\sigma_{ij}^{\mathrm{N}^n\mathrm{LO},\mathrm{HTL}}$ with $n=0,\,1,\,2,\,3$ of Eq.~\eqref{eq:ggf_master} are computed using an effective theory in which the heavy top-quark is integrated out, see Refs.~\cite{Kramer:1996iq,Chetyrkin:1997un,Schroder:2005hy,Chetyrkin:2005ia}.
These corrections are known exactly at NLO~\cite{Dawson:1990zj,Djouadi:1991tka}, NNLO~\cite{Anastasiou:2002yz,Harlander:2002wh,Ravindran:2003um} and at N$^3$LO~\cite{Anastasiou:2016cez,Anastasiou:2015vya,Mistlberger:2018etf}.
The $\mathrm{R}_\mathrm{LO}$ ratio accounts for top-quark mass effects at LO, cf.\ eq.~\eqref{eq:RLO}.

The HTL results presented in this work were produced using the \texttt{iHixs2} code~\cite{Dulat:2018rbf}.
These corrections are also available in the \texttt{n3loxs}~\cite{Baglio:2022wzu} and  \texttt{ggHiggs}~\cite{Ball:2013bra,Bonvini:2014jma,Bonvini:2016frm,Ahmed:2016otz,Bonvini:2018ixe,Bonvini:2018iwt} tools.
The theoretical uncertainty associated with the truncation at N$^3$LO corrections is estimated via a 3-point scale variation defined by varying the common perturbative scale $\mu=\mu_R=\mu_F$ by a factor of 2 and 1/2 around the central scale $\mu=\MH/2$.
This scale variation prescription is at variance with the default implemented in the \texttt{iHixs2} code~\cite{Dulat:2018rbf} where $\mu = 0.4\ \MH$ is used for the lower scale variation, motivated by the fact that it was observed to give a larger scale variation than the usual 3-point prescription.
Furthermore, we note that the scale uncertainty of the pure HTL is rather insensitive to the choice of top-quark mass scheme, which enters only at NNLO via the Wilson coefficient $C$.
However, due to the re-weighting factor $\mathrm{R}_\mathrm{LO}$, the rescaled EFT is sensitive to the top-mass scheme.
In the on-shell scheme, which we adopt as our central recommendation, the top-quark mass in $\mathrm{R}_\mathrm{LO}$ is fixed for all values of the renormalisation scale, while in the $\overline{\mathrm{MS}}$ scheme, the value of the top-quark mass entering $\mathrm{R}_\mathrm{LO}$ is varied.
The variation of $\mathrm{R}_\mathrm{LO}$ due to the running of the top-quark mass in the $\overline{\mathrm{MS}}$ scheme is anti-correlated with the scale variation of the pure HTL.
As a consequence, the scale uncertainty of the rescaled HTL in the on-shell scheme and the pure HTL are approximately equal and larger than the scale uncertainty in the $\overline{\mathrm{MS}}$ scheme.
As a further consistency check, a comparison to approximate N$^4$LO HTL results obtained via the soft-virtual approximation is presented in Section~\ref{sec:ggf_n4lo_sv}.

\paragraph{Recommendation}
The rescaled HTL results are included to order N$^3$LO.
The corrections are evaluated with the top-quark mass renormalised in the on-shell scheme, see Section~\ref{sec:ggf_mq} for details of the inclusion of further quark mass corrections.
The theoretical uncertainty, $\delta(\mathrm{scale})$, associated with the missing higher-order rescaled HTL corrections, is derived via a 3-point scale variation of the Born rescaled heavy-top limit prediction (i.e.\ using the first line of Eq.~\eqref{eq:ggf_master}).


\subsection{EW and Mixed QCD--EW Corrections}
\label{sec:ggf_ew}

The combined Wilson coefficient, $C$, entering Eq.~\eqref{eq:ggf_master} is defined as,
\begin{align}
C = C_\mathrm{QCD} + \lambda_\mathrm{EWK} 
\Bigl( 1 + \frac{\alpha_s}{\pi} C_{1w} + \ldots \Bigr).
\end{align}
Here, $C_\mathrm{QCD}$ is the QCD Wilson coefficient required to match the HTL effective theory to the Standard Model, it is known at NNLO~\cite{Kramer:1996iq,Chetyrkin:1997un} and N$^3$LO~\cite{Schroder:2005hy,Chetyrkin:2005ia} accuracy.
The combined Wilson coefficient also incorporates the leading electro-weak corrections~\cite{Aglietti:2004nj,Actis:2008ug} via $\lambda_\mathrm{EWK}$, which is defined as the ratio of the NLO electroweak correction to the Born cross section.
The term $C_{1w}$ incorporates the mixed QCD--EW corrections.
The dominant electroweak correction proceeds via a massless quark loop coupled to the Higgs boson via massive gauge bosons, this part of the correction is known in the large $W$ and $Z$ boson mass limit~\cite{Anastasiou:2008tj}, the small $W$ and $Z$ boson mass limit~\cite{Anastasiou:2018adr}, and with exact $W$ and $Z$ boson masses for the gluon-induced channel~\cite{Bonetti:2016brm,Bonetti:2017ovy,Bonetti:2018ukf,Bonetti:2020hqh,Becchetti:2020wof,Bonetti:2022lrk}.
The calculation performed in the limit of large gauge boson mass yields $C_\mathrm{1w} =7/6$, while the calculation utilising finite gauge boson masses gives $C_\mathrm{1w} = -1.7$ for our central scale choice.
The use of finite gauge boson masses amounts to a $-0.53\%$ correction to the total cross section for a Higgs boson mass of $\MH=125.09$\,GeV for all collider energies between $7$\,TeV and $14$\,TeV, compatible with the $\delta(\mathrm{EWK})=1\%$ assigned in YR4~\cite{LHCHiggsCrossSectionWorkingGroup:2016ypw}.

\paragraph{Recommendation}
The recommendation for the central prediction includes the EW corrections using the calculation of Refs.~\cite{Aglietti:2004nj,Actis:2008ug}.
The mixed QCD--EW corrections for the gluon-induced channel are computed in the heavy boson approximation~\cite{Anastasiou:2008tj}. 
The treatment of the mixed corrections is motivated by the lack of knowledge of the quark channel contributions and results for the gluon-induced channel using the set of input parameters and settings recommended in this report.
However, the calculation of the corrections to the gluon-fusion channel with finite boson masses increases our confidence that the $\delta(\mathrm{EWK})=1\%$ error assigned to the missing/approximated QCD--EW corrections is reliable~\cite{Becchetti:2020wof}.


\subsection{Finite Quark Mass Corrections}
\label{sec:ggf_mq}

Using the notation $\left[\ldots \right]_{\alpha_s^i}$ to indicate only terms proportional to $\alpha_s^i$, we can define the contributions
\begin{align}
\delta \sigma_{ij}^{\mathrm{LO}, (t,b,c)} = 
& \sigma_{ij}^{\mathrm{LO}, (t,b,c)} - \left[C_\mathrm{QCD}^2 \mathrm{R}_\mathrm{LO} \sigma_{ij}^\mathrm{HTL} \right]_{\alpha_s^2}, \\
\delta \sigma_{ij}^{\mathrm{NLO}, (t,b,c)} = &
\sigma_{ij}^{\mathrm{NLO}, (t,b,c)} - \left[C_\mathrm{QCD}^2 \mathrm{R}_\mathrm{LO} \sigma_{ij}^\mathrm{HTL} \right]_{\alpha_s^3}, \\
\delta \sigma_{ij}^{\mathrm{NNLO}, (t,b,c)} = 
& \sigma_{ij}^{\mathrm{NNLO}, (t,b,c)} - \left[C_\mathrm{QCD}^2 \mathrm{R}_\mathrm{LO} \sigma_{ij}^\mathrm{HTL} \right]_{\alpha_s^4}, \label{eq:ggf_delta_tbc} &
\end{align}
which account for the dependence on the finite top, bottom and charm quark masses at LO, NLO and NNLO, respectively.
The exact QCD corrections with full dependence on the quark masses are known at LO~\cite{Wilczek:1977zn} and NLO~\cite{Graudenz:1992pv,Spira:1995rr}.
These corrections can be further decomposed into pieces involving only top, bottom or charm quarks coupling to the Higgs boson and their interferences,
\begin{align}
\delta\sigma_{ij}^{\mathrm{N}^n\mathrm{LO}, (t,b,c)} = \delta\sigma_{ij}^{\mathrm{N}^n\mathrm{LO}, (t)} 
+ \sigma_{ij}^{\mathrm{N}^n\mathrm{LO}, (b)} 
+ \sigma_{ij}^{\mathrm{N}^n\mathrm{LO}, (c)} 
+ \sigma_{ij}^{\mathrm{N}^n\mathrm{LO}, (t \times b)}
+ \sigma_{ij}^{\mathrm{N}^n\mathrm{LO}, (t \times c)}
+ \sigma_{ij}^{\mathrm{N}^n\mathrm{LO}, (b \times c)},
\end{align}
where $\delta\sigma_{ij}^{\mathrm{N}^n\mathrm{LO},(t)} $ is the difference between the top-only contribution and the rescaled HTL result, defined analogously to Eq.~\eqref{eq:ggf_delta_tbc}.
At NNLO not all terms of the decomposition are known, the top-only contribution $\sigma_{ij}^{\mathrm{NNLO},(t)}$ was computed in Refs.~\cite{Czakon:2020vql,Czakon:2021yub,Czakon:2024ywb}, the top--bottom interference was computed in Refs.~\cite{Czakon:2023kqm,Czakon:2024ywb}, the bottom-only, charm-only and charm interference contributions are known only to NLO~\cite{Graudenz:1992pv,Spira:1995rr}.
At N$^3$LO, the top-only virtual corrections are known to third order in a $1/m_t^2$ expansion~\cite{Davies:2019wmk}.

\begin{table}[h!]
\centering \small
\begin{tabular}{|c|rrrrrrr|}
\hline
\multicolumn{8}{|c|}{$\sqrt{s}=13.6$\,TeV} \\ \hline
& $\sigma^{(t,b,c)}$ [pb] & $\sigma^{(t)}$ [pb] & $\sigma^{(b)}$ [pb] & $\sigma^{(c)}$ [pb] & $\sigma^{(t \times b)}$ [pb] & $\sigma^{(t \times c)}$ [pb] & $\sigma^{(b \times c)}$ [pb] \\ \hline
$\sigma^\mathrm{LO}$  & $+15.865$ & $+17.115$ & $+0.045$ & $+0.000$ & $-1.172$ & $-0.131$ & $+0.008$ \\
$\sigma^\mathrm{NLO}$  & $+21.358$ & $+22.036$ & $+0.063$ & $+0.001$ & $-0.663$ & $-0.089$ & $+0.012$ \\
$\sigma^\mathrm{NNLO}$ & --- & $+10.103$ & --- & --- & $+0.04$ & --- & --- \\
\hline
Total & --- & $+49.254$ & --- & --- & $-1.795$ & --- & --- \\
\hline
\end{tabular}
\caption{The top, bottom and charm quark contributions to the gluon-fusion total cross section including finite quark mass corrections.
The top-quark is renormalised in the on-shell scheme, with the remaining quarks renormalised in the $\overline{\mathrm{MS}}$ scheme.
The LO and NLO corrections are computed using \texttt{iHixs2}~\cite{Dulat:2018rbf} which implements the calculations of Refs.~\cite{Graudenz:1992pv,Spira:1995rr}.
The available NNLO corrections are based on the calculations presented in Refs.~\cite{Czakon:2020vql,Czakon:2021yub,Czakon:2023kqm,Czakon:2024ywb}.
}
\label{tab:ggf_tbc}
\end{table}

In Table~\ref{tab:ggf_tbc}, we show the impact of finite top, bottom and charm masses at LO and NLO on the total cross section.
In this table, and the following sections, the omission of the subscripts $ij$ implies a sum over the contributions of all partonic channels.
We observe that the cross section is dominated by the top-only contribution, $\sigma_{ij}^{\mathrm{N^nLO}, (t)}$, which is discussed in  Section~\ref{sec:ggf_mt}.
The destructive top--bottom interference is the next largest contribution, amounting to a $-5\%$ correction at NLO, it is discussed in Section~\ref{sec:ggf_mbmt}.
The remaining finite quark contributions are all sub-percent corrections at NLO, they are discussed in Section~\ref{sec:ggf_mbmc}.

\subsubsection{Top Mass Corrections ($\sigma^{(t)}$)}
\label{sec:ggf_mt}

\begin{table}[h!]
\centering
\begin{tabular}{|c|r|r|r|r|r|}
\hline
& $7\,\mathrm{TeV}$ [pb] & $8\,\mathrm{TeV}$ [pb] & $13\,\mathrm{TeV}$ [pb] & $13.6\,\mathrm{TeV}$ [pb] & $14\,\mathrm{TeV}$ [pb] \\ \hline
$\delta \sigma^{\mathrm{NLO}, (t)}$  & $-0.0630$           & $-0.0907$           & $-0.3108$            & $-0.346$               & $-0.3707$            \\
$\delta \sigma^{\mathrm{NNLO}, (t)}$ & $+0.0395(2)$        & $+0.0538(3)$        & $+0.1488(9)$         & $+0.1634(8)$           & $+0.173(1)$          \\ \hline
\end{tabular}
\caption{
Top-quark mass corrections computed with the top-quark mass renormalised in the on-shell scheme and $\MH=125.09$\,GeV (except for the finite real--virtual corrections where $\MH=125$\,GeV), $M_t=172.5$\,GeV (except for the finite real-virtual corrections where $M_t = 173.055$\,GeV) for $\mu_R = \mu_F = \MH/2$.
The corrections are based on the calculations of Refs.~\cite{Czakon:2020vql,Czakon:2021yub,Czakon:2024ywb}.
}
\label{tab:ggf_mtcorr}
\end{table}

In Table~\ref{tab:ggf_mtcorr}, we report corrections due to a finite top-quark mass at NLO and NNLO computed using the setup of Section~\ref{sec:setup}.
The corrections are obtained using the original calculations presented in Refs.~\cite{Czakon:2020vql,Czakon:2021yub,Czakon:2024ywb}.
The numbers are provided for a fixed value of the Higgs boson mass, $\MH^\mathrm{ref}=125.09$\,GeV.

\paragraph{Different values of $\MH$}
Values for the top-quark mass corrections have not been re-evaluated for all Higgs boson masses required.
The corrections are extrapolated to different values of the Higgs boson mass by multiplying with the Higgs mass dependence of the corresponding reweighted HTL correction.
Fixing $\MH^\mathrm{ref}=125.09$\,GeV, our extrapolation formula reads,
\begin{align}
&R^H_n(\MH) = & \frac{\left[C_\mathrm{QCD}^2 \mathrm{R}_\mathrm{LO} \sigma_{ij}^\mathrm{HTL}\right]_{\alpha_s^n}(\MH)}{\left[C_\mathrm{QCD}^2 \mathrm{R}_\mathrm{LO} \sigma_{ij}^\mathrm{HTL}\right]_{\alpha_s^n}(\MH^\text{ref})},&
&\delta \sigma^{\mathrm{NNLO}, (t), \text{ext}}(\MH) = & R^H_4(\MH)\  \delta \sigma^{\mathrm{NNLO}, (t)}(\MH^\text{ref}).&
\end{align}
As a conservative error estimate, we assign a $100\%$ uncertainty to the shift induced by the $\MH$ extrapolation,
\begin{align}
\delta^\text{ext}(t) = & \delta \sigma^{\mathrm{NNLO}, (t), \text{ext}}(\MH) - \delta \sigma^{\mathrm{NNLO}, (t)}(\MH^\text{ref}).
\end{align}
Our extrapolation procedure is validated at NLO by comparing the extrapolation to the true NLO result, we observe that our extrapolated numbers and the true NLO numbers differ by at most $0.02\%$ of the total cross section, while our conservative error estimate is typically larger by a factor of 2 or more.
At NNLO we validate the extrapolation procedure using the $1/m_t^2$ expansion, we observe that our extrapolated numbers and the true NNLO $1/m_t^2$ numbers differ by at most $0.07\%$ of the total cross section, while our conservative error estimate is approximately equal to this difference.
Computing the extrapolation uncertainty at NNLO across all collider energies and Higgs boson masses in the recommendation, we observe that the uncertainty related to the extrapolation procedure never exceeds $ 0.03\%$ of the total cross section.
We consider this uncertainty to be negligible compared to the other remaining uncertainties and it is neglected in our recommendation.

\paragraph{Scheme Uncertainty of $m_t$} 
Results for the top-only contribution with the top-quark mass renormalised in the $\overline{\mathrm{MS}}$ scheme and the OS scheme are presented at NNLO in Ref.~\cite{Czakon:2024ywb} for $\sqrt{s}=13$\,TeV.
The difference between the two schemes is found to be very small at NNLO amounting to just $-0.01$ pb, approximately half of that observed at NLO.
Expressing the NNLO scheme uncertainty as a percentage of the total cross section and symmetrising, we obtain,
\begin{align}
\delta^\mathrm{scheme}(t) = \pm 0.01\%.
\end{align}
We consider this scheme uncertainty to be negligible compared to the other remaining uncertainties and it is neglected in our recommendation.

\paragraph{Missing Higher Orders}
We can estimate the impact of missing higher orders, beyond NNLO, of $\delta \sigma^{\mathrm{NNLO}, (t)}$ using a scale variation.
In Ref.~\cite{Czakon:2023kqm}, the impact of a 7-point scale variation on this correction was reported for $\sqrt{s}=13$\,TeV and found to amount to $^{+0.13}_{-0.03}$ pb.
Expressing this uncertainty as a percentage of the total cross section and symmetrising, we obtain,
\begin{align}
\delta^{\mathrm{MHOU}}(t) = \pm 0.17 \%. \label{eq:ggf_t_mhou}
\end{align}
Considering collider energies between $7$ and $14$\,TeV the missing higher order uncertainty varies between $\pm 0.14$ and $\pm 0.18\%$~\cite{Czakon:2023kqm}. 
We take Eq.~\eqref{eq:ggf_t_mhou} as a flat relative uncertainty across all collider energies and Higgs boson masses.

\paragraph{Recommendation}
We include the $\delta \sigma^{(t)}$ corrections, known to NNLO, in the central recommendation.
The uncertainties related to extrapolating the known results to different values of $\MH$ are found to be negligible (for the range considered here), and the remaining scheme uncertainty stemming from the treatment of the top-quark mass is also found to be negligible.
The uncertainty related to missing higher orders $\delta^{\mathrm{MHOU}}(t)$ should be considered and is included in the total finite quark mass correction uncertainty $\delta(t,b,c)$ quoted in the tables of Appendix~\ref{app:tables}.


\subsubsection{Top--Bottom Interference ($\sigma^{(t \times b)}$)}
\label{sec:ggf_mbmt}

\begin{table}[h!]
\centering
\begin{tabular}{|c|r|r|r|r|r|}
\hline
& $7\,\mathrm{TeV}$ [pb] & $8\,\mathrm{TeV}$ [pb] & $13\,\mathrm{TeV}$ [pb] & $13.6\,\mathrm{TeV}$ [pb] & $14\,\mathrm{TeV}$ [pb] \\ \hline
$\sigma^{\mathrm{LO}, (t \times b)}$   & $-0.39$                  & $-0.50$                   & $-1.09$                    & $-1.17$                      & $-1.22$                    \\
$\sigma^{\mathrm{NLO}, (t \times b)}$  & $-0.27$                  & $-0.32$                   & $-0.63$                    & $-0.66$                      & $-0.69$                   \\
$\sigma^{\mathrm{NNLO}, (t \times b)}$ & $-0.02$                   & $-0.02$                    & $+0.02$                    & $+0.04$                      & $+0.03$                    \\ \hline 
Total & $-0.68$ & $-0.84$ & $-1.70$ & $-1.80$ & $-1.88$ \\
\hline
\end{tabular}
\caption{
Top--bottom interference corrections computed with the top-quark mass renormalised in the on-shell scheme and the bottom-quark renormalised in the 5FS $\overline{\mathrm{MS}}$ scheme.
The mass of the Higgs boson is taken to be $\MH=125.09$\,GeV (except for the finite real--virtual corrections where $\MH=125$\,GeV), $M_t=172.5$\,GeV (except for the finite real--virtual corrections where $M_t = 173.055$\,GeV) for $\mu_R = \mu_F = \MH/2$. 
The corrections are based on the calculations of Refs.~\cite{Czakon:2023kqm,Czakon:2024ywb}.
}
\label{tab:ggf_mbmtcorr}
\end{table}

In Table~\ref{tab:ggf_mbmtcorr} we state top--bottom interference corrections at LO, NLO and NNLO evaluated with the setup of Section~\ref{sec:setup}.
The corrections were obtained using the original calculations presented in Refs.~\cite{Czakon:2023kqm,Czakon:2024ywb}.
The numbers are provided for a fixed value of the Higgs boson mass, $\MH^\text{ref}=125.09$\,GeV.

\paragraph{Different values of $\MH$}
We extrapolate the top--bottom interference contribution to different values of the Higgs boson mass again using the reweighted HTL correction, 
\begin{align}
\delta \sigma^{\mathrm{NNLO}, (t \times b), \text{ext}}(\MH) = & R^H_4(\MH)\  \delta \sigma^{\mathrm{NNLO}, (t \times b)}(\MH^\text{ref})
\end{align}
The error associated with the extrapolation to new values of $\MH$ is estimated using,
\begin{align}
\delta^\text{ext}(t \times b) = & \sigma^{\mathrm{NNLO}, (t \times b), \text{ext}}(\MH) - \sigma^{\mathrm{NNLO}, (t \times b)}(\MH^\mathrm{ref}). 
\end{align}
Our extrapolation procedure is validated at NLO by comparing the extrapolation to the true NLO result, we observe that our extrapolated numbers and the true NLO numbers differ by at most $0.06\%$ of the total cross section, while our conservative error estimate is typically larger by a factor of 3.
Computing the extrapolation uncertainty at NNLO across all collider energies and Higgs boson masses in the recommendation, we observe that it never exceeds $0.01\%$.
We consider this uncertainty to be negligible compared to the other remaining uncertainties and it is neglected in our recommendation.

\paragraph{Scheme Uncertainty of $m_b$}
For our central recommendation we take the 5-flavour-scheme (5FS) with the bottom quark mass renormalised in the $\overline{\mathrm{MS}}$ scheme.
In this scheme the perturbative series is found to converge well and the NNLO corrections are small~\cite{Czakon:2023kqm,Czakon:2024ywb}. 
We estimate the scheme uncertainty related to the bottom quark mass for the top--bottom interference by taking the difference between the 5FS result and the 4-flavour-scheme (4FS) result with the bottom quark renormalised in the $\overline{\mathrm{MS}}$ scheme,
\begin{align}
\delta^\text{scheme}(M_b) = & 
\left[
\sigma^\mathrm{LO,(t \times b)} (M_b^{\overline{\mathrm{MS}},\text{5FS}})
+\sigma^\mathrm{NLO,(t \times b)} (M_b^{\overline{\mathrm{MS}},\text{5FS}})
+\sigma^\mathrm{NNLO,(t \times b)} (M_b^{\overline{\mathrm{MS}},\text{5FS}})
\right] \nonumber \\
& - 
\left[
\sigma^\mathrm{LO,(t \times b)} (M_b^{\overline{\mathrm{MS}},\text{4FS}})
+\sigma^\mathrm{NLO,(t \times b)} (M_b^{\overline{\mathrm{MS}},\text{4FS}})
+\sigma^\mathrm{NNLO,(t \times b)} (M_b^{\overline{\mathrm{MS}},\text{4FS}})
\right],
\end{align}
as a percentage of the total cross section.
Results for the top--bottom interference in the 5FS and 4FS with the bottom quark renormalised in the $\overline{\mathrm{MS}}$ scheme at NNLO for $\sqrt{s}=  13$\,TeV are provided in Ref.~\cite{Czakon:2024ywb}, the difference between the two schemes amounts to $-0.09$ pb.
Expressing the NNLO scheme uncertainty as a percentage of the total cross section and symmetrising, we obtain,
\begin{align}
\delta^\mathrm{scheme}(t\times b) = \pm 0.09\%.
\end{align}
which we take as a flat relative uncertainty across all collider energies and Higgs boson masses.
This uncertainty is small compared to other uncertainties related to missing higher orders in the HTL, top-only and top--bottom contributions and we neglect it in the recommendation.

Repeating the above exercise in the 5FS but with the bottom mass renormalised in the $\overline{\mathrm{MS}}$ v.s.\ the on-shell scheme, we find a difference of $-0.25$ pb or $\pm 0.26 \%$ of the total cross section (after symmetrising).
However, the use of the on-shell scheme for the bottom quark in this process is poorly justified and the perturbative convergence in this scheme is observed to be poor, with the NNLO corrections found to be similar in size to the NLO corrections and with a larger scale uncertainty.


\paragraph{Missing Higher Orders}
We can estimate the impact of missing higher orders, beyond NNLO, of $\delta \sigma^{\mathrm{NNLO}, (t \times b)}$ using a scale variation.
In Ref.~\cite{Czakon:2024ywb}, the impact of a 7-point scale variation on this correction was reported for $\sqrt{s}=13$\,TeV and found to amount to $^{+0.13}_{-0.03}$ pb.
Expressing this uncertainty as a percentage of the total cross section and symmetrising, we obtain an uncertainty,
\begin{align}
\delta^{\mathrm{MHOU}}(t \times b) = \pm 0.17 \%
\end{align}
which we take as a flat relative uncertainty across all collider energies and Higgs boson masses.

\paragraph{Recommendation}
We include the $\delta \sigma^{(t\times b)}$ corrections, known to NNLO, in the central recommendation, utilising the 5FS with the bottom-quark mass renormalised in the $\overline{\mathrm{MS}}$ scheme.
The uncertainties related to extrapolating the known results to different values of $\MH$ are found to be negligible (for the range considered here).
The scheme uncertainty is assessed by comparing to results with the bottom-quark renormalised in 5FS and 4FS using the $\overline{\mathrm{MS}}$ renormalisation scheme at NNLO, the corresponding uncertainty, $\delta^\mathrm{scheme}(t \times b)$ is found to be small and is neglected.  
The uncertainty related to missing higher orders $\delta^{\mathrm{MHOU}}(t \times b)$, is estimated via a scale variation, it is included in the total finite quark mass correction uncertainty $\delta(t,b,c)$.


\subsubsection{Bottom \& Charm Contributions ($\sigma^{(b)},\sigma^{(c)},\sigma^{(t \times c)},\sigma^{(b \times c)}$)}
\label{sec:ggf_mbmc}

In Table~\ref{tab:ggf_tbc} we present the various interferences at LO and NLO.
The bottom-only, charm-only and charm interference contributions are known to NLO~\cite{Graudenz:1992pv,Spira:1995rr}.
We note that the contributions $\sigma^{(b)}, \sigma^{(c)}$ and $\sigma^{(b \times c)}$ are small, accounting for $+0.29\%$, $0.003\%$ and $0.05\%$ of the NLO cross section, respectively.
The interference $\sigma^{(t \times c)}$ is slightly larger, accounting for $-0.59\%$ of the NLO cross section.
We include all contributions to NLO using the $\overline{\mathrm{MS}}$ scheme for the light quarks.

\paragraph{Missing Higher Orders}
We can estimate the impact of missing higher orders, beyond NLO, using a scale variation.
Computing the 3-point scale variation of the bottom quark only contribution $\sigma^{(b)}$ at $\sqrt{s}= 13$\,TeV for the reference Higgs boson mass, we find  $\sigma^{(b)} = 0.1^{+0.03}_{-0.02}$ pb, which amounts to an uncertainty of $\delta^{\mathrm{MHOU}}(b)=\pm 0.06\%$ on the total cross section.
Considering, at NLO, the combination $\sigma^{(c)}+\sigma^{(t \times c)} + \sigma^{(b \times c)}= -0.20^{+0.03}_{-0.02}$ pb, this also amounts to an uncertainty at the level of $\delta^{\mathrm{MHOU}}(c,t \times c, b \times c)=\pm 0.06\%$ on the total cross section.
Considering collider energies of between $7$ and $14$\,TeV and the range of Higgs boson masses presented in this report, these uncertainties remain around the per-mille level.
We consider each of these uncertainties to be negligible compared to other remaining uncertainties and they are neglected in our recommendation.
 
\paragraph{Recommendation}
We include the $\sigma^{(b)},\sigma^{(c)},\sigma^{(t \times c)}$ and $\sigma^{(b \times c)}$ contributions at NLO in the central recommendation.
The uncertainty related to missing higher orders $\delta^{\mathrm{MHOU}}(b,c,t\times c,b \times c)$ is assessed via scale variation at NLO and found to be negligible compared to other sources of uncertainty, it is neglected in our recommendation.


\subsection{PDF Theory Uncertainty and Impact of aN$^3$LO PDFs}
\label{sec:ggf_an3lo_pdf}

At present, the PDF4LHC Working Group provide a combination NNLO set, PDF4LHC21~\cite{PDF4LHCWorkingGroup:2022cjn}, designed for Run 3 of the LHC.
The PDF4LHC21 set is based on a Monte Carlo combination of three global PDF sets  CT18~\cite{Hou:2019efy}, MSHT20~\cite{Bailey:2020ooq}, and NNPDF3.1~\cite{NNPDF:2017mvq}, all of which are available at NNLO accuracy.
For precision comparisons between data and theory or for systematic exploration of differences between PDF fits, the PDF4LHC working group recommends the use of as many modern individual PDF sets as possible, for example those presented in Refs.~\cite{Hou:2019efy,Alekhin:2024bhs,Bailey:2020ooq,NNPDF:2021njg}.

In the heavy-top limit, matrix elements for Higgs boson production are available at N$^3$LO.
To account for this mismatch, we can take the difference between the NNLO rescaled HTL prediction when computed with a NNLO PDF and a NLO PDF and use this as an estimate of the uncertainty,
\begin{align}
\delta(\mathrm{PDF\text{--}TH}) = \pm \left| \sigma^{(2),\mathrm{HTL},\mathrm{NNLO}} - \sigma^{(2),\mathrm{HTL},\mathrm{NLO}} \right|.
\label{eq:ggf_pdf_th}
\end{align}
We note that this definition of $\delta(\mathrm{PDF\text{--}TH})$ is twice that reported in the previous recommendation, this change is motivated by the study of the impact of approximate N$^3$LO PDFs discussed below.
For the computation of $\delta(\mathrm{PDF\text{--}TH})$ we rely on the PDF4LHC15~\cite{Butterworth:2015oua} combination PDF, which is available at both NLO and NNLO.

Beyond  NNLO, at present, two approximate N$^3$LO PDF sets are available, produced by the MSHT~\cite{McGowan:2022nag} and NNPDF~\cite{NNPDF:2024nan} collaborations.
A combined aN$^3$LO PDF set, MSHT20xNNPDF40\_aN3LO, 
has also been produced by the two fitting collaborations and was presented in Ref.~\cite{Cridge:2024icl}.
The approximate N$^3$LO PDF sets contain exact N$^3$LO information for the massless DIS coefficient functions~\cite{Vermaseren:2005qc}, while the massive neutral-current DIS coefficients are approximated.
The fits approximate the four-loop splitting functions using the Mellin moments known at the time they were performed supplemented by large/small-$x$ limits~\cite{Davies:2016jie,Moch:2017uml,Moch:2018wjh,Moch:2021qrk,Falcioni:2023luc,Gehrmann:2023cqm,Gehrmann:2023iah,Falcioni:2023tzp,Moch:2023tdj,Falcioni:2024xyt,Falcioni:2023vqq,Falcioni:2024xav,Falcioni:2024qpd,Kniehl:2025ttz}. 
Each fit includes the massive operator matrix elements (a.k.a transition matrix elements) known at the time it was performed~\cite{Bierenbaum:2009mv,Kawamura:2012cr,Ablinger:2014uka,Ablinger:2014lka,Ablinger:2014vwa,Ablinger:2014nga,Ablinger:2014tla,Blumlein:2021enk,Ablinger:2022wbb,Ablinger:2023ahe,Ablinger:2024xtt}, with the remaining matrix elements approximated.
The N$^3$LO hadronic coefficient functions are largely unknown and are approximated in the fits using either K-factors with nuisance parameters (for MSHT) or NNLO scale variations (for NNPDF).
In Ref.~\cite{Cooper-Sarkar:2024crx}, a benchmarking exercise was undertaken to understand the impact of the different splitting function approximations used in the global fits.
Recently, additional Mellin moments of the splitting functions have been computed and used to produce a new estimate of the four-loop splitting functions~\cite{Falcioni:2024qpd}.
Furthermore, the computation of all operator matrix elements has been completed~\cite{Ablinger:2025nnq}.
Including this new information in the aN$^3$LO PDF fits reduces the NNPDF prediction~\cite{NNPDF:2024nan} for the gluon-fusion Higgs total cross section by around 0.5\% and increases the MSHT prediction~\cite{McGowan:2022nag} by around 0.5\% relative to the fits used in the combination set MSHT20xNNPDF40~\cite{Cridge:2024icl} which enters this recommendation, bringing the two fits closer together by a shift that falls within their respective PDF uncertainties~\cite{Thorne:2024npj,Hekhorn:2025xke}.

\begin{table}[]
  \small
  \centering
\makebox[\textwidth][c]{%
\begin{tabular}{|rrrrrrr}
\hline
\multicolumn{7}{|c|}{(aN3LO PDF $\otimes$ N3LO HTL) vs (NNLO PDF $\otimes$ N3LO HTL)}                                                                                                       \\ \hline
& \multicolumn{3}{|c|}{aN3LO vs NNLO} &
\multicolumn{3}{c|}{aN3LO vs PDF4LHC21 NNLO} \\
\multicolumn{1}{|l|}{$\sqrt{s}$ [TeV]} & MSHT20xNNPDF40 & MSHT20 & \multicolumn{1}{l|}{NNPDF40} & MSHT20xNNPDF40 & MSHT20 & \multicolumn{1}{l|}{NNPDF40} \\ \hline
\multicolumn{1}{|r|}{7}    & $-4.2\%$   & $-6.0\%$ & \multicolumn{1}{r|}{$-2.0\%$}              & $-5.2\%$     & $-6.8\%$ & \multicolumn{1}{r|}{$-3.2\%$}  \\
\multicolumn{1}{|r|}{13.6} & $-3.8\%$   & $-5.2\%$ & \multicolumn{1}{r|}{$-2.1\%$}              & $-4.2\%$     & $-5.9\%$ & \multicolumn{1}{r|}{$-2.2\%$}  \\
\multicolumn{1}{|r|}{100}  & $-0.7\%$   & $+0.6\%$ & \multicolumn{1}{r|}{$-1.6\%$}              & $-1.0\%$     & $-0.7\%$ & \multicolumn{1}{r|}{$-1.0\%$} \\
\hline
\multicolumn{7}{|c|}{(aN3LO PDF $\otimes$ N3LO HTL) vs (NNLO PDF $\otimes$ NNLO HTL)}                                                                                                       \\ \hline
& \multicolumn{3}{|c|}{aN3LO vs NNLO} &
\multicolumn{3}{c|}{aN3LO vs PDF4LHC21 NNLO} \\
\multicolumn{1}{|l|}{$\sqrt{s}$ [TeV]} & MSHT20xNNPDF40 & MSHT20 & \multicolumn{1}{l|}{NNPDF40} & MSHT20xNNPDF40 & MSHT20 & \multicolumn{1}{l|}{NNPDF40} \\ \hline
\multicolumn{1}{|r|}{7} & $-0.8\%$      & $-2.6\%$ & \multicolumn{1}{r|}{$+1.5\%$}              & $-1.8\%$     & $-3.5\%$ & \multicolumn{1}{r|}{$+0.3\%$}  \\
\multicolumn{1}{|r|}{13.6} & $-0.5\%$   & $-1.9\%$ & \multicolumn{1}{r|}{$+1.3\%$}              & $-0.9\%$     & $-2.6\%$ & \multicolumn{1}{r|}{$+1.2\%$}  \\
\multicolumn{1}{|r|}{100} & $+3.1\%$    & $+4.4\%$ & \multicolumn{1}{r|}{$+2.1\%$}              & $+2.8\%$     & $+3.1\%$ & \multicolumn{1}{r|}{$+2.8\%$} \\ \hline
\end{tabular}
}
\caption{Impact of the aN$^3$LO PDF sets vs NNLO PDF sets on the total cross section when using the N$^3$LO HTL matrix element for (top) all predictions or (bottom) the N$^3$LO HTL matrix element for N$^3$LO predictions and the NNLO HTL matrix element for NNLO predictions. 
In the left columns the combination and individual aN$^3$LO PDF sets are compared to their corresponding NNLO variants.
In the right columns the aN$^3$LO PDF sets are compared to the PDF4LHC21 NNLO baseline prediction.}
\label{tab:ggf_an3lopdf_n3lo}
\end{table}

We observe that the difference between predictions produced using the N$^3$LO HTL matrix element with NNLO PDFs or aN$^3$LO PDFs is significant.
Comparing the combined set MSHT20xNNPDF40\_aN3LO to MSHT20xNNPDF40\_NNLO it amounts to a $-4.2\%$ suppression at $\sqrt{s}=7$\,TeV, a $-3.8\%$ suppression at $\sqrt{s}=13.6$\,TeV and a $-0.7\%$ suppression at $\sqrt{s}=100$\,TeV.
Comparing the combination aN$^3$LO PDF to our baseline PDF4LHC21 NNLO set we observe a $-5.2\%$ suppression at $\sqrt{s}=7$\,TeV, a $-4.2\%$ suppression at $\sqrt{s}=13.6$\,TeV and a $-1.0\%$ suppression at $\sqrt{s}=100$\,TeV.
The PDF uncertainty for the MSHT20xNNPDF40\_aN3LO combination set is $\pm 2.17\%$ at $\sqrt{s}=7$ TeV, $\pm 2.07\%$ at $\sqrt{s}=13.6$ TeV and $\pm 2.21\%$ at $\sqrt{s}=100$ TeV.
These values are obtained with the PDF sets that do not include QED evolution; the corresponding uncertainties for the QED-extended sets used in the recommendation are smaller, see Section~\ref{sec:ggf_qed_pdf}.
In Table~\ref{tab:ggf_an3lopdf_n3lo}, we provide a comparison of the individual PDF sets entering the aN$^3$LO combination, we observe that the MSHT20 set predicts a significantly larger impact of the aN$^3$LO PDFs than the NNPDF40 set.

Alternatively, we can compare results produced using the N$^3$LO HTL matrix element with aN$^3$LO PDFs to results produced using the NNLO HTL matrix element with NNLO PDFs.
For the combined set, the use of the N$^3$LO HTL matrix element with the aN$^3$LO PDFs gives a $-0.8\%$ suppression at $\sqrt{s}=7$\,TeV, a $-0.5\%$ suppression at $\sqrt{s}=13.6$\,TeV and a $+3.1\%$ enhancement at $\sqrt{s}=100$\,TeV.
In Table~\ref{tab:ggf_an3lopdf_n3lo}, we provide a comparison to the individual PDF sets, we observe that for collider energies relevant for the LHC the NNPDF40 aN$^3$LO PDFs predict an enhancement of the total cross section, while the MSHT20 aN$^3$LO PDFs predict a suppression.

In summary, using $\sqrt{s}=13.6$\,TeV, we obtain the smallest prediction for the total cross section using MSHT20 aN$^3$LO PDFs with the N$^3$LO HTL matrix element and the largest prediction using PDF4LHC21 NNLO PDFs with the N$^3$LO HTL matrix element, the MSHT20 aN$^3$LO result is $-5.9\%$ smaller than the largest prediction.

In Appendix~\ref{app:tables}, we provide tables which can be used to convert the recommendations produced with the NNLO PDF4LHC21 set to the aN$^3$LO MSHT20xNNPDF40\_aN3LO combination set.
The tables present the conversion summand,
\begin{align}
\Delta^\text{aN3LO} &=
\sigma( \text{MSHT20xNNPDF40\_aN3LO\_qed}) 
-\bigl[
\sigma(\text{PDF4LHC21}) + \Delta^\text{NNLO}_\text{QED}
\bigr],
\label{eq:ggf_pdf_conversion}
\end{align}
which corrects the PDF set used to compute the rescaled N$^3$LO HTL result (see first line of Eq.~\eqref{eq:ggf_master}), as well as a $\delta(\mathrm{PDF}+\alpha_s)$ uncertainty obtained using the aN$^3$LO set.
The use of the QED set for MSHT20xNNPDF40\_aN3LO and the addition of $\Delta^\text{NNLO}_\text{QED}$ for the PDF4LHC21 set accounts also for impact of QED evolution, as defined in Section~\ref{sec:ggf_qed_pdf}.

When utilising the MSHT20xNNPDF40\_aN3LO set the $\delta(\mathrm{PDF}+\alpha_s)$ uncertainty should be taken from the aN$^3$LO PDF set tables and the $\delta(\mathrm{PDF\text{--}TH})$ should not be included, as this term is only necessary when there exists a mismatch between the matrix element and PDF set order.

\paragraph{Recommendation}
The central recommendation is produced using the N$^3$LO HTL matrix elements with the PDF4LHC21 NNLO PDF set, in line with the general LHCHWG WG1 common setup outlined in Section~\ref{sec:setup}.
The PDF+$\alpha_s$ uncertainty is estimated according to the usual Hessian prescription via the LHAPDF~\cite{Buckley:2014ana} interface.
An additional uncertainty related to the mismatch between the perturbative order of the matrix element and the PDF, $\delta(\mathrm{PDF\text{--}TH})$, computed using Eq.~\eqref{eq:ggf_pdf_th}, is also reported.
Additionally, we provide tables which allow the central prediction to be converted to a prediction computed using the combination aN$^3$LO PDF set presented in Ref.~\cite{Cridge:2024icl}, the conversion and error prescription is outlined in Eq.~\eqref{eq:ggf_pdf_conversion} and surrounding text.


\subsection{Impact of QED DGLAP Evolution in PDFs}
\label{sec:ggf_qed_pdf}

\begin{table}[h!]
{\small
\centering
\makebox[\textwidth][c]{%
\renewcommand{\arraystretch}{1.2}
\begin{tabular}{|l|cccc|ccc|}
\hline
& \multicolumn{4}{|c|}{$\Delta^\text{NNLO}_\text{QED}$} &
\multicolumn{3}{c|}{$\Delta^\text{aN3LO}_\text{QED}$} \\
$\sqrt{s}$ [TeV] & PDF4LHC21 & $\langle$CT18$\rangle$ & MSHT20   & NNPDF31  & MSHT20xNNPDF40 & MSHT20     & NNPDF40    \\ \hline
7                & -0.77\% & -0.43\% & -0.69\% & -1.20\% & -1.55\% & -1.09\% & -1.74\% \\
8                & -0.77\% & -0.43\% & -0.66\% & -1.22\% & -1.53\% & -1.07\% & -1.73\% \\
13               & -0.74\% & -0.40\% & -0.58\% & -1.24\% & -1.46\% & -1.01\% & -1.66\% \\
13.6             & -0.74\% & -0.40\% & -0.58\% & -1.23\% & -1.45\% & -1.00\% & -1.65\% \\
14               & -0.74\% & -0.40\% & -0.57\% & -1.23\% & -1.45\% & -1.00\% & -1.65\% \\
100              & -0.60\% & -0.33\% & -0.40\% & -1.07\% & -1.10\% & -0.92\% & -1.19\% \\ \hline
\end{tabular}
}
}
\caption{
Impact of including QED DGLAP evolution in the individual and combination NNLO and aN$^3$LO PDFs sets on the N$^3$LO gluon fusion total cross section, for a Higgs boson mass of $\MH = 125.09$\,GeV.
}
\label{tab:ggf_pdf_qed}
\end{table}

The PDF sets can be extended to include also a photon parton distribution function and to account for QED and mixed QED--QCD DGLAP evolution.
The CT, MSHT and NNPDF collaborations have each produced QED extended sets at NNLO utilising the LuxQED method~\cite{Manohar:2016nzj,Manohar:2017eqh} to constrain the photon PDF~\cite{
Harland-Lang:2019pla,Xie:2021equ,Cridge:2021pxm,Cridge:2023ryv,
Bertone:2017bme,Barontini:2024dyb,NNPDF:2024djq}.
The MSHT and NNPDF collaborations have also produced QED extended sets at aN$^3$LO~\cite{Cridge:2023ryv,Barontini:2024dyb}.
The inclusion of a photon PDF primarily affects predictions for Higgs production in gluon fusion by altering the momentum distribution of the gluon PDF.
A fully consistent treatment of QED effects requires also the inclusion of photon induced Higgs boson production, which is neglected here.


In the recommendation presented in Appendix~\ref{app:tables}, we incorporate the effect of QED evolution for both the NNLO and aN$^3$LO PDF sets.
For the PDF4LHC21 NNLO set, we follow Ref.~\cite{Cooper-Sarkar:2025sqw} and define
\begin{align}
\Delta^\text{NNLO}_\text{QED} &=
    \frac{1}{3}\,\bigl(
    \langle \Delta^\text{NNLO}_\text{QED}(\text{CT18}) \rangle + 
    \Delta^\text{NNLO}_\text{QED}(\text{MSHT20}) + 
    \Delta^\text{NNLO}_\text{QED}(\text{NNPDF3.1})
    \bigr),
\label{eq:ggf_qed}
\end{align}
with,
\begin{align}
    \langle \Delta^\text{NNLO}_\text{QED}(\text{CT18}) \rangle &=
    \frac{1}{3}\,\bigl(
    \sigma(\text{CT18lux}) +
    \sigma(\text{CT18qed\_proton}) +
    \sigma(\text{CT18qed})
    \bigr)
    - \sigma(\text{CT18NNLO})
    , \nonumber \\
    \Delta^\text{NNLO}_\text{QED}(\text{MSHT20}) &=
    \sigma(\text{MSHT20qed\_nnlo}) - \sigma(\text{MSHT20nnlo\_as118})
    , \nonumber \\
    \Delta^\text{NNLO}_\text{QED}(\text{NNPDF3.1}) &=
    \sigma(\text{NNPDF31\_nnlo\_as\_0118\_luxqed}) - \sigma(\text{NNPDF31\_nnlo\_as\_0118}).
\label{eq:ggf_qed_ct18}
\end{align}
This shift is calculated using the rescaled N$^3$LO HTL result, i.e.\ the first line of Eq.~\eqref{eq:ggf_master}, expressed as a percentage then applied to the total cross section prediction.
Although the LuxQED method gives a constrained photon PDF, the impact of the inclusion of this photon PDF on the gluon distribution differs for each of the global sets, see Table~\ref{tab:ggf_pdf_qed}, in which $\langle$CT18$\rangle$ denotes the average over the three CT18 QED sets defined in the first line of Eq.~\eqref{eq:ggf_qed_ct18}.
The impact of the inclusion of QED effects on the PDF uncertainty of the PDF4LHC21 set has not yet been studied, in the recommendation, we report the PDF uncertainty obtained neglecting the QED effects.
For the MSHT20xNNPDF40 aN$^3$LO PDF set we define,
\begin{align}
\Delta^{\text{aN3LO}}_\text{QED} = \sigma(\text{MSHT20xNNPDF40\_aN3LO\_qed}) - \sigma(\text{MSHT20xNNPDF40\_aN3LO}).
\end{align}

Using the N$^3$LO HTL matrix element with NNLO PDFs, including QED evolution decreases the total cross section when using any of the CT18, MSHT20, NNPDF31 or combination PDF sets.
The predictions made with the CT18 and MSHT20 global sets decrease by around $-0.5\%$ due to the inclusion of QED evolution, while the NNPDF31 prediction decreases by more than $-1\%$. 
For the PDF4LHC21 set, the QED effects reduce the cross section by $-0.77\%$ at $\sqrt{s} = 7$\,TeV, $-0.74\%$ at $\sqrt{s} = 13.6$\,TeV and $-0.60\%$ at $\sqrt{s} = 100$\,TeV.



Alternatively, using the N$^3$LO HTL matrix element with  aN$^3$LO PDFs, the QED evolution has a larger impact.
For the MSHT20xNNPDF40 aN$^3$LO combination set we observe a $-1.55\%$ suppression at $\sqrt{s} = 7$\,TeV,  $-1.45\%$ at $\sqrt{s} = 13.6$\,TeV and a $-1.10\%$ suppression at $\sqrt{s} = 100$\,TeV.
Examining the individual PDF sets, we see that the MSHT20 aN$^3$LO set predicts a suppression of about $-1\%$ for all collider energies, while the NNPDF40 aN$^3$LO set predicts a larger $-1.7\%$ suppression for collider energies relevant for the LHC.

In summary, for the PDF4LHC21 and MSHT20xNNPDF40 combination sets, the impact of including QED evolution on the total cross section is between that of the individual CT18, MSHT20, NNPDF31 and NNPDF40 sets, which give a suppression ranging from $-0.40\%$ to $-1.65\%$ at $\sqrt{s}= 13.6$\,TeV.
Although all global fits utilise the LuxQED method to incorporate a photon PDF, the impact of QED evolution on the total Higgs cross section varies for the different PDF sets as well as the PDF order, it consistently decreases the cross section for both LHC and $100$\,TeV collider energies.
For the total cross section, including the QED evolution gives effects similar in size to the current PDF uncertainties, as the experimental and theoretical precision improves it will become increasingly important to account for QED evolution effects.

\paragraph{Recommendation}
The central recommendation is produced using the PDF4LHC21 PDF set in line with the common setup outlined in Section~\ref{sec:setup}.
The QED evolution effects are included according to Eq.~\eqref{eq:ggf_qed} following Ref.~\cite{Cooper-Sarkar:2025sqw}.
The QED evolution effects are included in the prediction using aN$^3$LO PDFs through the use of the MSHT20xNNPDF40\_aN3LO\_qed PDF set.
Global PDF fits indicate that incorporating QED evolution effects suppresses the total cross section for collider energies relevant for the LHC at the level of $-0.4\%$ to $-1.7\%$.


\subsection{Comparison to Approximate N$^4$LO Results}
\label{sec:ggf_n4lo_sv}

\begin{table}[h!]
  \small
  \centering
  \makebox[\textwidth][c]{%
\begin{tabular}{|lrrrrrr}
\hline
\multicolumn{7}{|c|}{13.6\,TeV} \\ \hline
\multicolumn{1}{|l|}{PDF}                                                 & N$^3$LO HTL & $\Delta(\mathrm{N}^3\mathrm{LO})$ & \multicolumn{1}{l|}{$\delta(\mathrm{scale})$} & N$^4$LO$_\mathrm{sv}$ HTL & $\Delta(\mathrm{N}^4\mathrm{LO}_\mathrm{sv})$ & \multicolumn{1}{l|}{$\delta(\mathrm{scale})$} \\ \hline
\multicolumn{1}{|l|}{\texttt{ABMPtt\_5\_nnlo}}           & 53.3 pb                 & $+3.3\%$                          & \multicolumn{1}{l|}{$^{+0.2\%}_{-3.6\%}$}     & 53.2 pb                 & $-0.1\%$                          & \multicolumn{1}{l|}{$^{+0.5\%}_{-2.1\%}$}     \\
\multicolumn{1}{|l|}{\texttt{CT18NNLO}}                  & 51.3 pb                 & $+3.5\%$                          & \multicolumn{1}{l|}{$^{+0.3\%}_{-3.9\%}$}     & 51.3 pb                 & $-0.1\%$                          & \multicolumn{1}{l|}{$^{+0.5\%}_{-2.3\%}$}     \\
\multicolumn{1}{|l|}{\texttt{PDF4LHC21\_40}}             & 51.6 pb                 & $+3.5\%$                          & \multicolumn{1}{l|}{$^{+0.3\%}_{-3.9\%}$}     & 51.5 pb                 & $-0.1\%$                          & \multicolumn{1}{l|}{$^{+0.5\%}_{-2.3\%}$}     \\
\multicolumn{1}{|l|}{\texttt{MSHT20an3lo\_as118}}        & 48.7 pb                 & $+3.5\%$                          & \multicolumn{1}{l|}{$^{+0.3\%}_{-3.9\%}$}     & 48.7 pb                 & $-0.1\%$                          & \multicolumn{1}{l|}{$^{+0.5\%}_{-2.3\%}$}     \\
\multicolumn{1}{|l|}{\texttt{NNPDF40\_an3lo\_as\_01180}} & 50.6 pb                 & $+3.5\%$                          & \multicolumn{1}{l|}{$^{+0.3\%}_{-3.9\%}$}     & 50.5 pb                 & $-0.1\%$                          & \multicolumn{1}{l|}{$^{+0.5\%}_{-2.3\%}$}     \\ \hline
\end{tabular}
}
\caption{
Heavy-top limit results at N$^3$LO and N$^4$LO$_\mathrm{sv}$.
The mass of the Higgs boson is taken to be $\MH=125$\,GeV and the top-quark is renormalised in the on-shell scheme.
The scale uncertainty is computed via a 7-point scale variation about the central scale $\mu_R = \mu_F = \MH/2$ with fixed $\alpha_s(M_Z)=0.118$.
Results obtained using the calculation of Ref.~\cite{Das:2020adl}.
}
\label{tab:ggf_n4lo_sv}
\end{table}

As described in Section~\ref{sec:ggf_htl}, results for Higgs production in gluon-fusion are available at order N$^3$LO in the heavy top-quark limit.
Beyond this order, it is possible to produce an approximation of the N$^4$LO result using the soft-virtual approximation, which captures the leading term in the expansion near to the Higgs boson production threshold.
In Ref.~\cite{Das:2020adl}, an approximation to this was calculated and at $\sqrt{s}=14$\,TeV it was found to give a $-0.1\%$ correction for the scale $\mu_R=\mu_F=\MH/2$ and to reduce the remaining 7-point scale uncertainty from about 5\% to less than 3\%.
In Table~\ref{tab:ggf_n4lo_sv}, we present results obtained using the calculation of Ref.~\cite{Das:2020adl} for $\sqrt{s}=13.6$\,TeV with the recommended parameters of the Higgs working group, noting that the 7-point scale variation used here is at variance with the 3-point prescription adopted for the recommendation (see Section~\ref{sec:ggf_htl}).
The magnitude of the correction and reduction in the scale dependence is found to be largely independent of the PDF set choice and order of the PDF.

\paragraph{Recommendation}
Knowledge of the N$^4$LO$_\mathrm{sv}$ corrections serves as an important cross-check of the perturbative stability of the N$^3$LO result and of the reliability of the uncertainty estimated via a scale variation.
Pending a complete N$^4$LO HTL calculation performed without the soft-virtual approximation, our recommended numbers are produced based on the N$^3$LO result.


\subsection{Comparison to Yellow Report 4}
\label{sec:ggf_yr4}

\begin{table}[]
{\renewcommand{\arraystretch}{1.4}
  \centering\small
  \makebox[\textwidth][c]{%
\begin{tabular}{|l|l|l|l|}
\hline
$\sqrt{s}=13$ TeV & YR4           & R5          & Comments                                                                                                   \\ \hline
$\sigma$                         & 48.58 pb        & 48.04 pb     & Reduced by $1.11\%$                                                                                          \\ \hline
{\color{black!55} $\delta(\text{scale})$} & $^{+0.21\%}_{-2.37\%}$ & $^{+0.29\%}_{-3.33\%}$ & $M_t^{\overline{\mathrm{MS}}} \rightarrow M_t^\mathrm{OS}$, 3-point scale variation                                \\
{\color{black!55} $\delta(\text{EWK})$}   & $\pm 1\%$        & $\pm 1\%$      & Mixed QCD-EW approx. checked against full calculation~\cite{Becchetti:2020wof}                  \\
{\color{black!55} $\delta(t,b,c)$}        & $\pm0.83\%$     & $\pm0.34\%$   & Included $\sigma^{\mathrm{NNLO}, (t)}, \sigma^{\mathrm{NNLO}, (t \times b)}$                         \\
{\color{black!55} $\delta(1/m_t)$}        & $\pm1\%$        & 0           & Included $\sigma^{\mathrm{NNLO}, (t)}$, remaining uncertainty included in $\delta(t,b,c)$              \\
{\color{black!55} $\delta(\text{trunc})$} & $\pm 0.37\%$     & 0           & Included complete N$^3$LO HTL corrections~\cite{Mistlberger:2018etf}                               \\
$\delta(\text{theory})$                 & $^{+3.41\%}_{-5.57\%}$   & $^{+1.63\%}_{-4.67\%}$ & --                                                                                                            \\ \hline
$\delta(\text{PDF}+\alpha_s)$            & $\pm 3.2\%$      & $^{+2.69\%}_{-2.28\%}$ & $\text{PDF4LHC15} \rightarrow \text{PDF4LHC21}$                                                                             \\
$\delta(\text{PDF-TH})$                  & $\pm 1.16\%$     & $\pm 2.31\%$       & Increased by $\times 2$ \\ \hline
\end{tabular}
}
}
\caption{ Comparison of the recommendation and uncertainties of YR4~\cite{LHCHiggsCrossSectionWorkingGroup:2016ypw} and
  the current recommendation (R5) for collider energy $\sqrt{s} = 13$
  TeV and Higgs boson mass $\MH = 125$ GeV with all other parameters
  chosen as their default in the respective reports.  In the
  $\delta(\text{theory})$ uncertainty of each recommendation the
  $\delta(\text{PDF-TH})$ uncertainty is omitted and reported
  separately.  }
\label{tab:ggf_yr4_vs_yr5}
\end{table}

In this Section the uncertainty estimate of the current gluon fusion total cross section recommendation and that of the original YR4~\cite{LHCHiggsCrossSectionWorkingGroup:2016ypw} are compared.
In Table~\ref{tab:ggf_yr4_vs_yr5}, each of the uncertainties presented in the current and previous report are summarised and the main factors driving the change in the uncertainty estimates are recorded.
The uncertainties presented are for collider energy $\sqrt{s} = 13$ TeV and Higgs boson mass $\MH = 125$ GeV, all other choices and settings are chosen as originally presented in their respective report.

YR4~\cite{LHCHiggsCrossSectionWorkingGroup:2016ypw} contains a discussion and analysis of how the various uncertainties can be interpreted, namely, either as a flat 100\% confidence interval (procedure F) or a Gaussian one-sigma symmetric 68\% confidence interval (procedure G). 
We do not repeat this analysis here, however, we do record the relevant differences between how uncertainties are calculated in the two reports:
\begin{itemize}
\item $\delta(\text{scale})$ - In YR4 the F-uncertainty of $^{+0.2\%}_{-2.4\%}$ was estimated by performing a scale variation scan in the range $\MH/4 \le \mu_R = \mu_F \le \MH$, while the G-uncertainty of $^{+3.0\%}_{-3.0\%}$ was estimated via a symmetrised 7-point scale uncertainty. 
In the current recommendation the scale uncertainty is estimated via a 3-point scale uncertainty.
As described in Section~\ref{sec:ggf_htl}, the previous recommendations used the $\overline{\mathrm{MS}}$ top-quark mass with scale $\mu_R$, which was varied in the re-weighting factor, $\text{R}_\text{LO}$, leading to a smaller scale uncertainty estimate.
In the current recommendation a fixed on-shell top-quark mass is used.
\item $\delta(\text{EWK})$ - The EW and mixed QCD-EW corrections are included identically to those in YR4.
The uncertainty numbers quoted in the current recommendation are the same as those given for the F-uncertainty in the previous recommendation.
An additional check was performed to confirm that the EFT approximation used, in which the weak bosons are assumed to be heavier than the Higgs boson, is compatible with the full result~\cite{Becchetti:2020wof} within the $\pm 1\%$ uncertainty assigned.
\item $\delta(t,b,c)$ - In the current recommendation, the missing N$^3$LO top and top-bottom quark mass effects, and NNLO bottom, charm, top-charm and bottom-charm quark mass effects are estimated via 7- and 3-point scale variations, respectively.
This procedure differs from both the F-uncertainty and G-uncertainty presented in YR4.
\item $\delta(\text{PDF}+\alpha_s)$ - The current recommendation is based on the PDF4LHC21 NNLO PDF set, the PDF and $\alpha_s$ uncertainties are estimated via the usual procedure, which matches that used in the previous report.
Total cross section numbers and PDF uncertainties are also provided for the combined MSHT20xNNPDF40 aN$^3$LO set, these numbers include within the $\delta(\text{PDF})$ uncertainty also an estimate of the missing higher-order uncertainty associated with the fixed order matrix elements used within the PDF fits.
\item $\delta(\text{PDF-TH})$ - The uncertainty related to the mismatch between the N$^3$LO matrix element and the NNLO PDF is estimated using Eq.~\eqref{eq:ggf_pdf_th}, this estimate is twice that reported in the previous YR4.
As explained in Section~\ref{sec:ggf_an3lo_pdf}, approximate N$^3$LO PDFs have been produced by two global fitting groups and were combined in Ref.~\cite{Cridge:2024icl}.
Switching from the PDF4LHC21 NNLO set to the MSHT20xNNPDF40 aN$^3$LO combination set reduces the total cross section at $\sqrt{s}=13$ TeV by around $-4.70\%$, this discrepancy motivates the decision to inflate the PDF-TH uncertainty.
This difference would be smaller at LHC energies (and larger at a 100 TeV collider) if the NNLO PDF was convoluted with the NNLO matrix element instead of the N$^3$LO matrix element.
An alternative, favoured by some of the community, is to estimate the PDF-TH associated with the use of the PDF4LHC21 NNLO set by directly taking the difference between using NNLO or aN$^3$LO PDF sets.
Furthermore, if agreement within the community is reached on the use of aN$^3$LO PDF sets, it is conceivable that the current PDF-TH uncertainty can be eliminated or significantly reduced, along with an update of the central cross section prediction.
\end{itemize}

\section{VBF}
\label{sec:VBF}
\subsection{Summary of Theory Developments}
In this section we summarise the theory work carried out in recent years on
Higgs production via vector-boson fusion (VBF) at the LHC. A comparative study
of differential cross sections at fixed order, including parton-shower
corrections, has recently been made public~\cite{Barone:2025jey}. Besides
providing state-of-the-art differential predictions at fixed order and
recommendations for theory uncertainties in parton-shower simulations, it
contains a pedagogical description of higher-order corrections in VBF and a
review of the experimental and theoretical state of the art, with references
collected in its Tables~2 and~3, respectively. We therefore refer the interested
reader to Ref.~\cite{Barone:2025jey} and references therein. For fixed-order
calculations the Les Houches wishlist~\cite{Huss:2025nlt} is a further useful
resource. For completeness we nonetheless give a short review of recent and
state-of-the-art theory calculations for VBF at the LHC.

The electroweak (EW) production of a Higgs boson in association with two jets
(HJJ) comprises several contributions, of VBF and of VH type. The quality of
the VBF approximation, and the impact of higher-multiplicity contributions,
have been studied in the
literature~\cite{Figy:2003nv,Berger:2004pca,Ciccolini:2007jr,Bolzoni:2011cu,
Campanario:2013fsa,Campanario:2018ppz,Barone:2025jey}.

The state of the art for EW corrections is NLO
EW~\cite{Ciccolini:2007jr,Ciccolini:2007ec}, which are used in this article.
For QCD corrections the state of the art is NNLO QCD in the VBF approximation
at the differential level~\cite{Cacciari:2015jma,Cruz-Martinez:2018rod,Asteriadis:2021gpd}, while
inclusive corrections are available at N$^3$LO~\cite{Dreyer:2016oyx} in the structure function approximation~\cite{Han:1992hr,Bolzoni:2010xr,Bolzoni:2011cu} and are
used for the predictions presented here. Since the state-of-the-art QCD
computations are performed in the VBF approximation, which neglects the cross
talk between the two quark lines, the impact of non-factorisable corrections
has been studied in detail~\cite{Liu:2019tuy,Dreyer:2020urf,Asteriadis:2023nyl,
Bronnum-Hansen:2023vzh,Long:2023mvc,Gates:2023iiv}. In the predictions
presented here, non-factorisable corrections are accounted for as an uncertainty.

As mentioned above, the EW production of HJJ proceeds through several
mechanisms. In addition, separate contributions at different orders in
perturbation theory lead to the same final
state~\cite{Andersen:2006ag,Andersen:2007mp,Harlander:2008xn,Greiner:2016awe,
Greiner:2015jha,Andersen:2017kfc,Andersen:2018tnm,Andersen:2018kjg,
Chen:2021azt,Andersen:2022zte}; a summary of these irreducible backgrounds can
be found in Ref.~\cite{Barone:2025jey}. The characteristics of the VBF
production mechanism have also been studied in several phase-space regions,
including those in which the Higgs boson is
boosted~\cite{Becker:2020rjp,Buckley:2021gfw}. Finally, while historically the
focus of both theoretical and experimental work has been on a stable Higgs
boson, recent work~\cite{Asteriadis:2021gpd,Asteriadis:2024nbg} has shown that
considering the decay products of the Higgs boson along with its production
mechanism can induce non-trivial effects.

Beyond fixed order, significant work related to parton showers has been carried
out over the years. Parton-shower corrections matched to
QCD~\cite{Nason:2009ai,Frixione:2013mta} and even to EW~\cite{Jager:2022acp}
corrections are available at NLO accuracy, and multi-jet merged results have
been investigated~\cite{Hoche:2021mkv,Chen:2021phj}. Uncertainties related to
NLO QCD+PS simulations have been studied in
detail~\cite{Jager:2020hkz,Buckley:2021gfw,Hoche:2021mkv}, and recommendations
regarding such uncertainties are given in Ref.~\cite{Barone:2025jey}. Soft
physics effects such as multi-parton interactions and
hadronisation~\cite{Hoche:2021mkv,Bittrich:2021ztq} have been studied, as have
developments towards next-to-leading-logarithmic showers for
VBF~\cite{vanBeekveld:2023chs}.

\subsection{Inclusive Cross Section Predictions}
\label{sec:vbf_xs}

The results compiled in Tables~\ref{tab:vbf_XStot_7}-\ref{tab:vbf_XStot_14} are
combined according to
\begin{align}
  \label{eq:vbf-xsec-combined}
  \sigma^{\VBF} = \sigma_{\NNNLO}^{\DIS} (1+\delta_{\ELWK}) + \sigma_{\gamma} ,
\end{align}
where $\sigma_{\NNNLO}^{\DIS}$ is the $\NNNLO$ QCD cross section in the DIS
approximation. The relative EW corrections omitting the photon-induced
contributions are denoted by $\delta_{\ELWK}$, while the photon-induced
contributions are collected in $\sigma_{\gamma}$.

The $\NNNLO$ QCD cross section is obtained with
\texttt{proVBFH}~\cite{Cacciari:2015jma,Dreyer:2016oyx}. The NLO EW
corrections and the photon-induced contributions are computed with
{\sc Hawk}~\cite{Denner:2014cla}, as are the interferences between the $s$-,
$t$-, and $u$-channel contributions. It is worth mentioning that while
{\sc Hawk} takes all partonic channels into account at LO, at NLO the channels
featuring bottom quarks in the external state are omitted, which is a small
effect. The non-factorisable corrections are taken from
Refs.~\cite{Liu:2019tuy,Asteriadis:2023nyl}. The central renormalisation and
factorisation scales are set to the fixed value $\mu_F=\mu_R=M_W$; a fixed
scale is required for the EW corrections, and the same choice is retained for
the QCD contribution so that the two are combined consistently. The scale
uncertainty $\Delta_\mathrm{scale}$ quoted in the tables is obtained from a
3-point variation, in which $\mu_F$ and $\mu_R$ are varied together by a factor
of two up and down about this central value, the envelope of the three
resulting predictions being taken as the uncertainty. It is well below the
other uncertainties throughout, a consequence of the very mild scale dependence
of the inclusive cross section at N$^3$LO~\cite{Dreyer:2016oyx}.

The theory uncertainties are computed as
\begin{align}
  \label{eq:vbf-TU}
  \Delta_{\mathrm{TU}} = \max\left\{0.5\%,\delta_{\ELWK}^2\right\} + \frac{|\sigma_{\mbox{\scriptsize nf}}| + |\sigma_{\mbox{\scriptsize s/t/u}}|}{\sigma^{\VBF}}\%
\end{align}
where $\sigma_{\mbox{\scriptsize s/t/u}}$ refers to the size of the
interference contribution between $s$-, $t$- and $u$-channel diagrams for
$\sqrt{s}=\{13,13.6,14\}$ TeV. For the legacy numbers corresponding to
$\sqrt{s}=\{7,8\}$ TeV the non-factorisable contribution,
$\sigma_{\mbox{\scriptsize nf}}$, was not computed, and we instead set
\begin{align}
  \label{eq:vbf-TU-legacy}
  \Delta_{\mathrm{TU}} = \max\left[\max\left\{0.5\%,\delta_{\ELWK}^2\right\} + \frac{|\sigma_{\mbox{\scriptsize s/t/u}}|}{\sigma^{\VBF}}\%,1.0\%\right],
\end{align}
which in this case always corresponds to $1\%$.

\subsection{Parton distributions beyond NNLO}
\label{sec:vbf_an3lo}

The prediction of Eq.~\eqref{eq:vbf-xsec-combined} combines an N$^3$LO QCD
matrix element with the NNLO PDF4LHC21 parton distributions of
Section~\ref{sec:pdf}, and therefore carries the same mismatch in perturbative
accuracy as the gluon-fusion prediction, discussed in
Section~\ref{sec:ggf_an3lo_pdf}. To quantify it we have recomputed the
inclusive N$^3$LO cross section in the setup of this report, at
$\sqrt{s}=13$, $13.6$ and $14$~TeV, with the NNLO and approximate N$^3$LO
variants of the MSHT20xNNPDF40 combination of Ref.~\cite{Cridge:2024icl}.

Replacing PDF4LHC21 by the aN$^3$LO combination raises the cross
section by $3.3\%$ at $\sqrt{s}=13.6$~TeV. 
Writing $\sigma[S]$ for the N$^3$LO cross section evaluated with the PDF
set $S$, this shift separates exactly into
\begin{equation}
  \underbrace{\sigma[\mathrm{aN}^3\mathrm{LO}]-\sigma[\mathrm{PDF4LHC21}]}_{+3.3\%}
  = \underbrace{\sigma[\mathrm{NNLO}]-\sigma[\mathrm{PDF4LHC21}]}_{+1.1\%}
  + \underbrace{\sigma[\mathrm{aN}^3\mathrm{LO}]-\sigma[\mathrm{NNLO}]}_{+2.2\%}\,,
  \label{eq:vbf-pdf-th}
\end{equation}
where NNLO and aN$^3$LO denote the two corresponding members of the
MSHT20xNNPDF40 family and the percentages are relative to
$\sigma[\mathrm{PDF4LHC21}]$.
The first term is a change of PDF set at fixed NNLO accuracy, and is
already covered by the $\Delta_{\mathrm{PDF}\oplus\alphas}$ uncertainty
quoted in the tables; the second is a change of perturbative order within
one family, and is not.
Only the second term describes a genuine PDF-theory effect. Since the position of the working group is to use PDF4LHC21 at
NNLO as the baseline, we cannot advocate applying this shift directly
to the cross section. It is in any case comparable to the total
uncertainty of $^{+2.5}_{-2.4}\%$ already assigned to VBF at this
energy, and our central prediction and uncertainty therefore do not
include it. Should a more conservative uncertainty band be desired, we
recommend that the PDF-theory uncertainty of Eq.~\eqref{eq:vbf-pdf-th},
$\Delta_{\mathrm{PDF-TH}} = \pm 2.2\%$, be added in quadrature to the
total uncertainty, across collider energies and Higgs-boson masses,
which brings the latter to approximately $\pm 3.3\%$. The energy dependence of the underlying shift is
negligible over the range considered, varying by less than $0.03$
percentage points between $13$ and $14$~TeV.

This estimate is taken directly from the difference between NNLO and aN$^3$LO
sets, rather than from the one-order-down proxy of Eq.~\eqref{eq:ggf_pdf_th}
used for gluon fusion. The two prescriptions are appropriate in different
circumstances. A direct difference is only as reliable as the agreement between
the two independent aN$^3$LO PDF determinations, and for VBF that agreement is
good: the MSHT20 and NNPDF4.0 aN$^3$LO sets differ by $0.3\%$, against $1.4\%$
for the corresponding NNLO sets, so the two groups agree substantially better
at aN$^3$LO than at NNLO and the shift is well under control. For gluon fusion the
same two sets differ by around $3\%$ on an effect of comparable size, so the
direct difference is not a reliable estimator there; the prescription of
Section~\ref{sec:ggf_an3lo_pdf} is used instead, and the resulting uncertainty
is included in the gluon-fusion recommendation.

\section{VH}
\label{sec:VH}
\subsection{Summary of Theory Developments} 
In this section, we shortly collect results available for the theoretical description of the associated production of a Higgs boson and a weak vector boson (VH) at the LHC, and we outline recent developments in theory predictions. We recall that the dominant partonic contribution to the process comes from the $q\bar{q}$-initiated, Drell-Yan-like channel, while a subleading contribution due to a loop-induced $gg$-initiated channel arises at NNLO QCD for the $ZH$ final state.

Concerning fixed-order predictions, the QCD corrections for the $q \bar{q}$-initiated process have been known at NNLO~\cite{Glashow:1978ab, Eichten:1984eu, HAN1991167, Brein:2003wg}. For the Drell-Yan-like contribution, N$^3$LO QCD corrections have become available~\cite{Baglio:2022wzu}. Electroweak (EW) NLO corrections have been computed in Refs.~\cite{Ciccolini:2003jy, Denner:2011id} and are available from the code \texttt{HAWK}~\cite{Denner:2011id,Denner:2014cla}, including the leptonic decays of the weak bosons. Differential predictions are available in Refs.~\cite{Ferrera:2011bk, Ferrera:2013yga,Ferrera:2014lca,Campbell:2016jau,Gauld:2021ule}, including NNLO QCD and NLO EW effects.
In recent years, considerable interest has been devoted to the QCD corrections to the subdominant process $gg \to ZH$~\cite{Dicus:1988yh,Kniehl:1990iva}, with the aim of reducing the substantial impact of the associated scale uncertainties. The NLO QCD corrections have been obtained using multiple approaches~\cite{Altenkamp:2012sx,Hasselhuhn:2016rqt,Davies:2020drs,Chen:2020gae,Wang:2021rxu,Alasfar:2021ppe,Chen:2022rua,Degrassi:2022mro,Davies:2025out, Davies:2026uxl}. The Refs.~\cite{Chen:2022rua,Degrassi:2022mro, Davies:2026uxl} discuss an additional source of theoretical uncertainty due to the choice of renormalization scheme for the top-quark mass ($m_t$). More recently, the virtual QCD corrections at NNLO have been computed~\cite{Davies:2025otz} in the large-$m_t$ limit. 

Given the relevance of VH production as an experimental probe of the Higgs coupling to bottom quarks, several works have discussed the effects of supplementing VH predictions with the Higgs decay to a $b \bar{b}$ pair~\cite{Caola:2017xuq,Gauld:2019yng,Behring:2020uzq}.

The resummation of threshold logarithms for the Drell-Yan (DY) component of VH production has been carried out at N$^3$LL accuracy~\cite{Kumar:2014uwa, Das:2022zie}. Soft-gluon resummation in the $gg$-initiated channel has been studied in Refs.~\cite{Harlander:2014wda,Das:2025wbj,Das:2025qym, Banerjee:2026hzp}. The matching of fixed-order corrections to the parton shower (PS) has been obtained for VH production at NLO QCD~\cite{Luisoni:2013cuh,Granata:2017iod} and at NLO EW~\cite{Granata:2017iod} accuracy. Considering NNLO QCD corrections, merged predictions including one jet for $gg \to ZH$ have been presented~\cite{Hespel:2015zea, Goncalves:2015mfa}, while NNLO+PS predictions for the full VH process have been achieved in Refs.~\cite{Astill:2016hpa,Astill:2018ivh, Alioli:2019qzz}. Matched predictions including the $H\to b \bar{b}$ decay at full NNLO+PS accuracy have been obtained in Ref.~\cite{Zanoli:2021iyp}.

\subsection{Inclusive Cross Section Predictions}
\label{sec:vh_inclxs_breakdown}

The results compiled in Tables~\ref{tab:w+h_xs_7tev}-\ref{tab:zh_xs_14tev} are obtained from the following combination of the different contributions to the cross section
\begin{align}
    \sigma^{VH} &= \sigma_\mathrm{DY}^{VH}(1+ \delta_\mathrm{EW})+ \sigma_\text{top} + \delta_{\tiny VZ}~\sigma^\mathrm{NLO}_{ggZH},
    \label{eq:sigma_vh_breakdown}
\end{align}
where $\sigma_\mathrm{DY}^{VH}$ is the Drell-Yan-like contribution at NNLO QCD, $\delta_\mathrm{EW}$ is a multiplicative factor that accounts for the relative EW corrections at NLO, $\delta_{\tiny VZ}$ equals 1 when $V=Z$ and 0 otherwise, and $\sigma^\mathrm{NLO}_{ggZH}$ is the NLO QCD cross section for the $gg$-initiated channel. The term $\sigma_\text{top}$ is defined as
\begin{align}
     \sigma_\text{top} &= \sigma_\mathrm{I}^{VH} + \delta_{VZ}~\sigma_\mathrm{II}^{ZH},
\end{align}
where $\sigma_\mathrm{I}^{VH}$ and $\sigma_\mathrm{II}^{ZH}$ are contributions mediated by top-quark loops in the $q\bar{q}$-initiated channel.
The terms $\sigma^\mathrm{NLO}_{ggZH}$ and $\sigma_\mathrm{II}^{ZH}$  are present only in the $ZH$ final state. In the Tables~\ref{tab:w+h_xs_7tev}-\ref{tab:zh_xs_14tev} we provide results for on-shell bosons, whereas the Tables~\ref{tab:w+_lep_xs_7tev}-\ref{tab:zh_lep_xs_14tev}, contain the predictions including the leptonic decays. Besides the multiplication of the results by the respective branching ratios, the latter predictions account for an additional photon-induced contribution, denoted as $\sigma^\gamma$, which is computed with \texttt{HAWK} and is non-negligible in the case of $WH$ production.

The results for all the contributions in Eq.~\eqref{eq:sigma_vh_breakdown} except $\sigma^\mathrm{NLO}_{ggZH}$ are obtained using \texttt{vh@nnlo}~\cite{Brein:2012ne, Harlander:2018yio} and \texttt{HAWK}~\cite{Denner:2011id,Denner:2014cla}. The results for $\sigma^\mathrm{NLO}_{ggZH}$ at centre-of-mass (CM) energies $\sqrt{s}=\{13, 13.6, 14\}$~TeV are taken from Ref.~\cite{CampilloAveleira:2025rbh}, which includes NLO QCD corrections with top-quark mass effects at $\mathcal{O}(\%)$ accuracy. This reference provides a comparison of the virtual corrections reported in Refs.~\cite{Chen:2022rua,Degrassi:2022mro} and supplements the latter with the new results for the real corrections using \texttt{GoSam}~\cite{Braun:2025afl}. Results for values of the Higgs mass not present in Ref.~\cite{CampilloAveleira:2025rbh} have been obtained with a linear interpolation, which is accurate at $\mathcal{O}(10^{-4})$, given the mild dependence of the cross section on this parameter. For $\sqrt{s}=\{7, 8\}$~TeV, also the values of $\sigma^\mathrm{NLO}_{ggZH}$ are computed using \texttt{vh@nnlo}.  We remind that N$^3$LO QCD corrections for the DY-like contribution are available~\cite{Baglio:2022wzu}, and it was shown that they are comparable in magnitude to the NNLO corrections and that they do not lead to a reduction of the associated scale uncertainties. However, these effects are not included in the current inclusive predictions, since a consistent combination of all the $O(\alpha_s^3)$ contributions provided by the existing codes is complicated by ambiguities in the assignments of subsets of diagrams to the quark- or gluon-initiated channels. The dominant scale uncertainties from $\sigma^\mathrm{NLO}_{ggZH}$ are expected to cover the omission of the N$^3$LO DY corrections in $ZH$ production.

The uncertainties due to the factorization and renormalization scale variation are denoted as $\Delta_\text{scale}$ and computed following the recommendation of the YR4~\cite{LHCHiggsCrossSectionWorkingGroup:2016ypw} from a 7-point scale variation around the central scale $M_{VH}$ in the interval
\begin{align}
    M_{VH}/3 \leq \mu_F ,\mu_R \leq 3 M_{VH},
\end{align}
with $M_{VH}$ the invariant mass of the $VH$ system. Only in the case in which $\sigma^\mathrm{NLO}_{ggZH}$ is taken from Ref.~\cite{CampilloAveleira:2025rbh}, the scale uncertainties for the latter are obtained from a 7-point scale variation around the central scale $M_{ZH}/2$ in the interval
\begin{align}
    M_{ZH}/4 \leq \mu_F ,\mu_R \leq M_{ZH}.
\end{align}
The choice of central scale follows the observation of
Ref.~\cite{Harlander:2014wda} that this scale seems to minimize the
effects of soft-gluon resummation; the interval for the scale
variation follows the recommendations for $gg\rightarrow H$ and
$gg\rightarrow HH$, for which NLO QCD corrections have a similar
impact as $gg \rightarrow ZH$. It must be noted that these choices are
based on the approximation that the quark-initiated and the
gluon-initiated channels are independent, and correlations of scale
uncertainties due to the interference of subgroups of diagrams from
the two channels are neglected. The PDF and $\alpha_s$ uncertainties
(denoted as $\Delta_\text{PDF}$ ad $\Delta_{\alpha_s}$, respectively)
are calculated following Section~\ref{sec:pdf}.

\section{\texorpdfstring{$\boldsymbol{t\bar{t}H}$}{ttH} and \texorpdfstring{$\boldsymbol{tH}$}{tH}}
\label{sec:ttH}

\subsection{Cross-section predictions for \texorpdfstring{$t\bar t H$}{ttbar H}}
\label{sec:results-xsects-tth}
New predictions for the $t\bar t H$ total cross section have recently appeared in Ref.~\cite{Balsach:2025jcw}. These results
combine predictions accurate at NNLO in QCD~\cite{Catani:2022mfv,Devoto:2024nhl} together with soft-gluon resummation up to NNLL~\cite{Kulesza:2015vda,Broggio:2015lya,Broggio:2016lfj,
Kulesza:2017ukk,Kulesza:2018tqz,Broggio:2019ewu,Kulesza:2020nfh}, and include all effects of EW origin up to NLO~\cite{Frixione:2014qaa,Frixione:2015zaa,Frederix:2018nkq}.
Soft-gluon resummation is carried out in two independent frameworks (SCET and dQCD), which makes it possible to obtain a conservative estimate of missing higher-order
uncertainties via scale variations. In Ref.~\cite{Balsach:2025jcw}, a thorough estimate of many sources of theoretical uncertainties is carried out.
The only one which may be phenomenologically relevant (besides scale variations, PDF and $\alphas$ uncertainties) is the one stemming 
from the approximation of the two-loop virtual amplitude carried out in order to perform the NNLO computation. Such an uncertainty amounts to
\begin{equation}
    \Delta_{\rm virt} = 0.9\%\,
\end{equation}
for the numbers presented here. The total theory uncertainty, $\Delta_{\rm th}$, is computed by combining in quadrature this effect together with scale variations. Following Ref.~\cite{Balsach:2025jcw}, the central value of the renormalisation
and factorisation scales is $\mu_R=\mu_F=m_t+\MH/2$.
The values for the total cross sections at 13, 13.6, and 14 TeV can be found in Tables~\ref{tab:tth-tot13},~\ref{tab:tth-tot136}, and~\ref{tab:tth-tot14}, respectively. Cross sections
for the mass values
\begin{equation}
    \MH \in [ 124.6, 125.0, 125.09, 125.38, 125.6, 126.0] \,\rm GeV\,,
\end{equation}
have been computed. All other values, inside and outside this range, have been obtained
from the linear formula
\begin{equation}
    \sigma(\MH)=\sigma(124.6) +\frac{\sigma(126.0)-\sigma(124.6)}{1.4}(\MH/\rm GeV-124.6)\,,
    \label{eq:inttth}
\end{equation}
which interpolates for $124.6 \,\rm GeV< \MH < 126\,\rm GeV$ and extrapolates outside that interval.
The cross sections at the directly computed mass points carry a residual Monte Carlo
uncertainty at the level of a few per mille, which Eq.~\eqref{eq:inttth} does not smooth;
it is visible as a mild non-monotonicity in $\MH$ of some of the tabulated predictions,
and is negligible compared with the quoted uncertainties.
For other energies, we advise employing the numbers
presented in Ref.~\cite{Karlberg:2024zxx}.

\subsection{Cross-section predictions for \texorpdfstring{$t H$}{tH}}
\label{sec:results-xsects-singleth}
In this section, we provide predictions for $tH$ production. 
For this process, three main production mechanisms concur: 
$t$-channel, $s$-channel and $tWH$ associated production. Their characteristics are very similar to those of single-top (without Higgs) production processes. As is the case
for these processes, at LHC energies the $t$-channel mode is by far the dominant one. At variance with $t\bar tH$, where the cross section
is proportional to the square of the top-quark Yukawa, for $tH$ it breaks down into three terms: a term independent of $y_t$, a linear term and a quadratic
term. As such, $tH$ production is sensitive to the sign of $y_t$. 

State-of-the-art predictions
for these processes include corrections up to NLO QCD~\cite{Demartin:2015uha,Demartin:2016axk} as well as NLO EW~\cite{Pagani:2020mov}. When
the latter are considered, however, interferences between the three production channels cannot be neglected, and the corresponding breakdown 
therefore becomes ill defined. Thus, the impact of EW corrections can only be assessed
by considering all the channels together (possibly imposing the selection cuts which enhance a given channel). In this case, the effect on the inclusive
cross section has been found in Ref.~\cite{Pagani:2020mov} to be about $-3.5\%$. Given the rather low rate of these processes, together with
the issues related to their inclusion when keeping the three channels separate, EW
corrections are not included in the reference cross sections.\\

All single-top and Higgs cross-sections presented below are computed with {\sc MadGraph5\_aMC@NLO}~\cite{Alwall:2014hca,Frederix:2018nkq}. 

The central value $\mu$ of the renormalisation and factorisation scales is set in a process-dependent manner:
\begin{itemize}
    \item $tH$ ($t$-channel): $\mu= \frac{\MH+m_t}{4}\,$, both for the computation in the 4FS and in the 5FS;
    \item $tH$ ($s$-channel): $\mu= \frac{\MH+m_t}{2}\,$;
    \item $tWH$: $\mu= \frac{\MH+m_t+m_W}{2}\,$;
\end{itemize}
All numbers are computed in the five-flavour scheme. For the $t$-channel production mechanism, the uncertainty $\Delta_{\rm scale+\textrm{FS}}$
is obtained as the envelope of the scale variations for the
five- and four-flavour scheme computation, both at NLO. For the 4FS prediction, we employ $m_b=4.92$~GeV.\\
For $tWH$, contributions belonging to $t\bar t H$ entering at NLO via real emissions are removed using the so-called ``Diagram removal with interference'' 
scheme (known as DR2), as implemented in {\sc MadGraph5\_aMC@NLO}, see Refs.~\cite{Demartin:2016axk,Frixione:2019fxg}.

Tables~\ref{tab:th-t-tot13},~\ref{tab:th-t-tot136},~\ref{tab:th-t-tot14};~\ref{tab:th-t-top13},~\ref{tab:th-t-top136},~\ref{tab:th-t-top14}; and~\ref{tab:th-t-atop13},~\ref{tab:th-t-atop136},~\ref{tab:th-t-atop14}
show predictions for $t$-channel production of $tH\,+\,\bar tH$ at $\sqrt{s}=$13, 13.6 and 14 TeV respectively, $tH$ production alone, and $\bar tH$ production alone at the same energies. Predictions for 
$s$-channel production are organised similary in 
Tables~\ref{tab:th-s-tot13},~\ref{tab:th-s-tot136},~\ref{tab:th-s-tot14};~\ref{tab:th-s-top13},~\ref{tab:th-s-top136},~\ref{tab:th-s-top14}; and~\ref{tab:th-s-atop13},~\ref{tab:th-s-atop136},~\ref{tab:th-s-atop14}. 
Finally, Tables~\ref{tab:thw-tot13},~\ref{tab:thw-tot136}, and~\ref{tab:thw-tot14}
show predictions for $tW^-H \,+ \,\bar t W^+H$ associated production (the cross section of the two separate processes are equal to each other in this case) at the three energies.
As for $t\bar t H$, cross sections
for the mass values
\begin{equation}
    \MH \in [ 124.6, 125., 125.09, 125.38, 125.6, 126.] \rm GeV\,,
\end{equation}
have been computed. All other values have been obtained from Eq.~\eqref{eq:inttth},
which interpolates within the computed range and extrapolates outside it.

\section{\texorpdfstring{$\boldsymbol{b\bar{b}H}$}{bbH}}
\label{sec:bbH}
\newcommand{\bbH}{$b\bar{b}H$}
\newcommand{\citere}[1]{Ref.~\cite{#1}}
\newcommand{\citeres}[1]{Refs.~\cite{#1}}
\newcommand{\tab}[1]{Table~\ref{#1}}
\newcommand{\nnnlo}{$\text{N}^3\text{LO}$}
\newcommand{\noun}[1]{{\scshape #1}}
\newcommand{\minnlo}{{\noun{MiNNLO$_{\textrm{PS}}$}}}
\newcommand{\nlonnllpart}{NLO+NNLL\textsubscript{part}+$y_by_t$}
\newcommand{\nlonnllpartnointf}{NLO+NNLL\textsubscript{part}}

\subsection{Summary of Theory Developments}

In this section, we present a summary of state-of-the-art predictions for \bbH{} production. Although its production rate at the LHC is comparable to that of $t\bar{t}H$, Higgs boson production in association with bottom quarks poses greater experimental challenges due to the lack of distinct electroweak signatures. The \bbH{} process receives contributions from both top-quark and bottom-quark Yukawa interactions, including interference terms. The small but non-negligible value of the bottom-quark mass compared to LHC energies makes the modelling of this process particularly interesting.

In the zero-mass variable-flavour number scheme (ZM-VFNS), the bottom quark can be treated as a massless parton, allowing for the resummation of collinear logarithmic contributions into the parton distribution functions, while neglecting power-suppressed terms in the bottom-quark mass. This approach defines the five-flavour scheme (5FS), where the cross-section, first computed at LO in~\citeres{Barnett:1987jw,Dicus:1988cx} and at NLO in~\citeres{Dicus:1998hs,Balazs:1998sb}, is currently known at \nnnlo{} QCD accuracy from~\citere{Duhr:2019kwi}. Additionally, the transverse-momentum and rapidity spectra have been described by resummed analytic predictions in~\citeres{Cal:2023mib,Das:2024pac}. The NNLO predictions~\cite{Harlander:2003ai} have been matched to parton showers at the fully exclusive level in~\citeres{Biello:2024vdh,Gavardi:2025zpf}. These predictions are proportional to the squared bottom-quark Yukawa coupling, and the interference with top-quark contributions vanishes due to the massless treatment of the bottom quark.

Alternatively, \bbH{} production can be described within a four-flavour scheme (4FS), where bottom quarks are treated as massive and produced explicitly in the hard scattering together with the Higgs boson. This approach allows for a systematic inclusion of all the bottom-quark mass effects order by order in perturbative QCD and a complete knowledge of the bottom-quark kinematics, beyond the 5FS collinear approximation. Fixed-order predictions in the 4FS, first obtained at LO in~\citeres{Raitio:1978pt,Ng:1983jm,Kunszt:1984ri}, are currently known exactly at NLO accuracy, including the interference between top- and bottom-Yukawa induced contributions~\cite{Dittmaier:2003ej,Dawson:2003kb,Wiesemann:2014ioa}. The first determination of NNLO corrections in the bottom-quark Yukawa channel has been presented in~\citere{Biello:2024pgo} using the \minnlo{} method for a parton-shower matching. The calculation is exact except for the double virtual contribution, which is estimated in the small bottom-mass limit, making use of the analytic knowledge of the two-loop amplitude with massless bottom quarks~\cite{Badger:2021ega,Badger:2024mir}. Beyond the purely bottom-quark-Yukawa contribution, top-quark-loop effects, which enter for the first time at NLO, are numerically important. The NLO prediction for the $y_t^2$ component has therefore been computed separately in the heavy-top effective field theory limit~\cite{Deutschmann:2018avk}, establishing the relevance of this contribution for the description of QCD backgrounds in di-Higgs searches~\cite{Manzoni:2023qaf}. The corresponding NLO+PS results were compared with a ggF NNLOPS prediction~\cite{Hamilton:2013fea}, revealing a significant sensitivity to the modelling of bottom-quark splittings in the parton shower. More recently, a detailed study~\cite{ATLAS:2025dix} further investigated the differences between the two descriptions and proposed a combined modelling strategy, using the NLO+PS prediction for the $y_t^2$ component of \bbH{} together with a bottom-vetoed NNLOPS sample for the complementary ggF contribution. In the complete-NLO calculation, further contributions were also found to be comparable in size to the bottom-Yukawa-induced component~\cite{Pagani:2020rsg}.

The two flavour schemes capture different aspects of the heavy-quark mass effects, and a combination of the two approaches is needed for a faithful determination of the cross-section. Matched inclusive predictions were first performed using a phenomenological combination known as Santander matching~\cite{Harlander:2011aa}. Two systematic approaches have been developed to combine the 4FS and 5FS cross sections while avoiding double-counting: the FONLL method~\cite{Forte:2015hba,Forte:2016sja} and the \nlonnllpart{} approach~\cite{Bonvini:2015pxa,Bonvini:2016fgf}. The FONLL-B variant matches NNLO predictions in the 5FS with NLO predictions in the 4FS and agrees well with the \nlonnllpart{} results across different values of the Higgs boson mass, as shown in YR4~\cite{LHCHiggsCrossSectionWorkingGroup:2016ypw}. Recently, the \nnnlo{} 5FS prediction has been matched with NLO 4FS cross-sections in~\citere{Duhr:2020kzd} using the FONLL method, reaching a novel level of accuracy in matched fully-inclusive results. Alternative general-mass variable-flavour number scheme (GM-VFNS) approaches, such as that in~\citere{Gauld:2021zmq}, can be employed to obtain matched predictions at both inclusive and fully-differential levels.

\paragraph{Recommendation}
The advantage of the \nlonnllpart{} method is the presence of an explicit dependence on a resummation scale $\mu_b$ that enables the incorporation of a resummation uncertainty estimation in the predictions. For this reason, the current recommendation of the LHCHWG is based on the \nlonnllpart{} method.

\subsection{Total cross sections with flavour-scheme matching}

\begin{table}[h!]
\centering
\begin{tabular}{c|c}
\hline
\textbf{Calculations} & \textbf{References} \\ 
\hline
NNLO 5FS QCD & \cite{Harlander:2003ai} \\
NLO 4FS QCD  & \cite{Dittmaier:2003ej,Dawson:2003kb} \\
\nlonnllpart{} & \cite{Bonvini:2015pxa,Bonvini:2016fgf} \\
\hline
\textbf{PDFs} & \textbf{References} \\
\hline
NNLO PDFs & \cite{Butterworth:2015oua,Bonvini:2016fgf}\\
\hline
\textbf{Codes} & \textbf{References} \\ 
\hline
\nlonnllpart{} & \cite{Bonvini:2016fgf}
\end{tabular}
\caption{Summary of theory references for \bbH{} production at the LHC upon which this recommendation is based.
}
\label{tab:bbf_cite}
\end{table}

Table~\ref{tab:bbf_cite} displays the theoretical references on which the cross-section predictions presented here are based. The available samplings are obtained using the code developed in~\cite{Bonvini:2016fgf}, based on the \nlonnllpart{} method, which matches NNLO QCD 5FS predictions with NLO QCD 4FS results. The matching is performed in a resummation-like way for the $y_b^2$ component of the cross-section, schematically:
\begin{align}
	\sigma^{\text{\nlonnllpartnointf}}=\sigma^{\text{resum}}+\left( \sigma^{\text{FO}} - \left.\sigma^{\text{resum}}\right|_{\mu_b=\mu_F} \right)\,.	\label{eq:FSmatching}
\end{align}
Here, $\sigma^{\text{resum}}$ denotes the prediction in which collinear logarithms are resummed. It is based on the 5FS calculation and depends on the $b$-quark matching scale $\mu_b$, which is of the order of the bottom-quark mass $m_b$. The fixed-order counterpart, $\sigma^{\text{FO}}$, is obtained from the $y_b^2$ component of the 4FS computation. The overlapping logarithmic contributions are subtracted using the resummed result evaluated at the hard scale $\mu_b = \mu_F$, where effectively only the perturbative logarithms are present and no resummation is applied. The non-singular interference component proportional to $y_t y_b$, which is present in the 4FS computation, can be added without introducing any double-counting issues:
\begin{align}
	\sigma^{\text{\nlonnllpart}} = \sigma^{\text{\nlonnllpartnointf}} + \sigma^{y_b y_t}\,.
\end{align}
In Tables~\ref{tab:bbH7},~\ref{tab:bbH8},~\ref{tab:bbH13} and~\ref{tab:bbH14} we report the \nlonnllpart{} cross-sections at respectively 7, 8, 13, and 14 TeV. These numbers are obtained using the following central values for the renormalisation and factorisation scales:
\begin{align}
\mu_F=\frac{1}{4}(\MH+2m_b),\hspace{1cm}\mu_R=\frac{1}{2}\MH.
\end{align}
The standard choice for the resummation scale $\mu_b$ is the bottom-quark mass; however, in regions where $\mu_b/\mu_F > 0.3$, this scale is chosen according to a more general profile, as explained in~\citere{Bonvini:2016fgf}. For a reliable and consistent combination of the flavour schemes, we employ dedicated PDF sets as described in~\citere{Bonvini:2016fgf}. In these sets, the bottom-quark distributions are obtained through a re-evolution of the \texttt{PDF4LHC15\_nnlo\_mc} inputs from an initial scale of $2$~GeV, for all relevant values of $m_b$ and $\mu_b$. This procedure is essential because the different PDF sets entering the PDF4LHC combination are constructed with varying choices of $m_b$. A consistent matching of collinear logarithms, however, requires a controlled and uniform treatment of the evolution across the bottom-quark threshold. The use of self-generated bottom-quark PDFs induces an asymmetry in the parametric uncertainties associated with the PDF inputs, $\delta(\text{PDF})$, whereas the PDF4LHC prescription would lead to symmetric uncertainty bands. The same effect also renders the strong-coupling uncertainty, $\delta(\alpha_s)$, asymmetric in the cross-section tables.

Theoretical uncertainties from renormalisation and factorisation scale variations are estimated using the envelope of the standard 7-point method and are denoted as $\delta(\mu_R,\mu_F)$.
In addition, we assess the uncertainty associated with the resummation scale $\mu_b$ by varying it up and down by a factor of two, and denote this contribution as $\delta(\mu_b)$.
The two sources of theoretical uncertainty are then combined in quadrature to obtain the total theoretical uncertainty, $\delta(\text{theory})$. In the tables, we also report the parametric uncertainties associated with the parton distribution functions, $\delta(\text{PDF})$, the strong coupling constant, $\delta(\alpha_s)$, and the bottom-quark mass, $\delta(m_b)$. The latter has been included due to its particular relevance for \bbH{} predictions, especially through its impact on the determination of the bottom-quark Yukawa coupling. These contributions are combined in quadrature to obtain the total parametric uncertainty, $\delta(\text{PDF}+\alpha_s+m_b)$.

The predictions at 13.6 TeV, summarised in~\tab{tab:bbH136}, are obtained through linear interpolation from cross-section grids at 13 and 14 TeV, using the relation
\begin{align}
	\sigma_{b\bar{b}H}(\sqrt{s}=13.6\,\text{TeV})=0.4\,\sigma_{b\bar{b}H}(\sqrt{s}=13.0\,\text{TeV})+0.6\,\sigma_{b\bar{b}H}(\sqrt{s}=14.0\,\text{TeV})\,.
\end{align}
Tests on the accuracy of the linear interpolation have been performed by comparing the 13.6 TeV prediction obtained from 13 and 14 TeV cross sections with that obtained using the interpolation of existing 13.5 and 14 TeV predictions in~\citere{10.21468/SciPostPhysCommRep.18}. Given the sizeable theoretical perturbative uncertainties in this process, and the quality of the interpolation procedure, dedicated runs at 13.6 TeV are not required. 

\FloatBarrier
\section{Discussion and Outlook}
\label{sec:Outlook}
\begin{figure}[!htbp]
    \centering
    \includegraphics[width=0.95\textwidth]{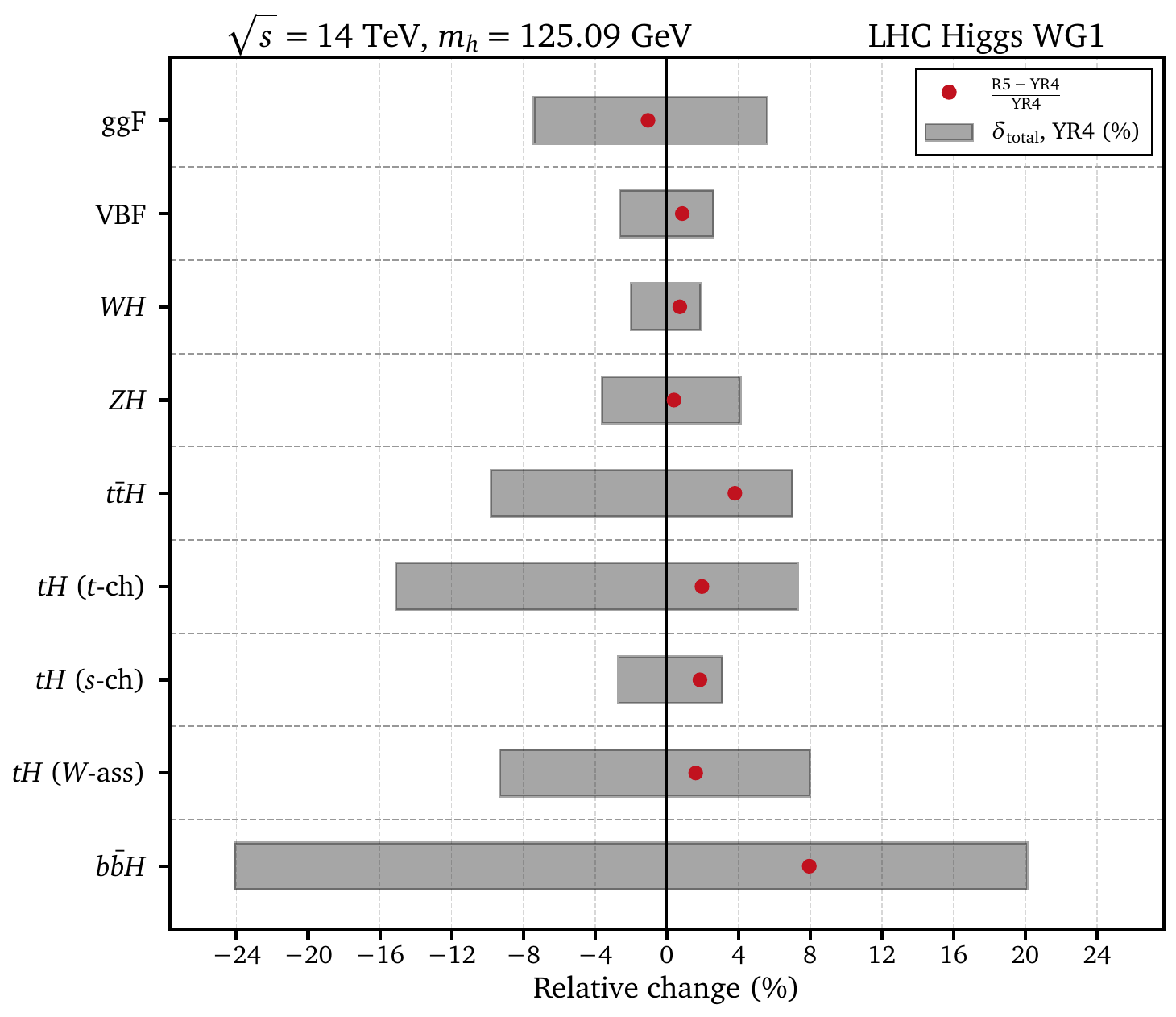}
    \caption{Change in the central value of the predictions of Higgs boson production cross-sections between the YR4 and R5, relative to the YR4 prediction, for collider energy $\sqrt{s} = 14$ TeV and Higgs boson mass $\MH = 125.09$ GeV. The relative uncertainty on the YR4 prediction is also shown. YR4 predictions for W-associated $tH$ production only exist for $\MH = 125.0$ GeV, so that Higgs mass is used instead.}
    \label{fig:central_values}
\end{figure}

\begin{figure}[!htbp]
    \centering
    \includegraphics[width=0.95\textwidth]{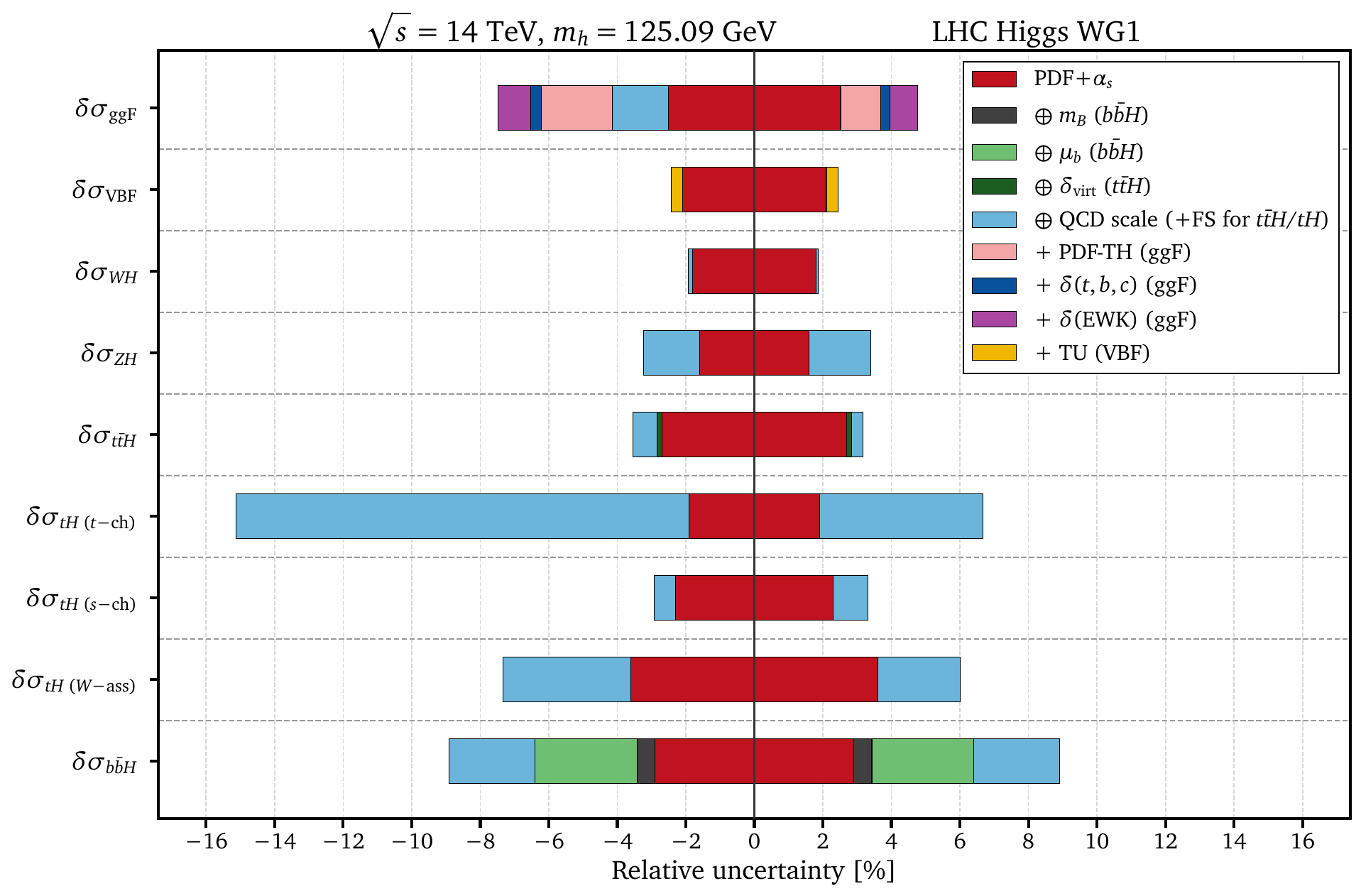}
    \caption{Contribution of various theoretical uncertainties on the Higgs boson production cross sections recommended in this report to the total relative uncertainty. The uncertainties are added one-at-a-time, and the various bars show the total uncertainty after adding each contribution. The symbols in the legend indicate where the source of uncertainty is added linearly ($+$) or in quadrature ($\oplus$).}
    \label{fig:R5_uncertainties_breakdown}
\end{figure}

\begin{figure}[!htbp]
    \centering
    \includegraphics[width=0.95\textwidth]{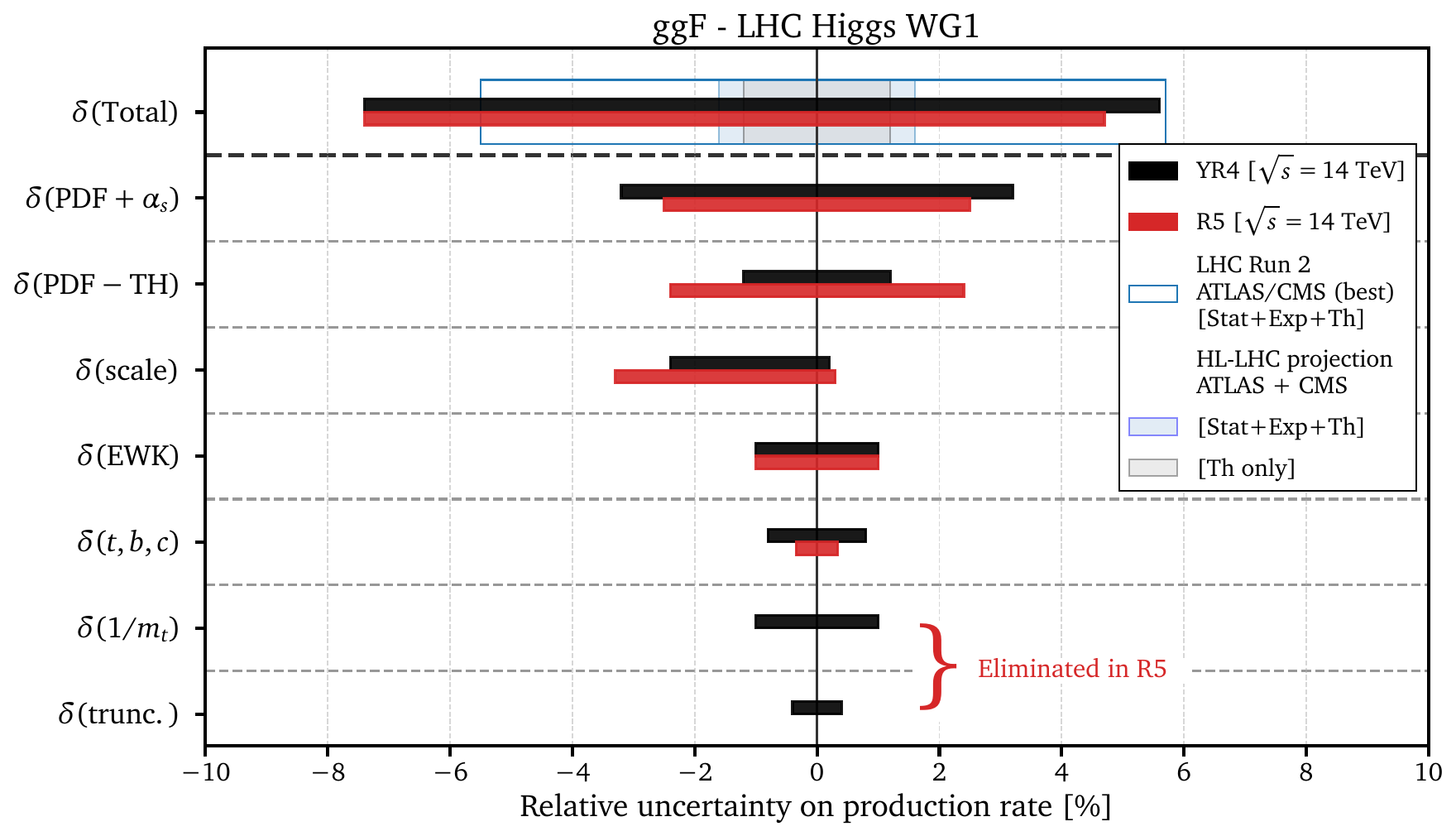}
    \caption{Comparison of the individual contributions to the total theoretical uncertainty on the ggF Higgs boson production cross section recommended in YR4 (black) and in R5 (red). This is compared with the best expected precision for LHC Run 2 Higgs boson ggF production cross section measurements (blue outlined box), whose total uncertainty includes statistical, experimental, and theoretical components, and with the projected HL-LHC ATLAS+CMS combined precision (filled blue box). The theoretical component of the HL-LHC uncertainty projection is also shown (filled gray box).}
    \label{fig:YR5_ggfsummary_uncertainties}
\end{figure}

The R5 update of Higgs boson cross-section predictions brings together many
theoretical advances made since YR4. The update modifies both the central
values and the uncertainties of the predictions with respect to the YR4.
Figure~\ref{fig:central_values} shows that the central values of the R5
predictions are consistent with those of YR4, within the YR4 uncertainties. Figure~\ref{fig:R5_uncertainties_breakdown} shows
the contributions of the various sources of theoretical uncertainty to the
total relative uncertainty for each production mode. The dominant
uncertainties generally arise from PDF-related sources
(PDF+$\alpha_s$, and PDF$-$TH for ggF) or from QCD scale variation.
Figure~\ref{fig:YR5_ggfsummary_uncertainties} illustrates the change in the
individual uncertainties on ggF since the YR4. Several sources of uncertainty
have been eliminated completely or reduced, while others, namely PDF$-$TH
and QCD scale, have increased for the reasons described in
Section~\ref{sec:ggf_yr4}.

\begin{figure}[!htbp]
    \centering
    \includegraphics[width=0.95\textwidth]{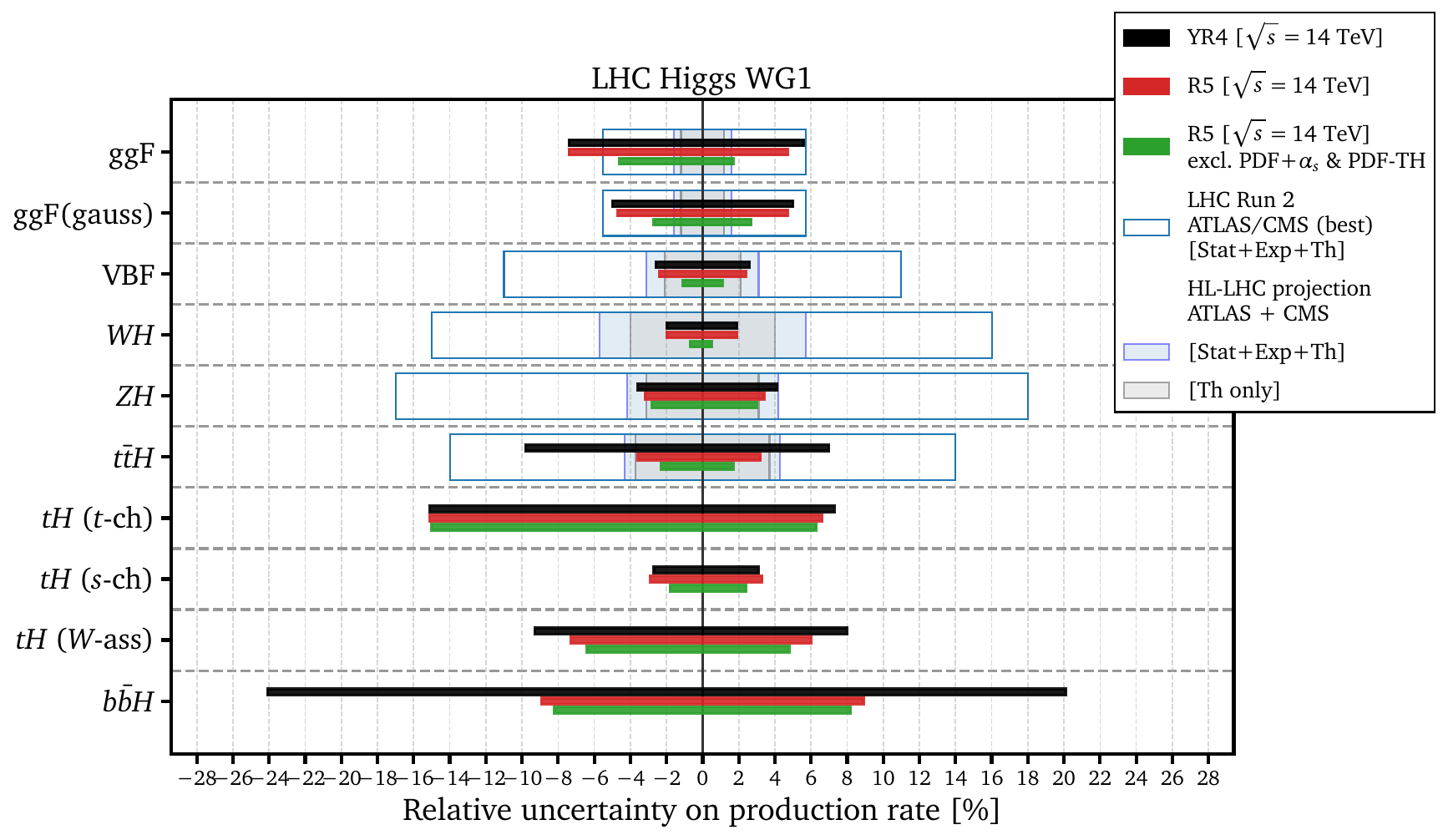}
    \caption{Comparison of the theoretical uncertainties on the Higgs boson production cross sections recommended in YR4 (black) and in R5 (red) for collider energy $\sqrt{s} = 14$ TeV and Higgs boson mass $\MH = 125.09$ GeV. The current recommendations excluding the PDF$+\alpha_{s}$ and PDF$-$TH contributions are also shown (green). These are compared with the best expected precision for LHC Run 2 Higgs boson production cross section measurements (blue outlined boxes), whose total uncertainties include statistical, experimental, and theoretical components, and with the projected HL-LHC ATLAS+CMS combined precision (filled blue box). The theoretical component of the HL-LHC uncertainty projection is also shown (filled gray box).}
    \label{fig:YR5_summary_uncertainties}
\end{figure}

Figure~\ref{fig:YR5_summary_uncertainties} expands this comparison by
illustrating the evolution of the theoretical uncertainties on the Higgs
boson production cross-sections between the YR4 and R5 for all production
modes, and compares these results to the expected precision for measurements of the
Higgs boson production cross-sections at the LHC in Run 2
\cite{ATLAS:2026comb,CMS:2026nce} and the projected precision of these
measurements at the HL-LHC~\cite{ATLAS:2022hsp}. The corresponding numerical
values are summarised in Table~\ref{tab:prod_uncertainties}. For the ggF,
VBF, ZH and $t\bar t H$ production modes, the anticipated precision of the
measurement at the HL-LHC approaches or exceeds the YR4 theory
precision, illustrating the need for the R5 cross-section update. With the
improved precision of the R5 cross-section predictions, only for ggF does the
projected HL-LHC measurement precision exceed that of the inclusive theory
prediction. Were the optional PDF-theory uncertainty discussed in
Section~\ref{sec:vbf_an3lo} to be included for VBF, bringing its total to
approximately $\pm3.3\%$, the same would become true of VBF. However, this is only part of the story; 
the impact of theory uncertainty extends beyond the uncertainty on the inclusive cross-section
prediction, and enters the projected measurement uncertainty itself, for instance via reliance on the predictions 
of Monte Carlo generators. Figure~\ref{fig:YR5_summary_uncertainties} therefore also shows the size of the projected uncertainty 
on the measurement arising only from theory sources. For all production modes, theory uncertainty is projected
to have the largest impact on the measurement precision. It should also
be noted that the HL-LHC projections themselves assume a reduction by a
factor of two in the theory uncertainties entering the measurement uncertainty. Overall, this
comparison makes clear that the theoretical
advancements brought together in the R5 cross-section predictions are
important for fully exploiting the potential of the HL-LHC, and that theory progress will determine
whether the projected precision of these measurements at the HL-LHC will be achieved. 

Figure~\ref{fig:YR5_summary_uncertainties} also makes clear that the
predictions for most production modes have seen an improvement in theory
precision. For several production modes, the relative uncertainty is reduced
by multiple percentage points; for others, the improvement is more moderate.
It should be noted that for \bbH{}, where the figure shows the largest change
in relative uncertainty, the comparison is made to the YR4 numbers that use
Santander matching, rather than to the updated numbers using
\nlonnllpart{} matching that were released very shortly after the YR4 and
have substantially smaller uncertainties. For all production modes, the
improved precision reflects advances in higher-order calculations,
improvements in PDFs, and refined treatments of theoretical and parametric
uncertainties. To illustrate the potential impact of further improvements in
PDFs, the theory precision without PDF-related uncertainties is also shown
in the figure. For several of the most precisely predicted production modes,
this results in a substantial reduction of the total uncertainty,
demonstrating that PDF-related uncertainties are an important limitation and
that further progress in this area will be crucial ahead of the HL-LHC.

Table~\ref{tab:summary_125p09} collects the central predictions of the R5
update at the three collider energies most relevant to the ongoing programme,
for the reference Higgs-boson mass $\MH=125.09$~GeV. It is intended as a
single-point summary of the recommendation; the complete results, for every
production mode, collider energy and Higgs-boson mass considered, are given in
Appendix~\ref{app:tables}. The uncertainty quoted for each mode is the total
uncertainty of that mode, and in the last column the cross sections are summed
over all production modes with the absolute uncertainties added linearly. That
sum is dominated by gluon fusion, which accounts for some $91\%$ of the upward
and $94\%$ of the downward uncertainty at each of the three energies. The
relative uncertainty on the summed cross section, $^{+4.5}_{-6.9}\%$, is
consequently close to that of gluon fusion alone and is, to the precision
quoted, independent of the collider energy.

\begin{table}[htbp]
\renewcommand{\arraystretch}{1.3}
\setlength{\belowcaptionskip}{10pt}
\centering
\caption{Relative uncertainties (\%) on Higgs boson production cross sections from YR4 and R5, including the R5 uncertainties excluding PDF$+\alpha_{s}$ and PDF$-$TH contributions. The YR4 and R5 numbers correspond to a collider energy $\sqrt{s} = 14$~TeV and a Higgs boson mass $\MH = 125.09$~GeV. These are compared with the best expected precision for LHC Run 2 Higgs boson production cross section measurements\cite{ATLAS:2026comb,CMS:2026nce}, whose total uncertainties include statistical, experimental, and theoretical components, and with the projected HL-LHC ATLAS+CMS combined precision~\cite{ATLAS:2022hsp}. The theoretical component of the HL-LHC uncertainty projection is also shown. For the Run~2 measurements and HL-LHC projections, \bbH{} is grouped with ggF and the $tH$ channels are not considered separately, so they are not shown.}
\label{tab:prod_uncertainties}
\begin{tabular}{lcccccc}
\toprule
Production mode &
YR4 &
R5 &
R5 (no PDF) &
Run 2 &
HL-LHC &
HL-LHC (Th only) \\
\midrule
ggF          & ${}^{+5.6}_{-7.4}$  & ${}^{+4.7}_{-7.4}$  & ${}^{+1.7}_{-4.6}$  & ${}^{+5.7}_{-5.5}$ & $\pm1.6$ & $\pm1.2$ \\
VBF          & $\pm2.6$            & $\pm2.4$            & $\pm1.1$            & $\pm11$            & $\pm3.1$ & $\pm2.1$  \\
WH           & ${}^{+1.9}_{-2.0}$  & ${}^{+1.9}_{-2.0}$  & ${}^{+0.5}_{-0.7}$  & ${}^{+16}_{-15}$   & $\pm5.7$ & $\pm4.0$  \\
ZH           & ${}^{+4.1}_{-3.6}$  & ${}^{+3.4}_{-3.2}$  & ${}^{+3.0}_{-2.8}$  & ${}^{+18}_{-17}$   & $\pm4.2$ & $\pm3.1$  \\
ttH          & ${}^{+7.0}_{-9.8}$  & ${}^{+3.2}_{-3.6}$  & ${}^{+1.7}_{-2.3}$  & $\pm14$            & $\pm4.3$ & $\pm3.7$  \\
tH ($t$-ch.) & ${}^{+7.3}_{-15.1}$ & ${}^{+6.6}_{-15.1}$ & ${}^{+6.3}_{-15.0}$ & --                  & --  & -- \\
tH ($s$-ch.) & ${}^{+3.1}_{-2.7}$  & ${}^{+3.3}_{-2.9}$  & ${}^{+2.4}_{-1.8}$  & --                  & --  & -- \\
tH ($W$-ass)  & ${}^{+8.0}_{-9.3}$  & ${}^{+6.0}_{-7.3}$  & ${}^{+4.8}_{-6.4}$  & --                  & --  & -- \\
bbH          & ${}^{+20.1}_{-24.1}$& ${}^{+8.9}_{-8.9}$  & ${}^{+8.2}_{-8.2}$  & --                  & --  & -- \\
\bottomrule
\end{tabular}
\end{table}

Looking beyond the present recommendations, continued progress in PDF
determinations will be particularly important. Approximate N$^3$LO PDF sets
are already available, and the perturbative ingredients entering N$^3$LO
evolution and matching continue to improve, as discussed in
Section~\ref{sec:an3lo-pdfs}. Establishing a robust common prescription for
PDFs at this accuracy will be important for processes whose hard-scattering
cross sections are already known at N$^3$LO. At the same time, as
percent-level precision becomes increasingly relevant, a more uniform
treatment of QED evolution and photon-initiated contributions will become
desirable. Further progress in higher-order perturbative calculations and
in the assessment of process-specific residual theoretical uncertainties
will remain important for production modes where these effects continue to
limit the prediction.

The recommendations presented in this report represent the current state of
the art. Continued progress in higher-order perturbative calculations, PDF
determinations, and the assessment of residual theoretical uncertainties
will be essential to match the experimental precision expected from the
HL-LHC programme.

\begin{landscape}
\begin{table}[ht!]
\caption{Summary of the recommended inclusive production cross sections for
$\MH = 125.09$~GeV at the three highest collider energies considered in this
report. The uncertainty quoted for each production mode is the total
uncertainty of that mode, as given in the corresponding table of
Appendix~\ref{app:tables}. In the last column the cross sections are summed
over all production modes and the corresponding absolute uncertainties are
added linearly.}
\label{tab:summary_125p09}
\begin{center}%
\begin{small}%
\tabcolsep4pt
\begin{tabular}{lccccccccccc}%
\toprule
$\sqrt{s}$ [TeV] & \multicolumn{11}{c}{Production cross section (in pb) for $\MH = 125.09$ GeV}
\\
\midrule
& ggF & VBF & $W^+H$ & $W^-H$ & $ZH$ & $t\bar{t}H$ & $tH$ ($t$-ch.) & $tH$ ($s$-ch.) & $tWH$ & $b\bar{b}H$ & total \\
\midrule
$13$ & $47.98^{+4.7\%}_{-7.4\%}$ & $3.811^{+2.5\%}_{-2.4\%}$ & $0.848^{+1.9\%}_{-2.0\%}$ & $0.532^{+2.0\%}_{-2.0\%}$ & $0.886^{+3.4\%}_{-3.1\%}$ & $0.529^{+3.4\%}_{-3.7\%}$ & $0.0760^{+6.7\%}_{-16.1\%}$ & $0.0029^{+3.4\%}_{-3.1\%}$ & $0.0154^{+6.0\%}_{-7.4\%}$ & $0.526^{+9.9\%}_{-10.0\%}$ & $55.21^{+4.5\%}_{-6.9\%}$ \\[0.2cm]
$13.6$ & $51.60^{+4.7\%}_{-7.4\%}$ & $4.109^{+2.5\%}_{-2.4\%}$ & $0.898^{+1.9\%}_{-2.0\%}$ & $0.567^{+1.9\%}_{-2.0\%}$ & $0.947^{+3.4\%}_{-3.1\%}$ & $0.592^{+3.3\%}_{-3.7\%}$ & $0.0853^{+6.7\%}_{-15.6\%}$ & $0.0031^{+3.3\%}_{-2.9\%}$ & $0.0174^{+6.0\%}_{-7.4\%}$ & $0.568^{+9.3\%}_{-9.3\%}$ & $59.39^{+4.5\%}_{-6.9\%}$ \\[0.2cm]
$14$ & $54.03^{+4.7\%}_{-7.4\%}$ & $4.312^{+2.4\%}_{-2.4\%}$ & $0.930^{+1.9\%}_{-2.0\%}$ & $0.591^{+1.9\%}_{-2.0\%}$ & $0.988^{+3.4\%}_{-3.2\%}$ & $0.636^{+3.2\%}_{-3.6\%}$ & $0.0919^{+6.6\%}_{-15.1\%}$ & $0.0033^{+3.3\%}_{-2.9\%}$ & $0.0188^{+6.0\%}_{-7.3\%}$ & $0.596^{+8.9\%}_{-8.9\%}$ & $62.20^{+4.5\%}_{-6.9\%}$ \\
\bottomrule
\end{tabular}%
\end{small}%
\end{center}%
\end{table}
\end{landscape}

\clearpage
\section{Conclusions}
\label{sec:conclusions}
We have presented updated LHC Higgs Working Group recommendations for the
inclusive production cross sections of a Standard Model Higgs boson at
$\sqrt{s}=7,\,8,\,13,\,13.6$ and $14$~TeV, and for Higgs-boson masses between
$120$ and $130$~GeV, covering gluon fusion, vector-boson fusion, associated
production with a vector boson, associated production with a top-quark pair or
a single top quark, and associated production with bottom quarks. Within this
scope the report supersedes Yellow Report~4
(YR4)~\cite{LHCHiggsCrossSectionWorkingGroup:2016ypw} and the ad interim
recommendations of Ref.~\cite{Karlberg:2024zxx}, and constitutes the inclusive
cross-section contribution to Report~5 of the working group.

For every production mode the central prediction remains consistent with YR4
within the uncertainties quoted there, while the precision has improved for
most channels. The improvements rest on specific calculations rather than on a
uniform advance: for gluon fusion, the complete N$^3$LO result in the heavy-top
limit~\cite{Mistlberger:2018etf} together with exact NNLO corrections for the
top-only contribution and the top--bottom
interference~\cite{Czakon:2024ywb}; for vector-boson fusion, the N$^3$LO
structure-function result~\cite{Dreyer:2016oyx} supplemented by explicit
calculations of the non-factorisable corrections; for $VH$, the completion of
the NLO QCD corrections to the loop-induced $gg\to ZH$
channel~\cite{CampilloAveleira:2025rbh}; for $t\bar tH$, NNLO QCD matched to
NNLL soft-gluon resummation and combined with the complete NLO electroweak
corrections~\cite{Balsach:2025jcw}; and for $b\bar bH$, the matched
\nlonnllpart{} prediction~\cite{Bonvini:2016fgf}, which carries an explicit
resummation-scale uncertainty. As a consequence the total theoretical
uncertainty, once parton-distribution uncertainties are set aside, is now
smaller than the precision projected for Higgs boson production
measurements at the HL-LHC in every channel considered except gluon fusion,
where the projected measurement remains the more precise of the two.

Parton distributions are the binding constraint on further improvement. They
are the leading source of uncertainty in the four most precisely predicted
channels -- gluon fusion, vector-boson fusion, $WH$ and $t\bar tH$ -- where
they account for between roughly a third and three quarters of the total; for
$ZH$, $b\bar bH$ and single-top associated production the residual
perturbative uncertainties dominate instead. The baseline of this
report is PDF4LHC21 at NNLO~\cite{PDF4LHCWorkingGroup:2022cjn}, in line with
the working group's practice of following the PDF4LHC recommendations. Since no
such recommendation exists beyond NNLO, the mismatch between an N$^3$LO matrix
element and NNLO distributions cannot be removed and has instead to be carried
as an uncertainty. How best to estimate that uncertainty proved to depend on
the process. The two independent approximate N$^3$LO determinations agree well
for vector-boson fusion, where they differ by $0.3\%$, but not for gluon
fusion, where they differ by around $3\%$ on an effect of comparable
size~\cite{Cridge:2024icl}. For gluon fusion we have accordingly doubled the
PDF-theory uncertainty relative to YR4, and even that does not fully bracket
the shift; for vector-boson fusion the shift is well determined and comparable
to the uncertainty already assigned, and is quantified but not applied to the
central prediction. Establishing a common prescription for parton distributions
at this accuracy is the single development that would most improve the
predictions collected here.

We have also quantified the effect of QED corrections to the DGLAP evolution,
which shifts the gluon-fusion cross section by up to about $1.7\%$ depending on
the parton distributions and the collider energy, and the vector-boson-fusion
cross section by about $0.7\%$~\cite{Cooper-Sarkar:2025sqw}. The treatment of
QED evolution and of photon-initiated production is not uniform across the
production modes considered here, for reasons of timing set out in
Section~\ref{sec:pdf}, and a uniform treatment is desirable in a future update.

The numerical predictions, together with a full breakdown of the associated
theoretical and parametric uncertainties, are tabulated in
Appendix~\ref{app:tables} for all production modes, collider energies and
Higgs-boson masses considered. The numbers are also available at
\url{https://gitlab.cern.ch/LHCHIGGSXS/LHCHXSWG1/crosssections} together with a set of \texttt{bibtex} citation records of all papers that should be cited when the cross section predictions from this report are used.\footnote{The numbers in that repository will be updated whenever new results are available for the various production modes. This includes, but is not limited to, updates when a PDF4LHC set becomes available at (a)N$^3$LO. The numbers presented in this report correspond to \texttt{v1.1.4} of the spreadsheet contained in the repository.} We hope that
this report will serve, as YR4 did before it, as the standard reference for
Higgs-boson production cross sections at the LHC and HL-LHC, and we anticipate
that further updates will be required as exact N$^3$LO ingredients, consistent
N$^3$LO parton distributions and refined uncertainty estimates become
available.

\section*{Acknowledgments}
This work was done on behalf of the LHCHWG.

G.B. acknowledges partial support from the U.S. Department of Energy, Office of
Science, under Award Number DE-SC0010010.
C.~Biello acknowledges support from the Swiss National Science Foundation under
grant 10001706.
J.C. acknowledges support from the Natural Sciences and Engineering Research
Council of Canada (NSERC).
S.F.R. and A.G. acknowledge support from the Italian Ministry of Universities
and Research (MUR) under the FIS grant (CUP: D53C24005480001, FLAME).
R.H. acknowledges support from the Dutch Research Council (NWO) under grant
\href{https://doi.org/10.61686/PGEJL69189}{10.61686/PGEJL69189}.
S.J. acknowledges support from the Royal Society under grant URF{\textbackslash}R1{\textbackslash}201268
and from the UK Science and Technology Facilities Council (STFC) under contracts
ST/X000745/1 and ST/X003167/1.
S.Y.K. receives funding from the Federal Ministry of Education and Research
(BMBF) under grant number 05H21PACCA.
J.M. acknowledges support from the Royal Society under grant
URF{\textbackslash}R1{\textbackslash}201519 and from the STFC under grant ST/W000512/1.
D.M. acknowledges support from the research project PID2023-147706NB-I00
funded by the Spanish MICIU/AEI/10.13039/501100011033 and FEDER/EU.
B.C., G.D., F.E., S.Y.K, A.M., C.T.P., M.V. and M.~Worek acknowledge support from the Deutsche Forschungsgemeinschaft
(DFG) under grant 396021762 -- TRR 257 ``Particle Physics Phenomenology after the
Higgs Discovery''.
D.V. acknowledges support from the Swiss National Science Foundation under
contract number 10003769.
S.Z. acknowledges support from the STFC under grant ST/X000761/1.
M.Z. acknowledges financial support from the Italian Ministry of University and
Research (MUR) through the PRIN2022 grant 2022EZ3S3F, funded by the European
Union -- NextGenerationEU.
L.Z. acknowledges partial support from the U.S. Department of Energy, Office of
Science, under Award Number DE-SC0010072.

A.~Cardini acknowledges funding from the ERC Grant StG -- 101115353: INTREPID,
10.3030/101115353.
S.-O.M. acknowledges funding from the ERC Advanced Grant 101095857:
Conformal-EIC.
L.S. acknowledges support by the ERC under grant 101041109 `BOSON'.
Funded by the European Union. Views and opinions expressed are however those of
the author(s) only and do not necessarily reflect those of the European Union or
the European Research Council Executive Agency. Neither the European Union nor
the granting authority can be held responsible for them.
\newpage
\appendix
\section{Reference tables}
\label{app:tables}
\subsection{$\bold{ggF}$}
\begin{landscape}
\begin{table}[ht!]
\caption{Total ggH cross sections in the SM for a LHC CM energy of $\sqrt{s}=7$\,TeV.}
\label{tab:ggH7}
\centering
\scriptsize
{\renewcommand{\arraystretch}{1.4}
\begin{tabular}{| cc | cccc | ccc | c |@{}c@{\;}| cc |@{}c@{\;}| ccc |}
\cline{1-10} \cline{12-13} \cline{15-17}
\multicolumn{2}{|c|}{\(\sqrt{s} = 7\,\mathrm{TeV}\)} &
\multicolumn{4}{l|}{} &
\multicolumn{4}{c|}{NNLO PDF} &
&
\multicolumn{2}{|c|}{aN${}^\text{3}$LO PDF} && \multicolumn{3}{l|}{} \\
\(\MH\,[\mathrm{GeV}]\) & \(\sigma\,[\mathrm{pb}]\) &
\(\delta(\mathrm{theory})\) &
{\color{black!55}\(\delta(\mathrm{scale})\)} & {\color{black!55}\(\delta(\mathrm{EWK})\)} & {\color{black!55}\(\delta(\mathrm{t,b,c})\)} &
\(\delta(\mathrm{PDF}+\alpha_s)\) &
{\color{black!55}\(\delta(\mathrm{PDF})\)} & {\color{black!55}\(\delta(\alpha_s)\)} &
\(\delta(\mathrm{PDF\text{--}TH})\) &
&
\(\Delta\,[\mathrm{pb}]\) & \(\delta(\mathrm{PDF}+\alpha_s)\) & &\(\delta(\mathrm{Gauss})\) & \(\delta(\mathrm{Total})\) & \shortstack{
  \(\mathrm{Gauss}\) \\
  \(\delta(\mathrm{Total})\)
  }
\\
\cline{1-10} \cline{12-13} \cline{15-17}
120.00 & 18.03 & \({}^{ +1.45}_{ -5.05}\%\) & {\color{black!55}\({}^{ +0.11}_{ -3.71}\%\)} & {\color{black!55}\(\pm{  1.00}\%\)} & {\color{black!55}\(\pm{  0.34}\%\)} & \(\pm 2.57\%\) & {\color{black!55}\({}^{ +1.70}_{ -1.70}\%\)} & {\color{black!55}\(\pm 1.92\%\)} & \(\pm{  1.84}\%\) &  & \(-0.95\) & \({}^{ +1.83}_{ -1.83}\%\) &  & \(\pm 3.98\%\) & \({}^{ +4.18}_{ -7.35}\%\) & \(\pm 4.74\%\)\\
122.00 & 17.43 & \({}^{ +1.45}_{ -5.05}\%\) & {\color{black!55}\({}^{ +0.11}_{ -3.71}\%\)} & {\color{black!55}\(\pm{  1.00}\%\)} & {\color{black!55}\(\pm{  0.34}\%\)} & \(\pm 2.58\%\) & {\color{black!55}\({}^{ +1.70}_{ -1.70}\%\)} & {\color{black!55}\(\pm 1.92\%\)} & \(\pm{  1.82}\%\) &  & \(-0.92\) & \({}^{ +1.82}_{ -1.82}\%\) &  & \(\pm 3.97\%\) & \({}^{ +4.16}_{ -7.34}\%\) & \(\pm 4.73\%\)\\
124.00 & 16.86 & \({}^{ +1.45}_{ -5.04}\%\) & {\color{black!55}\({}^{ +0.11}_{ -3.70}\%\)} & {\color{black!55}\(\pm{  1.00}\%\)} & {\color{black!55}\(\pm{  0.34}\%\)} & \(\pm 2.57\%\) & {\color{black!55}\({}^{ +1.70}_{ -1.70}\%\)} & {\color{black!55}\(\pm 1.92\%\)} & \(\pm{  1.80}\%\) &  & \(-0.89\) & \({}^{ +1.81}_{ -1.81}\%\) &  & \(\pm 3.95\%\) & \({}^{ +4.14}_{ -7.30}\%\) & \(\pm 4.71\%\)\\
124.60 & 16.70 & \({}^{ +1.45}_{ -5.03}\%\) & {\color{black!55}\({}^{ +0.11}_{ -3.69}\%\)} & {\color{black!55}\(\pm{  1.00}\%\)} & {\color{black!55}\(\pm{  0.34}\%\)} & \(\pm 2.57\%\) & {\color{black!55}\({}^{ +1.70}_{ -1.70}\%\)} & {\color{black!55}\(\pm 1.92\%\)} & \(\pm{  1.79}\%\) &  & \(-0.88\) & \({}^{ +1.81}_{ -1.81}\%\) &  & \(\pm 3.94\%\) & \({}^{ +4.14}_{ -7.30}\%\) & \(\pm 4.71\%\)\\
124.80 & 16.64 & \({}^{ +1.45}_{ -5.03}\%\) & {\color{black!55}\({}^{ +0.11}_{ -3.69}\%\)} & {\color{black!55}\(\pm{  1.00}\%\)} & {\color{black!55}\(\pm{  0.34}\%\)} & \(\pm 2.57\%\) & {\color{black!55}\({}^{ +1.70}_{ -1.70}\%\)} & {\color{black!55}\(\pm 1.92\%\)} & \(\pm{  1.79}\%\) &  & \(-0.88\) & \({}^{ +1.81}_{ -1.81}\%\) &  & \(\pm 3.94\%\) & \({}^{ +4.14}_{ -7.29}\%\) & \(\pm 4.71\%\)\\
125.00 & 16.59 & \({}^{ +1.45}_{ -5.03}\%\) & {\color{black!55}\({}^{ +0.11}_{ -3.69}\%\)} & {\color{black!55}\(\pm{  1.00}\%\)} & {\color{black!55}\(\pm{  0.34}\%\)} & \(\pm 2.57\%\) & {\color{black!55}\({}^{ +1.70}_{ -1.70}\%\)} & {\color{black!55}\(\pm 1.92\%\)} & \(\pm{  1.79}\%\) &  & \(-0.88\) & \({}^{ +1.81}_{ -1.81}\%\) &  & \(\pm 3.94\%\) & \({}^{ +4.14}_{ -7.29}\%\) & \(\pm 4.70\%\)\\
125.09 & 16.56 & \({}^{ +1.45}_{ -5.03}\%\) & {\color{black!55}\({}^{ +0.11}_{ -3.69}\%\)} & {\color{black!55}\(\pm{  1.00}\%\)} & {\color{black!55}\(\pm{  0.34}\%\)} & \(\pm 2.57\%\) & {\color{black!55}\({}^{ +1.70}_{ -1.70}\%\)} & {\color{black!55}\(\pm 1.92\%\)} & \(\pm{  1.79}\%\) &  & \(-0.87\) & \({}^{ +1.81}_{ -1.81}\%\) &  & \(\pm 3.94\%\) & \({}^{ +4.14}_{ -7.29}\%\) & \(\pm 4.70\%\)\\
125.20 & 16.53 & \({}^{ +1.45}_{ -5.03}\%\) & {\color{black!55}\({}^{ +0.11}_{ -3.69}\%\)} & {\color{black!55}\(\pm{  1.00}\%\)} & {\color{black!55}\(\pm{  0.34}\%\)} & \(\pm 2.57\%\) & {\color{black!55}\({}^{ +1.70}_{ -1.70}\%\)} & {\color{black!55}\(\pm 1.92\%\)} & \(\pm{  1.79}\%\) &  & \(-0.87\) & \({}^{ +1.81}_{ -1.81}\%\) &  & \(\pm 3.94\%\) & \({}^{ +4.13}_{ -7.29}\%\) & \(\pm 4.70\%\)\\
125.30 & 16.51 & \({}^{ +1.45}_{ -5.03}\%\) & {\color{black!55}\({}^{ +0.11}_{ -3.69}\%\)} & {\color{black!55}\(\pm{  1.00}\%\)} & {\color{black!55}\(\pm{  0.34}\%\)} & \(\pm 2.57\%\) & {\color{black!55}\({}^{ +1.70}_{ -1.70}\%\)} & {\color{black!55}\(\pm 1.92\%\)} & \(\pm{  1.79}\%\) &  & \(-0.87\) & \({}^{ +1.81}_{ -1.81}\%\) &  & \(\pm 3.94\%\) & \({}^{ +4.13}_{ -7.29}\%\) & \(\pm 4.70\%\)\\
125.38 & 16.48 & \({}^{ +1.45}_{ -5.03}\%\) & {\color{black!55}\({}^{ +0.11}_{ -3.69}\%\)} & {\color{black!55}\(\pm{  1.00}\%\)} & {\color{black!55}\(\pm{  0.34}\%\)} & \(\pm 2.57\%\) & {\color{black!55}\({}^{ +1.70}_{ -1.70}\%\)} & {\color{black!55}\(\pm 1.92\%\)} & \(\pm{  1.78}\%\) &  & \(-0.87\) & \({}^{ +1.81}_{ -1.81}\%\) &  & \(\pm 3.94\%\) & \({}^{ +4.13}_{ -7.29}\%\) & \(\pm 4.70\%\)\\
125.60 & 16.43 & \({}^{ +1.45}_{ -5.03}\%\) & {\color{black!55}\({}^{ +0.11}_{ -3.69}\%\)} & {\color{black!55}\(\pm{  1.00}\%\)} & {\color{black!55}\(\pm{  0.34}\%\)} & \(\pm 2.57\%\) & {\color{black!55}\({}^{ +1.70}_{ -1.70}\%\)} & {\color{black!55}\(\pm 1.92\%\)} & \(\pm{  1.78}\%\) &  & \(-0.87\) & \({}^{ +1.81}_{ -1.81}\%\) &  & \(\pm 3.93\%\) & \({}^{ +4.13}_{ -7.28}\%\) & \(\pm 4.70\%\)\\
126.00 & 16.32 & \({}^{ +1.45}_{ -5.03}\%\) & {\color{black!55}\({}^{ +0.11}_{ -3.69}\%\)} & {\color{black!55}\(\pm{  1.00}\%\)} & {\color{black!55}\(\pm{  0.34}\%\)} & \(\pm 2.57\%\) & {\color{black!55}\({}^{ +1.70}_{ -1.70}\%\)} & {\color{black!55}\(\pm 1.92\%\)} & \(\pm{  1.78}\%\) &  & \(-0.86\) & \({}^{ +1.81}_{ -1.81}\%\) &  & \(\pm 3.93\%\) & \({}^{ +4.13}_{ -7.28}\%\) & \(\pm 4.70\%\)\\
128.00 & 15.80 & \({}^{ +1.45}_{ -5.03}\%\) & {\color{black!55}\({}^{ +0.11}_{ -3.69}\%\)} & {\color{black!55}\(\pm{  1.00}\%\)} & {\color{black!55}\(\pm{  0.34}\%\)} & \(\pm 2.57\%\) & {\color{black!55}\({}^{ +1.70}_{ -1.70}\%\)} & {\color{black!55}\(\pm 1.92\%\)} & \(\pm{  1.76}\%\) &  & \(-0.83\) & \({}^{ +1.80}_{ -1.80}\%\) &  & \(\pm 3.92\%\) & \({}^{ +4.12}_{ -7.26}\%\) & \(\pm 4.69\%\)\\
130.00 & 15.30 & \({}^{ +1.46}_{ -5.02}\%\) & {\color{black!55}\({}^{ +0.12}_{ -3.68}\%\)} & {\color{black!55}\(\pm{  1.00}\%\)} & {\color{black!55}\(\pm{  0.34}\%\)} & \(\pm 2.57\%\) & {\color{black!55}\({}^{ +1.70}_{ -1.70}\%\)} & {\color{black!55}\(\pm 1.92\%\)} & \(\pm{  1.74}\%\) &  & \(-0.81\) & \({}^{ +1.79}_{ -1.79}\%\) &  & \(\pm 3.90\%\) & \({}^{ +4.10}_{ -7.23}\%\) & \(\pm 4.67\%\)\\
\cline{1-10} \cline{12-13} \cline{15-17}
\end{tabular}
}
\end{table}

\begin{table}[ht!]
\caption{Total ggH cross sections in the SM for a LHC CM energy of $\sqrt{s}=8$\,TeV.}
\label{tab:ggH8}
\centering
\scriptsize
{\renewcommand{\arraystretch}{1.4}
\begin{tabular}{| cc | cccc | ccc | c |@{}c@{\;}| cc |@{}c@{\;}| ccc |}
\cline{1-10} \cline{12-13} \cline{15-17}
\multicolumn{2}{|c|}{\(\sqrt{s} = 8\,\mathrm{TeV}\)} &
\multicolumn{4}{l|}{} &
\multicolumn{4}{c|}{NNLO PDF} &
&
\multicolumn{2}{|c|}{aN${}^\text{3}$LO PDF} && \multicolumn{3}{l|}{} \\
\(\MH\,[\mathrm{GeV}]\) & \(\sigma\,[\mathrm{pb}]\) &
\(\delta(\mathrm{theory})\) &
{\color{black!55}\(\delta(\mathrm{scale})\)} & {\color{black!55}\(\delta(\mathrm{EWK})\)} & {\color{black!55}\(\delta(\mathrm{t,b,c})\)} &
\(\delta(\mathrm{PDF}+\alpha_s)\) &
{\color{black!55}\(\delta(\mathrm{PDF})\)} & {\color{black!55}\(\delta(\alpha_s)\)} &
\(\delta(\mathrm{PDF\text{--}TH})\) &
&
\(\Delta\,[\mathrm{pb}]\) & \(\delta(\mathrm{PDF}+\alpha_s)\) & &\(\delta(\mathrm{Gauss})\) & \(\delta(\mathrm{Total})\) & \shortstack{
  \(\mathrm{Gauss}\) \\
  \(\delta(\mathrm{Total})\)
    }
\\
\cline{1-10} \cline{12-13} \cline{15-17}
120.00 & 22.88 & \({}^{ +1.48}_{ -4.97}\%\) & {\color{black!55}\({}^{ +0.14}_{ -3.63}\%\)} & {\color{black!55}\(\pm{  1.00}\%\)} & {\color{black!55}\(\pm{  0.34}\%\)} & \(\pm 2.55\%\) & {\color{black!55}\({}^{ +1.69}_{ -1.69}\%\)} & {\color{black!55}\(\pm 1.90\%\)} & \(\pm{  1.95}\%\) &  & \(-1.18\) & \({}^{ +1.83}_{ -1.83}\%\) &  & \(\pm 3.99\%\) & \({}^{ +4.27}_{ -7.37}\%\) & \(\pm 4.74\%\)\\
122.00 & 22.15 & \({}^{ +1.49}_{ -4.96}\%\) & {\color{black!55}\({}^{ +0.15}_{ -3.62}\%\)} & {\color{black!55}\(\pm{  1.00}\%\)} & {\color{black!55}\(\pm{  0.34}\%\)} & \(\pm 2.55\%\) & {\color{black!55}\({}^{ +1.68}_{ -1.68}\%\)} & {\color{black!55}\(\pm 1.90\%\)} & \(\pm{  1.93}\%\) &  & \(-1.14\) & \({}^{ +1.83}_{ -1.83}\%\) &  & \(\pm 3.98\%\) & \({}^{ +4.26}_{ -7.34}\%\) & \(\pm 4.72\%\)\\
124.00 & 21.45 & \({}^{ +1.49}_{ -4.95}\%\) & {\color{black!55}\({}^{ +0.15}_{ -3.61}\%\)} & {\color{black!55}\(\pm{  1.00}\%\)} & {\color{black!55}\(\pm{  0.34}\%\)} & \(\pm 2.54\%\) & {\color{black!55}\({}^{ +1.68}_{ -1.68}\%\)} & {\color{black!55}\(\pm 1.90\%\)} & \(\pm{  1.91}\%\) &  & \(-1.11\) & \({}^{ +1.82}_{ -1.82}\%\) &  & \(\pm 3.96\%\) & \({}^{ +4.24}_{ -7.31}\%\) & \(\pm 4.71\%\)\\
124.60 & 21.25 & \({}^{ +1.49}_{ -4.95}\%\) & {\color{black!55}\({}^{ +0.15}_{ -3.61}\%\)} & {\color{black!55}\(\pm{  1.00}\%\)} & {\color{black!55}\(\pm{  0.34}\%\)} & \(\pm 2.54\%\) & {\color{black!55}\({}^{ +1.68}_{ -1.68}\%\)} & {\color{black!55}\(\pm 1.90\%\)} & \(\pm{  1.90}\%\) &  & \(-1.09\) & \({}^{ +1.82}_{ -1.82}\%\) &  & \(\pm 3.95\%\) & \({}^{ +4.23}_{ -7.31}\%\) & \(\pm 4.70\%\)\\
124.80 & 21.18 & \({}^{ +1.49}_{ -4.95}\%\) & {\color{black!55}\({}^{ +0.15}_{ -3.61}\%\)} & {\color{black!55}\(\pm{  1.00}\%\)} & {\color{black!55}\(\pm{  0.34}\%\)} & \(\pm 2.54\%\) & {\color{black!55}\({}^{ +1.68}_{ -1.68}\%\)} & {\color{black!55}\(\pm 1.89\%\)} & \(\pm{  1.90}\%\) &  & \(-1.09\) & \({}^{ +1.82}_{ -1.82}\%\) &  & \(\pm 3.95\%\) & \({}^{ +4.23}_{ -7.30}\%\) & \(\pm 4.70\%\)\\
125.00 & 21.11 & \({}^{ +1.49}_{ -4.95}\%\) & {\color{black!55}\({}^{ +0.15}_{ -3.61}\%\)} & {\color{black!55}\(\pm{  1.00}\%\)} & {\color{black!55}\(\pm{  0.34}\%\)} & \(\pm 2.54\%\) & {\color{black!55}\({}^{ +1.68}_{ -1.68}\%\)} & {\color{black!55}\(\pm 1.89\%\)} & \(\pm{  1.89}\%\) &  & \(-1.09\) & \({}^{ +1.82}_{ -1.82}\%\) &  & \(\pm 3.95\%\) & \({}^{ +4.23}_{ -7.30}\%\) & \(\pm 4.70\%\)\\
125.09 & 21.08 & \({}^{ +1.49}_{ -4.95}\%\) & {\color{black!55}\({}^{ +0.15}_{ -3.61}\%\)} & {\color{black!55}\(\pm{  1.00}\%\)} & {\color{black!55}\(\pm{  0.34}\%\)} & \(\pm 2.54\%\) & {\color{black!55}\({}^{ +1.68}_{ -1.68}\%\)} & {\color{black!55}\(\pm 1.89\%\)} & \(\pm{  1.89}\%\) &  & \(-1.09\) & \({}^{ +1.82}_{ -1.82}\%\) &  & \(\pm 3.95\%\) & \({}^{ +4.23}_{ -7.30}\%\) & \(\pm 4.70\%\)\\
125.20 & 21.04 & \({}^{ +1.49}_{ -4.95}\%\) & {\color{black!55}\({}^{ +0.15}_{ -3.61}\%\)} & {\color{black!55}\(\pm{  1.00}\%\)} & {\color{black!55}\(\pm{  0.34}\%\)} & \(\pm 2.54\%\) & {\color{black!55}\({}^{ +1.68}_{ -1.68}\%\)} & {\color{black!55}\(\pm 1.89\%\)} & \(\pm{  1.89}\%\) &  & \(-1.08\) & \({}^{ +1.82}_{ -1.82}\%\) &  & \(\pm 3.95\%\) & \({}^{ +4.23}_{ -7.30}\%\) & \(\pm 4.70\%\)\\
125.30 & 21.01 & \({}^{ +1.49}_{ -4.95}\%\) & {\color{black!55}\({}^{ +0.15}_{ -3.61}\%\)} & {\color{black!55}\(\pm{  1.00}\%\)} & {\color{black!55}\(\pm{  0.34}\%\)} & \(\pm 2.54\%\) & {\color{black!55}\({}^{ +1.68}_{ -1.68}\%\)} & {\color{black!55}\(\pm 1.89\%\)} & \(\pm{  1.89}\%\) &  & \(-1.08\) & \({}^{ +1.82}_{ -1.82}\%\) &  & \(\pm 3.95\%\) & \({}^{ +4.23}_{ -7.29}\%\) & \(\pm 4.70\%\)\\
125.38 & 20.98 & \({}^{ +1.49}_{ -4.95}\%\) & {\color{black!55}\({}^{ +0.15}_{ -3.61}\%\)} & {\color{black!55}\(\pm{  1.00}\%\)} & {\color{black!55}\(\pm{  0.34}\%\)} & \(\pm 2.54\%\) & {\color{black!55}\({}^{ +1.68}_{ -1.68}\%\)} & {\color{black!55}\(\pm 1.89\%\)} & \(\pm{  1.89}\%\) &  & \(-1.08\) & \({}^{ +1.82}_{ -1.82}\%\) &  & \(\pm 3.95\%\) & \({}^{ +4.23}_{ -7.29}\%\) & \(\pm 4.69\%\)\\
125.60 & 20.91 & \({}^{ +1.49}_{ -4.95}\%\) & {\color{black!55}\({}^{ +0.15}_{ -3.61}\%\)} & {\color{black!55}\(\pm{  1.00}\%\)} & {\color{black!55}\(\pm{  0.34}\%\)} & \(\pm 2.54\%\) & {\color{black!55}\({}^{ +1.68}_{ -1.68}\%\)} & {\color{black!55}\(\pm 1.89\%\)} & \(\pm{  1.89}\%\) &  & \(-1.08\) & \({}^{ +1.82}_{ -1.82}\%\) &  & \(\pm 3.95\%\) & \({}^{ +4.22}_{ -7.29}\%\) & \(\pm 4.69\%\)\\
126.00 & 20.78 & \({}^{ +1.49}_{ -4.95}\%\) & {\color{black!55}\({}^{ +0.15}_{ -3.61}\%\)} & {\color{black!55}\(\pm{  1.00}\%\)} & {\color{black!55}\(\pm{  0.34}\%\)} & \(\pm 2.54\%\) & {\color{black!55}\({}^{ +1.68}_{ -1.68}\%\)} & {\color{black!55}\(\pm 1.89\%\)} & \(\pm{  1.88}\%\) &  & \(-1.07\) & \({}^{ +1.82}_{ -1.82}\%\) &  & \(\pm 3.94\%\) & \({}^{ +4.22}_{ -7.29}\%\) & \(\pm 4.69\%\)\\
128.00 & 20.14 & \({}^{ +1.49}_{ -4.94}\%\) & {\color{black!55}\({}^{ +0.15}_{ -3.60}\%\)} & {\color{black!55}\(\pm{  1.00}\%\)} & {\color{black!55}\(\pm{  0.34}\%\)} & \(\pm 2.54\%\) & {\color{black!55}\({}^{ +1.68}_{ -1.68}\%\)} & {\color{black!55}\(\pm 1.89\%\)} & \(\pm{  1.86}\%\) &  & \(-1.04\) & \({}^{ +1.81}_{ -1.81}\%\) &  & \(\pm 3.93\%\) & \({}^{ +4.20}_{ -7.26}\%\) & \(\pm 4.68\%\)\\
130.00 & 19.53 & \({}^{ +1.48}_{ -4.94}\%\) & {\color{black!55}\({}^{ +0.14}_{ -3.60}\%\)} & {\color{black!55}\(\pm{  1.00}\%\)} & {\color{black!55}\(\pm{  0.34}\%\)} & \(\pm 2.54\%\) & {\color{black!55}\({}^{ +1.68}_{ -1.68}\%\)} & {\color{black!55}\(\pm 1.89\%\)} & \(\pm{  1.84}\%\) &  & \(-1.01\) & \({}^{ +1.81}_{ -1.81}\%\) &  & \(\pm 3.91\%\) & \({}^{ +4.18}_{ -7.24}\%\) & \(\pm 4.66\%\)\\
\cline{1-10} \cline{12-13} \cline{15-17}
\end{tabular}
}
\end{table}

\begin{table}[ht!]
\caption{Total ggH cross sections in the SM for a LHC CM energy of $\sqrt{s}=13$\,TeV.}
\label{tab:ggH13}
\centering
\scriptsize
{\renewcommand{\arraystretch}{1.4}
\begin{tabular}{| cc | cccc | ccc | c |@{}c@{\;}| cc |@{}c@{\;}| ccc |}
\cline{1-10} \cline{12-13} \cline{15-17}
\multicolumn{2}{|c|}{\(\sqrt{s} = 13\,\mathrm{TeV}\)} &
\multicolumn{4}{l|}{} &
\multicolumn{4}{c|}{NNLO PDF} &
&
\multicolumn{2}{|c|}{aN${}^\text{3}$LO PDF} && \multicolumn{3}{l|}{} \\
\(\MH\,[\mathrm{GeV}]\) & \(\sigma\,[\mathrm{pb}]\) &
\(\delta(\mathrm{theory})\) &
{\color{black!55}\(\delta(\mathrm{scale})\)} & {\color{black!55}\(\delta(\mathrm{EWK})\)} & {\color{black!55}\(\delta(\mathrm{t,b,c})\)} &
\(\delta(\mathrm{PDF}+\alpha_s)\) &
{\color{black!55}\(\delta(\mathrm{PDF})\)} & {\color{black!55}\(\delta(\alpha_s)\)} &
\(\delta(\mathrm{PDF\text{--}TH})\) &
&
\(\Delta\,[\mathrm{pb}]\) & \(\delta(\mathrm{PDF}+\alpha_s)\) & &\(\delta(\mathrm{Gauss})\) & \(\delta(\mathrm{Total})\) & \shortstack{
  \(\mathrm{Gauss}\) \\
  \(\delta(\mathrm{Total})\)
    }
\\
\cline{1-10} \cline{12-13} \cline{15-17}
120.00 & 51.61 & \({}^{ +1.64}_{ -4.69}\%\) & {\color{black!55}\({}^{ +0.30}_{ -3.35}\%\)} & {\color{black!55}\(\pm{  1.00}\%\)} & {\color{black!55}\(\pm{  0.34}\%\)} & \(\pm 2.48\%\) & {\color{black!55}\({}^{ +1.65}_{ -1.65}\%\)} & {\color{black!55}\(\pm 1.84\%\)} & \(\pm{  2.38}\%\) &  & \(-2.36\) & \({}^{ +1.75}_{ -1.75}\%\) &  & \(\pm 4.08\%\) & \({}^{ +4.72}_{ -7.49}\%\) & \(\pm 4.78\%\)\\
122.00 & 50.14 & \({}^{ +1.63}_{ -4.68}\%\) & {\color{black!55}\({}^{ +0.29}_{ -3.34}\%\)} & {\color{black!55}\(\pm{  1.00}\%\)} & {\color{black!55}\(\pm{  0.34}\%\)} & \(\pm 2.48\%\) & {\color{black!55}\({}^{ +1.65}_{ -1.65}\%\)} & {\color{black!55}\(\pm 1.84\%\)} & \(\pm{  2.35}\%\) &  & \(-2.29\) & \({}^{ +1.75}_{ -1.75}\%\) &  & \(\pm 4.06\%\) & \({}^{ +4.69}_{ -7.46}\%\) & \(\pm 4.76\%\)\\
124.00 & 48.73 & \({}^{ +1.63}_{ -4.67}\%\) & {\color{black!55}\({}^{ +0.29}_{ -3.33}\%\)} & {\color{black!55}\(\pm{  1.00}\%\)} & {\color{black!55}\(\pm{  0.34}\%\)} & \(\pm 2.47\%\) & {\color{black!55}\({}^{ +1.65}_{ -1.65}\%\)} & {\color{black!55}\(\pm 1.84\%\)} & \(\pm{  2.33}\%\) &  & \(-2.23\) & \({}^{ +1.75}_{ -1.75}\%\) &  & \(\pm 4.04\%\) & \({}^{ +4.67}_{ -7.42}\%\) & \(\pm 4.74\%\)\\
124.60 & 48.31 & \({}^{ +1.63}_{ -4.67}\%\) & {\color{black!55}\({}^{ +0.29}_{ -3.33}\%\)} & {\color{black!55}\(\pm{  1.00}\%\)} & {\color{black!55}\(\pm{  0.34}\%\)} & \(\pm 2.47\%\) & {\color{black!55}\({}^{ +1.65}_{ -1.65}\%\)} & {\color{black!55}\(\pm 1.84\%\)} & \(\pm{  2.32}\%\) &  & \(-2.22\) & \({}^{ +1.75}_{ -1.75}\%\) &  & \(\pm 4.03\%\) & \({}^{ +4.66}_{ -7.41}\%\) & \(\pm 4.73\%\)\\
124.80 & 48.17 & \({}^{ +1.63}_{ -4.67}\%\) & {\color{black!55}\({}^{ +0.29}_{ -3.33}\%\)} & {\color{black!55}\(\pm{  1.00}\%\)} & {\color{black!55}\(\pm{  0.34}\%\)} & \(\pm 2.47\%\) & {\color{black!55}\({}^{ +1.65}_{ -1.65}\%\)} & {\color{black!55}\(\pm 1.83\%\)} & \(\pm{  2.32}\%\) &  & \(-2.21\) & \({}^{ +1.75}_{ -1.75}\%\) &  & \(\pm 4.03\%\) & \({}^{ +4.66}_{ -7.41}\%\) & \(\pm 4.73\%\)\\
125.00 & 48.04 & \({}^{ +1.63}_{ -4.67}\%\) & {\color{black!55}\({}^{ +0.29}_{ -3.33}\%\)} & {\color{black!55}\(\pm{  1.00}\%\)} & {\color{black!55}\(\pm{  0.34}\%\)} & \(\pm 2.47\%\) & {\color{black!55}\({}^{ +1.65}_{ -1.65}\%\)} & {\color{black!55}\(\pm 1.83\%\)} & \(\pm{  2.31}\%\) &  & \(-2.20\) & \({}^{ +1.75}_{ -1.75}\%\) &  & \(\pm 4.03\%\) & \({}^{ +4.65}_{ -7.40}\%\) & \(\pm 4.73\%\)\\
125.09 & 47.98 & \({}^{ +1.63}_{ -4.67}\%\) & {\color{black!55}\({}^{ +0.29}_{ -3.33}\%\)} & {\color{black!55}\(\pm{  1.00}\%\)} & {\color{black!55}\(\pm{  0.34}\%\)} & \(\pm 2.47\%\) & {\color{black!55}\({}^{ +1.65}_{ -1.65}\%\)} & {\color{black!55}\(\pm 1.83\%\)} & \(\pm{  2.31}\%\) &  & \(-2.20\) & \({}^{ +1.75}_{ -1.75}\%\) &  & \(\pm 4.03\%\) & \({}^{ +4.65}_{ -7.40}\%\) & \(\pm 4.73\%\)\\
125.20 & 47.90 & \({}^{ +1.63}_{ -4.67}\%\) & {\color{black!55}\({}^{ +0.29}_{ -3.33}\%\)} & {\color{black!55}\(\pm{  1.00}\%\)} & {\color{black!55}\(\pm{  0.34}\%\)} & \(\pm 2.47\%\) & {\color{black!55}\({}^{ +1.65}_{ -1.65}\%\)} & {\color{black!55}\(\pm 1.83\%\)} & \(\pm{  2.31}\%\) &  & \(-2.20\) & \({}^{ +1.75}_{ -1.75}\%\) &  & \(\pm 4.03\%\) & \({}^{ +4.65}_{ -7.40}\%\) & \(\pm 4.73\%\)\\
125.30 & 47.83 & \({}^{ +1.63}_{ -4.66}\%\) & {\color{black!55}\({}^{ +0.29}_{ -3.32}\%\)} & {\color{black!55}\(\pm{  1.00}\%\)} & {\color{black!55}\(\pm{  0.34}\%\)} & \(\pm 2.47\%\) & {\color{black!55}\({}^{ +1.65}_{ -1.65}\%\)} & {\color{black!55}\(\pm 1.83\%\)} & \(\pm{  2.31}\%\) &  & \(-2.20\) & \({}^{ +1.75}_{ -1.75}\%\) &  & \(\pm 4.03\%\) & \({}^{ +4.65}_{ -7.40}\%\) & \(\pm 4.72\%\)\\
125.38 & 47.78 & \({}^{ +1.63}_{ -4.67}\%\) & {\color{black!55}\({}^{ +0.29}_{ -3.33}\%\)} & {\color{black!55}\(\pm{  1.00}\%\)} & {\color{black!55}\(\pm{  0.34}\%\)} & \(\pm 2.47\%\) & {\color{black!55}\({}^{ +1.65}_{ -1.65}\%\)} & {\color{black!55}\(\pm 1.83\%\)} & \(\pm{  2.31}\%\) &  & \(-2.19\) & \({}^{ +1.75}_{ -1.75}\%\) &  & \(\pm 4.03\%\) & \({}^{ +4.65}_{ -7.40}\%\) & \(\pm 4.72\%\)\\
125.60 & 47.63 & \({}^{ +1.63}_{ -4.66}\%\) & {\color{black!55}\({}^{ +0.29}_{ -3.32}\%\)} & {\color{black!55}\(\pm{  1.00}\%\)} & {\color{black!55}\(\pm{  0.34}\%\)} & \(\pm 2.47\%\) & {\color{black!55}\({}^{ +1.65}_{ -1.65}\%\)} & {\color{black!55}\(\pm 1.83\%\)} & \(\pm{  2.31}\%\) &  & \(-2.19\) & \({}^{ +1.75}_{ -1.75}\%\) &  & \(\pm 4.03\%\) & \({}^{ +4.65}_{ -7.40}\%\) & \(\pm 4.72\%\)\\
126.00 & 47.36 & \({}^{ +1.63}_{ -4.66}\%\) & {\color{black!55}\({}^{ +0.29}_{ -3.32}\%\)} & {\color{black!55}\(\pm{  1.00}\%\)} & {\color{black!55}\(\pm{  0.34}\%\)} & \(\pm 2.47\%\) & {\color{black!55}\({}^{ +1.65}_{ -1.65}\%\)} & {\color{black!55}\(\pm 1.83\%\)} & \(\pm{  2.30}\%\) &  & \(-2.18\) & \({}^{ +1.75}_{ -1.75}\%\) &  & \(\pm 4.02\%\) & \({}^{ +4.64}_{ -7.39}\%\) & \(\pm 4.72\%\)\\
128.00 & 46.06 & \({}^{ +1.62}_{ -4.65}\%\) & {\color{black!55}\({}^{ +0.28}_{ -3.31}\%\)} & {\color{black!55}\(\pm{  1.00}\%\)} & {\color{black!55}\(\pm{  0.34}\%\)} & \(\pm 2.47\%\) & {\color{black!55}\({}^{ +1.64}_{ -1.64}\%\)} & {\color{black!55}\(\pm 1.83\%\)} & \(\pm{  2.28}\%\) &  & \(-2.12\) & \({}^{ +1.75}_{ -1.75}\%\) &  & \(\pm 4.00\%\) & \({}^{ +4.61}_{ -7.36}\%\) & \(\pm 4.70\%\)\\
130.00 & 44.81 & \({}^{ +1.62}_{ -4.65}\%\) & {\color{black!55}\({}^{ +0.28}_{ -3.31}\%\)} & {\color{black!55}\(\pm{  1.00}\%\)} & {\color{black!55}\(\pm{  0.34}\%\)} & \(\pm 2.46\%\) & {\color{black!55}\({}^{ +1.64}_{ -1.64}\%\)} & {\color{black!55}\(\pm 1.82\%\)} & \(\pm{  2.25}\%\) &  & \(-2.07\) & \({}^{ +1.76}_{ -1.76}\%\) &  & \(\pm 3.99\%\) & \({}^{ +4.59}_{ -7.33}\%\) & \(\pm 4.68\%\)\\
\cline{1-10} \cline{12-13} \cline{15-17}
\end{tabular}
}
\end{table}

\begin{table}[ht!]
\caption{Total ggH cross sections in the SM for a LHC CM energy of $\sqrt{s}=13.6$\,TeV.}
\label{tab:ggH136}
\centering
\scriptsize
{\renewcommand{\arraystretch}{1.4}
\begin{tabular}{| cc | cccc | ccc | c |@{}c@{\;}| cc |@{}c@{\;}| ccc |}
\cline{1-10} \cline{12-13} \cline{15-17}
\multicolumn{2}{|c|}{\(\sqrt{s} = 13.6\,\mathrm{TeV}\)} &
\multicolumn{4}{l|}{} &
\multicolumn{4}{c|}{NNLO PDF} &
&
\multicolumn{2}{|c|}{aN${}^\text{3}$LO PDF} && \multicolumn{3}{l|}{} \\
\(\MH\,[\mathrm{GeV}]\) & \(\sigma\,[\mathrm{pb}]\) &
\(\delta(\mathrm{theory})\) &
{\color{black!55}\(\delta(\mathrm{scale})\)} & {\color{black!55}\(\delta(\mathrm{EWK})\)} & {\color{black!55}\(\delta(\mathrm{t,b,c})\)} &
\(\delta(\mathrm{PDF}+\alpha_s)\) &
{\color{black!55}\(\delta(\mathrm{PDF})\)} & {\color{black!55}\(\delta(\alpha_s)\)} &
\(\delta(\mathrm{PDF\text{--}TH})\) &
&
\(\Delta\,[\mathrm{pb}]\) & \(\delta(\mathrm{PDF}+\alpha_s)\) & &\(\delta(\mathrm{Gauss})\) & \(\delta(\mathrm{Total})\) & \shortstack{
  \(\mathrm{Gauss}\) \\
  \(\delta(\mathrm{Total})\)
    }
\\
\cline{1-10} \cline{12-13} \cline{15-17}
120.00 & 55.48 & \({}^{ +1.65}_{ -4.66}\%\) & {\color{black!55}\({}^{ +0.31}_{ -3.32}\%\)} & {\color{black!55}\(\pm{  1.00}\%\)} & {\color{black!55}\(\pm{  0.34}\%\)} & \(\pm 2.48\%\) & {\color{black!55}\({}^{ +1.65}_{ -1.65}\%\)} & {\color{black!55}\(\pm 1.84\%\)} & \(\pm{  2.42}\%\) &  & \(-2.50\) & \({}^{ +1.73}_{ -1.73}\%\) &  & \(\pm 4.09\%\) & \({}^{ +4.77}_{ -7.51}\%\) & \(\pm 4.78\%\)\\
122.00 & 53.91 & \({}^{ +1.66}_{ -4.65}\%\) & {\color{black!55}\({}^{ +0.32}_{ -3.31}\%\)} & {\color{black!55}\(\pm{  1.00}\%\)} & {\color{black!55}\(\pm{  0.34}\%\)} & \(\pm 2.47\%\) & {\color{black!55}\({}^{ +1.65}_{ -1.65}\%\)} & {\color{black!55}\(\pm 1.84\%\)} & \(\pm{  2.40}\%\) &  & \(-2.43\) & \({}^{ +1.73}_{ -1.73}\%\) &  & \(\pm 4.07\%\) & \({}^{ +4.75}_{ -7.47}\%\) & \(\pm 4.76\%\)\\
124.00 & 52.40 & \({}^{ +1.65}_{ -4.65}\%\) & {\color{black!55}\({}^{ +0.31}_{ -3.31}\%\)} & {\color{black!55}\(\pm{  1.00}\%\)} & {\color{black!55}\(\pm{  0.34}\%\)} & \(\pm 2.47\%\) & {\color{black!55}\({}^{ +1.64}_{ -1.64}\%\)} & {\color{black!55}\(\pm 1.83\%\)} & \(\pm{  2.37}\%\) &  & \(-2.37\) & \({}^{ +1.74}_{ -1.74}\%\) &  & \(\pm 4.05\%\) & \({}^{ +4.72}_{ -7.44}\%\) & \(\pm 4.75\%\)\\
124.60 & 51.95 & \({}^{ +1.65}_{ -4.65}\%\) & {\color{black!55}\({}^{ +0.31}_{ -3.31}\%\)} & {\color{black!55}\(\pm{  1.00}\%\)} & {\color{black!55}\(\pm{  0.34}\%\)} & \(\pm 2.47\%\) & {\color{black!55}\({}^{ +1.64}_{ -1.64}\%\)} & {\color{black!55}\(\pm 1.83\%\)} & \(\pm{  2.36}\%\) &  & \(-2.35\) & \({}^{ +1.74}_{ -1.74}\%\) &  & \(\pm 4.05\%\) & \({}^{ +4.71}_{ -7.43}\%\) & \(\pm 4.74\%\)\\
124.80 & 51.81 & \({}^{ +1.65}_{ -4.65}\%\) & {\color{black!55}\({}^{ +0.31}_{ -3.31}\%\)} & {\color{black!55}\(\pm{  1.00}\%\)} & {\color{black!55}\(\pm{  0.34}\%\)} & \(\pm 2.47\%\) & {\color{black!55}\({}^{ +1.64}_{ -1.64}\%\)} & {\color{black!55}\(\pm 1.83\%\)} & \(\pm{  2.36}\%\) &  & \(-2.35\) & \({}^{ +1.74}_{ -1.74}\%\) &  & \(\pm 4.05\%\) & \({}^{ +4.71}_{ -7.43}\%\) & \(\pm 4.74\%\)\\
125.00 & 51.66 & \({}^{ +1.65}_{ -4.65}\%\) & {\color{black!55}\({}^{ +0.31}_{ -3.31}\%\)} & {\color{black!55}\(\pm{  1.00}\%\)} & {\color{black!55}\(\pm{  0.34}\%\)} & \(\pm 2.47\%\) & {\color{black!55}\({}^{ +1.64}_{ -1.64}\%\)} & {\color{black!55}\(\pm 1.83\%\)} & \(\pm{  2.36}\%\) &  & \(-2.34\) & \({}^{ +1.74}_{ -1.74}\%\) &  & \(\pm 4.04\%\) & \({}^{ +4.70}_{ -7.42}\%\) & \(\pm 4.74\%\)\\
125.09 & 51.60 & \({}^{ +1.65}_{ -4.65}\%\) & {\color{black!55}\({}^{ +0.31}_{ -3.31}\%\)} & {\color{black!55}\(\pm{  1.00}\%\)} & {\color{black!55}\(\pm{  0.34}\%\)} & \(\pm 2.47\%\) & {\color{black!55}\({}^{ +1.64}_{ -1.64}\%\)} & {\color{black!55}\(\pm 1.83\%\)} & \(\pm{  2.36}\%\) &  & \(-2.34\) & \({}^{ +1.74}_{ -1.74}\%\) &  & \(\pm 4.04\%\) & \({}^{ +4.70}_{ -7.42}\%\) & \(\pm 4.74\%\)\\
125.20 & 51.52 & \({}^{ +1.65}_{ -4.65}\%\) & {\color{black!55}\({}^{ +0.31}_{ -3.31}\%\)} & {\color{black!55}\(\pm{  1.00}\%\)} & {\color{black!55}\(\pm{  0.34}\%\)} & \(\pm 2.47\%\) & {\color{black!55}\({}^{ +1.64}_{ -1.64}\%\)} & {\color{black!55}\(\pm 1.83\%\)} & \(\pm{  2.35}\%\) &  & \(-2.33\) & \({}^{ +1.74}_{ -1.74}\%\) &  & \(\pm 4.04\%\) & \({}^{ +4.70}_{ -7.42}\%\) & \(\pm 4.74\%\)\\
125.30 & 51.45 & \({}^{ +1.65}_{ -4.64}\%\) & {\color{black!55}\({}^{ +0.31}_{ -3.30}\%\)} & {\color{black!55}\(\pm{  1.00}\%\)} & {\color{black!55}\(\pm{  0.34}\%\)} & \(\pm 2.47\%\) & {\color{black!55}\({}^{ +1.64}_{ -1.64}\%\)} & {\color{black!55}\(\pm 1.83\%\)} & \(\pm{  2.35}\%\) &  & \(-2.33\) & \({}^{ +1.74}_{ -1.74}\%\) &  & \(\pm 4.04\%\) & \({}^{ +4.70}_{ -7.42}\%\) & \(\pm 4.73\%\)\\
125.38 & 51.39 & \({}^{ +1.65}_{ -4.64}\%\) & {\color{black!55}\({}^{ +0.31}_{ -3.30}\%\)} & {\color{black!55}\(\pm{  1.00}\%\)} & {\color{black!55}\(\pm{  0.34}\%\)} & \(\pm 2.47\%\) & {\color{black!55}\({}^{ +1.64}_{ -1.64}\%\)} & {\color{black!55}\(\pm 1.83\%\)} & \(\pm{  2.35}\%\) &  & \(-2.33\) & \({}^{ +1.74}_{ -1.74}\%\) &  & \(\pm 4.04\%\) & \({}^{ +4.70}_{ -7.42}\%\) & \(\pm 4.73\%\)\\
125.60 & 51.23 & \({}^{ +1.65}_{ -4.64}\%\) & {\color{black!55}\({}^{ +0.31}_{ -3.30}\%\)} & {\color{black!55}\(\pm{  1.00}\%\)} & {\color{black!55}\(\pm{  0.34}\%\)} & \(\pm 2.47\%\) & {\color{black!55}\({}^{ +1.64}_{ -1.64}\%\)} & {\color{black!55}\(\pm 1.83\%\)} & \(\pm{  2.35}\%\) &  & \(-2.32\) & \({}^{ +1.74}_{ -1.74}\%\) &  & \(\pm 4.04\%\) & \({}^{ +4.69}_{ -7.42}\%\) & \(\pm 4.73\%\)\\
126.00 & 50.95 & \({}^{ +1.65}_{ -4.64}\%\) & {\color{black!55}\({}^{ +0.31}_{ -3.30}\%\)} & {\color{black!55}\(\pm{  1.00}\%\)} & {\color{black!55}\(\pm{  0.34}\%\)} & \(\pm 2.47\%\) & {\color{black!55}\({}^{ +1.64}_{ -1.64}\%\)} & {\color{black!55}\(\pm 1.83\%\)} & \(\pm{  2.34}\%\) &  & \(-2.31\) & \({}^{ +1.74}_{ -1.74}\%\) &  & \(\pm 4.03\%\) & \({}^{ +4.69}_{ -7.41}\%\) & \(\pm 4.73\%\)\\
128.00 & 49.56 & \({}^{ +1.64}_{ -4.63}\%\) & {\color{black!55}\({}^{ +0.30}_{ -3.29}\%\)} & {\color{black!55}\(\pm{  1.00}\%\)} & {\color{black!55}\(\pm{  0.34}\%\)} & \(\pm 2.46\%\) & {\color{black!55}\({}^{ +1.64}_{ -1.64}\%\)} & {\color{black!55}\(\pm 1.83\%\)} & \(\pm{  2.32}\%\) &  & \(-2.25\) & \({}^{ +1.74}_{ -1.74}\%\) &  & \(\pm 4.01\%\) & \({}^{ +4.67}_{ -7.37}\%\) & \(\pm 4.71\%\)\\
130.00 & 48.23 & \({}^{ +1.64}_{ -4.62}\%\) & {\color{black!55}\({}^{ +0.30}_{ -3.28}\%\)} & {\color{black!55}\(\pm{  1.00}\%\)} & {\color{black!55}\(\pm{  0.34}\%\)} & \(\pm 2.46\%\) & {\color{black!55}\({}^{ +1.64}_{ -1.64}\%\)} & {\color{black!55}\(\pm 1.82\%\)} & \(\pm{  2.30}\%\) &  & \(-2.20\) & \({}^{ +1.74}_{ -1.74}\%\) &  & \(\pm 3.99\%\) & \({}^{ +4.64}_{ -7.34}\%\) & \(\pm 4.69\%\)\\
\cline{1-10} \cline{12-13} \cline{15-17}
\end{tabular}
}
\end{table}

\begin{table}[ht!]
\caption{Total ggH cross sections in the SM for a LHC CM energy of $\sqrt{s}=14$\,TeV.}
\label{tab:ggH14}
\centering
\scriptsize
{\renewcommand{\arraystretch}{1.4}
\begin{tabular}{| cc | cccc | ccc | c |@{}c@{\;}| cc |@{}c@{\;}| ccc |}
\cline{1-10} \cline{12-13} \cline{15-17}
\multicolumn{2}{|c|}{\(\sqrt{s} = 14\,\mathrm{TeV}\)} &
\multicolumn{4}{l|}{} &
\multicolumn{4}{c|}{NNLO PDF} &
&
\multicolumn{2}{|c|}{aN${}^\text{3}$LO PDF} && \multicolumn{3}{l|}{} \\
\(\MH\,[\mathrm{GeV}]\) & \(\sigma\,[\mathrm{pb}]\) &
\(\delta(\mathrm{theory})\) &
{\color{black!55}\(\delta(\mathrm{scale})\)} & {\color{black!55}\(\delta(\mathrm{EWK})\)} & {\color{black!55}\(\delta(\mathrm{t,b,c})\)} &
\(\delta(\mathrm{PDF}+\alpha_s)\) &
{\color{black!55}\(\delta(\mathrm{PDF})\)} & {\color{black!55}\(\delta(\alpha_s)\)} &
\(\delta(\mathrm{PDF\text{--}TH})\) &
&
\(\Delta\,[\mathrm{pb}]\) & \(\delta(\mathrm{PDF}+\alpha_s)\) & &\(\delta(\mathrm{Gauss})\) & \(\delta(\mathrm{Total})\) & \shortstack{
  \(\mathrm{Gauss}\) \\
  \(\delta(\mathrm{Total})\)
    }
\\
\cline{1-10} \cline{12-13} \cline{15-17}
120.00 & 58.06 & \({}^{ +1.66}_{ -4.65}\%\) & {\color{black!55}\({}^{ +0.32}_{ -3.31}\%\)} & {\color{black!55}\(\pm{  1.00}\%\)} & {\color{black!55}\(\pm{  0.34}\%\)} & \(\pm 2.48\%\) & {\color{black!55}\({}^{ +1.65}_{ -1.65}\%\)} & {\color{black!55}\(\pm 1.84\%\)} & \(\pm{  2.45}\%\) &  & \(-2.59\) & \({}^{ +1.72}_{ -1.72}\%\) &  & \(\pm 4.10\%\) & \({}^{ +4.80}_{ -7.52}\%\) & \(\pm 4.79\%\)\\
122.00 & 56.43 & \({}^{ +1.66}_{ -4.64}\%\) & {\color{black!55}\({}^{ +0.32}_{ -3.30}\%\)} & {\color{black!55}\(\pm{  1.00}\%\)} & {\color{black!55}\(\pm{  0.34}\%\)} & \(\pm 2.47\%\) & {\color{black!55}\({}^{ +1.65}_{ -1.65}\%\)} & {\color{black!55}\(\pm 1.84\%\)} & \(\pm{  2.42}\%\) &  & \(-2.53\) & \({}^{ +1.72}_{ -1.72}\%\) &  & \(\pm 4.08\%\) & \({}^{ +4.77}_{ -7.48}\%\) & \(\pm 4.77\%\)\\
124.00 & 54.86 & \({}^{ +1.66}_{ -4.63}\%\) & {\color{black!55}\({}^{ +0.32}_{ -3.29}\%\)} & {\color{black!55}\(\pm{  1.00}\%\)} & {\color{black!55}\(\pm{  0.34}\%\)} & \(\pm 2.47\%\) & {\color{black!55}\({}^{ +1.64}_{ -1.64}\%\)} & {\color{black!55}\(\pm 1.83\%\)} & \(\pm{  2.40}\%\) &  & \(-2.46\) & \({}^{ +1.73}_{ -1.73}\%\) &  & \(\pm 4.06\%\) & \({}^{ +4.75}_{ -7.45}\%\) & \(\pm 4.75\%\)\\
124.60 & 54.40 & \({}^{ +1.66}_{ -4.63}\%\) & {\color{black!55}\({}^{ +0.32}_{ -3.29}\%\)} & {\color{black!55}\(\pm{  1.00}\%\)} & {\color{black!55}\(\pm{  0.34}\%\)} & \(\pm 2.47\%\) & {\color{black!55}\({}^{ +1.64}_{ -1.64}\%\)} & {\color{black!55}\(\pm 1.83\%\)} & \(\pm{  2.39}\%\) &  & \(-2.44\) & \({}^{ +1.73}_{ -1.73}\%\) &  & \(\pm 4.05\%\) & \({}^{ +4.74}_{ -7.44}\%\) & \(\pm 4.74\%\)\\
124.80 & 54.25 & \({}^{ +1.66}_{ -4.63}\%\) & {\color{black!55}\({}^{ +0.32}_{ -3.29}\%\)} & {\color{black!55}\(\pm{  1.00}\%\)} & {\color{black!55}\(\pm{  0.34}\%\)} & \(\pm 2.47\%\) & {\color{black!55}\({}^{ +1.64}_{ -1.64}\%\)} & {\color{black!55}\(\pm 1.83\%\)} & \(\pm{  2.39}\%\) &  & \(-2.44\) & \({}^{ +1.73}_{ -1.73}\%\) &  & \(\pm 4.05\%\) & \({}^{ +4.74}_{ -7.44}\%\) & \(\pm 4.74\%\)\\
125.00 & 54.10 & \({}^{ +1.66}_{ -4.63}\%\) & {\color{black!55}\({}^{ +0.32}_{ -3.29}\%\)} & {\color{black!55}\(\pm{  1.00}\%\)} & {\color{black!55}\(\pm{  0.34}\%\)} & \(\pm 2.47\%\) & {\color{black!55}\({}^{ +1.64}_{ -1.64}\%\)} & {\color{black!55}\(\pm 1.83\%\)} & \(\pm{  2.39}\%\) &  & \(-2.43\) & \({}^{ +1.73}_{ -1.73}\%\) &  & \(\pm 4.05\%\) & \({}^{ +4.74}_{ -7.43}\%\) & \(\pm 4.74\%\)\\
125.09 & 54.03 & \({}^{ +1.66}_{ -4.63}\%\) & {\color{black!55}\({}^{ +0.32}_{ -3.29}\%\)} & {\color{black!55}\(\pm{  1.00}\%\)} & {\color{black!55}\(\pm{  0.34}\%\)} & \(\pm 2.47\%\) & {\color{black!55}\({}^{ +1.64}_{ -1.64}\%\)} & {\color{black!55}\(\pm 1.83\%\)} & \(\pm{  2.38}\%\) &  & \(-2.43\) & \({}^{ +1.73}_{ -1.73}\%\) &  & \(\pm 4.05\%\) & \({}^{ +4.74}_{ -7.43}\%\) & \(\pm 4.74\%\)\\
125.20 & 53.95 & \({}^{ +1.66}_{ -4.63}\%\) & {\color{black!55}\({}^{ +0.32}_{ -3.29}\%\)} & {\color{black!55}\(\pm{  1.00}\%\)} & {\color{black!55}\(\pm{  0.34}\%\)} & \(\pm 2.47\%\) & {\color{black!55}\({}^{ +1.64}_{ -1.64}\%\)} & {\color{black!55}\(\pm 1.83\%\)} & \(\pm{  2.38}\%\) &  & \(-2.42\) & \({}^{ +1.73}_{ -1.73}\%\) &  & \(\pm 4.05\%\) & \({}^{ +4.74}_{ -7.43}\%\) & \(\pm 4.74\%\)\\
125.30 & 53.87 & \({}^{ +1.66}_{ -4.63}\%\) & {\color{black!55}\({}^{ +0.32}_{ -3.29}\%\)} & {\color{black!55}\(\pm{  1.00}\%\)} & {\color{black!55}\(\pm{  0.34}\%\)} & \(\pm 2.47\%\) & {\color{black!55}\({}^{ +1.64}_{ -1.64}\%\)} & {\color{black!55}\(\pm 1.83\%\)} & \(\pm{  2.38}\%\) &  & \(-2.42\) & \({}^{ +1.73}_{ -1.73}\%\) &  & \(\pm 4.05\%\) & \({}^{ +4.74}_{ -7.43}\%\) & \(\pm 4.74\%\)\\
125.38 & 53.81 & \({}^{ +1.66}_{ -4.62}\%\) & {\color{black!55}\({}^{ +0.32}_{ -3.28}\%\)} & {\color{black!55}\(\pm{  1.00}\%\)} & {\color{black!55}\(\pm{  0.34}\%\)} & \(\pm 2.47\%\) & {\color{black!55}\({}^{ +1.64}_{ -1.64}\%\)} & {\color{black!55}\(\pm 1.83\%\)} & \(\pm{  2.38}\%\) &  & \(-2.42\) & \({}^{ +1.73}_{ -1.73}\%\) &  & \(\pm 4.04\%\) & \({}^{ +4.73}_{ -7.43}\%\) & \(\pm 4.74\%\)\\
125.60 & 53.65 & \({}^{ +1.66}_{ -4.62}\%\) & {\color{black!55}\({}^{ +0.32}_{ -3.28}\%\)} & {\color{black!55}\(\pm{  1.00}\%\)} & {\color{black!55}\(\pm{  0.34}\%\)} & \(\pm 2.47\%\) & {\color{black!55}\({}^{ +1.64}_{ -1.64}\%\)} & {\color{black!55}\(\pm 1.83\%\)} & \(\pm{  2.38}\%\) &  & \(-2.41\) & \({}^{ +1.73}_{ -1.73}\%\) &  & \(\pm 4.04\%\) & \({}^{ +4.73}_{ -7.42}\%\) & \(\pm 4.73\%\)\\
126.00 & 53.35 & \({}^{ +1.66}_{ -4.62}\%\) & {\color{black!55}\({}^{ +0.32}_{ -3.28}\%\)} & {\color{black!55}\(\pm{  1.00}\%\)} & {\color{black!55}\(\pm{  0.34}\%\)} & \(\pm 2.46\%\) & {\color{black!55}\({}^{ +1.64}_{ -1.64}\%\)} & {\color{black!55}\(\pm 1.83\%\)} & \(\pm{  2.37}\%\) &  & \(-2.40\) & \({}^{ +1.73}_{ -1.73}\%\) &  & \(\pm 4.04\%\) & \({}^{ +4.72}_{ -7.42}\%\) & \(\pm 4.73\%\)\\
128.00 & 51.90 & \({}^{ +1.65}_{ -4.62}\%\) & {\color{black!55}\({}^{ +0.31}_{ -3.28}\%\)} & {\color{black!55}\(\pm{  1.00}\%\)} & {\color{black!55}\(\pm{  0.34}\%\)} & \(\pm 2.46\%\) & {\color{black!55}\({}^{ +1.64}_{ -1.64}\%\)} & {\color{black!55}\(\pm 1.82\%\)} & \(\pm{  2.35}\%\) &  & \(-2.34\) & \({}^{ +1.73}_{ -1.73}\%\) &  & \(\pm 4.02\%\) & \({}^{ +4.70}_{ -7.39}\%\) & \(\pm 4.71\%\)\\
130.00 & 50.52 & \({}^{ +1.65}_{ -4.61}\%\) & {\color{black!55}\({}^{ +0.31}_{ -3.27}\%\)} & {\color{black!55}\(\pm{  1.00}\%\)} & {\color{black!55}\(\pm{  0.34}\%\)} & \(\pm 2.46\%\) & {\color{black!55}\({}^{ +1.64}_{ -1.64}\%\)} & {\color{black!55}\(\pm 1.82\%\)} & \(\pm{  2.32}\%\) &  & \(-2.28\) & \({}^{ +1.73}_{ -1.73}\%\) &  & \(\pm 4.00\%\) & \({}^{ +4.67}_{ -7.36}\%\) & \(\pm 4.70\%\)\\
\cline{1-10} \cline{12-13} \cline{15-17}
\end{tabular}
}
\end{table}

\end{landscape}

\FloatBarrier
\subsection{$\bold{VBF}$}
\begin{landscape}
\begin{table}[ht!]
\caption{Total VBF cross sections in the SM for a LHC CM energy of $\sqrt{s}=7$ TeV, including QCD and EW corrections
and their uncertainties for different Higgs-boson masses \texorpdfstring{$\MH$}\MH. 
The cross sections $\sigma_{\NNNLO}^{\DIS}$, $\sigma_{\gamma}$, $\sigma_{\mbox{\scriptsize nf}}$, $\sigma_{\mbox{\scriptsize s/t/u}}$ refer to the NNNLO QCD cross section in the DIS approximation, the photon-induced contributions, the non-factorisable corrections, and the interference between the $s$, $t$, and $u$ diagrams, respectively.
The corrections $\delta_{\ELWK}$ represent the EW corrections excluding the photon-induced contributions.
The uncertainties $\Delta_{\mathrm{PDF}/\alphas/\mathrm{PDF\oplus\alphas}}$ refer to the PDF uncertainty, uncertainty related to the strong coupling, and the quadratically combined uncertainty.
For more details see Section~\ref{sec:VBF}. The uncertainty $\Delta_{\mathrm{theory}}$ is the linear sum of the scale and $\Delta_{\mathrm{TU}}$ uncertainties, separately in the up and down directions.}
\label{tab:vbf_XStot_7}
\centering
\scriptsize
{\renewcommand{\arraystretch}{1.4}
\begin{tabular}{| cc | ccc | ccc | ccc | cc | c |}
\cline{1-14}
\multicolumn{2}{|c|}{\(\sqrt{s} = 7\,\mathrm{TeV}\)} &
\multicolumn{3}{|c|}{Theory uncertainty} &
\multicolumn{3}{|c|}{Parametric uncertainty} &
\multicolumn{5}{|c|}{Cross-section decomposition} &
\multicolumn{1}{c|}{Total uncertainty} \\
\(\MH\,[\mathrm{GeV}]\) & \(\sigma^{\VBF}\,[\mathrm{fb}]\) &
\(\Delta_{\rm theory}\) &
{\color{black!55}\(\Delta_{\rm scale}\)} & {\color{black!55}\(\Delta_{\rm TU}\)} &
\(\Delta_{\rm PDF\oplus\alphas}\) &
{\color{black!55}\(\Delta_{\rm PDF}\)} & {\color{black!55}\(\Delta_{\alphas}\)} &
\(\sigma_{\NNNLO}^{\DIS}\,[\mathrm{fb}]\) & \(\delta_{\ELWK}\,[\%]\) & \(\sigma_\gamma\,[\mathrm{fb}]\) & \(\sigma_{\mathrm{nf}}\,[\mathrm{fb}]\) & \(\sigma_{s/t/u}\,[\mathrm{fb}]\) &
\(\Delta_{\rm total}\)
\\
\cline{1-14}
$120.00$ & $1310$ & $^{+1.1}_{-1.1}\%$ & {\color{black!55}$^{+0.067}_{-0.050}\%$} & {\color{black!55}$\pm 1.0\%$} & $\pm 2.3\%$ & {\color{black!55}$\pm 2.3\%$} & {\color{black!55}$\pm 0.3\%$} & $1360$ & $-4.4\%$ & $9.7$ & $-$ & $-5.3$ & $^{+2.6}_{-2.6}\%$ \\
$122.00$ & $1285$ & $^{+1.1}_{-1.1}\%$ & {\color{black!55}$^{+0.065}_{-0.051}\%$} & {\color{black!55}$\pm 1.0\%$} & $\pm 2.3\%$ & {\color{black!55}$\pm 2.3\%$} & {\color{black!55}$\pm 0.3\%$} & $1333$ & $-4.4\%$ & $9.6$ & $-$ & $-5.0$ & $^{+2.6}_{-2.6}\%$ \\
$124.00$ & $1261$ & $^{+1.1}_{-1.1}\%$ & {\color{black!55}$^{+0.064}_{-0.051}\%$} & {\color{black!55}$\pm 1.0\%$} & $\pm 2.4\%$ & {\color{black!55}$\pm 2.3\%$} & {\color{black!55}$\pm 0.3\%$} & $1309$ & $-4.4\%$ & $9.5$ & $-$ & $-4.7$ & $^{+2.6}_{-2.6}\%$ \\
$124.60$ & $1254$ & $^{+1.1}_{-1.1}\%$ & {\color{black!55}$^{+0.064}_{-0.051}\%$} & {\color{black!55}$\pm 1.0\%$} & $\pm 2.4\%$ & {\color{black!55}$\pm 2.3\%$} & {\color{black!55}$\pm 0.3\%$} & $1301$ & $-4.3\%$ & $9.5$ & $-$ & $-4.6$ & $^{+2.6}_{-2.6}\%$ \\
$124.80$ & $1252$ & $^{+1.1}_{-1.1}\%$ & {\color{black!55}$^{+0.064}_{-0.051}\%$} & {\color{black!55}$\pm 1.0\%$} & $\pm 2.4\%$ & {\color{black!55}$\pm 2.3\%$} & {\color{black!55}$\pm 0.3\%$} & $1299$ & $-4.3\%$ & $9.4$ & $-$ & $-4.6$ & $^{+2.6}_{-2.6}\%$ \\
$125.00$ & $1249$ & $^{+1.1}_{-1.1}\%$ & {\color{black!55}$^{+0.064}_{-0.051}\%$} & {\color{black!55}$\pm 1.0\%$} & $\pm 2.4\%$ & {\color{black!55}$\pm 2.3\%$} & {\color{black!55}$\pm 0.3\%$} & $1296$ & $-4.3\%$ & $9.4$ & $-$ & $-4.5$ & $^{+2.6}_{-2.6}\%$ \\
$125.09$ & $1248$ & $^{+1.1}_{-1.1}\%$ & {\color{black!55}$^{+0.063}_{-0.051}\%$} & {\color{black!55}$\pm 1.0\%$} & $\pm 2.4\%$ & {\color{black!55}$\pm 2.3\%$} & {\color{black!55}$\pm 0.3\%$} & $1295$ & $-4.3\%$ & $9.4$ & $-$ & $-4.5$ & $^{+2.6}_{-2.6}\%$ \\
$125.20$ & $1247$ & $^{+1.1}_{-1.1}\%$ & {\color{black!55}$^{+0.063}_{-0.051}\%$} & {\color{black!55}$\pm 1.0\%$} & $\pm 2.4\%$ & {\color{black!55}$\pm 2.3\%$} & {\color{black!55}$\pm 0.3\%$} & $1294$ & $-4.3\%$ & $9.4$ & $-$ & $-4.5$ & $^{+2.6}_{-2.6}\%$ \\
$125.30$ & $1246$ & $^{+1.1}_{-1.1}\%$ & {\color{black!55}$^{+0.063}_{-0.051}\%$} & {\color{black!55}$\pm 1.0\%$} & $\pm 2.4\%$ & {\color{black!55}$\pm 2.3\%$} & {\color{black!55}$\pm 0.3\%$} & $1293$ & $-4.3\%$ & $9.4$ & $-$ & $-4.4$ & $^{+2.6}_{-2.6}\%$ \\
$125.38$ & $1245$ & $^{+1.1}_{-1.1}\%$ & {\color{black!55}$^{+0.063}_{-0.051}\%$} & {\color{black!55}$\pm 1.0\%$} & $\pm 2.4\%$ & {\color{black!55}$\pm 2.3\%$} & {\color{black!55}$\pm 0.3\%$} & $1292$ & $-4.3\%$ & $9.4$ & $-$ & $-4.4$ & $^{+2.6}_{-2.6}\%$ \\
$125.60$ & $1242$ & $^{+1.1}_{-1.1}\%$ & {\color{black!55}$^{+0.063}_{-0.051}\%$} & {\color{black!55}$\pm 1.0\%$} & $\pm 2.4\%$ & {\color{black!55}$\pm 2.3\%$} & {\color{black!55}$\pm 0.3\%$} & $1289$ & $-4.3\%$ & $9.4$ & $-$ & $-4.4$ & $^{+2.6}_{-2.6}\%$ \\
$126.00$ & $1238$ & $^{+1.1}_{-1.1}\%$ & {\color{black!55}$^{+0.063}_{-0.051}\%$} & {\color{black!55}$\pm 1.0\%$} & $\pm 2.4\%$ & {\color{black!55}$\pm 2.3\%$} & {\color{black!55}$\pm 0.3\%$} & $1284$ & $-4.3\%$ & $9.4$ & $-$ & $-4.4$ & $^{+2.6}_{-2.6}\%$ \\
$128.00$ & $1215$ & $^{+1.1}_{-1.1}\%$ & {\color{black!55}$^{+0.061}_{-0.051}\%$} & {\color{black!55}$\pm 1.0\%$} & $\pm 2.4\%$ & {\color{black!55}$\pm 2.3\%$} & {\color{black!55}$\pm 0.3\%$} & $1260$ & $-4.3\%$ & $9.2$ & $-$ & $-4.1$ & $^{+2.6}_{-2.6}\%$ \\
$130.00$ & $1192$ & $^{+1.1}_{-1.1}\%$ & {\color{black!55}$^{+0.060}_{-0.051}\%$} & {\color{black!55}$\pm 1.0\%$} & $\pm 2.4\%$ & {\color{black!55}$\pm 2.3\%$} & {\color{black!55}$\pm 0.3\%$} & $1237$ & $-4.3\%$ & $9.1$ & $-$ & $-3.8$ & $^{+2.6}_{-2.6}\%$ \\
\cline{1-14}
\end{tabular}
}
\end{table}

\begin{table}[ht!]
\caption{Total VBF cross sections in the SM for a LHC CM energy of $\sqrt{s}=8$ TeV, including QCD and EW corrections
and their uncertainties for different Higgs-boson masses \texorpdfstring{$\MH$}\MH. 
The cross sections $\sigma_{\NNNLO}^{\DIS}$, $\sigma_{\gamma}$, $\sigma_{\mbox{\scriptsize nf}}$, $\sigma_{\mbox{\scriptsize s/t/u}}$ refer to the NNNLO QCD cross section in the DIS approximation, the photon-induced contributions, the non-factorisable corrections, and the interference between the $s$, $t$, and $u$ diagrams, respectively.
The corrections $\delta_{\ELWK}$ represent the EW corrections excluding the photon-induced contributions.
The uncertainties $\Delta_{\mathrm{PDF}/\alphas/\mathrm{PDF\oplus\alphas}}$ refer to the PDF uncertainty, uncertainty related to the strong coupling, and the quadratically combined uncertainty.
For more details see Section~\ref{sec:VBF}. The uncertainty $\Delta_{\mathrm{theory}}$ is the linear sum of the scale and $\Delta_{\mathrm{TU}}$ uncertainties, separately in the up and down directions.}
\label{tab:vbf_XStot_8}
\centering
\scriptsize
{\renewcommand{\arraystretch}{1.4}
\begin{tabular}{| cc | ccc | ccc | ccc | cc | c |}
\cline{1-14}
\multicolumn{2}{|c|}{\(\sqrt{s} = 8\,\mathrm{TeV}\)} &
\multicolumn{3}{|c|}{Theory uncertainty} &
\multicolumn{3}{|c|}{Parametric uncertainty} &
\multicolumn{5}{|c|}{Cross-section decomposition} &
\multicolumn{1}{c|}{Total uncertainty} \\
\(\MH\,[\mathrm{GeV}]\) & \(\sigma^{\VBF}\,[\mathrm{fb}]\) &
\(\Delta_{\rm theory}\) &
{\color{black!55}\(\Delta_{\rm scale}\)} & {\color{black!55}\(\Delta_{\rm TU}\)} &
\(\Delta_{\rm PDF\oplus\alphas}\) &
{\color{black!55}\(\Delta_{\rm PDF}\)} & {\color{black!55}\(\Delta_{\alphas}\)} &
\(\sigma_{\NNNLO}^{\DIS}\,[\mathrm{fb}]\) & \(\delta_{\ELWK}\,[\%]\) & \(\sigma_\gamma\,[\mathrm{fb}]\) & \(\sigma_{\mathrm{nf}}\,[\mathrm{fb}]\) & \(\sigma_{s/t/u}\,[\mathrm{fb}]\) &
\(\Delta_{\rm total}\)
\\
\cline{1-14}
$120.00$ & $1687$ & $^{+1.1}_{-1.1}\%$ & {\color{black!55}$^{+0.082}_{-0.061}\%$} & {\color{black!55}$\pm 1.0\%$} & $\pm 2.3\%$ & {\color{black!55}$\pm 2.3\%$} & {\color{black!55}$\pm 0.3\%$} & $1754$ & $-4.6\%$ & $13.2$ & $-$ & $-6.2$ & $^{+2.5}_{-2.5}\%$ \\
$122.00$ & $1657$ & $^{+1.1}_{-1.1}\%$ & {\color{black!55}$^{+0.081}_{-0.061}\%$} & {\color{black!55}$\pm 1.0\%$} & $\pm 2.3\%$ & {\color{black!55}$\pm 2.3\%$} & {\color{black!55}$\pm 0.3\%$} & $1722$ & $-4.6\%$ & $13.0$ & $-$ & $-5.9$ & $^{+2.5}_{-2.5}\%$ \\
$124.00$ & $1627$ & $^{+1.1}_{-1.1}\%$ & {\color{black!55}$^{+0.081}_{-0.061}\%$} & {\color{black!55}$\pm 1.0\%$} & $\pm 2.3\%$ & {\color{black!55}$\pm 2.3\%$} & {\color{black!55}$\pm 0.3\%$} & $1691$ & $-4.5\%$ & $12.9$ & $-$ & $-5.5$ & $^{+2.5}_{-2.5}\%$ \\
$124.60$ & $1618$ & $^{+1.1}_{-1.1}\%$ & {\color{black!55}$^{+0.081}_{-0.061}\%$} & {\color{black!55}$\pm 1.0\%$} & $\pm 2.3\%$ & {\color{black!55}$\pm 2.3\%$} & {\color{black!55}$\pm 0.3\%$} & $1681$ & $-4.5\%$ & $12.8$ & $-$ & $-5.5$ & $^{+2.5}_{-2.5}\%$ \\
$124.80$ & $1615$ & $^{+1.1}_{-1.1}\%$ & {\color{black!55}$^{+0.081}_{-0.061}\%$} & {\color{black!55}$\pm 1.0\%$} & $\pm 2.3\%$ & {\color{black!55}$\pm 2.3\%$} & {\color{black!55}$\pm 0.3\%$} & $1678$ & $-4.5\%$ & $12.8$ & $-$ & $-5.5$ & $^{+2.5}_{-2.5}\%$ \\
$125.00$ & $1612$ & $^{+1.1}_{-1.1}\%$ & {\color{black!55}$^{+0.081}_{-0.061}\%$} & {\color{black!55}$\pm 1.0\%$} & $\pm 2.3\%$ & {\color{black!55}$\pm 2.3\%$} & {\color{black!55}$\pm 0.3\%$} & $1675$ & $-4.5\%$ & $12.8$ & $-$ & $-5.4$ & $^{+2.5}_{-2.5}\%$ \\
$125.09$ & $1611$ & $^{+1.1}_{-1.1}\%$ & {\color{black!55}$^{+0.081}_{-0.061}\%$} & {\color{black!55}$\pm 1.0\%$} & $\pm 2.3\%$ & {\color{black!55}$\pm 2.3\%$} & {\color{black!55}$\pm 0.3\%$} & $1674$ & $-4.5\%$ & $12.8$ & $-$ & $-5.4$ & $^{+2.5}_{-2.5}\%$ \\
$125.20$ & $1609$ & $^{+1.1}_{-1.1}\%$ & {\color{black!55}$^{+0.081}_{-0.061}\%$} & {\color{black!55}$\pm 1.0\%$} & $\pm 2.3\%$ & {\color{black!55}$\pm 2.3\%$} & {\color{black!55}$\pm 0.3\%$} & $1672$ & $-4.5\%$ & $12.8$ & $-$ & $-5.4$ & $^{+2.5}_{-2.5}\%$ \\
$125.30$ & $1608$ & $^{+1.1}_{-1.1}\%$ & {\color{black!55}$^{+0.081}_{-0.061}\%$} & {\color{black!55}$\pm 1.0\%$} & $\pm 2.3\%$ & {\color{black!55}$\pm 2.3\%$} & {\color{black!55}$\pm 0.3\%$} & $1671$ & $-4.5\%$ & $12.8$ & $-$ & $-5.4$ & $^{+2.6}_{-2.5}\%$ \\
$125.38$ & $1607$ & $^{+1.1}_{-1.1}\%$ & {\color{black!55}$^{+0.080}_{-0.061}\%$} & {\color{black!55}$\pm 1.0\%$} & $\pm 2.3\%$ & {\color{black!55}$\pm 2.3\%$} & {\color{black!55}$\pm 0.3\%$} & $1669$ & $-4.5\%$ & $12.8$ & $-$ & $-5.4$ & $^{+2.5}_{-2.5}\%$ \\
$125.60$ & $1604$ & $^{+1.1}_{-1.1}\%$ & {\color{black!55}$^{+0.080}_{-0.061}\%$} & {\color{black!55}$\pm 1.0\%$} & $\pm 2.3\%$ & {\color{black!55}$\pm 2.3\%$} & {\color{black!55}$\pm 0.3\%$} & $1666$ & $-4.5\%$ & $12.8$ & $-$ & $-5.3$ & $^{+2.6}_{-2.5}\%$ \\
$126.00$ & $1598$ & $^{+1.1}_{-1.1}\%$ & {\color{black!55}$^{+0.080}_{-0.061}\%$} & {\color{black!55}$\pm 1.0\%$} & $\pm 2.3\%$ & {\color{black!55}$\pm 2.3\%$} & {\color{black!55}$\pm 0.3\%$} & $1660$ & $-4.5\%$ & $12.7$ & $-$ & $-5.2$ & $^{+2.6}_{-2.5}\%$ \\
$128.00$ & $1569$ & $^{+1.1}_{-1.1}\%$ & {\color{black!55}$^{+0.079}_{-0.061}\%$} & {\color{black!55}$\pm 1.0\%$} & $\pm 2.3\%$ & {\color{black!55}$\pm 2.3\%$} & {\color{black!55}$\pm 0.3\%$} & $1630$ & $-4.5\%$ & $12.6$ & $-$ & $-4.9$ & $^{+2.6}_{-2.5}\%$ \\
$130.00$ & $1542$ & $^{+1.1}_{-1.1}\%$ & {\color{black!55}$^{+0.079}_{-0.061}\%$} & {\color{black!55}$\pm 1.0\%$} & $\pm 2.3\%$ & {\color{black!55}$\pm 2.3\%$} & {\color{black!55}$\pm 0.3\%$} & $1601$ & $-4.5\%$ & $12.4$ & $-$ & $-4.6$ & $^{+2.6}_{-2.5}\%$ \\
\cline{1-14}
\end{tabular}
}
\end{table}

\begin{table}[ht!]
\caption{Total VBF cross sections in the SM for a LHC CM energy of $\sqrt{s}=13$ TeV, including QCD and EW corrections
and their uncertainties for different Higgs-boson masses \texorpdfstring{$\MH$}\MH. 
The cross sections $\sigma_{\NNNLO}^{\DIS}$, $\sigma_{\gamma}$, $\sigma_{\mbox{\scriptsize nf}}$, $\sigma_{\mbox{\scriptsize s/t/u}}$ refer to the NNNLO QCD cross section in the DIS approximation, the photon-induced contributions, the non-factorisable corrections, and the interference between the $s$, $t$, and $u$ diagrams, respectively.
The corrections $\delta_{\ELWK}$ represent the EW corrections excluding the photon-induced contributions.
The uncertainties $\Delta_{\mathrm{PDF}/\alphas/\mathrm{PDF\oplus\alphas}}$ refer to the PDF uncertainty, uncertainty related to the strong coupling, and the quadratically combined uncertainty.
For more details see Section~\ref{sec:VBF}. The uncertainty $\Delta_{\mathrm{theory}}$ is the linear sum of the scale and $\Delta_{\mathrm{TU}}$ uncertainties, separately in the up and down directions.}
\label{tab:vbf_XStot_13}
\centering
\scriptsize
{\renewcommand{\arraystretch}{1.4}
\begin{tabular}{| cc | ccc | ccc | ccc | cc | c |}
\cline{1-14}
\multicolumn{2}{|c|}{\(\sqrt{s} = 13\,\mathrm{TeV}\)} &
\multicolumn{3}{|c|}{Theory uncertainty} &
\multicolumn{3}{|c|}{Parametric uncertainty} &
\multicolumn{5}{|c|}{Cross-section decomposition} &
\multicolumn{1}{c|}{Total uncertainty} \\
\(\MH\,[\mathrm{GeV}]\) & \(\sigma^{\VBF}\,[\mathrm{fb}]\) &
\(\Delta_{\rm theory}\) &
{\color{black!55}\(\Delta_{\rm scale}\)} & {\color{black!55}\(\Delta_{\rm TU}\)} &
\(\Delta_{\rm PDF\oplus\alphas}\) &
{\color{black!55}\(\Delta_{\rm PDF}\)} & {\color{black!55}\(\Delta_{\alphas}\)} &
\(\sigma_{\NNNLO}^{\DIS}\,[\mathrm{fb}]\) & \(\delta_{\ELWK}\,[\%]\) & \(\sigma_\gamma\,[\mathrm{fb}]\) & \(\sigma_{\mathrm{nf}}\,[\mathrm{fb}]\) & \(\sigma_{s/t/u}\,[\mathrm{fb}]\) &
\(\Delta_{\rm total}\)
\\
\cline{1-14}
$120.00$ & $3967$ & $^{+1.1}_{-1.1}\%$ & {\color{black!55}$^{+0.13}_{-0.091}\%$} & {\color{black!55}$\pm 1.0\%$} & $\pm 2.2\%$ & {\color{black!55}$\pm 2.1\%$} & {\color{black!55}$\pm 0.4\%$} & $4148$ & $-5.2\%$ & $36.1$ & $-8.9$ & $-11.5$ & $^{+2.5}_{-2.5}\%$ \\
$122.00$ & $3905$ & $^{+1.1}_{-1.1}\%$ & {\color{black!55}$^{+0.13}_{-0.092}\%$} & {\color{black!55}$\pm 1.0\%$} & $\pm 2.2\%$ & {\color{black!55}$\pm 2.1\%$} & {\color{black!55}$\pm 0.4\%$} & $4082$ & $-5.2\%$ & $35.8$ & $-8.5$ & $-10.6$ & $^{+2.5}_{-2.5}\%$ \\
$124.00$ & $3844$ & $^{+1.1}_{-1.1}\%$ & {\color{black!55}$^{+0.13}_{-0.092}\%$} & {\color{black!55}$\pm 1.0\%$} & $\pm 2.2\%$ & {\color{black!55}$\pm 2.1\%$} & {\color{black!55}$\pm 0.4\%$} & $4017$ & $-5.2\%$ & $35.4$ & $-8.2$ & $-10.2$ & $^{+2.5}_{-2.4}\%$ \\
$124.60$ & $3825$ & $^{+1.1}_{-1.1}\%$ & {\color{black!55}$^{+0.13}_{-0.093}\%$} & {\color{black!55}$\pm 1.0\%$} & $\pm 2.2\%$ & {\color{black!55}$\pm 2.1\%$} & {\color{black!55}$\pm 0.4\%$} & $3998$ & $-5.2\%$ & $35.3$ & $-8.1$ & $-10.0$ & $^{+2.5}_{-2.4}\%$ \\
$124.80$ & $3819$ & $^{+1.1}_{-1.1}\%$ & {\color{black!55}$^{+0.13}_{-0.093}\%$} & {\color{black!55}$\pm 1.0\%$} & $\pm 2.2\%$ & {\color{black!55}$\pm 2.1\%$} & {\color{black!55}$\pm 0.4\%$} & $3992$ & $-5.2\%$ & $35.3$ & $-8.1$ & $-10.0$ & $^{+2.5}_{-2.4}\%$ \\
$125.00$ & $3813$ & $^{+1.1}_{-1.1}\%$ & {\color{black!55}$^{+0.13}_{-0.093}\%$} & {\color{black!55}$\pm 1.0\%$} & $\pm 2.2\%$ & {\color{black!55}$\pm 2.1\%$} & {\color{black!55}$\pm 0.4\%$} & $3985$ & $-5.2\%$ & $35.2$ & $-8.0$ & $-10.0$ & $^{+2.5}_{-2.4}\%$ \\
$125.09$ & $3811$ & $^{+1.1}_{-1.1}\%$ & {\color{black!55}$^{+0.13}_{-0.093}\%$} & {\color{black!55}$\pm 1.0\%$} & $\pm 2.2\%$ & {\color{black!55}$\pm 2.1\%$} & {\color{black!55}$\pm 0.4\%$} & $3982$ & $-5.2\%$ & $35.2$ & $-8.0$ & $-10.0$ & $^{+2.5}_{-2.4}\%$ \\
$125.20$ & $3807$ & $^{+1.1}_{-1.1}\%$ & {\color{black!55}$^{+0.13}_{-0.093}\%$} & {\color{black!55}$\pm 1.0\%$} & $\pm 2.2\%$ & {\color{black!55}$\pm 2.1\%$} & {\color{black!55}$\pm 0.4\%$} & $3979$ & $-5.2\%$ & $35.2$ & $-8.0$ & $-10.0$ & $^{+2.5}_{-2.4}\%$ \\
$125.30$ & $3804$ & $^{+1.1}_{-1.1}\%$ & {\color{black!55}$^{+0.13}_{-0.093}\%$} & {\color{black!55}$\pm 1.0\%$} & $\pm 2.2\%$ & {\color{black!55}$\pm 2.1\%$} & {\color{black!55}$\pm 0.4\%$} & $3976$ & $-5.2\%$ & $35.2$ & $-8.0$ & $-9.9$ & $^{+2.5}_{-2.4}\%$ \\
$125.38$ & $3802$ & $^{+1.1}_{-1.1}\%$ & {\color{black!55}$^{+0.13}_{-0.093}\%$} & {\color{black!55}$\pm 1.0\%$} & $\pm 2.2\%$ & {\color{black!55}$\pm 2.1\%$} & {\color{black!55}$\pm 0.4\%$} & $3973$ & $-5.2\%$ & $35.2$ & $-8.0$ & $-9.8$ & $^{+2.5}_{-2.4}\%$ \\
$125.60$ & $3795$ & $^{+1.1}_{-1.1}\%$ & {\color{black!55}$^{+0.13}_{-0.093}\%$} & {\color{black!55}$\pm 1.0\%$} & $\pm 2.2\%$ & {\color{black!55}$\pm 2.1\%$} & {\color{black!55}$\pm 0.4\%$} & $3966$ & $-5.2\%$ & $35.1$ & $-8.0$ & $-9.7$ & $^{+2.5}_{-2.4}\%$ \\
$126.00$ & $3784$ & $^{+1.1}_{-1.1}\%$ & {\color{black!55}$^{+0.13}_{-0.093}\%$} & {\color{black!55}$\pm 1.0\%$} & $\pm 2.2\%$ & {\color{black!55}$\pm 2.1\%$} & {\color{black!55}$\pm 0.4\%$} & $3954$ & $-5.2\%$ & $35.1$ & $-7.9$ & $-9.6$ & $^{+2.5}_{-2.4}\%$ \\
$128.00$ & $3725$ & $^{+1.1}_{-1.0}\%$ & {\color{black!55}$^{+0.13}_{-0.093}\%$} & {\color{black!55}$\pm 1.0\%$} & $\pm 2.2\%$ & {\color{black!55}$\pm 2.2\%$} & {\color{black!55}$\pm 0.4\%$} & $3892$ & $-5.2\%$ & $34.7$ & $-7.7$ & $-9.2$ & $^{+2.5}_{-2.4}\%$ \\
$130.00$ & $3667$ & $^{+1.1}_{-1.0}\%$ & {\color{black!55}$^{+0.13}_{-0.094}\%$} & {\color{black!55}$\pm 0.9\%$} & $\pm 2.2\%$ & {\color{black!55}$\pm 2.2\%$} & {\color{black!55}$\pm 0.3\%$} & $3831$ & $-5.2\%$ & $34.3$ & $-7.5$ & $-8.6$ & $^{+2.5}_{-2.4}\%$ \\
\cline{1-14}
\end{tabular}
}
\end{table}

\begin{table}[ht!]
\caption{Total VBF cross sections in the SM for a LHC CM energy of $\sqrt{s}=13.6$ TeV, including QCD and EW corrections
and their uncertainties for different Higgs-boson masses \texorpdfstring{$\MH$}\MH. 
The cross sections $\sigma_{\NNNLO}^{\DIS}$, $\sigma_{\gamma}$, $\sigma_{\mbox{\scriptsize nf}}$, $\sigma_{\mbox{\scriptsize s/t/u}}$ refer to the NNNLO QCD cross section in the DIS approximation, the photon-induced contributions, the non-factorisable corrections, and the interference between the $s$, $t$, and $u$ diagrams, respectively.
The corrections $\delta_{\ELWK}$ represent the EW corrections excluding the photon-induced contributions.
The uncertainties $\Delta_{\mathrm{PDF}/\alphas/\mathrm{PDF\oplus\alphas}}$ refer to the PDF uncertainty, uncertainty related to the strong coupling, and the quadratically combined uncertainty.
For more details see Section~\ref{sec:VBF}. The uncertainty $\Delta_{\mathrm{theory}}$ is the linear sum of the scale and $\Delta_{\mathrm{TU}}$ uncertainties, separately in the up and down directions.}
\label{tab:vbf_XStot_136}
\centering
\scriptsize
{\renewcommand{\arraystretch}{1.4}
\begin{tabular}{| cc | ccc | ccc | ccc | cc | c |}
\cline{1-14}
\multicolumn{2}{|c|}{\(\sqrt{s} = 13.6\,\mathrm{TeV}\)} &
\multicolumn{3}{|c|}{Theory uncertainty} &
\multicolumn{3}{|c|}{Parametric uncertainty} &
\multicolumn{5}{|c|}{Cross-section decomposition} &
\multicolumn{1}{c|}{Total uncertainty} \\
\(\MH\,[\mathrm{GeV}]\) & \(\sigma^{\VBF}\,[\mathrm{fb}]\) &
\(\Delta_{\rm theory}\) &
{\color{black!55}\(\Delta_{\rm scale}\)} & {\color{black!55}\(\Delta_{\rm TU}\)} &
\(\Delta_{\rm PDF\oplus\alphas}\) &
{\color{black!55}\(\Delta_{\rm PDF}\)} & {\color{black!55}\(\Delta_{\alphas}\)} &
\(\sigma_{\NNNLO}^{\DIS}\,[\mathrm{fb}]\) & \(\delta_{\ELWK}\,[\%]\) & \(\sigma_\gamma\,[\mathrm{fb}]\) & \(\sigma_{\mathrm{nf}}\,[\mathrm{fb}]\) & \(\sigma_{s/t/u}\,[\mathrm{fb}]\) &
\(\Delta_{\rm total}\)
\\
\cline{1-14}
$120.00$ & $4276$ & $^{+1.1}_{-1.1}\%$ & {\color{black!55}$^{+0.13}_{-0.093}\%$} & {\color{black!55}$\pm 1.0\%$} & $\pm 2.2\%$ & {\color{black!55}$\pm 2.1\%$} & {\color{black!55}$\pm 0.4\%$} & $4473$ & $-5.3\%$ & $39.4$ & $-9.2$ & $-11.9$ & $^{+2.5}_{-2.4}\%$ \\
$122.00$ & $4210$ & $^{+1.1}_{-1.1}\%$ & {\color{black!55}$^{+0.13}_{-0.094}\%$} & {\color{black!55}$\pm 1.0\%$} & $\pm 2.2\%$ & {\color{black!55}$\pm 2.1\%$} & {\color{black!55}$\pm 0.4\%$} & $4403$ & $-5.3\%$ & $39.0$ & $-8.8$ & $-11.4$ & $^{+2.5}_{-2.4}\%$ \\
$124.00$ & $4144$ & $^{+1.1}_{-1.1}\%$ & {\color{black!55}$^{+0.13}_{-0.095}\%$} & {\color{black!55}$\pm 1.0\%$} & $\pm 2.2\%$ & {\color{black!55}$\pm 2.1\%$} & {\color{black!55}$\pm 0.4\%$} & $4334$ & $-5.3\%$ & $38.6$ & $-8.5$ & $-10.9$ & $^{+2.5}_{-2.4}\%$ \\
$124.60$ & $4125$ & $^{+1.1}_{-1.1}\%$ & {\color{black!55}$^{+0.13}_{-0.095}\%$} & {\color{black!55}$\pm 1.0\%$} & $\pm 2.2\%$ & {\color{black!55}$\pm 2.1\%$} & {\color{black!55}$\pm 0.4\%$} & $4313$ & $-5.3\%$ & $38.5$ & $-8.4$ & $-10.8$ & $^{+2.5}_{-2.4}\%$ \\
$124.80$ & $4118$ & $^{+1.1}_{-1.1}\%$ & {\color{black!55}$^{+0.13}_{-0.095}\%$} & {\color{black!55}$\pm 1.0\%$} & $\pm 2.2\%$ & {\color{black!55}$\pm 2.1\%$} & {\color{black!55}$\pm 0.4\%$} & $4307$ & $-5.3\%$ & $38.5$ & $-8.4$ & $-10.7$ & $^{+2.5}_{-2.4}\%$ \\
$125.00$ & $4112$ & $^{+1.1}_{-1.1}\%$ & {\color{black!55}$^{+0.13}_{-0.095}\%$} & {\color{black!55}$\pm 1.0\%$} & $\pm 2.2\%$ & {\color{black!55}$\pm 2.1\%$} & {\color{black!55}$\pm 0.4\%$} & $4300$ & $-5.3\%$ & $38.5$ & $-8.3$ & $-10.7$ & $^{+2.5}_{-2.4}\%$ \\
$125.09$ & $4109$ & $^{+1.1}_{-1.1}\%$ & {\color{black!55}$^{+0.13}_{-0.095}\%$} & {\color{black!55}$\pm 1.0\%$} & $\pm 2.2\%$ & {\color{black!55}$\pm 2.1\%$} & {\color{black!55}$\pm 0.4\%$} & $4297$ & $-5.3\%$ & $38.4$ & $-8.3$ & $-10.7$ & $^{+2.5}_{-2.4}\%$ \\
$125.20$ & $4106$ & $^{+1.1}_{-1.1}\%$ & {\color{black!55}$^{+0.13}_{-0.095}\%$} & {\color{black!55}$\pm 1.0\%$} & $\pm 2.2\%$ & {\color{black!55}$\pm 2.1\%$} & {\color{black!55}$\pm 0.4\%$} & $4293$ & $-5.3\%$ & $38.4$ & $-8.3$ & $-10.6$ & $^{+2.5}_{-2.4}\%$ \\
$125.30$ & $4102$ & $^{+1.1}_{-1.1}\%$ & {\color{black!55}$^{+0.13}_{-0.095}\%$} & {\color{black!55}$\pm 1.0\%$} & $\pm 2.2\%$ & {\color{black!55}$\pm 2.1\%$} & {\color{black!55}$\pm 0.4\%$} & $4290$ & $-5.3\%$ & $38.4$ & $-8.3$ & $-10.5$ & $^{+2.5}_{-2.4}\%$ \\
$125.38$ & $4100$ & $^{+1.1}_{-1.1}\%$ & {\color{black!55}$^{+0.13}_{-0.095}\%$} & {\color{black!55}$\pm 1.0\%$} & $\pm 2.2\%$ & {\color{black!55}$\pm 2.1\%$} & {\color{black!55}$\pm 0.4\%$} & $4287$ & $-5.3\%$ & $38.4$ & $-8.3$ & $-10.4$ & $^{+2.5}_{-2.4}\%$ \\
$125.60$ & $4093$ & $^{+1.1}_{-1.1}\%$ & {\color{black!55}$^{+0.13}_{-0.095}\%$} & {\color{black!55}$\pm 1.0\%$} & $\pm 2.2\%$ & {\color{black!55}$\pm 2.1\%$} & {\color{black!55}$\pm 0.4\%$} & $4280$ & $-5.3\%$ & $38.3$ & $-8.3$ & $-10.4$ & $^{+2.5}_{-2.4}\%$ \\
$126.00$ & $4080$ & $^{+1.1}_{-1.0}\%$ & {\color{black!55}$^{+0.13}_{-0.095}\%$} & {\color{black!55}$\pm 1.0\%$} & $\pm 2.2\%$ & {\color{black!55}$\pm 2.1\%$} & {\color{black!55}$\pm 0.4\%$} & $4266$ & $-5.3\%$ & $38.3$ & $-8.2$ & $-10.3$ & $^{+2.5}_{-2.4}\%$ \\
$128.00$ & $4018$ & $^{+1.1}_{-1.0}\%$ & {\color{black!55}$^{+0.13}_{-0.096}\%$} & {\color{black!55}$\pm 0.9\%$} & $\pm 2.2\%$ & {\color{black!55}$\pm 2.1\%$} & {\color{black!55}$\pm 0.4\%$} & $4200$ & $-5.2\%$ & $37.9$ & $-8.0$ & $-9.8$ & $^{+2.4}_{-2.4}\%$ \\
$130.00$ & $3956$ & $^{+1.1}_{-1.0}\%$ & {\color{black!55}$^{+0.13}_{-0.097}\%$} & {\color{black!55}$\pm 0.9\%$} & $\pm 2.2\%$ & {\color{black!55}$\pm 2.1\%$} & {\color{black!55}$\pm 0.4\%$} & $4135$ & $-5.2\%$ & $37.5$ & $-7.8$ & $-9.1$ & $^{+2.4}_{-2.4}\%$ \\
\cline{1-14}
\end{tabular}
}
\end{table}

\begin{table}[ht!]
\caption{Total VBF cross sections in the SM for a LHC CM energy of $\sqrt{s}=14$ TeV, including QCD and EW corrections
and their uncertainties for different Higgs-boson masses \texorpdfstring{$\MH$}\MH. 
The cross sections $\sigma_{\NNNLO}^{\DIS}$, $\sigma_{\gamma}$, $\sigma_{\mbox{\scriptsize nf}}$, $\sigma_{\mbox{\scriptsize s/t/u}}$ refer to the NNNLO QCD cross section in the DIS approximation, the photon-induced contributions, the non-factorisable corrections, and the interference between the $s$, $t$, and $u$ diagrams, respectively.
The corrections $\delta_{\ELWK}$ represent the EW corrections excluding the photon-induced contributions.
For more details see Section~\ref{sec:VBF}. The uncertainty $\Delta_{\mathrm{theory}}$ is the linear sum of the scale and $\Delta_{\mathrm{TU}}$ uncertainties, separately in the up and down directions.}
\label{tab:vbf_XStot_14}
\centering
\scriptsize
{\renewcommand{\arraystretch}{1.4}
\begin{tabular}{| cc | ccc | ccc | ccc | cc | c |}
\cline{1-14}
\multicolumn{2}{|c|}{\(\sqrt{s} = 14\,\mathrm{TeV}\)} &
\multicolumn{3}{|c|}{Theory uncertainty} &
\multicolumn{3}{|c|}{Parametric uncertainty} &
\multicolumn{5}{|c|}{Cross-section decomposition} &
\multicolumn{1}{c|}{Total uncertainty} \\
\(\MH\,[\mathrm{GeV}]\) & \(\sigma^{\VBF}\,[\mathrm{fb}]\) &
\(\Delta_{\rm theory}\) &
{\color{black!55}\(\Delta_{\rm scale}\)} & {\color{black!55}\(\Delta_{\rm TU}\)} &
\(\Delta_{\rm PDF\oplus\alphas}\) &
{\color{black!55}\(\Delta_{\rm PDF}\)} & {\color{black!55}\(\Delta_{\alphas}\)} &
\(\sigma_{\NNNLO}^{\DIS}\,[\mathrm{fb}]\) & \(\delta_{\ELWK}\,[\%]\) & \(\sigma_\gamma\,[\mathrm{fb}]\) & \(\sigma_{\mathrm{nf}}\,[\mathrm{fb}]\) & \(\sigma_{s/t/u}\,[\mathrm{fb}]\) &
\(\Delta_{\rm total}\)
\\
\cline{1-14}
$120.00$ & $4486$ & $^{+1.1}_{-1.1}\%$ & {\color{black!55}$^{+0.14}_{-0.094}\%$} & {\color{black!55}$\pm 1.0\%$} & $\pm 2.2\%$ & {\color{black!55}$\pm 2.1\%$} & {\color{black!55}$\pm 0.4\%$} & $4694$ & $-5.3\%$ & $41.7$ & $-9.9$ & $-12.4$ & $^{+2.5}_{-2.4}\%$ \\
$122.00$ & $4416$ & $^{+1.1}_{-1.1}\%$ & {\color{black!55}$^{+0.14}_{-0.095}\%$} & {\color{black!55}$\pm 1.0\%$} & $\pm 2.2\%$ & {\color{black!55}$\pm 2.1\%$} & {\color{black!55}$\pm 0.4\%$} & $4620$ & $-5.3\%$ & $41.3$ & $-9.5$ & $-11.9$ & $^{+2.5}_{-2.4}\%$ \\
$124.00$ & $4348$ & $^{+1.1}_{-1.1}\%$ & {\color{black!55}$^{+0.14}_{-0.096}\%$} & {\color{black!55}$\pm 1.0\%$} & $\pm 2.2\%$ & {\color{black!55}$\pm 2.1\%$} & {\color{black!55}$\pm 0.4\%$} & $4549$ & $-5.3\%$ & $40.8$ & $-9.1$ & $-11.2$ & $^{+2.5}_{-2.4}\%$ \\
$124.60$ & $4328$ & $^{+1.1}_{-1.1}\%$ & {\color{black!55}$^{+0.14}_{-0.097}\%$} & {\color{black!55}$\pm 1.0\%$} & $\pm 2.2\%$ & {\color{black!55}$\pm 2.1\%$} & {\color{black!55}$\pm 0.4\%$} & $4527$ & $-5.3\%$ & $40.7$ & $-9.0$ & $-11.0$ & $^{+2.5}_{-2.4}\%$ \\
\end{tabular}
}
\end{table}
\end{landscape}

\FloatBarrier
\subsection{$\bold{VH}$}
\FloatBarrier
\subsubsection{$W^+H$ cross sections}

\begin{table}[ht!]
\caption{Total cross section and relative uncertainties for the $W^+H$ production at a CM energy of  $\sqrt{s}=7$~TeV and for different values of the Higgs-boson mass. Corrections due to individual contributions are included, see section~\ref{sec:vh_inclxs_breakdown}.}
\label{tab:w+h_xs_7tev}
\makebox[\textwidth][c]{
\centering
\scriptsize
{\renewcommand{\arraystretch}{1.4}
\begin{tabular}{| cc | c | ccc | ccc | c |}
\cline{1-10}
\multicolumn{2}{|c|}{\(\sqrt{s} = 7\,\mathrm{TeV}\)} &
\multicolumn{1}{|c|}{Theory uncertainty} &
\multicolumn{3}{|c|}{Parametric uncertainty} &
\multicolumn{3}{|c|}{Cross-section decomposition} &
\multicolumn{1}{c|}{Total uncertainty} \\
\(\MH\,[\mathrm{GeV}]\) & \(\sigma\,[\mathrm{fb}]\) &
\(\Delta_{\rm scale}\) &
\(\Delta_{\rm PDF\oplus\alpha_s}\) &
{\color{black!55}\(\Delta_{\rm PDF}\)} & {\color{black!55}\(\Delta_{\alpha_s}\)} &
\(\sigma_{\rm DY}\,[\mathrm{fb}]\) & \(\sigma_{\rm top}\,[\mathrm{fb}]\) & \(\delta_{\rm EW}\,[\%]\) &
\(\Delta_{\rm total}\)
\\
\cline{1-10}
$120.00$ & $429.1$ & $^{+0.67}_{-0.92}\%$ & $\pm 2.3\%$ & {\color{black!55}$\pm 2.3\%$} & {\color{black!55}$\pm 0.4\%$} & $454.7$ & $4.3$ & $-6.58\%$ & $^{+2.4}_{-2.5}\%$ \\
$122.00$ & $406.3$ & $^{+0.68}_{-0.92}\%$ & $\pm 2.3\%$ & {\color{black!55}$\pm 2.3\%$} & {\color{black!55}$\pm 0.4\%$} & $430.9$ & $4.1$ & $-6.67\%$ & $^{+2.4}_{-2.5}\%$ \\
$124.00$ & $385.0$ & $^{+0.68}_{-0.93}\%$ & $\pm 2.4\%$ & {\color{black!55}$\pm 2.3\%$} & {\color{black!55}$\pm 0.4\%$} & $408.6$ & $4.0$ & $-6.76\%$ & $^{+2.5}_{-2.5}\%$ \\
$124.60$ & $378.8$ & $^{+0.69}_{-0.93}\%$ & $\pm 2.4\%$ & {\color{black!55}$\pm 2.3\%$} & {\color{black!55}$\pm 0.4\%$} & $402.2$ & $3.9$ & $-6.78\%$ & $^{+2.5}_{-2.5}\%$ \\
$124.80$ & $376.8$ & $^{+0.69}_{-0.93}\%$ & $\pm 2.4\%$ & {\color{black!55}$\pm 2.3\%$} & {\color{black!55}$\pm 0.4\%$} & $400.1$ & $3.9$ & $-6.79\%$ & $^{+2.5}_{-2.5}\%$ \\
$125.00$ & $374.8$ & $^{+0.69}_{-0.93}\%$ & $\pm 2.4\%$ & {\color{black!55}$\pm 2.3\%$} & {\color{black!55}$\pm 0.4\%$} & $398.0$ & $3.9$ & $-6.80\%$ & $^{+2.5}_{-2.5}\%$ \\
\end{tabular}
}
}
\end{table}

\begin{table}[ht!]
\caption{Total cross section and relative uncertainties for the $W^+H$ production at a CM energy of  $\sqrt{s}=8$~TeV and for different values of the Higgs-boson mass. Corrections due to individual contributions are included, see section~\ref{sec:vh_inclxs_breakdown}.}
\label{tab:w+h_xs_8tev}
\makebox[\textwidth][c]{
\centering
\scriptsize
{\renewcommand{\arraystretch}{1.4}
\begin{tabular}{| cc | c | ccc | ccc | c |}
\cline{1-10}
\multicolumn{2}{|c|}{\(\sqrt{s} = 8\,\mathrm{TeV}\)} &
\multicolumn{1}{|c|}{Theory uncertainty} &
\multicolumn{3}{|c|}{Parametric uncertainty} &
\multicolumn{3}{|c|}{Cross-section decomposition} &
\multicolumn{1}{c|}{Total uncertainty} \\
\(\MH\,[\mathrm{GeV}]\) & \(\sigma\,[\mathrm{fb}]\) &
\(\Delta_{\rm scale}\) &
\(\Delta_{\rm PDF\oplus\alpha_s}\) &
{\color{black!55}\(\Delta_{\rm PDF}\)} & {\color{black!55}\(\Delta_{\alpha_s}\)} &
\(\sigma_{\rm DY}\,[\mathrm{fb}]\) & \(\sigma_{\rm top}\,[\mathrm{fb}]\) & \(\delta_{\rm EW}\,[\%]\) &
\(\Delta_{\rm total}\)
\\
\cline{1-10}
$120.00$ & $515.7$ & $^{+0.62}_{-0.88}\%$ & $\pm 2.2\%$ & {\color{black!55}$\pm 2.2\%$} & {\color{black!55}$\pm 0.5\%$} & $546.4$ & $5.5$ & $-6.63\%$ & $^{+2.3}_{-2.4}\%$ \\
$122.00$ & $488.6$ & $^{+0.63}_{-0.88}\%$ & $\pm 2.2\%$ & {\color{black!55}$\pm 2.2\%$} & {\color{black!55}$\pm 0.5\%$} & $518.1$ & $5.2$ & $-6.72\%$ & $^{+2.3}_{-2.4}\%$ \\
$124.00$ & $463.2$ & $^{+0.64}_{-0.89}\%$ & $\pm 2.2\%$ & {\color{black!55}$\pm 2.2\%$} & {\color{black!55}$\pm 0.4\%$} & $491.6$ & $5.1$ & $-6.80\%$ & $^{+2.3}_{-2.4}\%$ \\
$124.60$ & $455.9$ & $^{+0.63}_{-0.89}\%$ & $\pm 2.2\%$ & {\color{black!55}$\pm 2.2\%$} & {\color{black!55}$\pm 0.4\%$} & $484.0$ & $5.0$ & $-6.83\%$ & $^{+2.3}_{-2.4}\%$ \\
$124.80$ & $453.4$ & $^{+0.65}_{-0.88}\%$ & $\pm 2.2\%$ & {\color{black!55}$\pm 2.2\%$} & {\color{black!55}$\pm 0.4\%$} & $481.5$ & $4.9$ & $-6.85\%$ & $^{+2.3}_{-2.4}\%$ \\
$125.00$ & $451.1$ & $^{+0.65}_{-0.89}\%$ & $\pm 2.2\%$ & {\color{black!55}$\pm 2.2\%$} & {\color{black!55}$\pm 0.5\%$} & $479.0$ & $4.9$ & $-6.86\%$ & $^{+2.3}_{-2.4}\%$ \\
\end{tabular}
}
}
\end{table}

\begin{table}[ht!]
\caption{Total cross section and relative uncertainties for the $W^+H$ production at a CM energy of  $\sqrt{s}=13$~TeV and for different values of the Higgs-boson mass. Corrections due to individual contributions are included, see section~\ref{sec:vh_inclxs_breakdown}.}
\label{tab:w+h_xs_13tev}
\makebox[\textwidth][c]{
\centering
\scriptsize
{\renewcommand{\arraystretch}{1.4}
\begin{tabular}{| cc | c | ccc | ccc | c |}
\cline{1-10}
\multicolumn{2}{|c|}{\(\sqrt{s} = 13\,\mathrm{TeV}\)} &
\multicolumn{1}{|c|}{Theory uncertainty} &
\multicolumn{3}{|c|}{Parametric uncertainty} &
\multicolumn{3}{|c|}{Cross-section decomposition} &
\multicolumn{1}{c|}{Total uncertainty} \\
\(\MH\,[\mathrm{GeV}]\) & \(\sigma\,[\mathrm{fb}]\) &
\(\Delta_{\rm scale}\) &
\(\Delta_{\rm PDF\oplus\alpha_s}\) &
{\color{black!55}\(\Delta_{\rm PDF}\)} & {\color{black!55}\(\Delta_{\alpha_s}\)} &
\(\sigma_{\rm DY}\,[\mathrm{fb}]\) & \(\sigma_{\rm top}\,[\mathrm{fb}]\) & \(\delta_{\rm EW}\,[\%]\) &
\(\Delta_{\rm total}\)
\\
\cline{1-10}
$120.00$ & $967.6$ & $^{+0.49}_{-0.76}\%$ & $\pm 1.9\%$ & {\color{black!55}$\pm 1.8\%$} & {\color{black!55}$\pm 0.6\%$} & $1024.7$ & $12.4$ & $-6.78\%$ & $^{+1.9}_{-2.0}\%$ \\
$122.00$ & $918.4$ & $^{+0.51}_{-0.76}\%$ & $\pm 1.9\%$ & {\color{black!55}$\pm 1.8\%$} & {\color{black!55}$\pm 0.6\%$} & $973.5$ & $11.8$ & $-6.87\%$ & $^{+1.9}_{-2.0}\%$ \\
$124.00$ & $872.3$ & $^{+0.53}_{-0.76}\%$ & $\pm 1.9\%$ & {\color{black!55}$\pm 1.8\%$} & {\color{black!55}$\pm 0.6\%$} & $925.3$ & $11.3$ & $-6.95\%$ & $^{+1.9}_{-2.0}\%$ \\
$124.60$ & $859.1$ & $^{+0.51}_{-0.78}\%$ & $\pm 1.9\%$ & {\color{black!55}$\pm 1.8\%$} & {\color{black!55}$\pm 0.6\%$} & $911.4$ & $11.3$ & $-6.98\%$ & $^{+1.9}_{-2.0}\%$ \\
$124.80$ & $854.6$ & $^{+0.51}_{-0.77}\%$ & $\pm 1.9\%$ & {\color{black!55}$\pm 1.8\%$} & {\color{black!55}$\pm 0.6\%$} & $906.9$ & $11.2$ & $-6.99\%$ & $^{+1.9}_{-2.0}\%$ \\
$125.00$ & $850.2$ & $^{+0.52}_{-0.77}\%$ & $\pm 1.9\%$ & {\color{black!55}$\pm 1.8\%$} & {\color{black!55}$\pm 0.6\%$} & $902.3$ & $11.1$ & $-7.01\%$ & $^{+1.9}_{-2.0}\%$ \\
\end{tabular}
}
}
\end{table}

\begin{table}[ht!]
\caption{Total cross section and relative uncertainties for the $W^+H$ production at a CM energy of  $\sqrt{s}=13.6$~TeV and for different values of the Higgs-boson mass. Corrections due to individual contributions are included, see section~\ref{sec:vh_inclxs_breakdown}.}
\label{tab:w+h_xs_13.6tev}
\makebox[\textwidth][c]{
\centering
\scriptsize
{\renewcommand{\arraystretch}{1.4}
\begin{tabular}{| cc | c | ccc | ccc | c |}
\cline{1-10}
\multicolumn{2}{|c|}{\(\sqrt{s} = 13.6\,\mathrm{TeV}\)} &
\multicolumn{1}{|c|}{Theory uncertainty} &
\multicolumn{3}{|c|}{Parametric uncertainty} &
\multicolumn{3}{|c|}{Cross-section decomposition} &
\multicolumn{1}{c|}{Total uncertainty} \\
\(\MH\,[\mathrm{GeV}]\) & \(\sigma\,[\mathrm{fb}]\) &
\(\Delta_{\rm scale}\) &
\(\Delta_{\rm PDF\oplus\alpha_s}\) &
{\color{black!55}\(\Delta_{\rm PDF}\)} & {\color{black!55}\(\Delta_{\alpha_s}\)} &
\(\sigma_{\rm DY}\,[\mathrm{fb}]\) & \(\sigma_{\rm top}\,[\mathrm{fb}]\) & \(\delta_{\rm EW}\,[\%]\) &
\(\Delta_{\rm total}\)
\\
\cline{1-10}
$120.00$ & $1023.3$ & $^{+0.49}_{-0.75}\%$ & $\pm 1.8\%$ & {\color{black!55}$\pm 1.7\%$} & {\color{black!55}$\pm 0.6\%$} & $1083.6$ & $13.2$ & $-6.78\%$ & $^{+1.9}_{-2.0}\%$ \\
$122.00$ & $971.4$ & $^{+0.49}_{-0.75}\%$ & $\pm 1.8\%$ & {\color{black!55}$\pm 1.7\%$} & {\color{black!55}$\pm 0.6\%$} & $1029.6$ & $12.6$ & $-6.88\%$ & $^{+1.9}_{-2.0}\%$ \\
$124.00$ & $922.7$ & $^{+0.51}_{-0.76}\%$ & $\pm 1.8\%$ & {\color{black!55}$\pm 1.8\%$} & {\color{black!55}$\pm 0.6\%$} & $978.8$ & $12.2$ & $-6.97\%$ & $^{+1.9}_{-2.0}\%$ \\
$124.60$ & $908.8$ & $^{+0.49}_{-0.77}\%$ & $\pm 1.8\%$ & {\color{black!55}$\pm 1.8\%$} & {\color{black!55}$\pm 0.6\%$} & $964.2$ & $12.1$ & $-7.00\%$ & $^{+1.9}_{-2.0}\%$ \\
$124.80$ & $904.3$ & $^{+0.49}_{-0.77}\%$ & $\pm 1.9\%$ & {\color{black!55}$\pm 1.8\%$} & {\color{black!55}$\pm 0.6\%$} & $959.4$ & $12.1$ & $-7.00\%$ & $^{+1.9}_{-2.0}\%$ \\
$125.00$ & $899.5$ & $^{+0.51}_{-0.76}\%$ & $\pm 1.9\%$ & {\color{black!55}$\pm 1.8\%$} & {\color{black!55}$\pm 0.6\%$} & $954.6$ & $11.9$ & $-7.01\%$ & $^{+1.9}_{-2.0}\%$ \\
\end{tabular}
}
}
\end{table}

\begin{table}[ht!]
\caption{Total cross section and relative uncertainties for the $W^+H$ production at a CM energy of  $\sqrt{s}=14$~TeV and for different values of the Higgs-boson mass. Corrections due to individual contributions are included, see section~\ref{sec:vh_inclxs_breakdown}.}
\label{tab:w+h_xs_14tev}
\makebox[\textwidth][c]{
\centering
\scriptsize
{\renewcommand{\arraystretch}{1.4}
\begin{tabular}{| cc | c | ccc | ccc | c |}
\cline{1-10}
\multicolumn{2}{|c|}{\(\sqrt{s} = 14\,\mathrm{TeV}\)} &
\multicolumn{1}{|c|}{Theory uncertainty} &
\multicolumn{3}{|c|}{Parametric uncertainty} &
\multicolumn{3}{|c|}{Cross-section decomposition} &
\multicolumn{1}{c|}{Total uncertainty} \\
\(\MH\,[\mathrm{GeV}]\) & \(\sigma\,[\mathrm{fb}]\) &
\(\Delta_{\rm scale}\) &
\(\Delta_{\rm PDF\oplus\alpha_s}\) &
{\color{black!55}\(\Delta_{\rm PDF}\)} & {\color{black!55}\(\Delta_{\alpha_s}\)} &
\(\sigma_{\rm DY}\,[\mathrm{fb}]\) & \(\sigma_{\rm top}\,[\mathrm{fb}]\) & \(\delta_{\rm EW}\,[\%]\) &
\(\Delta_{\rm total}\)
\\
\cline{1-10}
$120.00$ & $1060.6$ & $^{+0.48}_{-0.74}\%$ & $\pm 1.8\%$ & {\color{black!55}$\pm 1.7\%$} & {\color{black!55}$\pm 0.6\%$} & $1123.0$ & $13.8$ & $-6.79\%$ & $^{+1.9}_{-2.0}\%$ \\
$122.00$ & $1007.0$ & $^{+0.48}_{-0.75}\%$ & $\pm 1.8\%$ & {\color{black!55}$\pm 1.7\%$} & {\color{black!55}$\pm 0.6\%$} & $1067.2$ & $13.3$ & $-6.88\%$ & $^{+1.9}_{-2.0}\%$ \\
$124.00$ & $956.6$ & $^{+0.51}_{-0.75}\%$ & $\pm 1.8\%$ & {\color{black!55}$\pm 1.7\%$} & {\color{black!55}$\pm 0.6\%$} & $1014.6$ & $12.7$ & $-6.98\%$ & $^{+1.9}_{-2.0}\%$ \\
$124.60$ & $942.1$ & $^{+0.50}_{-0.76}\%$ & $\pm 1.8\%$ & {\color{black!55}$\pm 1.7\%$} & {\color{black!55}$\pm 0.6\%$} & $999.5$ & $12.7$ & $-7.01\%$ & $^{+1.9}_{-2.0}\%$ \\
$124.80$ & $937.2$ & $^{+0.50}_{-0.75}\%$ & $\pm 1.8\%$ & {\color{black!55}$\pm 1.7\%$} & {\color{black!55}$\pm 0.6\%$} & $994.5$ & $12.5$ & $-7.02\%$ & $^{+1.9}_{-2.0}\%$ \\
$125.00$ & $932.5$ & $^{+0.50}_{-0.77}\%$ & $\pm 1.8\%$ & {\color{black!55}$\pm 1.7\%$} & {\color{black!55}$\pm 0.6\%$} & $989.5$ & $12.6$ & $-7.03\%$ & $^{+1.9}_{-2.0}\%$ \\
\end{tabular}
}
}
\end{table}

\FloatBarrier
\subsubsection{$W^-H$ cross sections}

\begin{table}[ht!]
\caption{Total cross section and relative uncertainties for the $W^-H$ production at a CM energy of  $\sqrt{s}=7$~TeV and for different values of the Higgs-boson mass. Corrections due to individual contributions are included, see section~\ref{sec:vh_inclxs_breakdown}.}
\label{tab:w-h_xs_7tev}
\makebox[\textwidth][c]{
\centering
\scriptsize
{\renewcommand{\arraystretch}{1.4}
\begin{tabular}{| cc | c | ccc | ccc | c |}
\cline{1-10}
\multicolumn{2}{|c|}{\(\sqrt{s} = 7\,\mathrm{TeV}\)} &
\multicolumn{1}{|c|}{Theory uncertainty} &
\multicolumn{3}{|c|}{Parametric uncertainty} &
\multicolumn{3}{|c|}{Cross-section decomposition} &
\multicolumn{1}{c|}{Total uncertainty} \\
\(\MH\,[\mathrm{GeV}]\) & \(\sigma\,[\mathrm{fb}]\) &
\(\Delta_{\rm scale}\) &
\(\Delta_{\rm PDF\oplus\alpha_s}\) &
{\color{black!55}\(\Delta_{\rm PDF}\)} & {\color{black!55}\(\Delta_{\alpha_s}\)} &
\(\sigma_{\rm DY}\,[\mathrm{fb}]\) & \(\sigma_{\rm top}\,[\mathrm{fb}]\) & \(\delta_{\rm EW}\,[\%]\) &
\(\Delta_{\rm total}\)
\\
\cline{1-10}
$120.00$ & $238.4$ & $^{+0.60}_{-0.88}\%$ & $\pm 2.4\%$ & {\color{black!55}$\pm 2.4\%$} & {\color{black!55}$\pm 0.4\%$} & $252.5$ & $2.5$ & $-6.58\%$ & $^{+2.5}_{-2.6}\%$ \\
$122.00$ & $225.1$ & $^{+0.61}_{-0.88}\%$ & $\pm 2.4\%$ & {\color{black!55}$\pm 2.4\%$} & {\color{black!55}$\pm 0.4\%$} & $238.7$ & $2.4$ & $-6.67\%$ & $^{+2.5}_{-2.6}\%$ \\
$124.00$ & $212.7$ & $^{+0.64}_{-0.89}\%$ & $\pm 2.4\%$ & {\color{black!55}$\pm 2.4\%$} & {\color{black!55}$\pm 0.4\%$} & $225.7$ & $2.2$ & $-6.76\%$ & $^{+2.5}_{-2.6}\%$ \\
$124.60$ & $209.1$ & $^{+0.63}_{-0.88}\%$ & $\pm 2.4\%$ & {\color{black!55}$\pm 2.4\%$} & {\color{black!55}$\pm 0.4\%$} & $222.0$ & $2.2$ & $-6.78\%$ & $^{+2.5}_{-2.6}\%$ \\
$124.80$ & $208.0$ & $^{+0.63}_{-0.89}\%$ & $\pm 2.4\%$ & {\color{black!55}$\pm 2.4\%$} & {\color{black!55}$\pm 0.4\%$} & $220.7$ & $2.2$ & $-6.79\%$ & $^{+2.5}_{-2.6}\%$ \\
$125.00$ & $206.8$ & $^{+0.63}_{-0.89}\%$ & $\pm 2.4\%$ & {\color{black!55}$\pm 2.4\%$} & {\color{black!55}$\pm 0.4\%$} & $219.5$ & $2.2$ & $-6.80\%$ & $^{+2.5}_{-2.6}\%$ \\
\end{tabular}
}
}
\end{table}

\begin{table}[ht!]
\caption{Total cross section and relative uncertainties for $W^-H$ production at a CM energy of  $\sqrt{s}=8$~TeV and for different values of the Higgs-boson mass. Corrections due to individual contributions are included, see section~\ref{sec:vh_inclxs_breakdown}.}
\label{tab:w-h_xs_8tev}
\makebox[\textwidth][c]{
\centering
\scriptsize
{\renewcommand{\arraystretch}{1.4}
\begin{tabular}{| cc | c | ccc | ccc | c |}
\cline{1-10}
\multicolumn{2}{|c|}{\(\sqrt{s} = 8\,\mathrm{TeV}\)} &
\multicolumn{1}{|c|}{Theory uncertainty} &
\multicolumn{3}{|c|}{Parametric uncertainty} &
\multicolumn{3}{|c|}{Cross-section decomposition} &
\multicolumn{1}{c|}{Total uncertainty} \\
\(\MH\,[\mathrm{GeV}]\) & \(\sigma\,[\mathrm{fb}]\) &
\(\Delta_{\rm scale}\) &
\(\Delta_{\rm PDF\oplus\alpha_s}\) &
{\color{black!55}\(\Delta_{\rm PDF}\)} & {\color{black!55}\(\Delta_{\alpha_s}\)} &
\(\sigma_{\rm DY}\,[\mathrm{fb}]\) & \(\sigma_{\rm top}\,[\mathrm{fb}]\) & \(\delta_{\rm EW}\,[\%]\) &
\(\Delta_{\rm total}\)
\\
\cline{1-10}
$120.00$ & $295.3$ & $^{+0.57}_{-0.83}\%$ & $\pm 2.3\%$ & {\color{black!55}$\pm 2.2\%$} & {\color{black!55}$\pm 0.4\%$} & $312.8$ & $3.2$ & $-6.63\%$ & $^{+2.3}_{-2.4}\%$ \\
$122.00$ & $279.1$ & $^{+0.56}_{-0.84}\%$ & $\pm 2.3\%$ & {\color{black!55}$\pm 2.2\%$} & {\color{black!55}$\pm 0.4\%$} & $295.9$ & $3.1$ & $-6.72\%$ & $^{+2.3}_{-2.4}\%$ \\
$124.00$ & $263.9$ & $^{+0.58}_{-0.84}\%$ & $\pm 2.3\%$ & {\color{black!55}$\pm 2.2\%$} & {\color{black!55}$\pm 0.4\%$} & $280.0$ & $2.9$ & $-6.80\%$ & $^{+2.4}_{-2.4}\%$ \\
$124.60$ & $259.5$ & $^{+0.58}_{-0.84}\%$ & $\pm 2.3\%$ & {\color{black!55}$\pm 2.3\%$} & {\color{black!55}$\pm 0.4\%$} & $275.5$ & $2.9$ & $-6.83\%$ & $^{+2.4}_{-2.5}\%$ \\
$124.80$ & $258.1$ & $^{+0.58}_{-0.84}\%$ & $\pm 2.3\%$ & {\color{black!55}$\pm 2.3\%$} & {\color{black!55}$\pm 0.4\%$} & $274.0$ & $2.9$ & $-6.85\%$ & $^{+2.4}_{-2.4}\%$ \\
$125.00$ & $256.7$ & $^{+0.58}_{-0.84}\%$ & $\pm 2.3\%$ & {\color{black!55}$\pm 2.3\%$} & {\color{black!55}$\pm 0.4\%$} & $272.5$ & $2.9$ & $-6.86\%$ & $^{+2.4}_{-2.4}\%$ \\
\end{tabular}
}
}
\end{table}

\begin{table}[ht!]
\caption{Total cross section and relative uncertainties for the $W^-H$ production at a CM energy of  $\sqrt{s}=13$~TeV and for different values of the Higgs-boson mass. Corrections due to individual contributions are included, see section~\ref{sec:vh_inclxs_breakdown}.}
\label{tab:w-h_xs_13tev}
\makebox[\textwidth][c]{
\centering
\scriptsize
{\renewcommand{\arraystretch}{1.4}
\begin{tabular}{| cc | c | ccc | ccc | c |}
\cline{1-10}
\multicolumn{2}{|c|}{\(\sqrt{s} = 13\,\mathrm{TeV}\)} &
\multicolumn{1}{|c|}{Theory uncertainty} &
\multicolumn{3}{|c|}{Parametric uncertainty} &
\multicolumn{3}{|c|}{Cross-section decomposition} &
\multicolumn{1}{c|}{Total uncertainty} \\
\(\MH\,[\mathrm{GeV}]\) & \(\sigma\,[\mathrm{fb}]\) &
\(\Delta_{\rm scale}\) &
\(\Delta_{\rm PDF\oplus\alpha_s}\) &
{\color{black!55}\(\Delta_{\rm PDF}\)} & {\color{black!55}\(\Delta_{\alpha_s}\)} &
\(\sigma_{\rm DY}\,[\mathrm{fb}]\) & \(\sigma_{\rm top}\,[\mathrm{fb}]\) & \(\delta_{\rm EW}\,[\%]\) &
\(\Delta_{\rm total}\)
\\
\cline{1-10}
$120.00$ & $609.4$ & $^{+0.42}_{-0.70}\%$ & $\pm 1.9\%$ & {\color{black!55}$\pm 1.8\%$} & {\color{black!55}$\pm 0.6\%$} & $645.2$ & $7.9$ & $-6.78\%$ & $^{+2.0}_{-2.0}\%$ \\
$122.00$ & $577.3$ & $^{+0.43}_{-0.70}\%$ & $\pm 1.9\%$ & {\color{black!55}$\pm 1.8\%$} & {\color{black!55}$\pm 0.6\%$} & $611.7$ & $7.6$ & $-6.87\%$ & $^{+2.0}_{-2.0}\%$ \\
$124.00$ & $547.2$ & $^{+0.44}_{-0.70}\%$ & $\pm 1.9\%$ & {\color{black!55}$\pm 1.8\%$} & {\color{black!55}$\pm 0.6\%$} & $580.3$ & $7.3$ & $-6.95\%$ & $^{+2.0}_{-2.0}\%$ \\
$124.60$ & $538.6$ & $^{+0.44}_{-0.71}\%$ & $\pm 1.9\%$ & {\color{black!55}$\pm 1.8\%$} & {\color{black!55}$\pm 0.6\%$} & $571.3$ & $7.2$ & $-6.98\%$ & $^{+2.0}_{-2.0}\%$ \\
$124.80$ & $535.7$ & $^{+0.45}_{-0.71}\%$ & $\pm 1.9\%$ & {\color{black!55}$\pm 1.8\%$} & {\color{black!55}$\pm 0.6\%$} & $568.3$ & $7.1$ & $-6.99\%$ & $^{+2.0}_{-2.0}\%$ \\
$125.00$ & $532.8$ & $^{+0.45}_{-0.70}\%$ & $\pm 1.9\%$ & {\color{black!55}$\pm 1.8\%$} & {\color{black!55}$\pm 0.6\%$} & $565.3$ & $7.1$ & $-7.01\%$ & $^{+2.0}_{-2.0}\%$ \\
\end{tabular}
}
}
\end{table}

\begin{table}[ht!]
\caption{Total cross section and relative uncertainties for the $W^-H$ production at a CM energy of  $\sqrt{s}=13.6$~TeV and for different values of the Higgs-boson mass. Corrections due to individual contributions are included, see section~\ref{sec:vh_inclxs_breakdown}.}
\label{tab:w-h_xs_13.6tev}
\makebox[\textwidth][c]{
\centering
\scriptsize
{\renewcommand{\arraystretch}{1.4}
\begin{tabular}{| cc | c | ccc | ccc | c |}
\cline{1-10}
\multicolumn{2}{|c|}{\(\sqrt{s} = 13.6\,\mathrm{TeV}\)} &
\multicolumn{1}{|c|}{Theory uncertainty} &
\multicolumn{3}{|c|}{Parametric uncertainty} &
\multicolumn{3}{|c|}{Cross-section decomposition} &
\multicolumn{1}{c|}{Total uncertainty} \\
\(\MH\,[\mathrm{GeV}]\) & \(\sigma\,[\mathrm{fb}]\) &
\(\Delta_{\rm scale}\) &
\(\Delta_{\rm PDF\oplus\alpha_s}\) &
{\color{black!55}\(\Delta_{\rm PDF}\)} & {\color{black!55}\(\Delta_{\alpha_s}\)} &
\(\sigma_{\rm DY}\,[\mathrm{fb}]\) & \(\sigma_{\rm top}\,[\mathrm{fb}]\) & \(\delta_{\rm EW}\,[\%]\) &
\(\Delta_{\rm total}\)
\\
\cline{1-10}
$120.00$ & $649.6$ & $^{+0.41}_{-0.68}\%$ & $\pm 1.9\%$ & {\color{black!55}$\pm 1.8\%$} & {\color{black!55}$\pm 0.6\%$} & $687.6$ & $8.6$ & $-6.78\%$ & $^{+1.9}_{-2.0}\%$ \\
$122.00$ & $615.4$ & $^{+0.43}_{-0.69}\%$ & $\pm 1.9\%$ & {\color{black!55}$\pm 1.8\%$} & {\color{black!55}$\pm 0.6\%$} & $652.0$ & $8.2$ & $-6.88\%$ & $^{+1.9}_{-2.0}\%$ \\
$124.00$ & $583.4$ & $^{+0.43}_{-0.70}\%$ & $\pm 1.9\%$ & {\color{black!55}$\pm 1.8\%$} & {\color{black!55}$\pm 0.6\%$} & $618.7$ & $7.9$ & $-6.97\%$ & $^{+1.9}_{-2.0}\%$ \\
$124.60$ & $574.2$ & $^{+0.43}_{-0.70}\%$ & $\pm 1.9\%$ & {\color{black!55}$\pm 1.8\%$} & {\color{black!55}$\pm 0.6\%$} & $609.1$ & $7.8$ & $-7.00\%$ & $^{+1.9}_{-2.0}\%$ \\
$124.80$ & $571.2$ & $^{+0.43}_{-0.70}\%$ & $\pm 1.9\%$ & {\color{black!55}$\pm 1.8\%$} & {\color{black!55}$\pm 0.6\%$} & $605.9$ & $7.7$ & $-7.00\%$ & $^{+1.9}_{-2.0}\%$ \\
$125.00$ & $568.2$ & $^{+0.43}_{-0.71}\%$ & $\pm 1.9\%$ & {\color{black!55}$\pm 1.8\%$} & {\color{black!55}$\pm 0.6\%$} & $602.8$ & $7.7$ & $-7.01\%$ & $^{+1.9}_{-2.0}\%$ \\
\end{tabular}
}
}
\end{table}

\begin{table}[ht!]
\caption{Total cross section and relative uncertainties for the $W^-H$ production at a CM energy of  $\sqrt{s}=14$~TeV and for different values of the Higgs-boson mass. Corrections due to individual contributions are included, see section~\ref{sec:vh_inclxs_breakdown}.}
\label{tab:w-h_xs_14tev}
\makebox[\textwidth][c]{
\centering
\scriptsize
{\renewcommand{\arraystretch}{1.4}
\begin{tabular}{| cc | c | ccc | ccc | c |}
\cline{1-10}
\multicolumn{2}{|c|}{\(\sqrt{s} = 14\,\mathrm{TeV}\)} &
\multicolumn{1}{|c|}{Theory uncertainty} &
\multicolumn{3}{|c|}{Parametric uncertainty} &
\multicolumn{3}{|c|}{Cross-section decomposition} &
\multicolumn{1}{c|}{Total uncertainty} \\
\(\MH\,[\mathrm{GeV}]\) & \(\sigma\,[\mathrm{fb}]\) &
\(\Delta_{\rm scale}\) &
\(\Delta_{\rm PDF\oplus\alpha_s}\) &
{\color{black!55}\(\Delta_{\rm PDF}\)} & {\color{black!55}\(\Delta_{\alpha_s}\)} &
\(\sigma_{\rm DY}\,[\mathrm{fb}]\) & \(\sigma_{\rm top}\,[\mathrm{fb}]\) & \(\delta_{\rm EW}\,[\%]\) &
\(\Delta_{\rm total}\)
\\
\cline{1-10}
$120.00$ & $676.5$ & $^{+0.41}_{-0.68}\%$ & $\pm 1.9\%$ & {\color{black!55}$\pm 1.8\%$} & {\color{black!55}$\pm 0.6\%$} & $716.1$ & $9.0$ & $-6.79\%$ & $^{+1.9}_{-2.0}\%$ \\
$122.00$ & $641.0$ & $^{+0.42}_{-0.68}\%$ & $\pm 1.9\%$ & {\color{black!55}$\pm 1.8\%$} & {\color{black!55}$\pm 0.6\%$} & $679.1$ & $8.6$ & $-6.88\%$ & $^{+1.9}_{-2.0}\%$ \\
$124.00$ & $607.8$ & $^{+0.43}_{-0.68}\%$ & $\pm 1.9\%$ & {\color{black!55}$\pm 1.8\%$} & {\color{black!55}$\pm 0.6\%$} & $644.5$ & $8.3$ & $-6.98\%$ & $^{+1.9}_{-2.0}\%$ \\
$124.60$ & $598.3$ & $^{+0.42}_{-0.69}\%$ & $\pm 1.9\%$ & {\color{black!55}$\pm 1.8\%$} & {\color{black!55}$\pm 0.6\%$} & $634.5$ & $8.2$ & $-7.01\%$ & $^{+1.9}_{-2.0}\%$ \\
$124.80$ & $595.1$ & $^{+0.43}_{-0.69}\%$ & $\pm 1.9\%$ & {\color{black!55}$\pm 1.8\%$} & {\color{black!55}$\pm 0.6\%$} & $631.2$ & $8.2$ & $-7.02\%$ & $^{+1.9}_{-2.0}\%$ \\
$125.00$ & $591.9$ & $^{+0.43}_{-0.70}\%$ & $\pm 1.9\%$ & {\color{black!55}$\pm 1.8\%$} & {\color{black!55}$\pm 0.6\%$} & $627.9$ & $8.1$ & $-7.03\%$ & $^{+1.9}_{-2.0}\%$ \\
\end{tabular}
}
}
\end{table}

\FloatBarrier
\subsubsection{$ZH$ cross section}

\begin{table}[ht!]
\caption{Total cross section and relative uncertainties for the $ZH$ production at a CM energy of  $\sqrt{s}=7$~TeV and for different values of the Higgs-boson mass. Corrections due to individual contributions are included, see section~\ref{sec:vh_inclxs_breakdown}.}
\label{tab:zh_xs_7tev}
\makebox[\textwidth][c]{
\centering
\scriptsize
{\renewcommand{\arraystretch}{1.4}
\begin{tabular}{| cc | c | ccc | cccc | c |}
\cline{1-11}
\multicolumn{2}{|c|}{\(\sqrt{s} = 7\,\mathrm{TeV}\)} &
\multicolumn{1}{|c|}{Theory uncertainty} &
\multicolumn{3}{|c|}{Parametric uncertainty} &
\multicolumn{4}{|c|}{Cross-section decomposition} &
\multicolumn{1}{c|}{Total uncertainty} \\
\(\MH\,[\mathrm{GeV}]\) & \(\sigma\,[\mathrm{fb}]\) &
\(\Delta_{\rm scale}\) &
\(\Delta_{\rm PDF\oplus\alpha_s}\) &
{\color{black!55}\(\Delta_{\rm PDF}\)} & {\color{black!55}\(\Delta_{\alpha_s}\)} &
\(\sigma_{\rm DY}\,[\mathrm{fb}]\) & \(\sigma_{\rm top}\,[\mathrm{fb}]\) & \(\sigma^{\rm NLO}_{ggZH}\,[\mathrm{fb}]\) & \(\delta_{\rm EW}\,[\%]\) &
\(\Delta_{\rm total}\)
\\
\cline{1-11}
$120.00$ & $388.4$ & $^{+2.5}_{-2.3}\%$ & $\pm 2.0\%$ & {\color{black!55}$\pm 2.0\%$} & {\color{black!55}$\pm 0.4\%$} & $374.4$ & $3.5$ & $29.5$ & $-5.07\%$ & $^{+3.2}_{-3.1}\%$ \\
$122.00$ & $369.1$ & $^{+2.6}_{-2.3}\%$ & $\pm 2.1\%$ & {\color{black!55}$\pm 2.0\%$} & {\color{black!55}$\pm 0.4\%$} & $354.8$ & $3.4$ & $28.9$ & $-5.09\%$ & $^{+3.3}_{-3.1}\%$ \\
$124.00$ & $350.8$ & $^{+2.6}_{-2.4}\%$ & $\pm 2.1\%$ & {\color{black!55}$\pm 2.0\%$} & {\color{black!55}$\pm 0.3\%$} & $336.5$ & $3.2$ & $28.3$ & $-5.11\%$ & $^{+3.4}_{-3.1}\%$ \\
$124.60$ & $345.5$ & $^{+2.7}_{-2.4}\%$ & $\pm 2.1\%$ & {\color{black!55}$\pm 2.0\%$} & {\color{black!55}$\pm 0.3\%$} & $331.2$ & $3.2$ & $28.0$ & $-5.12\%$ & $^{+3.4}_{-3.2}\%$ \\
$124.80$ & $343.8$ & $^{+2.7}_{-2.4}\%$ & $\pm 2.1\%$ & {\color{black!55}$\pm 2.0\%$} & {\color{black!55}$\pm 0.4\%$} & $329.5$ & $3.2$ & $28.0$ & $-5.12\%$ & $^{+3.4}_{-3.2}\%$ \\
$125.00$ & $342.1$ & $^{+2.7}_{-2.4}\%$ & $\pm 2.1\%$ & {\color{black!55}$\pm 2.0\%$} & {\color{black!55}$\pm 0.4\%$} & $327.7$ & $3.2$ & $27.9$ & $-5.12\%$ & $^{+3.4}_{-3.2}\%$ \\
\end{tabular}
}
}
\end{table}

\begin{table}[ht!]
\caption{Total cross section and relative uncertainties for the $ZH$ production at a CM energy of  $\sqrt{s}=8$~TeV and for different values of the Higgs-boson mass. Corrections due to individual contributions are included, see section~\ref{sec:vh_inclxs_breakdown}.}
\label{tab:zh_xs_8tev}
\makebox[\textwidth][c]{
\centering
\scriptsize
{\renewcommand{\arraystretch}{1.4}
\begin{tabular}{| cc | c | ccc | cccc | c |}
\cline{1-11}
\multicolumn{2}{|c|}{\(\sqrt{s} = 8\,\mathrm{TeV}\)} &
\multicolumn{1}{|c|}{Theory uncertainty} &
\multicolumn{3}{|c|}{Parametric uncertainty} &
\multicolumn{4}{|c|}{Cross-section decomposition} &
\multicolumn{1}{c|}{Total uncertainty} \\
\(\MH\,[\mathrm{GeV}]\) & \(\sigma\,[\mathrm{fb}]\) &
\(\Delta_{\rm scale}\) &
\(\Delta_{\rm PDF\oplus\alpha_s}\) &
{\color{black!55}\(\Delta_{\rm PDF}\)} & {\color{black!55}\(\Delta_{\alpha_s}\)} &
\(\sigma_{\rm DY}\,[\mathrm{fb}]\) & \(\sigma_{\rm top}\,[\mathrm{fb}]\) & \(\sigma^{\rm NLO}_{ggZH}\,[\mathrm{fb}]\) & \(\delta_{\rm EW}\,[\%]\) &
\(\Delta_{\rm total}\)
\\
\cline{1-11}
$120.00$ & $481.0$ & $^{+2.7}_{-2.4}\%$ & $\pm 1.9\%$ & {\color{black!55}$\pm 1.9\%$} & {\color{black!55}$\pm 0.4\%$} & $458.1$ & $4.5$ & $41.7$ & $-5.09\%$ & $^{+3.3}_{-3.1}\%$ \\
$122.00$ & $457.2$ & $^{+2.8}_{-2.5}\%$ & $\pm 1.9\%$ & {\color{black!55}$\pm 1.9\%$} & {\color{black!55}$\pm 0.4\%$} & $434.5$ & $4.3$ & $40.7$ & $-5.11\%$ & $^{+3.4}_{-3.1}\%$ \\
$124.00$ & $435.1$ & $^{+2.8}_{-2.5}\%$ & $\pm 1.9\%$ & {\color{black!55}$\pm 1.9\%$} & {\color{black!55}$\pm 0.4\%$} & $412.3$ & $4.2$ & $39.8$ & $-5.14\%$ & $^{+3.4}_{-3.2}\%$ \\
$124.60$ & $428.7$ & $^{+2.9}_{-2.6}\%$ & $\pm 1.9\%$ & {\color{black!55}$\pm 1.9\%$} & {\color{black!55}$\pm 0.4\%$} & $405.9$ & $4.1$ & $39.5$ & $-5.15\%$ & $^{+3.5}_{-3.2}\%$ \\
$124.80$ & $426.5$ & $^{+2.9}_{-2.5}\%$ & $\pm 1.9\%$ & {\color{black!55}$\pm 1.9\%$} & {\color{black!55}$\pm 0.4\%$} & $403.8$ & $4.1$ & $39.4$ & $-5.15\%$ & $^{+3.5}_{-3.2}\%$ \\
$125.00$ & $424.4$ & $^{+2.9}_{-2.5}\%$ & $\pm 1.9\%$ & {\color{black!55}$\pm 1.9\%$} & {\color{black!55}$\pm 0.4\%$} & $401.7$ & $4.1$ & $39.3$ & $-5.15\%$ & $^{+3.5}_{-3.2}\%$ \\
\end{tabular}
}
}
\end{table}

\begin{table}[ht!]
\caption{Total cross section and relative uncertainties for the $ZH$ production at a CM energy of  $\sqrt{s}=13$~TeV and for different values of the Higgs-boson mass. Corrections due to individual contributions are included, see section~\ref{sec:vh_inclxs_breakdown}.}
\label{tab:zh_xs_13tev}
\makebox[\textwidth][c]{
\centering
\scriptsize
{\renewcommand{\arraystretch}{1.4}
\begin{tabular}{| cc | c | ccc | cccc | c |}
\cline{1-11}
\multicolumn{2}{|c|}{\(\sqrt{s} = 13\,\mathrm{TeV}\)} &
\multicolumn{1}{|c|}{Theory uncertainty} &
\multicolumn{3}{|c|}{Parametric uncertainty} &
\multicolumn{4}{|c|}{Cross-section decomposition} &
\multicolumn{1}{c|}{Total uncertainty} \\
\(\MH\,[\mathrm{GeV}]\) & \(\sigma\,[\mathrm{fb}]\) &
\(\Delta_{\rm scale}\) &
\(\Delta_{\rm PDF\oplus\alpha_s}\) &
{\color{black!55}\(\Delta_{\rm PDF}\)} & {\color{black!55}\(\Delta_{\alpha_s}\)} &
\(\sigma_{\rm DY}\,[\mathrm{fb}]\) & \(\sigma_{\rm top}\,[\mathrm{fb}]\) & \(\sigma^{\rm NLO}_{ggZH}\,[\mathrm{fb}]\) & \(\delta_{\rm EW}\,[\%]\) &
\(\Delta_{\rm total}\)
\\
\cline{1-11}
$120.00$ & $998.8$ & $^{+2.8}_{-2.5}\%$ & $\pm 1.7\%$ & {\color{black!55}$\pm 1.6\%$} & {\color{black!55}$\pm 0.5\%$} & $910.3$ & $10.3$ & $125.2$ & $-5.17\%$ & $^{+3.3}_{-3.0}\%$ \\
$122.00$ & $952.6$ & $^{+2.9}_{-2.5}\%$ & $\pm 1.7\%$ & {\color{black!55}$\pm 1.6\%$} & {\color{black!55}$\pm 0.5\%$} & $865.0$ & $10.0$ & $122.5$ & $-5.19\%$ & $^{+3.4}_{-3.0}\%$ \\
$124.00$ & $909.0$ & $^{+3.0}_{-2.6}\%$ & $\pm 1.7\%$ & {\color{black!55}$\pm 1.6\%$} & {\color{black!55}$\pm 0.5\%$} & $822.6$ & $9.5$ & $119.8$ & $-5.22\%$ & $^{+3.4}_{-3.1}\%$ \\
$124.60$ & $896.3$ & $^{+3.0}_{-2.6}\%$ & $\pm 1.7\%$ & {\color{black!55}$\pm 1.6\%$} & {\color{black!55}$\pm 0.5\%$} & $810.2$ & $9.4$ & $119.0$ & $-5.23\%$ & $^{+3.4}_{-3.1}\%$ \\
$124.80$ & $892.1$ & $^{+3.0}_{-2.6}\%$ & $\pm 1.7\%$ & {\color{black!55}$\pm 1.6\%$} & {\color{black!55}$\pm 0.5\%$} & $806.2$ & $9.4$ & $118.7$ & $-5.23\%$ & $^{+3.5}_{-3.1}\%$ \\
$125.00$ & $888.2$ & $^{+3.0}_{-2.6}\%$ & $\pm 1.6\%$ & {\color{black!55}$\pm 1.6\%$} & {\color{black!55}$\pm 0.5\%$} & $802.3$ & $9.4$ & $118.4$ & $-5.23\%$ & $^{+3.4}_{-3.1}\%$ \\
\end{tabular}
}
}
\end{table}

\begin{table}[ht!]
\caption{Total cross section and relative uncertainties for the $ZH$ production at a CM energy of  $\sqrt{s}=13.6$~TeV and for different values of the Higgs-boson mass. Corrections due to individual contributions are included, see section~\ref{sec:vh_inclxs_breakdown}.}
\label{tab:zh_xs_13.6tev}
\makebox[\textwidth][c]{
\centering
\scriptsize
{\renewcommand{\arraystretch}{1.4}
\begin{tabular}{| cc | c | ccc | cccc | c |}
\cline{1-11}
\multicolumn{2}{|c|}{\(\sqrt{s} = 13.6\,\mathrm{TeV}\)} &
\multicolumn{1}{|c|}{Theory uncertainty} &
\multicolumn{3}{|c|}{Parametric uncertainty} &
\multicolumn{4}{|c|}{Cross-section decomposition} &
\multicolumn{1}{c|}{Total uncertainty} \\
\(\MH\,[\mathrm{GeV}]\) & \(\sigma\,[\mathrm{fb}]\) &
\(\Delta_{\rm scale}\) &
\(\Delta_{\rm PDF\oplus\alpha_s}\) &
{\color{black!55}\(\Delta_{\rm PDF}\)} & {\color{black!55}\(\Delta_{\alpha_s}\)} &
\(\sigma_{\rm DY}\,[\mathrm{fb}]\) & \(\sigma_{\rm top}\,[\mathrm{fb}]\) & \(\sigma^{\rm NLO}_{ggZH}\,[\mathrm{fb}]\) & \(\delta_{\rm EW}\,[\%]\) &
\(\Delta_{\rm total}\)
\\
\cline{1-11}
$120.00$ & $1065.9$ & $^{+2.8}_{-2.6}\%$ & $\pm 1.6\%$ & {\color{black!55}$\pm 1.5\%$} & {\color{black!55}$\pm 0.5\%$} & $967.3$ & $11.1$ & $137.6$ & $-5.17\%$ & $^{+3.2}_{-3.1}\%$ \\
$122.00$ & $1017.0$ & $^{+2.8}_{-2.6}\%$ & $\pm 1.6\%$ & {\color{black!55}$\pm 1.6\%$} & {\color{black!55}$\pm 0.5\%$} & $919.2$ & $10.7$ & $134.8$ & $-5.20\%$ & $^{+3.3}_{-3.1}\%$ \\
$124.00$ & $970.7$ & $^{+2.9}_{-2.7}\%$ & $\pm 1.6\%$ & {\color{black!55}$\pm 1.5\%$} & {\color{black!55}$\pm 0.5\%$} & $874.2$ & $10.3$ & $131.9$ & $-5.22\%$ & $^{+3.3}_{-3.1}\%$ \\
$124.60$ & $957.4$ & $^{+2.9}_{-2.7}\%$ & $\pm 1.7\%$ & {\color{black!55}$\pm 1.6\%$} & {\color{black!55}$\pm 0.5\%$} & $861.2$ & $10.2$ & $131.1$ & $-5.23\%$ & $^{+3.4}_{-3.2}\%$ \\
$124.80$ & $953.1$ & $^{+2.9}_{-2.7}\%$ & $\pm 1.6\%$ & {\color{black!55}$\pm 1.6\%$} & {\color{black!55}$\pm 0.5\%$} & $857.0$ & $10.1$ & $130.8$ & $-5.24\%$ & $^{+3.4}_{-3.1}\%$ \\
$125.00$ & $948.8$ & $^{+2.9}_{-2.7}\%$ & $\pm 1.6\%$ & {\color{black!55}$\pm 1.5\%$} & {\color{black!55}$\pm 0.5\%$} & $852.9$ & $10.1$ & $130.5$ & $-5.24\%$ & $^{+3.3}_{-3.1}\%$ \\
\end{tabular}
}
}
\end{table}

\begin{table}[ht!]
\caption{Total cross section and relative uncertainties for the $ZH$ production at a CM energy of  $\sqrt{s}=14$~TeV and for different values of the Higgs-boson mass. Corrections due to individual contributions are included, see section~\ref{sec:vh_inclxs_breakdown}.}
\label{tab:zh_xs_14tev}
\makebox[\textwidth][c]{
\centering
\scriptsize
{\renewcommand{\arraystretch}{1.4}
\begin{tabular}{| cc | c | ccc | cccc | c |}
\cline{1-11}
\multicolumn{2}{|c|}{\(\sqrt{s} = 14\,\mathrm{TeV}\)} &
\multicolumn{1}{|c|}{Theory uncertainty} &
\multicolumn{3}{|c|}{Parametric uncertainty} &
\multicolumn{4}{|c|}{Cross-section decomposition} &
\multicolumn{1}{c|}{Total uncertainty} \\
\(\MH\,[\mathrm{GeV}]\) & \(\sigma\,[\mathrm{fb}]\) &
\(\Delta_{\rm scale}\) &
\(\Delta_{\rm PDF\oplus\alpha_s}\) &
{\color{black!55}\(\Delta_{\rm PDF}\)} & {\color{black!55}\(\Delta_{\alpha_s}\)} &
\(\sigma_{\rm DY}\,[\mathrm{fb}]\) & \(\sigma_{\rm top}\,[\mathrm{fb}]\) & \(\sigma^{\rm NLO}_{ggZH}\,[\mathrm{fb}]\) & \(\delta_{\rm EW}\,[\%]\) &
\(\Delta_{\rm total}\)
\\
\cline{1-11}
$120.00$ & $1111.8$ & $^{+2.8}_{-2.6}\%$ & $\pm 1.6\%$ & {\color{black!55}$\pm 1.5\%$} & {\color{black!55}$\pm 0.6\%$} & $1005.5$ & $11.6$ & $146.7$ & $-5.17\%$ & $^{+3.2}_{-3.1}\%$ \\
$122.00$ & $1060.9$ & $^{+2.9}_{-2.7}\%$ & $\pm 1.6\%$ & {\color{black!55}$\pm 1.5\%$} & {\color{black!55}$\pm 0.5\%$} & $955.7$ & $11.3$ & $143.6$ & $-5.20\%$ & $^{+3.3}_{-3.2}\%$ \\
$124.00$ & $1012.7$ & $^{+2.9}_{-2.7}\%$ & $\pm 1.6\%$ & {\color{black!55}$\pm 1.6\%$} & {\color{black!55}$\pm 0.5\%$} & $908.9$ & $10.8$ & $140.5$ & $-5.23\%$ & $^{+3.4}_{-3.2}\%$ \\
$124.60$ & $998.7$ & $^{+3.0}_{-2.7}\%$ & $\pm 1.6\%$ & {\color{black!55}$\pm 1.5\%$} & {\color{black!55}$\pm 0.5\%$} & $895.4$ & $10.7$ & $139.5$ & $-5.24\%$ & $^{+3.4}_{-3.2}\%$ \\
$124.80$ & $994.1$ & $^{+3.0}_{-2.8}\%$ & $\pm 1.6\%$ & {\color{black!55}$\pm 1.5\%$} & {\color{black!55}$\pm 0.5\%$} & $891.0$ & $10.6$ & $139.2$ & $-5.24\%$ & $^{+3.4}_{-3.2}\%$ \\
$125.00$ & $989.5$ & $^{+3.0}_{-2.7}\%$ & $\pm 1.6\%$ & {\color{black!55}$\pm 1.6\%$} & {\color{black!55}$\pm 0.5\%$} & $886.5$ & $10.6$ & $138.9$ & $-5.24\%$ & $^{+3.4}_{-3.2}\%$ \\
\end{tabular}
}
}
\end{table}

\FloatBarrier

\FloatBarrier
\subsubsection{\texorpdfstring{$\ell^+ \nu_{\ell} H$}{l+ nul H} cross section}

\begin{table}[ht!]
\caption{Total cross section and relative uncertainties for the  $W^+ ( \rightarrow \ell^+ \nu_{\ell}) H$ production at a CM energy of  $\sqrt{s}=7$~TeV and for different values of the Higgs-boson mass. Corrections due to individual contributions are included, see section~\ref{sec:vh_inclxs_breakdown}.}
\label{tab:w+_lep_xs_7tev}
\makebox[\textwidth][c]{
\centering
\scriptsize
{\renewcommand{\arraystretch}{1.4}
\begin{tabular}{| cc | c | ccc | cccc | c |}
\cline{1-11}
\multicolumn{2}{|c|}{\(\sqrt{s} = 7\,\mathrm{TeV}\)} &
\multicolumn{1}{|c|}{Theory uncertainty} &
\multicolumn{3}{|c|}{Parametric uncertainty} &
\multicolumn{4}{|c|}{Cross-section decomposition} &
\multicolumn{1}{c|}{Total uncertainty} \\
\(\MH\,[\mathrm{GeV}]\) & \(\sigma\,[\mathrm{fb}]\) &
\(\Delta_{\rm scale}\) &
\(\Delta_{\rm PDF\oplus\alpha_s}\) &
{\color{black!55}\(\Delta_{\rm PDF}\)} & {\color{black!55}\(\Delta_{\alpha_s}\)} &
\(\sigma_{\rm DY}\,[\mathrm{fb}]\) & \(\sigma_{\rm top}\,[\mathrm{fb}]\) & \(\sigma^\gamma\,[\mathrm{fb}]\) & \(\delta_{\rm EW}\,[\%]\) &
\(\Delta_{\rm total}\)
\\
\cline{1-11}
$120.00$ & $47.68$ & $^{+0.7}_{-0.9}\%$ & $\pm 2.3\%$ & {\color{black!55}$\pm 2.3\%$} & {\color{black!55}$\pm 0.4\%$} & $49.51$ & $0.47$ & $1.15$ & $-6.98\%$ & $^{+2.4}_{-2.5}\%$ \\
$122.00$ & $45.19$ & $^{+0.7}_{-0.9}\%$ & $\pm 2.3\%$ & {\color{black!55}$\pm 2.3\%$} & {\color{black!55}$\pm 0.4\%$} & $46.92$ & $0.45$ & $1.13$ & $-7.07\%$ & $^{+2.4}_{-2.5}\%$ \\
$124.00$ & $42.86$ & $^{+0.7}_{-0.9}\%$ & $\pm 2.3\%$ & {\color{black!55}$\pm 2.3\%$} & {\color{black!55}$\pm 0.4\%$} & $44.50$ & $0.43$ & $1.12$ & $-7.15\%$ & $^{+2.4}_{-2.5}\%$ \\
$124.60$ & $42.19$ & $^{+0.7}_{-0.9}\%$ & $\pm 2.3\%$ & {\color{black!55}$\pm 2.3\%$} & {\color{black!55}$\pm 0.4\%$} & $43.80$ & $0.43$ & $1.11$ & $-7.18\%$ & $^{+2.4}_{-2.5}\%$ \\
$124.80$ & $41.97$ & $^{+0.7}_{-0.9}\%$ & $\pm 2.3\%$ & {\color{black!55}$\pm 2.3\%$} & {\color{black!55}$\pm 0.4\%$} & $43.57$ & $0.42$ & $1.11$ & $-7.19\%$ & $^{+2.4}_{-2.5}\%$ \\
$125.00$ & $41.75$ & $^{+0.7}_{-0.9}\%$ & $\pm 2.3\%$ & {\color{black!55}$\pm 2.3\%$} & {\color{black!55}$\pm 0.4\%$} & $43.34$ & $0.42$ & $1.11$ & $-7.20\%$ & $^{+2.4}_{-2.5}\%$ \\
\end{tabular}
}
}
\end{table}

\begin{table}[ht!]
\caption{Total cross section and relative uncertainties for the  $W^+ ( \rightarrow \ell^+ \nu_{\ell}) H$ production at a CM energy of  $\sqrt{s}=8$~TeV and for different values of the Higgs-boson mass. Corrections due to individual contributions are included, see section~\ref{sec:vh_inclxs_breakdown}.}
\label{tab:w+_lep_xs_8tev}
\makebox[\textwidth][c]{
\centering
\scriptsize
{\renewcommand{\arraystretch}{1.4}
\begin{tabular}{| cc | c | ccc | cccc | c |}
\cline{1-11}
\multicolumn{2}{|c|}{\(\sqrt{s} = 8\,\mathrm{TeV}\)} &
\multicolumn{1}{|c|}{Theory uncertainty} &
\multicolumn{3}{|c|}{Parametric uncertainty} &
\multicolumn{4}{|c|}{Cross-section decomposition} &
\multicolumn{1}{c|}{Total uncertainty} \\
\(\MH\,[\mathrm{GeV}]\) & \(\sigma\,[\mathrm{fb}]\) &
\(\Delta_{\rm scale}\) &
\(\Delta_{\rm PDF\oplus\alpha_s}\) &
{\color{black!55}\(\Delta_{\rm PDF}\)} & {\color{black!55}\(\Delta_{\alpha_s}\)} &
\(\sigma_{\rm DY}\,[\mathrm{fb}]\) & \(\sigma_{\rm top}\,[\mathrm{fb}]\) & \(\sigma^\gamma\,[\mathrm{fb}]\) & \(\delta_{\rm EW}\,[\%]\) &
\(\Delta_{\rm total}\)
\\
\cline{1-11}
$120.00$ & $57.47$ & $^{+0.6}_{-0.9}\%$ & $\pm 2.2\%$ & {\color{black!55}$\pm 2.1\%$} & {\color{black!55}$\pm 0.4\%$} & $59.50$ & $0.60$ & $1.56$ & $-7.02\%$ & $^{+2.2}_{-2.3}\%$ \\
$122.00$ & $54.52$ & $^{+0.6}_{-0.9}\%$ & $\pm 2.2\%$ & {\color{black!55}$\pm 2.1\%$} & {\color{black!55}$\pm 0.4\%$} & $56.42$ & $0.57$ & $1.53$ & $-7.11\%$ & $^{+2.3}_{-2.3}\%$ \\
$124.00$ & $51.75$ & $^{+0.6}_{-0.9}\%$ & $\pm 2.2\%$ & {\color{black!55}$\pm 2.1\%$} & {\color{black!55}$\pm 0.4\%$} & $53.53$ & $0.55$ & $1.51$ & $-7.20\%$ & $^{+2.2}_{-2.3}\%$ \\
$124.60$ & $50.95$ & $^{+0.6}_{-0.9}\%$ & $\pm 2.2\%$ & {\color{black!55}$\pm 2.1\%$} & {\color{black!55}$\pm 0.4\%$} & $52.70$ & $0.54$ & $1.51$ & $-7.23\%$ & $^{+2.2}_{-2.3}\%$ \\
$124.80$ & $50.68$ & $^{+0.6}_{-0.9}\%$ & $\pm 2.2\%$ & {\color{black!55}$\pm 2.1\%$} & {\color{black!55}$\pm 0.4\%$} & $52.43$ & $0.54$ & $1.51$ & $-7.24\%$ & $^{+2.3}_{-2.3}\%$ \\
$125.00$ & $50.42$ & $^{+0.6}_{-0.9}\%$ & $\pm 2.2\%$ & {\color{black!55}$\pm 2.1\%$} & {\color{black!55}$\pm 0.4\%$} & $52.16$ & $0.54$ & $1.51$ & $-7.26\%$ & $^{+2.3}_{-2.3}\%$ \\
\end{tabular}
}
}
\end{table}

\begin{table}[ht!]
\caption{Total cross section and relative uncertainties for the  $W^+ ( \rightarrow \ell^+ \nu_{\ell}) H$ production at a CM energy of  $\sqrt{s}=13$~TeV and for different values of the Higgs-boson mass. Corrections due to individual contributions are included, see section~\ref{sec:vh_inclxs_breakdown}.}
\label{tab:w+_lep_xs_13tev}
\makebox[\textwidth][c]{
\centering
\scriptsize
{\renewcommand{\arraystretch}{1.4}
\begin{tabular}{| cc | c | ccc | cccc | c |}
\cline{1-11}
\multicolumn{2}{|c|}{\(\sqrt{s} = 13\,\mathrm{TeV}\)} &
\multicolumn{1}{|c|}{Theory uncertainty} &
\multicolumn{3}{|c|}{Parametric uncertainty} &
\multicolumn{4}{|c|}{Cross-section decomposition} &
\multicolumn{1}{c|}{Total uncertainty} \\
\(\MH\,[\mathrm{GeV}]\) & \(\sigma\,[\mathrm{fb}]\) &
\(\Delta_{\rm scale}\) &
\(\Delta_{\rm PDF\oplus\alpha_s}\) &
{\color{black!55}\(\Delta_{\rm PDF}\)} & {\color{black!55}\(\Delta_{\alpha_s}\)} &
\(\sigma_{\rm DY}\,[\mathrm{fb}]\) & \(\sigma_{\rm top}\,[\mathrm{fb}]\) & \(\sigma^\gamma\,[\mathrm{fb}]\) & \(\delta_{\rm EW}\,[\%]\) &
\(\Delta_{\rm total}\)
\\
\cline{1-11}
$120.00$ & $109.15$ & $^{+0.5}_{-0.7}\%$ & $\pm 1.8\%$ & {\color{black!55}$\pm 1.7\%$} & {\color{black!55}$\pm 0.6\%$} & $111.58$ & $1.35$ & $4.22$ & $-7.16\%$ & $^{+1.9}_{-1.9}\%$ \\
$122.00$ & $103.77$ & $^{+0.5}_{-0.7}\%$ & $\pm 1.8\%$ & {\color{black!55}$\pm 1.7\%$} & {\color{black!55}$\pm 0.5\%$} & $106.00$ & $1.29$ & $4.18$ & $-7.26\%$ & $^{+1.9}_{-1.9}\%$ \\
$124.00$ & $98.73$ & $^{+0.5}_{-0.7}\%$ & $\pm 1.8\%$ & {\color{black!55}$\pm 1.7\%$} & {\color{black!55}$\pm 0.6\%$} & $100.76$ & $1.23$ & $4.13$ & $-7.34\%$ & $^{+1.9}_{-1.9}\%$ \\
$124.60$ & $97.29$ & $^{+0.5}_{-0.8}\%$ & $\pm 1.8\%$ & {\color{black!55}$\pm 1.7\%$} & {\color{black!55}$\pm 0.5\%$} & $99.25$ & $1.23$ & $4.12$ & $-7.36\%$ & $^{+1.9}_{-1.9}\%$ \\
$124.80$ & $96.80$ & $^{+0.5}_{-0.7}\%$ & $\pm 1.8\%$ & {\color{black!55}$\pm 1.7\%$} & {\color{black!55}$\pm 0.5\%$} & $98.75$ & $1.22$ & $4.11$ & $-7.38\%$ & $^{+1.8}_{-1.9}\%$ \\
$125.00$ & $96.31$ & $^{+0.5}_{-0.7}\%$ & $\pm 1.8\%$ & {\color{black!55}$\pm 1.7\%$} & {\color{black!55}$\pm 0.5\%$} & $98.26$ & $1.21$ & $4.11$ & $-7.40\%$ & $^{+1.8}_{-1.9}\%$ \\
\end{tabular}
}
}
\end{table}

\begin{table}[ht!]
\caption{Total cross section and relative uncertainties for the  $W^+ ( \rightarrow \ell^+ \nu_{\ell}) H$ production at a CM energy of  $\sqrt{s}=13.6$~TeV and for different values of the Higgs-boson mass. Corrections due to individual contributions are included, see section~\ref{sec:vh_inclxs_breakdown}.}
\label{tab:w+_lep_xs_13.6tev}
\makebox[\textwidth][c]{
\centering
\scriptsize
{\renewcommand{\arraystretch}{1.4}
\begin{tabular}{| cc | c | ccc | cccc | c |}
\cline{1-11}
\multicolumn{2}{|c|}{\(\sqrt{s} = 13.6\,\mathrm{TeV}\)} &
\multicolumn{1}{|c|}{Theory uncertainty} &
\multicolumn{3}{|c|}{Parametric uncertainty} &
\multicolumn{4}{|c|}{Cross-section decomposition} &
\multicolumn{1}{c|}{Total uncertainty} \\
\(\MH\,[\mathrm{GeV}]\) & \(\sigma\,[\mathrm{fb}]\) &
\(\Delta_{\rm scale}\) &
\(\Delta_{\rm PDF\oplus\alpha_s}\) &
{\color{black!55}\(\Delta_{\rm PDF}\)} & {\color{black!55}\(\Delta_{\alpha_s}\)} &
\(\sigma_{\rm DY}\,[\mathrm{fb}]\) & \(\sigma_{\rm top}\,[\mathrm{fb}]\) & \(\sigma^\gamma\,[\mathrm{fb}]\) & \(\delta_{\rm EW}\,[\%]\) &
\(\Delta_{\rm total}\)
\\
\cline{1-11}
$120.00$ & $115.60$ & $^{+0.5}_{-0.7}\%$ & $\pm 1.8\%$ & {\color{black!55}$\pm 1.7\%$} & {\color{black!55}$\pm 0.6\%$} & $118.00$ & $1.44$ & $4.60$ & $-7.16\%$ & $^{+1.8}_{-1.9}\%$ \\
$122.00$ & $109.90$ & $^{+0.5}_{-0.7}\%$ & $\pm 1.8\%$ & {\color{black!55}$\pm 1.7\%$} & {\color{black!55}$\pm 0.6\%$} & $112.11$ & $1.38$ & $4.55$ & $-7.27\%$ & $^{+1.8}_{-1.9}\%$ \\
$124.00$ & $104.57$ & $^{+0.5}_{-0.7}\%$ & $\pm 1.8\%$ & {\color{black!55}$\pm 1.7\%$} & {\color{black!55}$\pm 0.5\%$} & $106.59$ & $1.33$ & $4.50$ & $-7.36\%$ & $^{+1.8}_{-1.9}\%$ \\
$124.60$ & $103.05$ & $^{+0.5}_{-0.7}\%$ & $\pm 1.8\%$ & {\color{black!55}$\pm 1.7\%$} & {\color{black!55}$\pm 0.5\%$} & $104.99$ & $1.32$ & $4.49$ & $-7.38\%$ & $^{+1.8}_{-1.9}\%$ \\
$124.80$ & $102.55$ & $^{+0.5}_{-0.7}\%$ & $\pm 1.8\%$ & {\color{black!55}$\pm 1.7\%$} & {\color{black!55}$\pm 0.6\%$} & $104.47$ & $1.32$ & $4.49$ & $-7.39\%$ & $^{+1.8}_{-1.9}\%$ \\
$125.00$ & $102.02$ & $^{+0.5}_{-0.7}\%$ & $\pm 1.8\%$ & {\color{black!55}$\pm 1.7\%$} & {\color{black!55}$\pm 0.5\%$} & $103.95$ & $1.30$ & $4.48$ & $-7.41\%$ & $^{+1.9}_{-1.9}\%$ \\
\end{tabular}
}
}
\end{table}

\begin{table}[ht!]
\caption{Total cross section and relative uncertainties for the  $W^+ ( \rightarrow \ell^+ \nu_{\ell}) H$ production at a CM energy of  $\sqrt{s}=14$~TeV and for different values of the Higgs-boson mass. Corrections due to individual contributions are included, see section~\ref{sec:vh_inclxs_breakdown}.}
\label{tab:w+_lep_xs_14tev}
\makebox[\textwidth][c]{
\centering
\scriptsize
{\renewcommand{\arraystretch}{1.4}
\begin{tabular}{| cc | c | ccc | cccc | c |}
\cline{1-11}
\multicolumn{2}{|c|}{\(\sqrt{s} = 14\,\mathrm{TeV}\)} &
\multicolumn{1}{|c|}{Theory uncertainty} &
\multicolumn{3}{|c|}{Parametric uncertainty} &
\multicolumn{4}{|c|}{Cross-section decomposition} &
\multicolumn{1}{c|}{Total uncertainty} \\
\(\MH\,[\mathrm{GeV}]\) & \(\sigma\,[\mathrm{fb}]\) &
\(\Delta_{\rm scale}\) &
\(\Delta_{\rm PDF\oplus\alpha_s}\) &
{\color{black!55}\(\Delta_{\rm PDF}\)} & {\color{black!55}\(\Delta_{\alpha_s}\)} &
\(\sigma_{\rm DY}\,[\mathrm{fb}]\) & \(\sigma_{\rm top}\,[\mathrm{fb}]\) & \(\sigma^\gamma\,[\mathrm{fb}]\) & \(\delta_{\rm EW}\,[\%]\) &
\(\Delta_{\rm total}\)
\\
\cline{1-11}
$120.00$ & $119.89$ & $^{+0.5}_{-0.7}\%$ & $\pm 1.8\%$ & {\color{black!55}$\pm 1.7\%$} & {\color{black!55}$\pm 0.6\%$} & $122.29$ & $1.50$ & $4.86$ & $-7.17\%$ & $^{+1.8}_{-1.9}\%$ \\
$122.00$ & $114.01$ & $^{+0.5}_{-0.7}\%$ & $\pm 1.7\%$ & {\color{black!55}$\pm 1.7\%$} & {\color{black!55}$\pm 0.6\%$} & $116.21$ & $1.45$ & $4.81$ & $-7.28\%$ & $^{+1.8}_{-1.9}\%$ \\
$124.00$ & $108.49$ & $^{+0.5}_{-0.7}\%$ & $\pm 1.7\%$ & {\color{black!55}$\pm 1.7\%$} & {\color{black!55}$\pm 0.6\%$} & $110.48$ & $1.39$ & $4.76$ & $-7.36\%$ & $^{+1.8}_{-1.9}\%$ \\
$124.60$ & $106.92$ & $^{+0.5}_{-0.7}\%$ & $\pm 1.7\%$ & {\color{black!55}$\pm 1.7\%$} & {\color{black!55}$\pm 0.6\%$} & $108.83$ & $1.38$ & $4.74$ & $-7.39\%$ & $^{+1.8}_{-1.9}\%$ \\
$124.80$ & $106.37$ & $^{+0.5}_{-0.7}\%$ & $\pm 1.8\%$ & {\color{black!55}$\pm 1.7\%$} & {\color{black!55}$\pm 0.6\%$} & $108.29$ & $1.36$ & $4.74$ & $-7.40\%$ & $^{+1.8}_{-1.9}\%$ \\
$125.00$ & $105.86$ & $^{+0.5}_{-0.7}\%$ & $\pm 1.8\%$ & {\color{black!55}$\pm 1.7\%$} & {\color{black!55}$\pm 0.6\%$} & $107.75$ & $1.37$ & $4.74$ & $-7.42\%$ & $^{+1.8}_{-1.9}\%$ \\
\end{tabular}
}
}
\end{table}

\FloatBarrier
\subsubsection{\texorpdfstring{$\ell^- \bar{\nu}_{\ell} H$}{l- nul-bar H} cross section}

\begin{table}[ht!]
\caption{Total cross section and relative uncertainties for the $W^- ( \rightarrow \ell^- \bar{\nu}_{\ell}) H$ production at a CM energy of  $\sqrt{s}=7$~TeV and for different values of the Higgs-boson mass. Corrections due to individual contributions are included, see section~\ref{sec:vh_inclxs_breakdown}.}
\label{tab:w-_lep_xs_7tev}
\makebox[\textwidth][c]{
\centering
\scriptsize
{\renewcommand{\arraystretch}{1.4}
\begin{tabular}{| cc | c | ccc | cccc | c |}
\cline{1-11}
\multicolumn{2}{|c|}{\(\sqrt{s} = 7\,\mathrm{TeV}\)} &
\multicolumn{1}{|c|}{Theory uncertainty} &
\multicolumn{3}{|c|}{Parametric uncertainty} &
\multicolumn{4}{|c|}{Cross-section decomposition} &
\multicolumn{1}{c|}{Total uncertainty} \\
\(\MH\,[\mathrm{GeV}]\) & \(\sigma\,[\mathrm{fb}]\) &
\(\Delta_{\rm scale}\) &
\(\Delta_{\rm PDF\oplus\alpha_s}\) &
{\color{black!55}\(\Delta_{\rm PDF}\)} & {\color{black!55}\(\Delta_{\alpha_s}\)} &
\(\sigma_{\rm DY}\,[\mathrm{fb}]\) & \(\sigma_{\rm top}\,[\mathrm{fb}]\) & \(\sigma^\gamma\,[\mathrm{fb}]\) & \(\delta_{\rm EW}\,[\%]\) &
\(\Delta_{\rm total}\)
\\
\cline{1-11}
$120.00$ & $26.56$ & $^{+0.6}_{-0.9}\%$ & $\pm 2.3\%$ & {\color{black!55}$\pm 2.3\%$} & {\color{black!55}$\pm 0.4\%$} & $27.50$ & $0.27$ & $0.67$ & $-6.80\%$ & $^{+2.4}_{-2.5}\%$ \\
$122.00$ & $25.11$ & $^{+0.6}_{-0.9}\%$ & $\pm 2.4\%$ & {\color{black!55}$\pm 2.3\%$} & {\color{black!55}$\pm 0.4\%$} & $25.99$ & $0.26$ & $0.66$ & $-6.88\%$ & $^{+2.4}_{-2.5}\%$ \\
$124.00$ & $23.75$ & $^{+0.6}_{-0.9}\%$ & $\pm 2.4\%$ & {\color{black!55}$\pm 2.3\%$} & {\color{black!55}$\pm 0.4\%$} & $24.58$ & $0.24$ & $0.65$ & $-6.98\%$ & $^{+2.5}_{-2.5}\%$ \\
$124.60$ & $23.37$ & $^{+0.6}_{-0.9}\%$ & $\pm 2.4\%$ & {\color{black!55}$\pm 2.3\%$} & {\color{black!55}$\pm 0.4\%$} & $24.17$ & $0.24$ & $0.64$ & $-7.00\%$ & $^{+2.5}_{-2.5}\%$ \\
$124.80$ & $23.24$ & $^{+0.6}_{-0.9}\%$ & $\pm 2.4\%$ & {\color{black!55}$\pm 2.4\%$} & {\color{black!55}$\pm 0.4\%$} & $24.04$ & $0.24$ & $0.64$ & $-7.01\%$ & $^{+2.5}_{-2.5}\%$ \\
$125.00$ & $23.11$ & $^{+0.6}_{-0.9}\%$ & $\pm 2.4\%$ & {\color{black!55}$\pm 2.4\%$} & {\color{black!55}$\pm 0.4\%$} & $23.91$ & $0.24$ & $0.64$ & $-7.02\%$ & $^{+2.5}_{-2.5}\%$ \\
\end{tabular}
}
}
\end{table}

\begin{table}[ht!]
\caption{Total cross section and relative uncertainties for the $W^- ( \rightarrow \ell^- \bar{\nu}_{\ell}) H$ production at a CM energy of  $\sqrt{s}=8$~TeV and for different values of the Higgs-boson mass. Corrections due to individual contributions are included, see section~\ref{sec:vh_inclxs_breakdown}.}
\label{tab:w-_lep_xs_8tev}
\makebox[\textwidth][c]{
\centering
\scriptsize
{\renewcommand{\arraystretch}{1.4}
\begin{tabular}{| cc | c | ccc | cccc | c |}
\cline{1-11}
\multicolumn{2}{|c|}{\(\sqrt{s} = 8\,\mathrm{TeV}\)} &
\multicolumn{1}{|c|}{Theory uncertainty} &
\multicolumn{3}{|c|}{Parametric uncertainty} &
\multicolumn{4}{|c|}{Cross-section decomposition} &
\multicolumn{1}{c|}{Total uncertainty} \\
\(\MH\,[\mathrm{GeV}]\) & \(\sigma\,[\mathrm{fb}]\) &
\(\Delta_{\rm scale}\) &
\(\Delta_{\rm PDF\oplus\alpha_s}\) &
{\color{black!55}\(\Delta_{\rm PDF}\)} & {\color{black!55}\(\Delta_{\alpha_s}\)} &
\(\sigma_{\rm DY}\,[\mathrm{fb}]\) & \(\sigma_{\rm top}\,[\mathrm{fb}]\) & \(\sigma^\gamma\,[\mathrm{fb}]\) & \(\delta_{\rm EW}\,[\%]\) &
\(\Delta_{\rm total}\)
\\
\cline{1-11}
$120.00$ & $33.00$ & $^{+0.6}_{-0.8}\%$ & $\pm 2.2\%$ & {\color{black!55}$\pm 2.2\%$} & {\color{black!55}$\pm 0.4\%$} & $34.06$ & $0.35$ & $0.92$ & $-6.85\%$ & $^{+2.3}_{-2.3}\%$ \\
$122.00$ & $31.23$ & $^{+0.6}_{-0.8}\%$ & $\pm 2.2\%$ & {\color{black!55}$\pm 2.2\%$} & {\color{black!55}$\pm 0.4\%$} & $32.22$ & $0.34$ & $0.91$ & $-6.94\%$ & $^{+2.3}_{-2.4}\%$ \\
$124.00$ & $29.57$ & $^{+0.6}_{-0.8}\%$ & $\pm 2.2\%$ & {\color{black!55}$\pm 2.2\%$} & {\color{black!55}$\pm 0.4\%$} & $30.49$ & $0.32$ & $0.90$ & $-7.02\%$ & $^{+2.3}_{-2.4}\%$ \\
$124.60$ & $29.09$ & $^{+0.6}_{-0.8}\%$ & $\pm 2.2\%$ & {\color{black!55}$\pm 2.2\%$} & {\color{black!55}$\pm 0.4\%$} & $30.00$ & $0.31$ & $0.89$ & $-7.05\%$ & $^{+2.3}_{-2.4}\%$ \\
$124.80$ & $28.93$ & $^{+0.6}_{-0.8}\%$ & $\pm 2.2\%$ & {\color{black!55}$\pm 2.2\%$} & {\color{black!55}$\pm 0.4\%$} & $29.83$ & $0.31$ & $0.89$ & $-7.07\%$ & $^{+2.3}_{-2.4}\%$ \\
$125.00$ & $28.78$ & $^{+0.6}_{-0.8}\%$ & $\pm 2.2\%$ & {\color{black!55}$\pm 2.2\%$} & {\color{black!55}$\pm 0.4\%$} & $29.67$ & $0.31$ & $0.89$ & $-7.07\%$ & $^{+2.3}_{-2.4}\%$ \\
\end{tabular}
}
}
\end{table}

\begin{table}[ht!]
\caption{Total cross section and relative uncertainties for the $W^- ( \rightarrow \ell^- \bar{\nu}_{\ell}) H$ production at a CM energy of  $\sqrt{s}=13$~TeV and for different values of the Higgs-boson mass. Corrections due to individual contributions are included, see section~\ref{sec:vh_inclxs_breakdown}.}
\label{tab:w-_lep_xs_13tev}
\makebox[\textwidth][c]{
\centering
\scriptsize
{\renewcommand{\arraystretch}{1.4}
\begin{tabular}{| cc | c | ccc | cccc | c |}
\cline{1-11}
\multicolumn{2}{|c|}{\(\sqrt{s} = 13\,\mathrm{TeV}\)} &
\multicolumn{1}{|c|}{Theory uncertainty} &
\multicolumn{3}{|c|}{Parametric uncertainty} &
\multicolumn{4}{|c|}{Cross-section decomposition} &
\multicolumn{1}{c|}{Total uncertainty} \\
\(\MH\,[\mathrm{GeV}]\) & \(\sigma\,[\mathrm{fb}]\) &
\(\Delta_{\rm scale}\) &
\(\Delta_{\rm PDF\oplus\alpha_s}\) &
{\color{black!55}\(\Delta_{\rm PDF}\)} & {\color{black!55}\(\Delta_{\alpha_s}\)} &
\(\sigma_{\rm DY}\,[\mathrm{fb}]\) & \(\sigma_{\rm top}\,[\mathrm{fb}]\) & \(\sigma^\gamma\,[\mathrm{fb}]\) & \(\delta_{\rm EW}\,[\%]\) &
\(\Delta_{\rm total}\)
\\
\cline{1-11}
$120.00$ & $68.91$ & $^{+0.4}_{-0.7}\%$ & $\pm 1.8\%$ & {\color{black!55}$\pm 1.8\%$} & {\color{black!55}$\pm 0.5\%$} & $70.26$ & $0.86$ & $2.71$ & $-7.01\%$ & $^{+1.9}_{-2.0}\%$ \\
$122.00$ & $65.39$ & $^{+0.4}_{-0.7}\%$ & $\pm 1.8\%$ & {\color{black!55}$\pm 1.8\%$} & {\color{black!55}$\pm 0.5\%$} & $66.61$ & $0.83$ & $2.68$ & $-7.10\%$ & $^{+1.9}_{-2.0}\%$ \\
$124.00$ & $62.09$ & $^{+0.4}_{-0.7}\%$ & $\pm 1.8\%$ & {\color{black!55}$\pm 1.8\%$} & {\color{black!55}$\pm 0.5\%$} & $63.19$ & $0.79$ & $2.65$ & $-7.18\%$ & $^{+1.9}_{-1.9}\%$ \\
$124.60$ & $61.15$ & $^{+0.4}_{-0.7}\%$ & $\pm 1.8\%$ & {\color{black!55}$\pm 1.8\%$} & {\color{black!55}$\pm 0.5\%$} & $62.21$ & $0.78$ & $2.64$ & $-7.21\%$ & $^{+1.9}_{-2.0}\%$ \\
$124.80$ & $60.83$ & $^{+0.4}_{-0.7}\%$ & $\pm 1.8\%$ & {\color{black!55}$\pm 1.8\%$} & {\color{black!55}$\pm 0.5\%$} & $61.88$ & $0.78$ & $2.64$ & $-7.22\%$ & $^{+1.9}_{-2.0}\%$ \\
$125.00$ & $60.52$ & $^{+0.4}_{-0.7}\%$ & $\pm 1.8\%$ & {\color{black!55}$\pm 1.8\%$} & {\color{black!55}$\pm 0.5\%$} & $61.56$ & $0.77$ & $2.64$ & $-7.23\%$ & $^{+1.9}_{-2.0}\%$ \\
\end{tabular}
}
}
\end{table}

\begin{table}[ht!]
\caption{Total cross section and relative uncertainties for the $W^- ( \rightarrow \ell^- \bar{\nu}_{\ell}) H$ production at a CM energy of  $\sqrt{s}=13.6$~TeV and for different values of the Higgs-boson mass. Corrections due to individual contributions are included, see section~\ref{sec:vh_inclxs_breakdown}.}
\label{tab:w-_lep_xs_13.6tev}
\makebox[\textwidth][c]{
\centering
\scriptsize
{\renewcommand{\arraystretch}{1.4}
\begin{tabular}{| cc | c | ccc | cccc | c |}
\cline{1-11}
\multicolumn{2}{|c|}{\(\sqrt{s} = 13.6\,\mathrm{TeV}\)} &
\multicolumn{1}{|c|}{Theory uncertainty} &
\multicolumn{3}{|c|}{Parametric uncertainty} &
\multicolumn{4}{|c|}{Cross-section decomposition} &
\multicolumn{1}{c|}{Total uncertainty} \\
\(\MH\,[\mathrm{GeV}]\) & \(\sigma\,[\mathrm{fb}]\) &
\(\Delta_{\rm scale}\) &
\(\Delta_{\rm PDF\oplus\alpha_s}\) &
{\color{black!55}\(\Delta_{\rm PDF}\)} & {\color{black!55}\(\Delta_{\alpha_s}\)} &
\(\sigma_{\rm DY}\,[\mathrm{fb}]\) & \(\sigma_{\rm top}\,[\mathrm{fb}]\) & \(\sigma^\gamma\,[\mathrm{fb}]\) & \(\delta_{\rm EW}\,[\%]\) &
\(\Delta_{\rm total}\)
\\
\cline{1-11}
$120.00$ & $73.53$ & $^{+0.4}_{-0.7}\%$ & $\pm 1.8\%$ & {\color{black!55}$\pm 1.7\%$} & {\color{black!55}$\pm 0.6\%$} & $74.88$ & $0.94$ & $2.98$ & $-7.02\%$ & $^{+1.8}_{-1.9}\%$ \\
$122.00$ & $69.80$ & $^{+0.4}_{-0.7}\%$ & $\pm 1.8\%$ & {\color{black!55}$\pm 1.7\%$} & {\color{black!55}$\pm 0.6\%$} & $71.00$ & $0.89$ & $2.94$ & $-7.11\%$ & $^{+1.9}_{-1.9}\%$ \\
$124.00$ & $66.30$ & $^{+0.4}_{-0.7}\%$ & $\pm 1.8\%$ & {\color{black!55}$\pm 1.7\%$} & {\color{black!55}$\pm 0.5\%$} & $67.37$ & $0.86$ & $2.91$ & $-7.20\%$ & $^{+1.9}_{-1.9}\%$ \\
$124.60$ & $65.28$ & $^{+0.4}_{-0.7}\%$ & $\pm 1.8\%$ & {\color{black!55}$\pm 1.7\%$} & {\color{black!55}$\pm 0.5\%$} & $66.32$ & $0.85$ & $2.90$ & $-7.23\%$ & $^{+1.9}_{-1.9}\%$ \\
$124.80$ & $64.95$ & $^{+0.4}_{-0.7}\%$ & $\pm 1.8\%$ & {\color{black!55}$\pm 1.7\%$} & {\color{black!55}$\pm 0.5\%$} & $65.98$ & $0.84$ & $2.90$ & $-7.23\%$ & $^{+1.9}_{-1.9}\%$ \\
$125.00$ & $64.63$ & $^{+0.4}_{-0.7}\%$ & $\pm 1.8\%$ & {\color{black!55}$\pm 1.7\%$} & {\color{black!55}$\pm 0.5\%$} & $65.64$ & $0.84$ & $2.89$ & $-7.23\%$ & $^{+1.9}_{-1.9}\%$ \\
\end{tabular}
}
}
\end{table}

\begin{table}[ht!]
\caption{Total cross section and relative uncertainties for the $W^- ( \rightarrow \ell^- \bar{\nu}_{\ell}) H$ production at a CM energy of  $\sqrt{s}=14$~TeV and for different values of the Higgs-boson mass. Corrections due to individual contributions are included, see section~\ref{sec:vh_inclxs_breakdown}.}
\label{tab:w-_lep_xs_14tev}
\makebox[\textwidth][c]{
\centering
\scriptsize
{\renewcommand{\arraystretch}{1.4}
\begin{tabular}{| cc | c | ccc | cccc | c |}
\cline{1-11}
\multicolumn{2}{|c|}{\(\sqrt{s} = 14\,\mathrm{TeV}\)} &
\multicolumn{1}{|c|}{Theory uncertainty} &
\multicolumn{3}{|c|}{Parametric uncertainty} &
\multicolumn{4}{|c|}{Cross-section decomposition} &
\multicolumn{1}{c|}{Total uncertainty} \\
\(\MH\,[\mathrm{GeV}]\) & \(\sigma\,[\mathrm{fb}]\) &
\(\Delta_{\rm scale}\) &
\(\Delta_{\rm PDF\oplus\alpha_s}\) &
{\color{black!55}\(\Delta_{\rm PDF}\)} & {\color{black!55}\(\Delta_{\alpha_s}\)} &
\(\sigma_{\rm DY}\,[\mathrm{fb}]\) & \(\sigma_{\rm top}\,[\mathrm{fb}]\) & \(\sigma^\gamma\,[\mathrm{fb}]\) & \(\delta_{\rm EW}\,[\%]\) &
\(\Delta_{\rm total}\)
\\
\cline{1-11}
$120.00$ & $76.64$ & $^{+0.4}_{-0.7}\%$ & $\pm 1.8\%$ & {\color{black!55}$\pm 1.7\%$} & {\color{black!55}$\pm 0.6\%$} & $77.98$ & $0.98$ & $3.16$ & $-7.03\%$ & $^{+1.8}_{-1.9}\%$ \\
$122.00$ & $72.76$ & $^{+0.4}_{-0.7}\%$ & $\pm 1.8\%$ & {\color{black!55}$\pm 1.7\%$} & {\color{black!55}$\pm 0.6\%$} & $73.96$ & $0.94$ & $3.12$ & $-7.11\%$ & $^{+1.8}_{-1.9}\%$ \\
$124.00$ & $69.11$ & $^{+0.4}_{-0.7}\%$ & $\pm 1.8\%$ & {\color{black!55}$\pm 1.7\%$} & {\color{black!55}$\pm 0.5\%$} & $70.18$ & $0.90$ & $3.09$ & $-7.20\%$ & $^{+1.8}_{-1.9}\%$ \\
$124.60$ & $68.07$ & $^{+0.4}_{-0.7}\%$ & $\pm 1.8\%$ & {\color{black!55}$\pm 1.7\%$} & {\color{black!55}$\pm 0.6\%$} & $69.09$ & $0.90$ & $3.08$ & $-7.23\%$ & $^{+1.8}_{-1.9}\%$ \\
$124.80$ & $67.72$ & $^{+0.4}_{-0.7}\%$ & $\pm 1.8\%$ & {\color{black!55}$\pm 1.7\%$} & {\color{black!55}$\pm 0.5\%$} & $68.73$ & $0.89$ & $3.07$ & $-7.24\%$ & $^{+1.8}_{-1.9}\%$ \\
$125.00$ & $67.38$ & $^{+0.4}_{-0.7}\%$ & $\pm 1.8\%$ & {\color{black!55}$\pm 1.7\%$} & {\color{black!55}$\pm 0.6\%$} & $68.38$ & $0.88$ & $3.07$ & $-7.25\%$ & $^{+1.8}_{-1.9}\%$ \\
\end{tabular}
}
}
\end{table}

\FloatBarrier

\begin{landscape}
\FloatBarrier
\subsubsection{\texorpdfstring{$\nu \bar{\nu} H$}{nu nu-bar H} cross section}

\begin{table}[ht!]
\caption{Total cross section and relative uncertainties for the $Z ( \rightarrow 
\nu \bar{\nu}) H$ production at a CM energy of  $\sqrt{s}=7$~TeV and for different values of the Higgs-boson mass. Corrections due to individual contributions are included, see section~\ref{sec:vh_inclxs_breakdown}.}
\label{tab:zh_nu_xs_7tev}
\centering
\scriptsize
{\renewcommand{\arraystretch}{1.4}
\begin{tabular}{| cc | c | ccc | ccccc | c |}
\cline{1-12}
\multicolumn{2}{|c|}{\(\sqrt{s} = 7\,\mathrm{TeV}\)} &
\multicolumn{1}{|c|}{Theory uncertainty} &
\multicolumn{3}{|c|}{Parametric uncertainty} &
\multicolumn{5}{|c|}{Cross-section decomposition} &
\multicolumn{1}{c|}{Total uncertainty} \\
\(\MH\,[\mathrm{GeV}]\) & \(\sigma\,[\mathrm{fb}]\) &
\(\Delta_{\rm scale}\) &
\(\Delta_{\rm PDF\oplus\alpha_s}\) &
{\color{black!55}\(\Delta_{\rm PDF}\)} & {\color{black!55}\(\Delta_{\alpha_s}\)} &
\(\sigma_{\rm DY}\,[\mathrm{fb}]\) & \(\sigma_{\rm top}\,[\mathrm{fb}]\) & \(\sigma^{\rm NLO}_{ggZH}\,[\mathrm{fb}]\) & \(\sigma^\gamma\,[\mathrm{fb}]\) & \(\delta_{\rm EW}\,[\%]\) &
\(\Delta_{\rm total}\)
\\
\cline{1-12}
$120.00$ & $78.04$ & $^{+2.5}_{-2.3}\%$ & $\pm 2.0\%$ & {\color{black!55}$\pm 2.0\%$} & {\color{black!55}$\pm 0.4\%$} & $74.58$ & $0.71$ & $4.91$ & $0.00$ & $-4.19\%$ & $^{+3.2}_{-3.0}\%$ \\
$122.00$ & $74.15$ & $^{+2.5}_{-2.3}\%$ & $\pm 2.0\%$ & {\color{black!55}$\pm 2.0\%$} & {\color{black!55}$\pm 0.4\%$} & $70.69$ & $0.68$ & $4.81$ & $0.00$ & $-4.21\%$ & $^{+3.3}_{-3.1}\%$ \\
$124.00$ & $70.47$ & $^{+2.6}_{-2.4}\%$ & $\pm 2.0\%$ & {\color{black!55}$\pm 2.0\%$} & {\color{black!55}$\pm 0.3\%$} & $67.03$ & $0.65$ & $4.70$ & $0.00$ & $-4.24\%$ & $^{+3.3}_{-3.1}\%$ \\
$124.60$ & $69.40$ & $^{+2.7}_{-2.4}\%$ & $\pm 2.1\%$ & {\color{black!55}$\pm 2.0\%$} & {\color{black!55}$\pm 0.3\%$} & $65.98$ & $0.64$ & $4.66$ & $0.00$ & $-4.24\%$ & $^{+3.4}_{-3.1}\%$ \\
$124.80$ & $69.06$ & $^{+2.6}_{-2.4}\%$ & $\pm 2.1\%$ & {\color{black!55}$\pm 2.0\%$} & {\color{black!55}$\pm 0.3\%$} & $65.63$ & $0.64$ & $4.66$ & $0.00$ & $-4.24\%$ & $^{+3.3}_{-3.1}\%$ \\
$125.00$ & $68.71$ & $^{+2.6}_{-2.4}\%$ & $\pm 2.0\%$ & {\color{black!55}$\pm 2.0\%$} & {\color{black!55}$\pm 0.3\%$} & $65.29$ & $0.63$ & $4.65$ & $0.00$ & $-4.25\%$ & $^{+3.3}_{-3.1}\%$ \\
\end{tabular}
}
\end{table}

\begin{table}[ht!]
\caption{Total cross section and relative uncertainties for the $Z ( \rightarrow 
\nu \bar{\nu}) H$ production at a CM energy of  $\sqrt{s}=8$~TeV and for different values of the Higgs-boson mass. Corrections due to individual contributions are included, see section~\ref{sec:vh_inclxs_breakdown}.}
\label{tab:zh_nu_xs_8tev}
\centering
\scriptsize
{\renewcommand{\arraystretch}{1.4}
\begin{tabular}{| cc | c | ccc | ccccc | c |}
\cline{1-12}
\multicolumn{2}{|c|}{\(\sqrt{s} = 8\,\mathrm{TeV}\)} &
\multicolumn{1}{|c|}{Theory uncertainty} &
\multicolumn{3}{|c|}{Parametric uncertainty} &
\multicolumn{5}{|c|}{Cross-section decomposition} &
\multicolumn{1}{c|}{Total uncertainty} \\
\(\MH\,[\mathrm{GeV}]\) & \(\sigma\,[\mathrm{fb}]\) &
\(\Delta_{\rm scale}\) &
\(\Delta_{\rm PDF\oplus\alpha_s}\) &
{\color{black!55}\(\Delta_{\rm PDF}\)} & {\color{black!55}\(\Delta_{\alpha_s}\)} &
\(\sigma_{\rm DY}\,[\mathrm{fb}]\) & \(\sigma_{\rm top}\,[\mathrm{fb}]\) & \(\sigma^{\rm NLO}_{ggZH}\,[\mathrm{fb}]\) & \(\sigma^\gamma\,[\mathrm{fb}]\) & \(\delta_{\rm EW}\,[\%]\) &
\(\Delta_{\rm total}\)
\\
\cline{1-12}
$120.00$ & $96.63$ & $^{+2.7}_{-2.4}\%$ & $\pm 1.9\%$ & {\color{black!55}$\pm 1.9\%$} & {\color{black!55}$\pm 0.4\%$} & $91.26$ & $0.90$ & $7.02$ & $0.00$ & $-4.21\%$ & $^{+3.3}_{-3.1}\%$ \\
$122.00$ & $91.85$ & $^{+2.8}_{-2.4}\%$ & $\pm 1.9\%$ & {\color{black!55}$\pm 1.9\%$} & {\color{black!55}$\pm 0.4\%$} & $86.55$ & $0.86$ & $6.82$ & $0.00$ & $-4.24\%$ & $^{+3.4}_{-3.1}\%$ \\
$124.00$ & $87.40$ & $^{+2.8}_{-2.5}\%$ & $\pm 1.9\%$ & {\color{black!55}$\pm 1.9\%$} & {\color{black!55}$\pm 0.4\%$} & $82.14$ & $0.83$ & $6.69$ & $0.00$ & $-4.26\%$ & $^{+3.4}_{-3.2}\%$ \\
$124.60$ & $86.11$ & $^{+2.8}_{-2.5}\%$ & $\pm 1.9\%$ & {\color{black!55}$\pm 1.9\%$} & {\color{black!55}$\pm 0.4\%$} & $80.86$ & $0.82$ & $6.63$ & $0.00$ & $-4.27\%$ & $^{+3.4}_{-3.2}\%$ \\
$124.80$ & $85.68$ & $^{+2.9}_{-2.5}\%$ & $\pm 1.9\%$ & {\color{black!55}$\pm 1.9\%$} & {\color{black!55}$\pm 0.4\%$} & $80.45$ & $0.82$ & $6.62$ & $0.00$ & $-4.27\%$ & $^{+3.4}_{-3.2}\%$ \\
$125.00$ & $85.26$ & $^{+2.9}_{-2.5}\%$ & $\pm 1.9\%$ & {\color{black!55}$\pm 1.9\%$} & {\color{black!55}$\pm 0.4\%$} & $80.03$ & $0.81$ & $6.61$ & $0.00$ & $-4.27\%$ & $^{+3.5}_{-3.2}\%$ \\
\end{tabular}
}
\end{table}

\begin{table}[ht!]
\caption{Total cross section and relative uncertainties for the $Z ( \rightarrow 
\nu \bar{\nu}) H$ production at a CM energy of  $\sqrt{s}=13$~TeV and for different values of the Higgs-boson mass. Corrections due to individual contributions are included, see section~\ref{sec:vh_inclxs_breakdown}.}
\label{tab:zh_nu_xs_13tev}
\centering
\scriptsize
{\renewcommand{\arraystretch}{1.4}
%
}
\end{table}

\begin{table}[ht!]
\caption{Total cross section and relative uncertainties for the $Z ( \rightarrow 
\nu \bar{\nu}) H$ production at a CM energy of  $\sqrt{s}=13.6$~TeV and for different values of the Higgs-boson mass. Corrections due to individual contributions are included, see section~\ref{sec:vh_inclxs_breakdown}.}
\label{tab:zh_nu_xs_13.6tev}
\centering
\scriptsize
{\renewcommand{\arraystretch}{1.4}
%
}
\end{table}

\begin{table}[ht!]
\caption{Total cross section and relative uncertainties for the $Z ( \rightarrow 
\nu \bar{\nu}) H$ production at a CM energy of  $\sqrt{s}=14$~TeV and for different values of the Higgs-boson mass. Corrections due to individual contributions are included, see section~\ref{sec:vh_inclxs_breakdown}.}
\label{tab:zh_nu_xs_14tev}
\centering
\scriptsize
{\renewcommand{\arraystretch}{1.4}
%
}
\end{table}

\FloatBarrier
\subsubsection{\texorpdfstring{$\ell^+ \ell^- H$}{l+ l- H} cross section}

\begin{table}[ht!]
\caption{Total cross section and relative uncertainties for the $Z ( \rightarrow \ell^+ \ell^-) H$ production at a CM energy of  $\sqrt{s}=7$~TeV and for different values of the Higgs-boson mass. Corrections due to individual contributions are included, see section~\ref{sec:vh_inclxs_breakdown}.}
\label{tab:zh_lep_xs_7tev}
\centering
\scriptsize
{\renewcommand{\arraystretch}{1.4}
%
}
\end{table}

\begin{table}[ht!]
\caption{Total cross section and relative uncertainties for the $Z ( \rightarrow \ell^+ \ell^-) H$ production at a CM energy of  $\sqrt{s}=8$~TeV and for different values of the Higgs-boson mass. Corrections due to individual contributions are included, see section~\ref{sec:vh_inclxs_breakdown}.}
\label{tab:zh_lep_xs_8tev}
\centering
\scriptsize
{\renewcommand{\arraystretch}{1.4}
%
}
\end{table}

\begin{table}[ht!]
\caption{Total cross section and relative uncertainties for the $Z ( \rightarrow \ell^+ \ell^-) H$ production at a CM energy of  $\sqrt{s}=13$~TeV and for different values of the Higgs-boson mass. Corrections due to individual contributions are included, see section~\ref{sec:vh_inclxs_breakdown}.}
\label{tab:zh_lep_xs_13tev}
\centering
\scriptsize
{\renewcommand{\arraystretch}{1.4}
%
}
\end{table}

\begin{table}[ht!]
\caption{Total cross section and relative uncertainties for the $Z ( \rightarrow \ell^+ \ell^-) H$ production at a CM energy of  $\sqrt{s}=13.6$~TeV and for different values of the Higgs-boson mass. Corrections due to individual contributions are included, see section~\ref{sec:vh_inclxs_breakdown}.}
\label{tab:zh_lep_xs_13.6tev}
\centering
\scriptsize
{\renewcommand{\arraystretch}{1.4}
%
}
\end{table}

\begin{table}[ht!]
\caption{Total cross section and relative uncertainties for the $Z ( \rightarrow \ell^+ \ell^-) H$ production at a CM energy of  $\sqrt{s}=14$~TeV and for different values of the Higgs-boson mass. Corrections due to individual contributions are included, see section~\ref{sec:vh_inclxs_breakdown}.}
\label{tab:zh_lep_xs_14tev}
\centering
\scriptsize
{\renewcommand{\arraystretch}{1.4}
%
}
\end{table}

\FloatBarrier
\end{landscape}

\FloatBarrier
\subsection{\texorpdfstring{$\boldsymbol{t\bar{t}H}$}{ttH} and \texorpdfstring{$\boldsymbol{tH}$}{tH}}

\FloatBarrier
\subsubsection{\texorpdfstring{$\boldsymbol{t\bar{t}H}$}{ttH}}

\begin{table}[ht!]
\caption{\label{tab:tth-tot13} Predictions for the process $t\bar tH$ as from Ref.~\cite{Balsach:2025jcw}, at $\sqrt{s} = 13$ TeV. Predictions include effects at
NNLO QCD, supplemented by NNLL soft-gluon resummation, and NLO effects of EW origin~\cite{Frixione:2014qaa,Frixione:2015zaa,Kulesza:2015vda,Broggio:2015lya,Broggio:2016lfj,
Kulesza:2017ukk,Frederix:2018nkq,Kulesza:2018tqz,Broggio:2019ewu,Kulesza:2020nfh,
Catani:2022mfv,Devoto:2024nhl}. See Sect.~\ref{sec:results-xsects-tth} for details.}
\centering
\scriptsize
{\renewcommand{\arraystretch}{1.4}
\begin{tabular}{| cc | ccc | ccc | c |}
\cline{1-9}
\multicolumn{2}{|c|}{\(\sqrt{s} = 13\,\mathrm{TeV}\)} &
\multicolumn{3}{|c|}{Theory uncertainty} &
\multicolumn{3}{|c|}{Parametric uncertainty} &
\multicolumn{1}{c|}{Total uncertainty} \\
\(\MH\,[\mathrm{GeV}]\) & \(\sigma\,[\mathrm{fb}]\) &
\(\Delta_{\rm th}\) &
{\color{black!55}\(\Delta_{\rm scale}\)} & {\color{black!55}\(\Delta_{\rm virt}\)} &
\(\Delta_{{\rm PDF}\oplus\alphas}\) &
{\color{black!55}\(\Delta_{\rm PDF}\)} & {\color{black!55}\(\Delta_{\alphas}\)} &
\(\Delta_{\rm total}\)
\\
\cline{1-9}
$120.00$ & $588.7$ & $^{+2.0}_{-2.5}\%$ & {\color{black!55}$^{+1.8}_{-2.3}\%$} & {\color{black!55}$\pm 0.9\%$} & $\pm 2.9\%$ & {\color{black!55}$\pm 2.3\%$} & {\color{black!55}$\pm 1.7\%$} & $^{+3.5}_{-3.8}\%$ \\
$122.00$ & $565.0$ & $^{+1.9}_{-2.5}\%$ & {\color{black!55}$^{+1.7}_{-2.3}\%$} & {\color{black!55}$\pm 0.9\%$} & $\pm 2.9\%$ & {\color{black!55}$\pm 2.3\%$} & {\color{black!55}$\pm 1.7\%$} & $^{+3.4}_{-3.8}\%$ \\
$124.00$ & $541.3$ & $^{+1.9}_{-2.5}\%$ & {\color{black!55}$^{+1.7}_{-2.3}\%$} & {\color{black!55}$\pm 0.9\%$} & $\pm 2.9\%$ & {\color{black!55}$\pm 2.3\%$} & {\color{black!55}$\pm 1.7\%$} & $^{+3.4}_{-3.8}\%$ \\
$124.60$ & $534.2$ & $^{+1.8}_{-2.4}\%$ & {\color{black!55}$^{+1.6}_{-2.2}\%$} & {\color{black!55}$\pm 0.9\%$} & $\pm 2.9\%$ & {\color{black!55}$\pm 2.3\%$} & {\color{black!55}$\pm 1.7\%$} & $^{+3.4}_{-3.7}\%$ \\
$124.80$ & $531.8$ & $^{+1.8}_{-2.4}\%$ & {\color{black!55}$^{+1.6}_{-2.2}\%$} & {\color{black!55}$\pm 0.9\%$} & $\pm 2.9\%$ & {\color{black!55}$\pm 2.3\%$} & {\color{black!55}$\pm 1.7\%$} & $^{+3.4}_{-3.7}\%$ \\
$125.00$ & $530.6$ & $^{+1.9}_{-2.4}\%$ & {\color{black!55}$^{+1.7}_{-2.3}\%$} & {\color{black!55}$\pm 0.9\%$} & $\pm 2.9\%$ & {\color{black!55}$\pm 2.3\%$} & {\color{black!55}$\pm 1.7\%$} & $^{+3.4}_{-3.7}\%$ \\
$125.09$ & $528.8$ & $^{+1.8}_{-2.4}\%$ & {\color{black!55}$^{+1.6}_{-2.2}\%$} & {\color{black!55}$\pm 0.9\%$} & $\pm 2.9\%$ & {\color{black!55}$\pm 2.3\%$} & {\color{black!55}$\pm 1.7\%$} & $^{+3.4}_{-3.7}\%$ \\
$125.20$ & $527.1$ & $^{+1.8}_{-2.4}\%$ & {\color{black!55}$^{+1.6}_{-2.2}\%$} & {\color{black!55}$\pm 0.9\%$} & $\pm 2.9\%$ & {\color{black!55}$\pm 2.3\%$} & {\color{black!55}$\pm 1.7\%$} & $^{+3.4}_{-3.7}\%$ \\
$125.30$ & $525.9$ & $^{+1.8}_{-2.4}\%$ & {\color{black!55}$^{+1.6}_{-2.2}\%$} & {\color{black!55}$\pm 0.9\%$} & $\pm 2.9\%$ & {\color{black!55}$\pm 2.3\%$} & {\color{black!55}$\pm 1.7\%$} & $^{+3.4}_{-3.7}\%$ \\
$125.38$ & $524.9$ & $^{+1.8}_{-2.4}\%$ & {\color{black!55}$^{+1.6}_{-2.2}\%$} & {\color{black!55}$\pm 0.9\%$} & $\pm 2.9\%$ & {\color{black!55}$\pm 2.3\%$} & {\color{black!55}$\pm 1.7\%$} & $^{+3.4}_{-3.7}\%$ \\
$125.60$ & $522.1$ & $^{+1.8}_{-2.4}\%$ & {\color{black!55}$^{+1.6}_{-2.2}\%$} & {\color{black!55}$\pm 0.9\%$} & $\pm 2.9\%$ & {\color{black!55}$\pm 2.3\%$} & {\color{black!55}$\pm 1.7\%$} & $^{+3.4}_{-3.7}\%$ \\
$126.00$ & $517.6$ & $^{+1.8}_{-2.4}\%$ & {\color{black!55}$^{+1.6}_{-2.2}\%$} & {\color{black!55}$\pm 0.9\%$} & $\pm 2.9\%$ & {\color{black!55}$\pm 2.3\%$} & {\color{black!55}$\pm 1.7\%$} & $^{+3.4}_{-3.7}\%$ \\
$128.00$ & $493.9$ & $^{+1.7}_{-2.4}\%$ & {\color{black!55}$^{+1.5}_{-2.2}\%$} & {\color{black!55}$\pm 0.9\%$} & $\pm 2.9\%$ & {\color{black!55}$\pm 2.3\%$} & {\color{black!55}$\pm 1.7\%$} & $^{+3.4}_{-3.7}\%$ \\
$130.00$ & $470.2$ & $^{+1.7}_{-2.3}\%$ & {\color{black!55}$^{+1.5}_{-2.1}\%$} & {\color{black!55}$\pm 0.9\%$} & $\pm 2.9\%$ & {\color{black!55}$\pm 2.3\%$} & {\color{black!55}$\pm 1.7\%$} & $^{+3.4}_{-3.7}\%$
 \\
\cline{1-9}
\end{tabular}
}
\end{table}

\begin{table}[ht!]
\caption{\label{tab:tth-tot136} Predictions for the process $t\bar tH$ as from Ref.~\cite{Balsach:2025jcw}, at $\sqrt{s} = 13.6$ TeV. Predictions include effects at
NNLO QCD, supplemented by NNLL soft-gluon resummation, and NLO effects of EW origin~\cite{Frixione:2014qaa,Frixione:2015zaa,Kulesza:2015vda,Broggio:2015lya,Broggio:2016lfj,
Kulesza:2017ukk,Frederix:2018nkq,Kulesza:2018tqz,Broggio:2019ewu,Kulesza:2020nfh,
Catani:2022mfv,Devoto:2024nhl}. See Sect.~\ref{sec:results-xsects-tth} for details.}
\centering
\scriptsize
{\renewcommand{\arraystretch}{1.4}
\begin{tabular}{| cc | ccc | ccc | c |}
\cline{1-9}
\multicolumn{2}{|c|}{\(\sqrt{s} = 13.6\,\mathrm{TeV}\)} &
\multicolumn{3}{|c|}{Theory uncertainty} &
\multicolumn{3}{|c|}{Parametric uncertainty} &
\multicolumn{1}{c|}{Total uncertainty} \\
\(\MH\,[\mathrm{GeV}]\) & \(\sigma\,[\mathrm{fb}]\) &
\(\Delta_{\rm th}\) &
{\color{black!55}\(\Delta_{\rm scale}\)} & {\color{black!55}\(\Delta_{\rm virt}\)} &
\(\Delta_{{\rm PDF}\oplus\alphas}\) &
{\color{black!55}\(\Delta_{\rm PDF}\)} & {\color{black!55}\(\Delta_{\alphas}\)} &
\(\Delta_{\rm total}\)
\\
\cline{1-9}
$120.00$ & $661.1$ & $^{+1.4}_{-2.2}\%$ & {\color{black!55}$^{+1.1}_{-2.0}\%$} & {\color{black!55}$\pm 0.9\%$} & $\pm 2.8\%$ & {\color{black!55}$\pm 2.2\%$} & {\color{black!55}$\pm 1.7\%$} & $^{+3.1}_{-3.5}\%$ \\
$122.00$ & $634.2$ & $^{+1.6}_{-2.3}\%$ & {\color{black!55}$^{+1.3}_{-2.1}\%$} & {\color{black!55}$\pm 0.9\%$} & $\pm 2.8\%$ & {\color{black!55}$\pm 2.2\%$} & {\color{black!55}$\pm 1.7\%$} & $^{+3.2}_{-3.6}\%$ \\
$124.00$ & $607.4$ & $^{+1.7}_{-2.3}\%$ & {\color{black!55}$^{+1.4}_{-2.1}\%$} & {\color{black!55}$\pm 0.9\%$} & $\pm 2.8\%$ & {\color{black!55}$\pm 2.2\%$} & {\color{black!55}$\pm 1.7\%$} & $^{+3.2}_{-3.6}\%$ \\
$124.60$ & $599.3$ & $^{+1.7}_{-2.4}\%$ & {\color{black!55}$^{+1.5}_{-2.2}\%$} & {\color{black!55}$\pm 0.9\%$} & $\pm 2.8\%$ & {\color{black!55}$\pm 2.2\%$} & {\color{black!55}$\pm 1.7\%$} & $^{+3.3}_{-3.7}\%$ \\
$124.80$ & $596.6$ & $^{+1.7}_{-2.4}\%$ & {\color{black!55}$^{+1.5}_{-2.2}\%$} & {\color{black!55}$\pm 0.9\%$} & $\pm 2.8\%$ & {\color{black!55}$\pm 2.2\%$} & {\color{black!55}$\pm 1.7\%$} & $^{+3.3}_{-3.7}\%$ \\
$125.00$ & $592.5$ & $^{+1.7}_{-2.3}\%$ & {\color{black!55}$^{+1.4}_{-2.1}\%$} & {\color{black!55}$\pm 0.9\%$} & $\pm 2.8\%$ & {\color{black!55}$\pm 2.2\%$} & {\color{black!55}$\pm 1.7\%$} & $^{+3.2}_{-3.6}\%$ \\
$125.09$ & $592.1$ & $^{+1.7}_{-2.4}\%$ & {\color{black!55}$^{+1.5}_{-2.2}\%$} & {\color{black!55}$\pm 0.9\%$} & $\pm 2.8\%$ & {\color{black!55}$\pm 2.2\%$} & {\color{black!55}$\pm 1.7\%$} & $^{+3.3}_{-3.7}\%$ \\
$125.20$ & $591.2$ & $^{+1.7}_{-2.4}\%$ & {\color{black!55}$^{+1.5}_{-2.2}\%$} & {\color{black!55}$\pm 0.9\%$} & $\pm 2.8\%$ & {\color{black!55}$\pm 2.2\%$} & {\color{black!55}$\pm 1.7\%$} & $^{+3.3}_{-3.7}\%$ \\
$125.30$ & $589.9$ & $^{+1.7}_{-2.4}\%$ & {\color{black!55}$^{+1.5}_{-2.2}\%$} & {\color{black!55}$\pm 0.9\%$} & $\pm 2.8\%$ & {\color{black!55}$\pm 2.2\%$} & {\color{black!55}$\pm 1.7\%$} & $^{+3.3}_{-3.7}\%$ \\
$125.38$ & $588.8$ & $^{+1.7}_{-2.4}\%$ & {\color{black!55}$^{+1.5}_{-2.2}\%$} & {\color{black!55}$\pm 0.9\%$} & $\pm 2.8\%$ & {\color{black!55}$\pm 2.2\%$} & {\color{black!55}$\pm 1.7\%$} & $^{+3.3}_{-3.7}\%$ \\
$125.60$ & $586.0$ & $^{+1.8}_{-2.4}\%$ & {\color{black!55}$^{+1.6}_{-2.2}\%$} & {\color{black!55}$\pm 0.9\%$} & $\pm 2.8\%$ & {\color{black!55}$\pm 2.2\%$} & {\color{black!55}$\pm 1.7\%$} & $^{+3.3}_{-3.7}\%$ \\
$126.00$ & $580.5$ & $^{+1.8}_{-2.4}\%$ & {\color{black!55}$^{+1.6}_{-2.2}\%$} & {\color{black!55}$\pm 0.9\%$} & $\pm 2.8\%$ & {\color{black!55}$\pm 2.2\%$} & {\color{black!55}$\pm 1.7\%$} & $^{+3.3}_{-3.7}\%$ \\
$128.00$ & $553.6$ & $^{+1.9}_{-2.5}\%$ & {\color{black!55}$^{+1.7}_{-2.3}\%$} & {\color{black!55}$\pm 0.9\%$} & $\pm 2.8\%$ & {\color{black!55}$\pm 2.2\%$} & {\color{black!55}$\pm 1.7\%$} & $^{+3.4}_{-3.7}\%$ \\
$130.00$ & $526.8$ & $^{+2.1}_{-2.6}\%$ & {\color{black!55}$^{+1.9}_{-2.4}\%$} & {\color{black!55}$\pm 0.9\%$} & $\pm 2.8\%$ & {\color{black!55}$\pm 2.2\%$} & {\color{black!55}$\pm 1.7\%$} & $^{+3.5}_{-3.8}\%$
 \\
\cline{1-9}
\end{tabular}
}
\end{table}

\begin{table}[ht!]
\caption{\label{tab:tth-tot14} Predictions for the process $t\bar tH$ as from Ref.~\cite{Balsach:2025jcw}, at $\sqrt{s} = 14$ TeV. Predictions include effects at
NNLO QCD, supplemented by NNLL soft-gluon resummation, and NLO effects of EW origin~\cite{Frixione:2014qaa,Frixione:2015zaa,Kulesza:2015vda,Broggio:2015lya,Broggio:2016lfj,
Kulesza:2017ukk,Frederix:2018nkq,Kulesza:2018tqz,Broggio:2019ewu,Kulesza:2020nfh,
Catani:2022mfv,Devoto:2024nhl}. See Sect.~\ref{sec:results-xsects-tth} for details.}
\centering
\scriptsize
{\renewcommand{\arraystretch}{1.4}

}
\end{table}

\FloatBarrier
\subsubsection{\texorpdfstring{$\boldsymbol{tH}$}{tH} (t-channel)}

\begin{table}[ht!]
\caption{\label{tab:th-t-tot13} Predictions for the process $tH + \bar t H$ ($t$-channel), at $\sqrt{s}=$13 TeV, computed at NLO QCD accuracy following Ref.~\cite{Demartin:2015uha}. See Sect.~\ref{sec:results-xsects-singleth} for details.}
\centering
\scriptsize
{\renewcommand{\arraystretch}{1.4}
%
}
\end{table}

\begin{table}[ht!]
\caption{\label{tab:th-t-tot136} Predictions for the process $tH + \bar t H$ ($t$-channel), at $\sqrt{s}=$13.6 TeV, computed at NLO QCD accuracy following Ref.~\cite{Demartin:2015uha}. See Sect.~\ref{sec:results-xsects-singleth} for details.}
\centering
\scriptsize
{\renewcommand{\arraystretch}{1.4}
%
}
\end{table}

\begin{table}[ht!]
\caption{\label{tab:th-t-tot14} Predictions for the process $tH + \bar t H$ ($t$-channel), at $\sqrt{s}=$14 TeV, computed at NLO QCD accuracy following Ref.~\cite{Demartin:2015uha}. See Sect.~\ref{sec:results-xsects-singleth} for details.}
\centering
\scriptsize
{\renewcommand{\arraystretch}{1.4}
%
}
\end{table}
\begin{table}[ht!]
\caption{\label{tab:th-t-top13} Predictions for the process $tH$ ($t$-channel), at $\sqrt{s}=$13 TeV, computed at NLO QCD accuracy following Ref.~\cite{Demartin:2015uha}. See Sect.~\ref{sec:results-xsects-singleth} for details.}
\centering
\scriptsize
{\renewcommand{\arraystretch}{1.4}
%
}
\end{table}

\begin{table}[ht!]
\caption{\label{tab:th-t-top136} Predictions for the process $tH$ ($t$-channel), at $\sqrt{s}=$13.6 TeV, computed at NLO QCD accuracy following Ref.~\cite{Demartin:2015uha}. See Sect.~\ref{sec:results-xsects-singleth} for details.}
\centering
\scriptsize
{\renewcommand{\arraystretch}{1.4}
%
}
\end{table}

\begin{table}[ht!]
\caption{\label{tab:th-t-top14} Predictions for the process $tH$ ($t$-channel), at $\sqrt{s}=$14 TeV, computed at NLO QCD accuracy following Ref.~\cite{Demartin:2015uha}. See Sect.~\ref{sec:results-xsects-singleth} for details.}
\centering
\scriptsize
{\renewcommand{\arraystretch}{1.4}
%
}
\end{table}
\begin{table}[ht!]
\caption{\label{tab:th-t-atop13} Predictions for the process $\bar tH$ ($t$-channel), at $\sqrt{s}=$13 TeV, computed at NLO QCD accuracy following Ref.~\cite{Demartin:2015uha}. See Sect.~\ref{sec:results-xsects-singleth} for details.}
\centering
\scriptsize
{\renewcommand{\arraystretch}{1.4}
%
}
\end{table}

\begin{table}[ht!]
\caption{\label{tab:th-t-atop136} Predictions for the process $\bar tH$ ($t$-channel), at $\sqrt{s}=$13.6 TeV, computed at NLO QCD accuracy following Ref.~\cite{Demartin:2015uha}. See Sect.~\ref{sec:results-xsects-singleth} for details.}
\centering
\scriptsize
{\renewcommand{\arraystretch}{1.4}
%
}
\end{table}

\begin{table}[ht!]
\caption{\label{tab:th-t-atop14} Predictions for the process $\bar tH$ ($t$-channel), at $\sqrt{s}=$14 TeV, computed at NLO QCD accuracy following Ref.~\cite{Demartin:2015uha}. See Sect.~\ref{sec:results-xsects-singleth} for details.}
\centering
\scriptsize
{\renewcommand{\arraystretch}{1.4}
%
}
\end{table}

\FloatBarrier
\subsubsection{\texorpdfstring{$\boldsymbol{tH}$}{tH} (s-channel)}

\begin{table}[ht!]
\caption{\label{tab:th-s-tot13} Predictions for the process $tH + \bar t H$ ($s$-channel), at $\sqrt{s}=$13 TeV, computed at NLO QCD accuracy following Ref.~\cite{Demartin:2015uha}. See Sect.~\ref{sec:results-xsects-singleth} for details.}
\centering
\scriptsize
{\renewcommand{\arraystretch}{1.4}
%
}
\end{table}

\begin{table}[ht!]
\caption{\label{tab:th-s-tot136} Predictions for the process $tH + \bar t H$ ($s$-channel), at $\sqrt{s}=$13.6 TeV, computed at NLO QCD accuracy following Ref.~\cite{Demartin:2015uha}. See Sect.~\ref{sec:results-xsects-singleth} for details.}
\centering
\scriptsize
{\renewcommand{\arraystretch}{1.4}
%
}
\end{table}

\begin{table}[ht!]
\caption{\label{tab:th-s-tot14} Predictions for the process $tH + \bar t H$ ($s$-channel), at $\sqrt{s}=$14 TeV, computed at NLO QCD accuracy following Ref.~\cite{Demartin:2015uha}. See Sect.~\ref{sec:results-xsects-singleth} for details.}
\centering
\scriptsize
{\renewcommand{\arraystretch}{1.4}
%
}
\end{table}
\begin{table}[ht!]
\caption{\label{tab:th-s-top13} Predictions for the process $tH$ ($s$-channel), at $\sqrt{s}=$13 TeV, computed at NLO QCD accuracy following Ref.~\cite{Demartin:2015uha}. See Sect.~\ref{sec:results-xsects-singleth} for details.}
\centering
\scriptsize
{\renewcommand{\arraystretch}{1.4}
%
}
\end{table}

\begin{table}[ht!]
\caption{\label{tab:th-s-top136} Predictions for the process $tH$ ($s$-channel), at $\sqrt{s}=$13.6 TeV, computed at NLO QCD accuracy following Ref.~\cite{Demartin:2015uha}. See Sect.~\ref{sec:results-xsects-singleth} for details.}
\centering
\scriptsize
{\renewcommand{\arraystretch}{1.4}
%
}
\end{table}

\begin{table}[ht!]
\caption{\label{tab:th-s-top14} Predictions for the process $tH$ ($s$-channel), at $\sqrt{s}=$14 TeV, computed at NLO QCD accuracy following Ref.~\cite{Demartin:2015uha}. See Sect.~\ref{sec:results-xsects-singleth} for details.}
\centering
\scriptsize
{\renewcommand{\arraystretch}{1.4}
%
}
\end{table}
\begin{table}[ht!]
\caption{\label{tab:th-s-atop13} Predictions for the process $\bar tH$ ($s$-channel), at $\sqrt{s}=$13 TeV, computed at NLO QCD accuracy following Ref.~\cite{Demartin:2015uha}. See Sect.~\ref{sec:results-xsects-singleth} for details.}
\centering
\scriptsize
{\renewcommand{\arraystretch}{1.4}
%
}
\end{table}

\begin{table}[ht!]
\caption{\label{tab:th-s-atop136} Predictions for the process $\bar tH$ ($s$-channel), at $\sqrt{s}=$13.6 TeV, computed at NLO QCD accuracy following Ref.~\cite{Demartin:2015uha}. See Sect.~\ref{sec:results-xsects-singleth} for details.}
\centering
\scriptsize
{\renewcommand{\arraystretch}{1.4}
%
}
\end{table}

\begin{table}[ht!]
\caption{\label{tab:th-s-atop14} Predictions for the process $\bar tH$ ($s$-channel), at $\sqrt{s}=$14 TeV, computed at NLO QCD accuracy following Ref.~\cite{Demartin:2015uha}. See Sect.~\ref{sec:results-xsects-singleth} for details.}
\centering
\scriptsize
{\renewcommand{\arraystretch}{1.4}
%
}
\end{table}

\FloatBarrier
\subsubsection{\texorpdfstring{$\boldsymbol{tWH}$}{tWH}}

\begin{table}[ht!]
\caption{\label{tab:thw-tot13} Predictions for the process $tHW^- + \bar t HW^+$ at NLO QCD accuracy, at $\sqrt{s}=$13 TeV. The rate of each of the two processes taken alone is
    half of the rate of their sum. Predictions have been computed in the DR2 scheme, following Ref.~\cite{Demartin:2016axk}. See Sect.~\ref{sec:results-xsects-singleth} for details.}
\centering
\scriptsize
{\renewcommand{\arraystretch}{1.4}
%
}
\end{table}

\begin{table}[ht!]
\caption{\label{tab:thw-tot136} Predictions for the process $tHW^- + \bar t HW^+$ at NLO QCD accuracy, at $\sqrt{s}=$13.6 TeV. The rate of each of the two processes taken alone is
    half of the rate of their sum. Predictions have been computed in the DR2 scheme, following Ref.~\cite{Demartin:2016axk}. See Sect.~\ref{sec:results-xsects-singleth} for details.}
\centering
\scriptsize
{\renewcommand{\arraystretch}{1.4}
%
}
\end{table}

\begin{table}[ht!]
\caption{\label{tab:thw-tot14} Predictions for the process $tHW^- + \bar t HW^+$ at NLO QCD accuracy, at $\sqrt{s}=$14 TeV. The rate of each of the two processes taken alone is
    half of the rate of their sum. Predictions have been computed in the DR2 scheme, following Ref.~\cite{Demartin:2016axk}. See Sect.~\ref{sec:results-xsects-singleth} for details.}
\centering
\scriptsize
{\renewcommand{\arraystretch}{1.4}
%
}
\end{table}

\clearpage
\subsection{\texorpdfstring{$\boldsymbol{b\bar{b}H}$}{bbH}}
\begin{table}[ht!]
\caption{Total \bbH{} cross sections in the SM for a LHC CM energy of $\sqrt{s}=7$ TeV obtained using the \nlonnllpart{} method for matching NNLO 5FS and NLO 4FS corrections in QCD. The results are presented with both theoretical and parametric uncertainty. The sources of uncertainty are defined in the main text.\\}
\label{tab:bbH7}
\makebox[\textwidth][c]{%
\centering
\scriptsize
{\renewcommand{\arraystretch}{1.4}
%
}
}
\end{table}

\begin{table}[ht!]
\caption{Total \bbH{} cross sections in the SM for a LHC CM energy of $\sqrt{s}=8$ TeV obtained using the \nlonnllpart{} method for matching NNLO 5FS and NLO 4FS corrections in QCD. The results are presented with both theoretical and parametric uncertainty. The sources of uncertainty are defined in the main text.\\}
\label{tab:bbH8}
\makebox[\textwidth][c]{%
\centering
\scriptsize
{\renewcommand{\arraystretch}{1.4}
%
}
}
\end{table}

\begin{table}[ht!]
\caption{Total \bbH{} cross sections in the SM for a LHC CM energy of $\sqrt{s}=13$ TeV obtained using the \nlonnllpart{} method for matching NNLO 5FS and NLO 4FS corrections in QCD. The results are presented with both theoretical and parametric uncertainty. The sources of uncertainty are defined in the main text.\\}
\label{tab:bbH13}
\makebox[\textwidth][c]{%
\centering
\scriptsize
{\renewcommand{\arraystretch}{1.4}
%
}
}
\end{table}

\begin{table}[ht!]
\caption{Total \bbH{} cross sections in the SM for a LHC CM energy of $\sqrt{s}=14$ TeV obtained using the \nlonnllpart{} method for matching NNLO 5FS and NLO 4FS corrections in QCD. The results are presented with both theoretical and parametric uncertainty. The sources of uncertainty are defined in the main text.\\}
\label{tab:bbH14}
\makebox[\textwidth][c]{%
\centering
\scriptsize
{\renewcommand{\arraystretch}{1.4}
%
}
}
\end{table}

\begin{table}[ht!]
\caption{Total \bbH{} cross sections in the SM for a LHC CM energy of $\sqrt{s}=13.6$ TeV obtained via linear interpolation of results from Tables~\ref{tab:bbH13} and~\ref{tab:bbH14}.  The results are presented with both theoretical and parametric uncertainty. The sources of uncertainty are defined in the main text.\\}
\label{tab:bbH136}
\makebox[\textwidth][c]{%
\centering
\scriptsize
{\renewcommand{\arraystretch}{1.4}
%
}
}
\end{table}

\clearpage






\bibliography{refs.bib}

\begin{thebibliography}{100}
\providecommand{\url}[1]{\texttt{#1}}
\providecommand{\urlprefix}{URL }
\expandafter\ifx\csname urlstyle\endcsname\relax
  \providecommand{\doi}[1]{doi:\discretionary{}{}{}#1}\else
  \providecommand{\doi}{doi:\discretionary{}{}{}\begingroup
  \urlstyle{rm}\Url}\fi
\providecommand{\eprint}[2][]{\url{#2}}

\bibitem{LHCHiggsCrossSectionWorkingGroup:2016ypw}
D.~de~Florian \emph{et~al.},
\newblock \emph{{Handbook of LHC Higgs Cross Sections: 4. Deciphering the
  Nature of the Higgs Sector}},
\newblock CERN Yellow Rep. Monogr. \textbf{2}, 1 (2017),
\newblock \doi{10.23731/CYRM-2017-002},
\newblock \href{http://arxiv.org/abs/1610.07922}{{\tt 1610.07922}}.

\bibitem{Karlberg:2024zxx}
A.~Karlberg \emph{et~al.},
\newblock \emph{{Ad interim recommendations for the Higgs boson production
  cross sections at $\sqrt{s} = 13.6$ TeV}}  (2024),
\newblock \href{http://arxiv.org/abs/2402.09955}{{\tt 2402.09955}}.

\bibitem{ATLAS:2012yve}
G.~Aad \emph{et~al.},
\newblock \emph{{Observation of a new particle in the search for the Standard
  Model Higgs boson with the ATLAS detector at the LHC}},
\newblock Phys. Lett. B \textbf{716}, 1 (2012),
\newblock \doi{10.1016/j.physletb.2012.08.020},
\newblock \href{http://arxiv.org/abs/1207.7214}{{\tt 1207.7214}}.

\bibitem{CMS:2012qbp}
S.~Chatrchyan \emph{et~al.},
\newblock \emph{{Observation of a New Boson at a Mass of 125 GeV with the CMS
  Experiment at the LHC}},
\newblock Phys. Lett. B \textbf{716}, 30 (2012),
\newblock \doi{10.1016/j.physletb.2012.08.021},
\newblock \href{http://arxiv.org/abs/1207.7235}{{\tt 1207.7235}}.

\bibitem{ATLAS:2022vkf}
G.~Aad \emph{et~al.},
\newblock \emph{{A detailed map of Higgs boson interactions by the ATLAS
  experiment ten years after the discovery}},
\newblock Nature \textbf{607}(7917), 52 (2022),
\newblock \doi{10.1038/s41586-022-04893-w},
\newblock [Erratum: Nature 612, E24 (2022)],
\newblock \href{http://arxiv.org/abs/2207.00092}{{\tt 2207.00092}}.

\bibitem{CMS:2022dwd}
A.~Tumasyan \emph{et~al.},
\newblock \emph{{A portrait of the Higgs boson by the CMS experiment ten years
  after the discovery.}},
\newblock Nature \textbf{607}(7917), 60 (2022),
\newblock \doi{10.1038/s41586-022-04892-x},
\newblock [Erratum: Nature 623, (2023)],
\newblock \href{http://arxiv.org/abs/2207.00043}{{\tt 2207.00043}}.

\bibitem{ATLAS:2026comb}
{ATLAS Collaboration},
\newblock \emph{{Combination of Higgs boson measurements at $\sqrt{s} =$ 13 TeV
  and their interpretations by the ATLAS experiment}}  (2026),
\newblock \href{http://arxiv.org/abs/2608.07332}{{\tt 2608.07332}}.

\bibitem{CMS:2026nce}
{CMS Collaboration},
\newblock \emph{{Combined measurements and interpretations of Higgs boson
  production and decay in proton-proton collisions at $\sqrt{s}$ = 13 TeV}},
\newblock CMS-HIG-21-018, CERN-EP-2026-009  (2026),
\newblock \href{http://arxiv.org/abs/2602.18611}{{\tt 2602.18611}}.

\bibitem{Cepeda:2019klc}
M.~Cepeda \emph{et~al.},
\newblock \emph{{Report from Working Group 2}: {Higgs Physics at the HL-LHC and
  HE-LHC}},
\newblock CERN Yellow Rep. Monogr. \textbf{7}, 221 (2019),
\newblock \doi{10.23731/CYRM-2019-007.221},
\newblock \href{http://arxiv.org/abs/1902.00134}{{\tt 1902.00134}}.

\bibitem{Anastasiou:2015vya}
C.~Anastasiou, C.~Duhr, F.~Dulat, F.~Herzog and B.~Mistlberger,
\newblock \emph{{Higgs Boson Gluon-Fusion Production in QCD at Three Loops}},
\newblock Phys. Rev. Lett. \textbf{114}, 212001 (2015),
\newblock \doi{10.1103/PhysRevLett.114.212001},
\newblock \href{http://arxiv.org/abs/1503.06056}{{\tt 1503.06056}}.

\bibitem{Anastasiou:2016cez}
C.~Anastasiou, C.~Duhr, F.~Dulat, E.~Furlan, T.~Gehrmann, F.~Herzog,
  A.~Lazopoulos and B.~Mistlberger,
\newblock \emph{{High precision determination of the gluon fusion Higgs boson
  cross-section at the LHC}},
\newblock JHEP \textbf{05}, 058 (2016),
\newblock \doi{10.1007/JHEP05(2016)058},
\newblock \href{http://arxiv.org/abs/1602.00695}{{\tt 1602.00695}}.

\bibitem{Mistlberger:2018etf}
B.~Mistlberger,
\newblock \emph{{Higgs boson production at hadron colliders at N$^{3}$LO in
  QCD}},
\newblock JHEP \textbf{05}, 028 (2018),
\newblock \doi{10.1007/JHEP05(2018)028},
\newblock \href{http://arxiv.org/abs/1802.00833}{{\tt 1802.00833}}.

\bibitem{Czakon:2020vql}
M.~L. Czakon and M.~Niggetiedt,
\newblock \emph{{Exact quark-mass dependence of the Higgs-gluon form factor at
  three loops in QCD}},
\newblock JHEP \textbf{05}, 149 (2020),
\newblock \doi{10.1007/JHEP05(2020)149},
\newblock \href{http://arxiv.org/abs/2001.03008}{{\tt 2001.03008}}.

\bibitem{Czakon:2021yub}
M.~Czakon, R.~V. Harlander, J.~Klappert and M.~Niggetiedt,
\newblock \emph{{Exact Top-Quark Mass Dependence in Hadronic Higgs
  Production}},
\newblock Phys. Rev. Lett. \textbf{127}(16), 162002 (2021),
\newblock \doi{10.1103/PhysRevLett.127.162002},
\newblock [Erratum: Phys.Rev.Lett. 131, 179901 (2023)],
\newblock \href{http://arxiv.org/abs/2105.04436}{{\tt 2105.04436}}.

\bibitem{Czakon:2024ywb}
M.~Czakon, F.~Eschment, M.~Niggetiedt, R.~Poncelet and T.~Schellenberger,
\newblock \emph{{Quark mass effects in Higgs production}},
\newblock JHEP \textbf{10}, 210 (2024),
\newblock \doi{10.1007/JHEP10(2024)210},
\newblock \href{http://arxiv.org/abs/2407.12413}{{\tt 2407.12413}}.

\bibitem{Czakon:2023kqm}
M.~Czakon, F.~Eschment, M.~Niggetiedt, R.~Poncelet and T.~Schellenberger,
\newblock \emph{{Top-Bottom Interference Contribution to Fully-Inclusive Higgs
  Production}},
\newblock Phys. Rev. Lett. \textbf{132}(21), 211902 (2024),
\newblock \doi{10.1103/PhysRevLett.132.211902},
\newblock \href{http://arxiv.org/abs/2312.09896}{{\tt 2312.09896}}.

\bibitem{Bonetti:2016brm}
M.~Bonetti, K.~Melnikov and L.~Tancredi,
\newblock \emph{{Two-loop electroweak corrections to Higgs{\textendash}gluon
  couplings to higher orders in the dimensional regularization parameter}},
\newblock Nucl. Phys. B \textbf{916}, 709 (2017),
\newblock \doi{10.1016/j.nuclphysb.2017.01.020},
\newblock \href{http://arxiv.org/abs/1610.05497}{{\tt 1610.05497}}.

\bibitem{Bonetti:2017ovy}
M.~Bonetti, K.~Melnikov and L.~Tancredi,
\newblock \emph{{Three-loop mixed QCD-electroweak corrections to Higgs boson
  gluon fusion}},
\newblock Phys. Rev. \textbf{D97}(3), 034004 (2018),
\newblock \doi{10.1103/PhysRevD.97.034004},
\newblock \href{http://arxiv.org/abs/1711.11113}{{\tt 1711.11113}}.

\bibitem{Bonetti:2018ukf}
M.~Bonetti, K.~Melnikov and L.~Tancredi,
\newblock \emph{{Higher order corrections to mixed QCD-EW contributions to
  Higgs boson production in gluon fusion}},
\newblock Phys. Rev. D \textbf{97}(5), 056017 (2018),
\newblock \doi{10.1103/PhysRevD.97.056017},
\newblock [Erratum: Phys.Rev.D 97, 099906 (2018)],
\newblock \href{http://arxiv.org/abs/1801.10403}{{\tt 1801.10403}}.

\bibitem{Becchetti:2020wof}
M.~Becchetti, R.~Bonciani, V.~Del~Duca, V.~Hirschi, F.~Moriello and
  A.~Schweitzer,
\newblock \emph{{Next-to-leading order corrections to light-quark mixed QCD-EW
  contributions to Higgs boson production}},
\newblock Phys. Rev. D \textbf{103}(5), 054037 (2021),
\newblock \doi{10.1103/PhysRevD.103.054037},
\newblock \href{http://arxiv.org/abs/2010.09451}{{\tt 2010.09451}}.

\bibitem{Das:2020adl}
G.~Das, S.~Moch and A.~Vogt,
\newblock \emph{{Approximate four-loop QCD corrections to the Higgs-boson
  production cross section}},
\newblock Phys. Lett. B \textbf{807}, 135546 (2020),
\newblock \doi{10.1016/j.physletb.2020.135546},
\newblock \href{http://arxiv.org/abs/2004.00563}{{\tt 2004.00563}}.

\bibitem{Dreyer:2016oyx}
F.~A. Dreyer and A.~Karlberg,
\newblock \emph{{Vector-Boson Fusion Higgs Production at Three Loops in QCD}},
\newblock Phys. Rev. Lett. \textbf{117}(7), 072001 (2016),
\newblock \doi{10.1103/PhysRevLett.117.072001},
\newblock \href{http://arxiv.org/abs/1606.00840}{{\tt 1606.00840}}.

\bibitem{Liu:2019tuy}
T.~Liu, K.~Melnikov and A.~A. Penin,
\newblock \emph{{Nonfactorizable QCD Effects in Higgs Boson Production via
  Vector Boson Fusion}},
\newblock Phys. Rev. Lett. \textbf{123}(12), 122002 (2019),
\newblock \doi{10.1103/PhysRevLett.123.122002},
\newblock \href{http://arxiv.org/abs/1906.10899}{{\tt 1906.10899}}.

\bibitem{Dreyer:2020urf}
F.~A. Dreyer, A.~Karlberg and L.~Tancredi,
\newblock \emph{{On the impact of non-factorisable corrections in VBF single
  and double Higgs production}},
\newblock JHEP \textbf{10}, 131 (2020),
\newblock \doi{10.1007/JHEP10(2020)131},
\newblock \href{http://arxiv.org/abs/2005.11334}{{\tt 2005.11334}}.

\bibitem{Asteriadis:2023nyl}
K.~Asteriadis, C.~Br\o{}nnum-Hansen and K.~Melnikov,
\newblock \emph{{Nonfactorizable corrections to Higgs boson production in weak
  boson fusion}},
\newblock Phys. Rev. D \textbf{109}(1), 014031 (2024),
\newblock \doi{10.1103/PhysRevD.109.014031},
\newblock \href{http://arxiv.org/abs/2305.08016}{{\tt 2305.08016}}.

\bibitem{Baglio:2022wzu}
J.~Baglio, C.~Duhr, B.~Mistlberger and R.~Szafron,
\newblock \emph{{Inclusive production cross sections at N$^{3}$LO}},
\newblock JHEP \textbf{12}, 066 (2022),
\newblock \doi{10.1007/JHEP12(2022)066},
\newblock \href{http://arxiv.org/abs/2209.06138}{{\tt 2209.06138}}.

\bibitem{Chen:2020gae}
L.~Chen, G.~Heinrich, S.~P. Jones, M.~Kerner, J.~Klappert and J.~Schlenk,
\newblock \emph{{$ZH$ production in gluon fusion: two-loop amplitudes with full
  top quark mass dependence}},
\newblock JHEP \textbf{03}, 125 (2021),
\newblock \doi{10.1007/JHEP03(2021)125},
\newblock \href{http://arxiv.org/abs/2011.12325}{{\tt 2011.12325}}.

\bibitem{Wang:2021rxu}
G.~Wang, X.~Xu, Y.~Xu and L.~L. Yang,
\newblock \emph{{Next-to-leading order corrections for $gg \to ZH$ with top
  quark mass dependence}},
\newblock Phys. Lett. B \textbf{829}, 137087 (2022),
\newblock \doi{10.1016/j.physletb.2022.137087},
\newblock \href{http://arxiv.org/abs/2107.08206}{{\tt 2107.08206}}.

\bibitem{Chen:2022rua}
L.~Chen, J.~Davies, G.~Heinrich, S.~P. Jones, M.~Kerner, G.~Mishima, J.~Schlenk
  and M.~Steinhauser,
\newblock \emph{{ZH production in gluon fusion at NLO in QCD}},
\newblock JHEP \textbf{08}, 056 (2022),
\newblock \doi{10.1007/JHEP08(2022)056},
\newblock \href{http://arxiv.org/abs/2204.05225}{{\tt 2204.05225}}.

\bibitem{Degrassi:2022mro}
G.~Degrassi, R.~Gr\"ober, M.~Vitti and X.~Zhao,
\newblock \emph{{On the NLO QCD corrections to gluon-initiated ZH production}},
\newblock JHEP \textbf{08}, 009 (2022),
\newblock \doi{10.1007/JHEP08(2022)009},
\newblock \href{http://arxiv.org/abs/2205.02769}{{\tt 2205.02769}}.

\bibitem{Davies:2025out}
J.~Davies, D.~Grau, K.~Sch{\"o}nwald, M.~Steinhauser, D.~Stremmer and M.~Vitti,
\newblock \emph{{Two-loop QCD corrections to $ZH$ and off-shell $Z$ boson pair
  production in gluon fusion}},
\newblock JHEP \textbf{03}, 200 (2026),
\newblock \doi{10.1007/JHEP03(2026)200},
\newblock \href{http://arxiv.org/abs/2509.07072}{{\tt 2509.07072}}.

\bibitem{Davies:2025otz}
J.~Davies, D.~Grau, K.~Sch{\"o}nwald, M.~Steinhauser and D.~Stremmer,
\newblock \emph{{Three-loop corrections to $gg\to ZH$ in the large top quark
  mass limit}},
\newblock JHEP \textbf{02}, 191 (2026),
\newblock \doi{10.1007/JHEP02(2026)191},
\newblock \href{http://arxiv.org/abs/2512.00156}{{\tt 2512.00156}}.

\bibitem{Catani:2022mfv}
S.~Catani, S.~Devoto, M.~Grazzini, S.~Kallweit, J.~Mazzitelli and C.~Savoini,
\newblock \emph{{Higgs Boson Production in Association with a Top-Antitop Quark
  Pair in Next-to-Next-to-Leading Order QCD}},
\newblock Phys. Rev. Lett. \textbf{130}(11), 111902 (2023),
\newblock \doi{10.1103/PhysRevLett.130.111902},
\newblock \href{http://arxiv.org/abs/2210.07846}{{\tt 2210.07846}}.

\bibitem{Devoto:2024nhl}
S.~Devoto, M.~Grazzini, S.~Kallweit, J.~Mazzitelli and C.~Savoini,
\newblock \emph{{Precise predictions for $ t\overline{t}H $ production at the
  LHC: inclusive cross section and differential distributions}},
\newblock JHEP \textbf{03}, 189 (2025),
\newblock \doi{10.1007/JHEP03(2025)189},
\newblock \href{http://arxiv.org/abs/2411.15340}{{\tt 2411.15340}}.

\bibitem{Balsach:2025jcw}
R.~Balsach \emph{et~al.},
\newblock \emph{{State-of-the-art cross sections for $t\bar t H$: NNLO
  predictions matched with NNLL resummation and EW corrections}}  (2025),
\newblock \doi{10.21468/SciPostPhysCommRep.10},
\newblock \href{http://arxiv.org/abs/2503.15043}{{\tt 2503.15043}}.

\bibitem{Demartin:2015uha}
F.~Demartin, F.~Maltoni, K.~Mawatari and M.~Zaro,
\newblock \emph{{Higgs production in association with a single top quark at the
  LHC}},
\newblock Eur. Phys. J. C \textbf{75}(6), 267 (2015),
\newblock \doi{10.1140/epjc/s10052-015-3475-9},
\newblock \href{http://arxiv.org/abs/1504.00611}{{\tt 1504.00611}}.

\bibitem{Demartin:2016axk}
F.~Demartin, B.~Maier, F.~Maltoni, K.~Mawatari and M.~Zaro,
\newblock \emph{{tWH associated production at the LHC}},
\newblock Eur. Phys. J. C \textbf{77}(1), 34 (2017),
\newblock \doi{10.1140/epjc/s10052-017-4601-7},
\newblock \href{http://arxiv.org/abs/1607.05862}{{\tt 1607.05862}}.

\bibitem{Pagani:2020mov}
D.~Pagani, I.~Tsinikos and E.~Vryonidou,
\newblock \emph{{NLO QCD+EW predictions for $tHj$ and $tZj$ production at the
  LHC}},
\newblock JHEP \textbf{08}, 082 (2020),
\newblock \doi{10.1007/JHEP08(2020)082},
\newblock \href{http://arxiv.org/abs/2006.10086}{{\tt 2006.10086}}.

\bibitem{Duhr:2019kwi}
C.~Duhr, F.~Dulat and B.~Mistlberger,
\newblock \emph{{Higgs Boson Production in Bottom-Quark Fusion to Third Order
  in the Strong Coupling}},
\newblock Phys. Rev. Lett. \textbf{125}(5), 051804 (2020),
\newblock \doi{10.1103/PhysRevLett.125.051804},
\newblock \href{http://arxiv.org/abs/1904.09990}{{\tt 1904.09990}}.

\bibitem{Duhr:2020kzd}
C.~Duhr, F.~Dulat, V.~Hirschi and B.~Mistlberger,
\newblock \emph{{Higgs production in bottom quark fusion: matching the 4- and
  5-flavour schemes to third order in the strong coupling}},
\newblock JHEP \textbf{08}(08), 017 (2020),
\newblock \doi{10.1007/JHEP08(2020)017},
\newblock \href{http://arxiv.org/abs/2004.04752}{{\tt 2004.04752}}.

\bibitem{Bonvini:2015pxa}
M.~Bonvini, A.~S. Papanastasiou and F.~J. Tackmann,
\newblock \emph{{Resummation and matching of b-quark mass effects in $
  b\overline{b}H $ production}},
\newblock JHEP \textbf{11}, 196 (2015),
\newblock \doi{10.1007/JHEP11(2015)196},
\newblock \href{http://arxiv.org/abs/1508.03288}{{\tt 1508.03288}}.

\bibitem{Bonvini:2016fgf}
M.~Bonvini, A.~S. Papanastasiou and F.~J. Tackmann,
\newblock \emph{{Matched predictions for the $ b\overline{b}H $ cross section
  at the 13 TeV LHC}},
\newblock JHEP \textbf{10}, 053 (2016),
\newblock \doi{10.1007/JHEP10(2016)053},
\newblock \href{http://arxiv.org/abs/1605.01733}{{\tt 1605.01733}}.

\bibitem{PDF4LHCWorkingGroup:2022cjn}
R.~D. Ball \emph{et~al.},
\newblock \emph{{The PDF4LHC21 combination of global PDF fits for the LHC Run
  III}},
\newblock J. Phys. G \textbf{49}(8), 080501 (2022),
\newblock \doi{10.1088/1361-6471/ac7216},
\newblock \href{http://arxiv.org/abs/2203.05506}{{\tt 2203.05506}}.

\bibitem{McGowan:2022nag}
J.~McGowan, T.~Cridge, L.~A. Harland-Lang and R.~S. Thorne,
\newblock \emph{{Approximate N$^{3}$LO parton distribution functions with
  theoretical uncertainties: MSHT20aN$^3$LO PDFs}},
\newblock Eur. Phys. J. C \textbf{83}(3), 185 (2023),
\newblock \doi{10.1140/epjc/s10052-023-11236-0},
\newblock \href{http://arxiv.org/abs/2207.04739}{{\tt 2207.04739}}.

\bibitem{NNPDF:2024nan}
R.~D. Ball \emph{et~al.},
\newblock \emph{{The path to $\hbox {N}^3\hbox {LO}$ parton distributions}},
\newblock Eur. Phys. J. C \textbf{84}(7), 659 (2024),
\newblock \doi{10.1140/epjc/s10052-024-12891-7},
\newblock \href{http://arxiv.org/abs/2402.18635}{{\tt 2402.18635}}.

\bibitem{Cooper-Sarkar:2024crx}
A.~Cooper-Sarkar, T.~Cridge, F.~Giuli, L.~A. Harland-Lang, F.~Hekhorn,
  J.~Huston, G.~Magni, S.~Moch and R.~S. Thorne,
\newblock \emph{{A Benchmarking of QCD Evolution at Approximate $N^3LO$}}
  (2024),
\newblock \href{http://arxiv.org/abs/2406.16188}{{\tt 2406.16188}}.

\bibitem{Cridge:2024icl}
T.~Cridge \emph{et~al.},
\newblock \emph{{Combination of aN$^3$LO PDFs and implications for Higgs
  production cross-sections at the LHC}},
\newblock J. Phys. G \textbf{52}, 6 (2025),
\newblock \doi{10.1088/1361-6471/adde78},
\newblock \href{http://arxiv.org/abs/2411.05373}{{\tt 2411.05373}}.

\bibitem{Manohar:2016nzj}
A.~Manohar, P.~Nason, G.~P. Salam and G.~Zanderighi,
\newblock \emph{{How bright is the proton? A precise determination of the
  photon parton distribution function}},
\newblock Phys. Rev. Lett. \textbf{117}(24), 242002 (2016),
\newblock \doi{10.1103/PhysRevLett.117.242002},
\newblock \href{http://arxiv.org/abs/1607.04266}{{\tt 1607.04266}}.

\bibitem{Cooper-Sarkar:2025sqw}
A.~M. Cooper-Sarkar, T.~Cridge, T.~J. Hobbs, J.~Huston, P.~Nadolsky,
  M.~Ponce-Chavez and K.~Xie,
\newblock \emph{{QED-enhanced PDF implications for the Higgs sector}},
\newblock JHEP \textbf{05}, 228 (2026),
\newblock \doi{10.1007/JHEP05(2026)228},
\newblock \href{http://arxiv.org/abs/2508.06603}{{\tt 2508.06603}}.

\bibitem{Barone:2025jey}
G.~Barone \emph{et~al.},
\newblock \emph{{Higgs production via vector-boson fusion at the LHC}}  (2025),
\newblock \doi{10.21468/SciPostPhysCommRep.13},
\newblock \href{http://arxiv.org/abs/2507.22574}{{\tt 2507.22574}}.

\bibitem{CampilloAveleira:2025rbh}
B.~Campillo~Aveleira \emph{et~al.},
\newblock \emph{{$gg \to ZH$ : updated predictions at NLO QCD}},
\newblock \doi{10.21468/SciPostPhysCommRep.16} (2025),
  \href{http://arxiv.org/abs/2508.09905}{{\tt 2508.09905}}.

\bibitem{10.21468/SciPostPhysCommRep.18}
C.~Biello, A.~Gavardi, R.~von Kuk, M.~A. Lim, S.~Manzoni, E.~Mazzeo,
  J.~Mazzitelli, A.~Sankar, M.~Spira, F.~J. Tackmann, M.~Wiesemann,
  G.~Zanderighi \emph{et~al.},
\newblock \emph{{Modelling $b\bar b H$ production for the LHC at 13.6 TeV}},
\newblock SciPost Phys. Comm. Rep. p.~18 (2026),
\newblock \doi{10.21468/SciPostPhysCommRep.18}.

\bibitem{ATLAS:2022hsp}
\emph{{Snowmass White Paper Contribution: Physics with the Phase-2 ATLAS and
  CMS Detectors}},
\newblock Tech. Rep. ATL-PHYS-PUB-2022-018, CERN (2022).

\bibitem{ParticleDataGroup:2020ssz}
P.~A. Zyla \emph{et~al.},
\newblock \emph{{Review of Particle Physics}},
\newblock PTEP \textbf{2020}(8), 083C01 (2020),
\newblock \doi{10.1093/ptep/ptaa104}.

\bibitem{BARDIN1988539}
D.~Bardin, A.~Leike, T.~Riemann and M.~Sachwitz,
\newblock \emph{Energy-dependent width effects in $e^+e^-$-annihilation near
  the z-boson pole},
\newblock Physics Letters B \textbf{206}(3), 539 (1988),
\newblock \doi{https://doi.org/10.1016/0370-2693(88)91627-9}.

\bibitem{Denner:2000bj}
A.~Denner, S.~Dittmaier, M.~Roth and D.~Wackeroth,
\newblock \emph{{Electroweak radiative corrections to $e^+ e^- \to W W \to 4$
  fermions in double pole approximation: The RACOONWW approach}},
\newblock Nucl. Phys. B \textbf{587}, 67 (2000),
\newblock \doi{10.1016/S0550-3213(00)00511-3},
\newblock \href{http://arxiv.org/abs/hep-ph/0006307}{{\tt hep-ph/0006307}}.

\bibitem{ATLAS:2015yey}
G.~Aad \emph{et~al.},
\newblock \emph{{Combined Measurement of the Higgs Boson Mass in $pp$
  Collisions at $\sqrt{s}=7$ and 8 TeV with the ATLAS and CMS Experiments}},
\newblock Phys. Rev. Lett. \textbf{114}, 191803 (2015),
\newblock \doi{10.1103/PhysRevLett.114.191803},
\newblock \href{http://arxiv.org/abs/1503.07589}{{\tt 1503.07589}}.

\bibitem{CMS:2020xrn}
A.~M. Sirunyan \emph{et~al.},
\newblock \emph{{A measurement of the Higgs boson mass in the diphoton decay
  channel}},
\newblock Phys. Lett. B \textbf{805}, 135425 (2020),
\newblock \doi{10.1016/j.physletb.2020.135425},
\newblock \href{http://arxiv.org/abs/2002.06398}{{\tt 2002.06398}}.

\bibitem{Hou:2019efy}
T.-J. Hou \emph{et~al.},
\newblock \emph{{New CTEQ global analysis of quantum chromodynamics with
  high-precision data from the LHC}},
\newblock Phys. Rev. \textbf{D103}(1), 014013 (2021),
\newblock \doi{10.1103/PhysRevD.103.014013},
\newblock \href{http://arxiv.org/abs/1912.10053}{{\tt 1912.10053}}.

\bibitem{Bailey:2020ooq}
S.~Bailey, T.~Cridge, L.~A. Harland-Lang, A.~D. Martin and R.~S. Thorne,
\newblock \emph{{Parton distributions from LHC, HERA, Tevatron and fixed target
  data: MSHT20 PDFs}},
\newblock Eur. Phys. J. C \textbf{81}(4), 341 (2021),
\newblock \doi{10.1140/epjc/s10052-021-09057-0},
\newblock \href{http://arxiv.org/abs/2012.04684}{{\tt 2012.04684}}.

\bibitem{NNPDF:2017mvq}
R.~D. Ball \emph{et~al.},
\newblock \emph{{Parton distributions from high-precision collider data}},
\newblock Eur. Phys. J. C \textbf{77}(10), 663 (2017),
\newblock \doi{10.1140/epjc/s10052-017-5199-5},
\newblock \href{http://arxiv.org/abs/1706.00428}{{\tt 1706.00428}}.

\bibitem{Manohar:2017eqh}
A.~V. Manohar, P.~Nason, G.~P. Salam and G.~Zanderighi,
\newblock \emph{{The Photon Content of the Proton}},
\newblock JHEP \textbf{12}, 046 (2017),
\newblock \doi{10.1007/JHEP12(2017)046},
\newblock \href{http://arxiv.org/abs/1708.01256}{{\tt 1708.01256}}.

\bibitem{Falcioni:2024xyt}
G.~Falcioni, F.~Herzog, S.~Moch, A.~Pelloni and A.~Vogt,
\newblock \emph{{Four-loop splitting functions in QCD {\textendash} The
  quark-to-gluon case}},
\newblock Phys. Lett. B \textbf{856}, 138906 (2024),
\newblock \doi{10.1016/j.physletb.2024.138906},
\newblock \href{http://arxiv.org/abs/2404.09701}{{\tt 2404.09701}}.

\bibitem{Falcioni:2024qpd}
G.~Falcioni, F.~Herzog, S.~Moch, A.~Pelloni and A.~Vogt,
\newblock \emph{{Four-loop splitting functions in QCD {\textendash} the
  gluon-gluon case {\textendash}}},
\newblock Phys. Lett. B \textbf{860}, 139194 (2025),
\newblock \doi{10.1016/j.physletb.2024.139194},
\newblock \href{http://arxiv.org/abs/2410.08089}{{\tt 2410.08089}}.

\bibitem{Gehrmann:2026qbl}
T.~Gehrmann, A.~von Manteuffel, V.~Sotnikov and T.-Z. Yang,
\newblock \emph{{The four-loop non-singlet splitting functions in QCD}}
  (2026),
\newblock \href{http://arxiv.org/abs/2604.09534}{{\tt 2604.09534}}.

\bibitem{Bertone:2017gds}
V.~Bertone,
\newblock \emph{{APFEL++: A new PDF evolution library in C++}},
\newblock PoS \textbf{DIS2017}, 201 (2018),
\newblock \doi{10.22323/1.297.0201},
\newblock \href{http://arxiv.org/abs/1708.00911}{{\tt 1708.00911}}.

\bibitem{Candido:2022tld}
A.~Candido, F.~Hekhorn and G.~Magni,
\newblock \emph{{EKO: evolution kernel operators}},
\newblock Eur. Phys. J. C \textbf{82}(10), 976 (2022),
\newblock \doi{10.1140/epjc/s10052-022-10878-w},
\newblock \href{http://arxiv.org/abs/2202.02338}{{\tt 2202.02338}}.

\bibitem{Karlberg:2025hxk}
A.~Karlberg, P.~Nason, G.~Salam, G.~Zanderighi and F.~Dreyer,
\newblock \emph{{HOPPET v2.0.0 release note}},
\newblock Eur. Phys. J. C \textbf{86}(2), 157 (2026),
\newblock \doi{10.1140/epjc/s10052-025-15200-y},
\newblock \href{http://arxiv.org/abs/2510.09310}{{\tt 2510.09310}}.

\bibitem{Hampson:2025pvi}
C.~Hampson and M.~Guzzi,
\newblock \emph{{DGLAP evolution at \(\text{N}^{3}\text{LO}\) with the candia
  algorithm}},
\newblock Phys. Rev. D \textbf{113}(7), 074008 (2026),
\newblock \doi{10.1103/p6vv-8zph},
\newblock \href{http://arxiv.org/abs/2512.22667}{{\tt 2512.22667}}.

\bibitem{Bertone:2024dpm}
V.~Bertone and A.~Karlberg,
\newblock \emph{{Benchmark of deep-inelastic-scattering structure functions
  at~$\mathcal {O}\left( \alpha _s ^3\right) $}},
\newblock Eur. Phys. J. C \textbf{84}(8), 774 (2024),
\newblock \doi{10.1140/epjc/s10052-024-13133-6},
\newblock \href{http://arxiv.org/abs/2404.15711}{{\tt 2404.15711}}.

\bibitem{Blumlein:2016xcy}
J.~Bl{\"u}mlein, G.~Falcioni and A.~De~Freitas,
\newblock \emph{{The Complete $O(\alpha_s^2)$ Non-Singlet Heavy Flavor
  Corrections to the Structure Functions $g_{1,2}^{ep}(x,Q^2)$,
  $F_{1,2,L}^{ep}(x,Q^2)$, $F_{1,2,3}^{\nu(\bar{\nu})}(x,Q^2)$ and the
  Associated Sum Rules}},
\newblock Nucl. Phys. B \textbf{910}, 568 (2016),
\newblock \doi{10.1016/j.nuclphysb.2016.06.018},
\newblock \href{http://arxiv.org/abs/1605.05541}{{\tt 1605.05541}}.

\bibitem{Buza:1996wv}
M.~Buza, Y.~Matiounine, J.~Smith and W.~L. van Neerven,
\newblock \emph{{Charm electroproduction viewed in the variable flavor number
  scheme versus fixed order perturbation theory}},
\newblock Eur. Phys. J. C \textbf{1}, 301 (1998),
\newblock \doi{10.1007/BF01245820},
\newblock \href{http://arxiv.org/abs/hep-ph/9612398}{{\tt hep-ph/9612398}}.

\bibitem{Ablinger:2017err}
J.~Ablinger, J.~Bl{\"u}mlein, A.~De~Freitas, A.~Hasselhuhn, C.~Schneider and
  F.~Wi{\ss}brock,
\newblock \emph{{Three Loop Massive Operator Matrix Elements and Asymptotic
  Wilson Coefficients with Two Different Masses}},
\newblock Nucl. Phys. B \textbf{921}, 585 (2017),
\newblock \doi{10.1016/j.nuclphysb.2017.05.017},
\newblock \href{http://arxiv.org/abs/1705.07030}{{\tt 1705.07030}}.

\bibitem{Buza:1995ie}
M.~Buza, Y.~Matiounine, J.~Smith, R.~Migneron and W.~L. van Neerven,
\newblock \emph{{Heavy quark coefficient functions at asymptotic values Q**2
  {\ensuremath{>}}{\ensuremath{>}} m**2}},
\newblock Nucl. Phys. B \textbf{472}, 611 (1996),
\newblock \doi{10.1016/0550-3213(96)00228-3},
\newblock \href{http://arxiv.org/abs/hep-ph/9601302}{{\tt hep-ph/9601302}}.

\bibitem{Laenen:1992zk}
E.~Laenen, S.~Riemersma, J.~Smith and W.~L. van Neerven,
\newblock \emph{{Complete O (alpha-s) corrections to heavy flavor structure
  functions in electroproduction}},
\newblock Nucl. Phys. B \textbf{392}, 162 (1993),
\newblock \doi{10.1016/0550-3213(93)90201-Y}.

\bibitem{Laenen:1992xs}
E.~Laenen, S.~Riemersma, J.~Smith and W.~L. van Neerven,
\newblock \emph{{O(alpha-s) corrections to heavy flavor inclusive distributions
  in electroproduction}},
\newblock Nucl. Phys. B \textbf{392}, 229 (1993),
\newblock \doi{10.1016/0550-3213(93)90202-Z}.

\bibitem{Bierenbaum:2007qe}
I.~Bierenbaum, J.~Blumlein and S.~Klein,
\newblock \emph{{Two-Loop Massive Operator Matrix Elements and Unpolarized
  Heavy Flavor Production at Asymptotic Values Q**2
  {\ensuremath{>}}{\ensuremath{>}} m**2}},
\newblock Nucl. Phys. B \textbf{780}, 40 (2007),
\newblock \doi{10.1016/j.nuclphysb.2007.04.030},
\newblock \href{http://arxiv.org/abs/hep-ph/0703285}{{\tt hep-ph/0703285}}.

\bibitem{Bierenbaum:2008yu}
I.~Bierenbaum, J.~Blumlein, S.~Klein and C.~Schneider,
\newblock \emph{{Two-Loop Massive Operator Matrix Elements for Unpolarized
  Heavy Flavor Production to O(epsilon)}},
\newblock Nucl. Phys. B \textbf{803}, 1 (2008),
\newblock \doi{10.1016/j.nuclphysb.2008.05.016},
\newblock \href{http://arxiv.org/abs/0803.0273}{{\tt 0803.0273}}.

\bibitem{Bierenbaum:2009zt}
I.~Bierenbaum, J.~Blumlein and S.~Klein,
\newblock \emph{{The Gluonic Operator Matrix Elements at O(alpha(s)**2) for DIS
  Heavy Flavor Production}},
\newblock Phys. Lett. B \textbf{672}, 401 (2009),
\newblock \doi{10.1016/j.physletb.2009.01.057},
\newblock \href{http://arxiv.org/abs/0901.0669}{{\tt 0901.0669}}.

\bibitem{Bierenbaum:2022biv}
I.~Bierenbaum, J.~Bl{\"u}mlein, A.~De~Freitas, A.~Goedicke, S.~Klein and
  K.~Sch{\"o}nwald,
\newblock \emph{{$O(\alpha_s^2$) polarized heavy flavor corrections to
  deep-inelastic scattering at $Q^2 \gg m^2$}},
\newblock Nucl. Phys. B \textbf{988}, 116114 (2023),
\newblock \doi{10.1016/j.nuclphysb.2023.116114},
\newblock \href{http://arxiv.org/abs/2211.15337}{{\tt 2211.15337}}.

\bibitem{Bierenbaum:2009mv}
I.~Bierenbaum, J.~Blumlein and S.~Klein,
\newblock \emph{{Mellin Moments of the O(alpha**3(s)) Heavy Flavor
  Contributions to unpolarized Deep-Inelastic Scattering at Q**2
  {\ensuremath{>}}{\ensuremath{>}} m**2 and Anomalous Dimensions}},
\newblock Nucl. Phys. B \textbf{820}, 417 (2009),
\newblock \doi{10.1016/j.nuclphysb.2009.06.005},
\newblock \href{http://arxiv.org/abs/0904.3563}{{\tt 0904.3563}}.

\bibitem{Ablinger:2010ty}
J.~Ablinger, J.~Blumlein, S.~Klein, C.~Schneider and F.~Wissbrock,
\newblock \emph{{The $O(\alpha_s^3)$ Massive Operator Matrix Elements of
  $O(n_f)$ for the Structure Function $F_2(x,Q^2)$ and Transversity}},
\newblock Nucl. Phys. B \textbf{844}, 26 (2011),
\newblock \doi{10.1016/j.nuclphysb.2010.10.021},
\newblock \href{http://arxiv.org/abs/1008.3347}{{\tt 1008.3347}}.

\bibitem{Ablinger:2014vwa}
J.~Ablinger, A.~Behring, J.~Bl{\"u}mlein, A.~De~Freitas, A.~Hasselhuhn, A.~von
  Manteuffel, M.~Round, C.~Schneider and F.~Wi{\ss}brock,
\newblock \emph{{The 3-Loop Non-Singlet Heavy Flavor Contributions and
  Anomalous Dimensions for the Structure Function $F_2(x,Q^2)$ and
  Transversity}},
\newblock Nucl. Phys. B \textbf{886}, 733 (2014),
\newblock \doi{10.1016/j.nuclphysb.2014.07.010},
\newblock \href{http://arxiv.org/abs/1406.4654}{{\tt 1406.4654}}.

\bibitem{Ablinger:2014nga}
J.~Ablinger, A.~Behring, J.~Bl{\"u}mlein, A.~De~Freitas, A.~von Manteuffel and
  C.~Schneider,
\newblock \emph{{The 3-loop pure singlet heavy flavor contributions to the
  structure function $F_2(x,Q^2)$ and the anomalous dimension}},
\newblock Nucl. Phys. B \textbf{890}, 48 (2014),
\newblock \doi{10.1016/j.nuclphysb.2014.10.008},
\newblock \href{http://arxiv.org/abs/1409.1135}{{\tt 1409.1135}}.

\bibitem{Ablinger:2014lka}
J.~Ablinger, J.~Bl{\"u}mlein, A.~D. Freitas, A.~Hasselhuhn, A.~von Manteuffel,
  M.~Round, C.~Schneider and F.~Wissbrock,
\newblock \emph{The transition matrix element $a_{gq}(n)$ of the variable
  flavor number scheme at $o(\alpha_s^3)$},
\newblock DESY 13--185, DO-TH 13/27, MITP/13-063, SFB/CPP-13-74, LPN 13-072,
  Higgstools 14-001  (2014),
\newblock \doi{https://doi.org/10.1016/j.nuclphysb.2014.02.007},
\newblock \href{http://arxiv.org/abs/1402.0359}{{\tt 1402.0359}}.

\bibitem{Behring:2014eya}
A.~Behring, I.~Bierenbaum, J.~Bl{\"u}mlein, A.~De~Freitas, S.~Klein and
  F.~Wi{\ss}brock,
\newblock \emph{{The logarithmic contributions to the $O(\alpha^3_s)$
  asymptotic massive Wilson coefficients and operator matrix elements in deeply
  inelastic scattering}},
\newblock Eur. Phys. J. C \textbf{74}(9), 3033 (2014),
\newblock \doi{10.1140/epjc/s10052-014-3033-x},
\newblock \href{http://arxiv.org/abs/1403.6356}{{\tt 1403.6356}}.

\bibitem{Ablinger:2018brx}
J.~Ablinger, J.~Bl{\"u}mlein, A.~De~Freitas, A.~Goedicke, C.~Schneider and
  K.~Sch{\"o}nwald,
\newblock \emph{{The Two-mass Contribution to the Three-Loop Gluonic Operator
  Matrix Element $A_{gg,Q}^{(3)}$}},
\newblock Nucl. Phys. B \textbf{932}, 129 (2018),
\newblock \doi{10.1016/j.nuclphysb.2018.04.023},
\newblock \href{http://arxiv.org/abs/1804.02226}{{\tt 1804.02226}}.

\bibitem{Ablinger:2017xml}
J.~Ablinger, J.~Bl{\"u}mlein, A.~De~Freitas, C.~Schneider and K.~Sch{\"o}nwald,
\newblock \emph{{The two-mass contribution to the three-loop pure singlet
  operator matrix element}},
\newblock Nucl. Phys. B \textbf{927}, 339 (2018),
\newblock \doi{10.1016/j.nuclphysb.2017.12.018},
\newblock \href{http://arxiv.org/abs/1711.06717}{{\tt 1711.06717}}.

\bibitem{Ablinger:2022wbb}
J.~Ablinger, A.~Behring, J.~Bl{\"u}mlein, A.~De~Freitas, A.~Goedicke, A.~von
  Manteuffel, C.~Schneider and K.~Sch{\"o}nwald,
\newblock \emph{{The unpolarized and polarized single-mass three-loop heavy
  flavor operator matrix elements A$_{gg,Q}$ and
  {\ensuremath{\Delta}}A$_{gg,Q}$}},
\newblock JHEP \textbf{12}, 134 (2022),
\newblock \doi{10.1007/JHEP12(2022)134},
\newblock \href{http://arxiv.org/abs/2211.05462}{{\tt 2211.05462}}.

\bibitem{Ablinger:2023ahe}
J.~Ablinger, A.~Behring, J.~Bl{\"u}mlein, A.~De~Freitas, A.~von Manteuffel,
  C.~Schneider and K.~Sch{\"o}nwald,
\newblock \emph{{The first{\textendash}order factorizable contributions to the
  three{\textendash}loop massive operator matrix elements AQg(3) and
  {\ensuremath{\Delta}}AQg(3)}},
\newblock Nucl. Phys. B \textbf{999}, 116427 (2024),
\newblock \doi{10.1016/j.nuclphysb.2023.116427},
\newblock \href{http://arxiv.org/abs/2311.00644}{{\tt 2311.00644}}.

\bibitem{Ablinger:2024xtt}
J.~Ablinger, A.~Behring, J.~Bl{\"u}mlein, A.~De~Freitas, A.~von Manteuffel,
  C.~Schneider and K.~Sch{\"o}nwald,
\newblock \emph{{The non-first-order-factorizable contributions to the
  three-loop single-mass operator matrix elements AQg(3) and
  {\ensuremath{\Delta}}AQg(3)}},
\newblock Phys. Lett. B \textbf{854}, 138713 (2024),
\newblock \doi{10.1016/j.physletb.2024.138713},
\newblock \href{http://arxiv.org/abs/2403.00513}{{\tt 2403.00513}}.

\bibitem{Ablinger:2025nnq}
J.~Ablinger, J.~Bl{\"u}mlein, A.~De~Freitas, A.~von Manteuffel, C.~Schneider
  and K.~Sch{\"o}nwald,
\newblock \emph{{The two-mass contributions to the three-loop massive operator
  matrix elements ${\widetilde{A}}_{Qg}^{(3)}$ and ${\Delta
  \widetilde{A}}_{Qg}^{(3)}$}},
\newblock JHEP \textbf{01}, 111 (2026),
\newblock \doi{10.1007/JHEP01(2026)111},
\newblock \href{http://arxiv.org/abs/2510.09403}{{\tt 2510.09403}}.

\bibitem{Blumlein:2023aso}
J.~Bl{\"u}mlein,
\newblock \emph{{Deep-inelastic scattering: What do we know?}},
\newblock Int. J. Mod. Phys. A \textbf{39}(09n10), 2441004 (2024),
\newblock \doi{10.1142/S0217751X24410045},
\newblock \href{http://arxiv.org/abs/2306.01362}{{\tt 2306.01362}}.

\bibitem{Blumlein:2021lmf}
J.~Bl{\"u}mlein and M.~Saragnese,
\newblock \emph{{The N3LO scheme-invariant QCD evolution of the non-singlet
  structure functions F2NS(x,Q2) and g1NS(x,Q2)}},
\newblock Phys. Lett. B \textbf{820}, 136589 (2021),
\newblock \doi{10.1016/j.physletb.2021.136589},
\newblock \href{http://arxiv.org/abs/2107.01293}{{\tt 2107.01293}}.

\bibitem{Ablinger:2025joi}
J.~Ablinger, A.~Behring, J.~Bl{\"u}mlein, A.~De~Freitas, A.~von Manteuffel,
  C.~Schneider and K.~Sch{\"o}nwald,
\newblock \emph{{The single-mass variable flavor number scheme at three-loop
  order}},
\newblock JHEP \textbf{03}, 248 (2026),
\newblock \doi{10.1007/JHEP03(2026)248},
\newblock \href{http://arxiv.org/abs/2510.02175}{{\tt 2510.02175}}.

\bibitem{Cridge:2025oel}
T.~Cridge, L.~A. Harland-Lang and R.~S. Thorne,
\newblock \emph{{Approximate N3LO PDFs: Updates and Consequences for
  Phenomenology}},
\newblock PoS \textbf{DIS2025}, 027 (2025),
\newblock \doi{10.22323/1.512.0027},
\newblock \href{http://arxiv.org/abs/2510.09321}{{\tt 2510.09321}}.

\bibitem{Anastasiou:2002yz}
C.~Anastasiou and K.~Melnikov,
\newblock \emph{{Higgs boson production at hadron colliders in NNLO QCD}},
\newblock Nucl. Phys. B \textbf{646}, 220 (2002),
\newblock \doi{10.1016/S0550-3213(02)00837-4},
\newblock \href{http://arxiv.org/abs/hep-ph/0207004}{{\tt hep-ph/0207004}}.

\bibitem{Harlander:2002wh}
R.~V. Harlander and W.~B. Kilgore,
\newblock \emph{{Next-to-next-to-leading order Higgs production at hadron
  colliders}},
\newblock Phys. Rev. Lett. \textbf{88}, 201801 (2002),
\newblock \doi{10.1103/PhysRevLett.88.201801},
\newblock \href{http://arxiv.org/abs/hep-ph/0201206}{{\tt hep-ph/0201206}}.

\bibitem{Ravindran:2003um}
V.~Ravindran, J.~Smith and W.~L. van Neerven,
\newblock \emph{{NNLO corrections to the total cross-section for Higgs boson
  production in hadron hadron collisions}},
\newblock Nucl. Phys. \textbf{B665}, 325 (2003),
\newblock \doi{10.1016/S0550-3213(03)00457-7},
\newblock \href{http://arxiv.org/abs/hep-ph/0302135}{{\tt hep-ph/0302135}}.

\bibitem{Dawson:1990zj}
S.~Dawson,
\newblock \emph{{Radiative corrections to Higgs boson production}},
\newblock Nucl. Phys. \textbf{B359}, 283 (1991),
\newblock \doi{10.1016/0550-3213(91)90061-2}.

\bibitem{Djouadi:1991tka}
A.~Djouadi, M.~Spira and P.~M. Zerwas,
\newblock \emph{{Production of Higgs bosons in proton colliders: QCD
  corrections}},
\newblock Phys. Lett. \textbf{B264}, 440 (1991),
\newblock \doi{10.1016/0370-2693(91)90375-Z}.

\bibitem{Graudenz:1992pv}
D.~Graudenz, M.~Spira and P.~M. Zerwas,
\newblock \emph{{QCD corrections to Higgs boson production at proton proton
  colliders}},
\newblock Phys. Rev. Lett. \textbf{70}, 1372 (1993),
\newblock \doi{10.1103/PhysRevLett.70.1372}.

\bibitem{Spira:1995rr}
M.~Spira, A.~Djouadi, D.~Graudenz and P.~M. Zerwas,
\newblock \emph{{Higgs boson production at the LHC}},
\newblock Nucl. Phys. B \textbf{453}, 17 (1995),
\newblock \doi{10.1016/0550-3213(95)00379-7},
\newblock \href{http://arxiv.org/abs/hep-ph/9504378}{{\tt hep-ph/9504378}}.

\bibitem{Actis:2008ts}
S.~Actis, G.~Passarino, C.~Sturm and S.~Uccirati,
\newblock \emph{{NNLO Computational Techniques: The Cases H ---{\ensuremath{>}}
  gamma gamma and H ---{\ensuremath{>}} g g}},
\newblock Nucl. Phys. B \textbf{811}, 182 (2009),
\newblock \doi{10.1016/j.nuclphysb.2008.11.024},
\newblock \href{http://arxiv.org/abs/0809.3667}{{\tt 0809.3667}}.

\bibitem{Actis:2008ug}
S.~Actis, G.~Passarino, C.~Sturm and S.~Uccirati,
\newblock \emph{{NLO Electroweak Corrections to Higgs Boson Production at
  Hadron Colliders}},
\newblock Phys. Lett. B \textbf{670}, 12 (2008),
\newblock \doi{10.1016/j.physletb.2008.10.018},
\newblock \href{http://arxiv.org/abs/0809.1301}{{\tt 0809.1301}}.

\bibitem{Aglietti:2004nj}
U.~Aglietti, R.~Bonciani, G.~Degrassi and A.~Vicini,
\newblock \emph{{Two loop light fermion contribution to Higgs production and
  decays}},
\newblock Phys. Lett. B \textbf{595}, 432 (2004),
\newblock \doi{10.1016/j.physletb.2004.06.063},
\newblock \href{http://arxiv.org/abs/hep-ph/0404071}{{\tt hep-ph/0404071}}.

\bibitem{Bonvini:2014joa}
M.~Bonvini and S.~Marzani,
\newblock \emph{{Resummed Higgs cross section at N$^{3}$LL}},
\newblock JHEP \textbf{09}, 007 (2014),
\newblock \doi{10.1007/JHEP09(2014)007},
\newblock \href{http://arxiv.org/abs/1405.3654}{{\tt 1405.3654}}.

\bibitem{Catani:2014uta}
S.~Catani, L.~Cieri, D.~de~Florian, G.~Ferrera and M.~Grazzini,
\newblock \emph{{Threshold resummation at N$^3$LL accuracy and soft-virtual
  cross sections at N$^3$LO}},
\newblock Nucl. Phys. \textbf{B888}, 75 (2014),
\newblock \doi{10.1016/j.nuclphysb.2014.09.012},
\newblock \href{http://arxiv.org/abs/1405.4827}{{\tt 1405.4827}}.

\bibitem{Bertone:2017bme}
V.~Bertone, S.~Carrazza, N.~P. Hartland and J.~Rojo,
\newblock \emph{{Illuminating the photon content of the proton within a global
  PDF analysis}},
\newblock SciPost Phys. \textbf{5}(1), 008 (2018),
\newblock \doi{10.21468/SciPostPhys.5.1.008},
\newblock \href{http://arxiv.org/abs/1712.07053}{{\tt 1712.07053}}.

\bibitem{Harland-Lang:2019pla}
L.~A. Harland-Lang, A.~D. Martin, R.~Nathvani and R.~S. Thorne,
\newblock \emph{{Ad Lucem: QED Parton Distribution Functions in the MMHT
  Framework}},
\newblock Eur. Phys. J. C \textbf{79}(10), 811 (2019),
\newblock \doi{10.1140/epjc/s10052-019-7296-0},
\newblock \href{http://arxiv.org/abs/1907.02750}{{\tt 1907.02750}}.

\bibitem{Xie:2021equ}
K.~Xie, T.~J. Hobbs, T.-J. Hou, C.~Schmidt, M.~Yan and C.~P. Yuan,
\newblock \emph{{Photon PDF within the CT18 global analysis}},
\newblock Phys. Rev. D \textbf{105}(5), 054006 (2022),
\newblock \doi{10.1103/PhysRevD.105.054006},
\newblock \href{http://arxiv.org/abs/2106.10299}{{\tt 2106.10299}}.

\bibitem{Cridge:2021pxm}
T.~Cridge, L.~A. Harland-Lang, A.~D. Martin and R.~S. Thorne,
\newblock \emph{{QED parton distribution functions in the MSHT20 fit}},
\newblock Eur. Phys. J. C \textbf{82}(1), 90 (2022),
\newblock \doi{10.1140/epjc/s10052-022-10028-2},
\newblock \href{http://arxiv.org/abs/2111.05357}{{\tt 2111.05357}}.

\bibitem{Cridge:2023ryv}
T.~Cridge, L.~A. Harland-Lang and R.~S. Thorne,
\newblock \emph{{Combining QED and approximate ${\rm N}^3$LO QCD corrections in
  a global PDF fit: MSHT20qed{\_}an3lo PDFs}},
\newblock SciPost Phys. \textbf{17}(1), 026 (2024),
\newblock \doi{10.21468/SciPostPhys.17.1.026},
\newblock \href{http://arxiv.org/abs/2312.07665}{{\tt 2312.07665}}.

\bibitem{Barontini:2024dyb}
A.~Barontini, N.~Laurenti and J.~Rojo,
\newblock \emph{{NNPDF4.0 aN$^3$LO PDFs with QED corrections}},
\newblock In \emph{{31st International Workshop on Deep-Inelastic Scattering
  and Related Subjects}} (2024), \href{http://arxiv.org/abs/2406.01779}{{\tt
  2406.01779}}.

\bibitem{NNPDF:2024djq}
R.~D. Ball \emph{et~al.},
\newblock \emph{{Photons in the proton: implications for the LHC}},
\newblock Eur. Phys. J. C \textbf{84}(5), 540 (2024),
\newblock \doi{10.1140/epjc/s10052-024-12731-8},
\newblock \href{http://arxiv.org/abs/2401.08749}{{\tt 2401.08749}}.

\bibitem{Alekhin:2017kpj}
S.~Alekhin, J.~Bl{\"u}mlein, S.~Moch and R.~Placakyte,
\newblock \emph{{Parton distribution functions, $\alpha_s$, and heavy-quark
  masses for LHC Run II}},
\newblock Phys. Rev. \textbf{D96}(1), 014011 (2017),
\newblock \doi{10.1103/PhysRevD.96.014011},
\newblock \href{http://arxiv.org/abs/1701.05838}{{\tt 1701.05838}}.

\bibitem{Dulat:2018rbf}
F.~Dulat, A.~Lazopoulos and B.~Mistlberger,
\newblock \emph{{iHixs 2 Inclusive Higgs cross sections}},
\newblock Comput. Phys. Commun. \textbf{233}, 243 (2018),
\newblock \doi{10.1016/j.cpc.2018.06.025},
\newblock \href{http://arxiv.org/abs/1802.00827}{{\tt 1802.00827}}.

\bibitem{Georgi:1977gs}
H.~M. Georgi, S.~L. Glashow, M.~E. Machacek and D.~V. Nanopoulos,
\newblock \emph{{Higgs Bosons from Two Gluon Annihilation in Proton Proton
  Collisions}},
\newblock Phys. Rev. Lett. \textbf{40}, 692 (1978),
\newblock \doi{10.1103/PhysRevLett.40.692}.

\bibitem{Wilczek:1977zn}
F.~Wilczek,
\newblock \emph{{Decays of Heavy Vector Mesons Into Higgs Particles}},
\newblock Phys. Rev. Lett. \textbf{39}, 1304 (1977),
\newblock \doi{10.1103/PhysRevLett.39.1304}.

\bibitem{Kramer:1996iq}
M.~Kramer, E.~Laenen and M.~Spira,
\newblock \emph{{Soft gluon radiation in Higgs boson production at the LHC}},
\newblock Nucl. Phys. \textbf{B511}, 523 (1998),
\newblock \doi{10.1016/S0550-3213(97)00679-2},
\newblock \href{http://arxiv.org/abs/hep-ph/9611272}{{\tt hep-ph/9611272}}.

\bibitem{Chetyrkin:1997un}
K.~G. Chetyrkin, B.~A. Kniehl and M.~Steinhauser,
\newblock \emph{{Decoupling relations to $\mathcal{O}(\alpha_s^3)$ and their
  connection to low-energy theorems}},
\newblock Nucl. Phys. \textbf{B510}, 61 (1998),
\newblock \doi{10.1016/S0550-3213(98)81004-3, 10.1016/S0550-3213(97)00649-4},
\newblock \href{http://arxiv.org/abs/hep-ph/9708255}{{\tt hep-ph/9708255}}.

\bibitem{Schroder:2005hy}
Y.~Schroder and M.~Steinhauser,
\newblock \emph{{Four-loop decoupling relations for the strong coupling}},
\newblock JHEP \textbf{01}, 051 (2006),
\newblock \doi{10.1088/1126-6708/2006/01/051},
\newblock \href{http://arxiv.org/abs/hep-ph/0512058}{{\tt hep-ph/0512058}}.

\bibitem{Chetyrkin:2005ia}
K.~Chetyrkin, J.~K{\"{u}}hn and C.~Sturm,
\newblock \emph{{QCD decoupling at four loops}},
\newblock Nuclear Physics B \textbf{744}(1-2), 121 (2006),
\newblock \doi{10.1016/j.nuclphysb.2006.03.020}.

\bibitem{Ball:2013bra}
R.~D. Ball, M.~Bonvini, S.~Forte, S.~Marzani and G.~Ridolfi,
\newblock \emph{{Higgs production in gluon fusion beyond NNLO}},
\newblock Nucl. Phys. B \textbf{874}, 746 (2013),
\newblock \doi{10.1016/j.nuclphysb.2013.06.012},
\newblock \href{http://arxiv.org/abs/1303.3590}{{\tt 1303.3590}}.

\bibitem{Bonvini:2014jma}
M.~Bonvini, R.~D. Ball, S.~Forte, S.~Marzani and G.~Ridolfi,
\newblock \emph{{Updated Higgs cross section at approximate N$^3$LO}},
\newblock J. Phys. G \textbf{41}, 095002 (2014),
\newblock \doi{10.1088/0954-3899/41/9/095002},
\newblock \href{http://arxiv.org/abs/1404.3204}{{\tt 1404.3204}}.

\bibitem{Bonvini:2016frm}
M.~Bonvini, S.~Marzani, C.~Muselli and L.~Rottoli,
\newblock \emph{{On the Higgs cross section at N$^{3}$LO+N$^{3}$LL and its
  uncertainty}},
\newblock JHEP \textbf{08}, 105 (2016),
\newblock \doi{10.1007/JHEP08(2016)105},
\newblock \href{http://arxiv.org/abs/1603.08000}{{\tt 1603.08000}}.

\bibitem{Ahmed:2016otz}
T.~Ahmed, M.~Bonvini, M.~C. Kumar, P.~Mathews, N.~Rana, V.~Ravindran and
  L.~Rottoli,
\newblock \emph{{Pseudo-scalar Higgs boson production at N$^3$ LO$_{\text {A}}$
  +N$^3$ LL $'$}},
\newblock Eur. Phys. J. C \textbf{76}(12), 663 (2016),
\newblock \doi{10.1140/epjc/s10052-016-4510-1},
\newblock \href{http://arxiv.org/abs/1606.00837}{{\tt 1606.00837}}.

\bibitem{Bonvini:2018ixe}
M.~Bonvini and S.~Marzani,
\newblock \emph{{Double resummation for Higgs production}},
\newblock Phys. Rev. Lett. \textbf{120}(20), 202003 (2018),
\newblock \doi{10.1103/PhysRevLett.120.202003},
\newblock \href{http://arxiv.org/abs/1802.07758}{{\tt 1802.07758}}.

\bibitem{Bonvini:2018iwt}
M.~Bonvini,
\newblock \emph{{Small-$x$ phenomenology at the LHC and beyond: HELL 3.0 and
  the case of the Higgs cross section}},
\newblock Eur. Phys. J. C \textbf{78}(10), 834 (2018),
\newblock \doi{10.1140/epjc/s10052-018-6315-x},
\newblock \href{http://arxiv.org/abs/1805.08785}{{\tt 1805.08785}}.

\bibitem{Anastasiou:2008tj}
C.~Anastasiou, R.~Boughezal and F.~Petriello,
\newblock \emph{{Mixed QCD-electroweak corrections to Higgs boson production in
  gluon fusion}},
\newblock JHEP \textbf{04}, 003 (2009),
\newblock \doi{10.1088/1126-6708/2009/04/003},
\newblock \href{http://arxiv.org/abs/0811.3458}{{\tt 0811.3458}}.

\bibitem{Anastasiou:2018adr}
C.~Anastasiou, V.~del Duca, E.~Furlan, B.~Mistlberger, F.~Moriello,
  A.~Schweitzer and C.~Specchia,
\newblock \emph{{Mixed QCD-electroweak corrections to Higgs production via
  gluon fusion in the small mass approximation}},
\newblock JHEP \textbf{03}, 162 (2019),
\newblock \doi{10.1007/JHEP03(2019)162},
\newblock \href{http://arxiv.org/abs/1811.11211}{{\tt 1811.11211}}.

\bibitem{Bonetti:2020hqh}
M.~Bonetti, E.~Panzer, V.~A. Smirnov and L.~Tancredi,
\newblock \emph{{Two-loop mixed QCD-EW corrections to $gg \to Hg$}},
\newblock JHEP \textbf{11}, 045 (2020),
\newblock \doi{10.1007/JHEP11(2020)045},
\newblock \href{http://arxiv.org/abs/2007.09813}{{\tt 2007.09813}}.

\bibitem{Bonetti:2022lrk}
M.~Bonetti, E.~Panzer and L.~Tancredi,
\newblock \emph{{Two-loop mixed QCD-EW corrections to $ q\overline{q} $
  {\textrightarrow} Hg, qg {\textrightarrow} Hq, and $ \overline{q}g $
  {\textrightarrow} $ H\overline{q} $}},
\newblock JHEP \textbf{06}, 115 (2022),
\newblock \doi{10.1007/JHEP06(2022)115},
\newblock \href{http://arxiv.org/abs/2203.17202}{{\tt 2203.17202}}.

\bibitem{Davies:2019wmk}
J.~Davies, F.~Herren and M.~Steinhauser,
\newblock \emph{{Top Quark Mass Effects in Higgs Boson Production at Four-Loop
  Order: Virtual Corrections}},
\newblock Phys. Rev. Lett. \textbf{124}(11), 112002 (2020),
\newblock \doi{10.1103/PhysRevLett.124.112002},
\newblock \href{http://arxiv.org/abs/1911.10214}{{\tt 1911.10214}}.

\bibitem{Alekhin:2024bhs}
S.~Alekhin, M.~V. Garzelli, S.~O. Moch and O.~Zenaiev,
\newblock \emph{{NNLO PDFs driven by top-quark data}},
\newblock Eur. Phys. J. C \textbf{85}(2), 162 (2025),
\newblock \doi{10.1140/epjc/s10052-025-13832-8},
\newblock \href{http://arxiv.org/abs/2407.00545}{{\tt 2407.00545}}.

\bibitem{NNPDF:2021njg}
R.~D. Ball \emph{et~al.},
\newblock \emph{{The path to proton structure at 1\% accuracy}},
\newblock Eur. Phys. J. C \textbf{82}(5), 428 (2022),
\newblock \doi{10.1140/epjc/s10052-022-10328-7},
\newblock \href{http://arxiv.org/abs/2109.02653}{{\tt 2109.02653}}.

\bibitem{Butterworth:2015oua}
J.~Butterworth \emph{et~al.},
\newblock \emph{{PDF4LHC recommendations for LHC Run II}},
\newblock J. Phys. \textbf{G43}, 023001 (2016),
\newblock \doi{10.1088/0954-3899/43/2/023001},
\newblock \href{http://arxiv.org/abs/1510.03865}{{\tt 1510.03865}}.

\bibitem{Vermaseren:2005qc}
J.~A.~M. Vermaseren, A.~Vogt and S.~Moch,
\newblock \emph{{The Third-order QCD corrections to deep-inelastic scattering
  by photon exchange}},
\newblock Nucl. Phys. B \textbf{724}, 3 (2005),
\newblock \doi{10.1016/j.nuclphysb.2005.06.020},
\newblock \href{http://arxiv.org/abs/hep-ph/0504242}{{\tt hep-ph/0504242}}.

\bibitem{Davies:2016jie}
J.~Davies, A.~Vogt, B.~Ruijl, T.~Ueda and J.~A.~M. Vermaseren,
\newblock \emph{{Large-nf contributions to the four-loop splitting functions in
  QCD}},
\newblock Nucl. Phys. B \textbf{915}, 335 (2017),
\newblock \doi{10.1016/j.nuclphysb.2016.12.012},
\newblock \href{http://arxiv.org/abs/1610.07477}{{\tt 1610.07477}}.

\bibitem{Moch:2017uml}
S.~Moch, B.~Ruijl, T.~Ueda, J.~A.~M. Vermaseren and A.~Vogt,
\newblock \emph{{Four-Loop Non-Singlet Splitting Functions in the Planar Limit
  and Beyond}},
\newblock JHEP \textbf{10}, 041 (2017),
\newblock \doi{10.1007/JHEP10(2017)041},
\newblock \href{http://arxiv.org/abs/1707.08315}{{\tt 1707.08315}}.

\bibitem{Moch:2018wjh}
S.~Moch, B.~Ruijl, T.~Ueda, J.~A.~M. Vermaseren and A.~Vogt,
\newblock \emph{{On quartic colour factors in splitting functions and the gluon
  cusp anomalous dimension}},
\newblock Phys. Lett. B \textbf{782}, 627 (2018),
\newblock \doi{10.1016/j.physletb.2018.06.017},
\newblock \href{http://arxiv.org/abs/1805.09638}{{\tt 1805.09638}}.

\bibitem{Moch:2021qrk}
S.~Moch, B.~Ruijl, T.~Ueda, J.~A.~M. Vermaseren and A.~Vogt,
\newblock \emph{{Low moments of the four-loop splitting functions in QCD}},
\newblock Phys. Lett. B \textbf{825}, 136853 (2022),
\newblock \doi{10.1016/j.physletb.2021.136853},
\newblock \href{http://arxiv.org/abs/2111.15561}{{\tt 2111.15561}}.

\bibitem{Falcioni:2023luc}
G.~Falcioni, F.~Herzog, S.~Moch and A.~Vogt,
\newblock \emph{{Four-loop splitting functions in QCD \textendash{} The
  quark-quark case}},
\newblock Phys. Lett. B \textbf{842}, 137944 (2023),
\newblock \doi{10.1016/j.physletb.2023.137944},
\newblock \href{http://arxiv.org/abs/2302.07593}{{\tt 2302.07593}}.

\bibitem{Gehrmann:2023cqm}
T.~Gehrmann, A.~von Manteuffel, V.~Sotnikov and T.-Z. Yang,
\newblock \emph{{Complete $ {N}_f^2 $ contributions to four-loop pure-singlet
  splitting functions}},
\newblock JHEP \textbf{01}, 029 (2024),
\newblock \doi{10.1007/JHEP01(2024)029},
\newblock \href{http://arxiv.org/abs/2308.07958}{{\tt 2308.07958}}.

\bibitem{Gehrmann:2023iah}
T.~Gehrmann, A.~von Manteuffel, V.~Sotnikov and T.-Z. Yang,
\newblock \emph{{The NfCF3 contribution to the non-singlet splitting function
  at four-loop order}},
\newblock Phys. Lett. B \textbf{849}, 138427 (2024),
\newblock \doi{10.1016/j.physletb.2023.138427},
\newblock \href{http://arxiv.org/abs/2310.12240}{{\tt 2310.12240}}.

\bibitem{Falcioni:2023tzp}
G.~Falcioni, F.~Herzog, S.~Moch, J.~Vermaseren and A.~Vogt,
\newblock \emph{{The double fermionic contribution to the four-loop
  quark-to-gluon splitting function}},
\newblock Phys. Lett. B \textbf{848}, 138351 (2024),
\newblock \doi{10.1016/j.physletb.2023.138351},
\newblock \href{http://arxiv.org/abs/2310.01245}{{\tt 2310.01245}}.

\bibitem{Moch:2023tdj}
S.~Moch, B.~Ruijl, T.~Ueda, J.~Vermaseren and A.~Vogt,
\newblock \emph{{Additional moments and x-space approximations of four-loop
  splitting functions in QCD}},
\newblock Phys. Lett. B \textbf{849}, 138468 (2024),
\newblock \doi{10.1016/j.physletb.2024.138468},
\newblock \href{http://arxiv.org/abs/2310.05744}{{\tt 2310.05744}}.

\bibitem{Falcioni:2023vqq}
G.~Falcioni, F.~Herzog, S.~Moch and A.~Vogt,
\newblock \emph{{Four-loop splitting functions in QCD \textendash{} The
  gluon-to-quark case}},
\newblock Phys. Lett. B \textbf{846}, 138215 (2023),
\newblock \doi{10.1016/j.physletb.2023.138215},
\newblock \href{http://arxiv.org/abs/2307.04158}{{\tt 2307.04158}}.

\bibitem{Falcioni:2024xav}
G.~Falcioni, F.~Herzog, S.~Moch and S.~Van~Thurenhout,
\newblock \emph{{Constraints for twist-two alien operators in QCD}},
\newblock JHEP \textbf{11}, 080 (2024),
\newblock \doi{10.1007/JHEP11(2024)080},
\newblock \href{http://arxiv.org/abs/2409.02870}{{\tt 2409.02870}}.

\bibitem{Kniehl:2025ttz}
B.~A. Kniehl, S.~Moch, V.~N. Velizhanin and A.~Vogt,
\newblock \emph{{Flavor Nonsinglet Splitting Functions at Four Loops in QCD:
  Fermionic Contributions}},
\newblock Phys. Rev. Lett. \textbf{135}(7), 071902 (2025),
\newblock \doi{10.1103/hkg5-88hr},
\newblock \href{http://arxiv.org/abs/2505.09381}{{\tt 2505.09381}}.

\bibitem{Kawamura:2012cr}
H.~Kawamura, N.~A. Lo~Presti, S.~Moch and A.~Vogt,
\newblock \emph{{On the next-to-next-to-leading order QCD corrections to
  heavy-quark production in deep-inelastic scattering}},
\newblock Nucl. Phys. B \textbf{864}, 399 (2012),
\newblock \doi{10.1016/j.nuclphysb.2012.07.001},
\newblock \href{http://arxiv.org/abs/1205.5727}{{\tt 1205.5727}}.

\bibitem{Ablinger:2014uka}
J.~Ablinger, J.~Bl{\"u}mlein, A.~De~Freitas, A.~Hasselhuhn, A.~von Manteuffel,
  M.~Round and C.~Schneider,
\newblock \emph{{The $O(\alpha_s^3 T_F^2)$ Contributions to the Gluonic
  Operator Matrix Element}},
\newblock Nucl. Phys. B \textbf{885}, 280 (2014),
\newblock \doi{10.1016/j.nuclphysb.2014.05.028},
\newblock \href{http://arxiv.org/abs/1405.4259}{{\tt 1405.4259}}.

\bibitem{Ablinger:2014tla}
J.~Ablinger, J.~Bl{\"u}mlein, A.~De~Freitas, A.~Hasselhuhn, A.~von Manteuffel,
  M.~Round and C.~Schneider,
\newblock \emph{{3-loop Massive $O(T_F^2)$ Contributions to the DIS Operator
  Matrix Element $A_{gg}$}},
\newblock Nucl. Part. Phys. Proc. \textbf{258-259}, 37 (2015),
\newblock \doi{10.1016/j.nuclphysbps.2015.01.009},
\newblock \href{http://arxiv.org/abs/1409.1435}{{\tt 1409.1435}}.

\bibitem{Blumlein:2021enk}
J.~Bl{\"u}mlein, P.~Marquard, C.~Schneider and K.~Sch{\"o}nwald,
\newblock \emph{{The three-loop unpolarized and polarized non-singlet anomalous
  dimensions from off shell operator matrix elements}},
\newblock Nucl. Phys. B \textbf{971}, 115542 (2021),
\newblock \doi{10.1016/j.nuclphysb.2021.115542},
\newblock \href{http://arxiv.org/abs/2107.06267}{{\tt 2107.06267}}.

\bibitem{Thorne:2024npj}
R.~S. Thorne, T.~Cridge and L.~Harland-Lang,
\newblock \emph{{MSHT Updates 2024}},
\newblock In \emph{{31st International Workshop on Deep-Inelastic Scattering
  and Related Subjects}} (2024), \href{http://arxiv.org/abs/2408.10008}{{\tt
  2408.10008}}.

\bibitem{Hekhorn:2025xke}
F.~Hekhorn,
\newblock \emph{{Towards N$^3$LO PDFs and their implications for Higgs
  production cross-sections}},
\newblock In \emph{{59th Rencontres de Moriond on QCD and High Energy
  Interactions}: {Moriond QCD 2025}} (2025),
  \href{http://arxiv.org/abs/2505.10369}{{\tt 2505.10369}}.

\bibitem{Buckley:2014ana}
A.~Buckley, J.~Ferrando, S.~Lloyd, K.~Nordstr{\"o}m, B.~Page, M.~R{\"u}fenacht,
  M.~Sch{\"o}nherr and G.~Watt,
\newblock \emph{{LHAPDF6: parton density access in the LHC precision era}},
\newblock Eur. Phys. J. C \textbf{75}, 132 (2015),
\newblock \doi{10.1140/epjc/s10052-015-3318-8},
\newblock \href{http://arxiv.org/abs/1412.7420}{{\tt 1412.7420}}.

\bibitem{Huss:2025nlt}
A.~Huss, J.~Huston, S.~Jones, M.~Pellen and R.~R{\"o}ntsch,
\newblock \emph{{Les Houches 2023 -- Physics at TeV Colliders: Report on the
  Standard Model Precision Wishlist}}  (2025),
\newblock \doi{10.21468/SciPostPhysCommRep.21},
\newblock \href{http://arxiv.org/abs/2504.06689}{{\tt 2504.06689}}.

\bibitem{Figy:2003nv}
T.~Figy, C.~Oleari and D.~Zeppenfeld,
\newblock \emph{{Next-to-leading order jet distributions for Higgs boson
  production via weak boson fusion}},
\newblock Phys. Rev. D \textbf{68}, 073005 (2003),
\newblock \doi{10.1103/PhysRevD.68.073005},
\newblock \href{http://arxiv.org/abs/hep-ph/0306109}{{\tt hep-ph/0306109}}.

\bibitem{Berger:2004pca}
E.~L. Berger and J.~M. Campbell,
\newblock \emph{{Higgs boson production in weak boson fusion at next-to-leading
  order}},
\newblock Phys. Rev. D \textbf{70}, 073011 (2004),
\newblock \doi{10.1103/PhysRevD.70.073011},
\newblock \href{http://arxiv.org/abs/hep-ph/0403194}{{\tt hep-ph/0403194}}.

\bibitem{Ciccolini:2007jr}
M.~Ciccolini, A.~Denner and S.~Dittmaier,
\newblock \emph{{Strong and electroweak corrections to the production of Higgs
  + 2jets via weak interactions at the LHC}},
\newblock Phys. Rev. Lett. \textbf{99}, 161803 (2007),
\newblock \doi{10.1103/PhysRevLett.99.161803},
\newblock \href{http://arxiv.org/abs/0707.0381}{{\tt 0707.0381}}.

\bibitem{Bolzoni:2011cu}
P.~Bolzoni, F.~Maltoni, S.-O. Moch and M.~Zaro,
\newblock \emph{{Vector boson fusion at NNLO in QCD: SM Higgs and beyond}},
\newblock Phys. Rev. D \textbf{85}, 035002 (2012),
\newblock \doi{10.1103/PhysRevD.85.035002},
\newblock \href{http://arxiv.org/abs/1109.3717}{{\tt 1109.3717}}.

\bibitem{Campanario:2013fsa}
F.~Campanario, T.~M. Figy, S.~Pl\"atzer and M.~Sj\"odahl,
\newblock \emph{{Electroweak Higgs Boson Plus Three Jet Production at
  Next-to-Leading-Order QCD}},
\newblock Phys. Rev. Lett. \textbf{111}(21), 211802 (2013),
\newblock \doi{10.1103/PhysRevLett.111.211802},
\newblock \href{http://arxiv.org/abs/1308.2932}{{\tt 1308.2932}}.

\bibitem{Campanario:2018ppz}
F.~Campanario, T.~M. Figy, S.~Pl\"atzer, M.~Rauch, P.~Schichtel and
  M.~Sj\"odahl,
\newblock \emph{{Stress testing the vector-boson-fusion approximation in
  multijet final states}},
\newblock Phys. Rev. D \textbf{98}(3), 033003 (2018),
\newblock \doi{10.1103/PhysRevD.98.033003},
\newblock \href{http://arxiv.org/abs/1802.09955}{{\tt 1802.09955}}.

\bibitem{Ciccolini:2007ec}
M.~Ciccolini, A.~Denner and S.~Dittmaier,
\newblock \emph{{Electroweak and QCD corrections to Higgs production via
  vector-boson fusion at the LHC}},
\newblock Phys. Rev. D \textbf{77}, 013002 (2008),
\newblock \doi{10.1103/PhysRevD.77.013002},
\newblock \href{http://arxiv.org/abs/0710.4749}{{\tt 0710.4749}}.

\bibitem{Cacciari:2015jma}
M.~Cacciari, F.~A. Dreyer, A.~Karlberg, G.~P. Salam and G.~Zanderighi,
\newblock \emph{{Fully Differential Vector-Boson-Fusion Higgs Production at
  Next-to-Next-to-Leading Order}},
\newblock Phys. Rev. Lett. \textbf{115}(8), 082002 (2015),
\newblock \doi{10.1103/PhysRevLett.115.082002},
\newblock \href{http://arxiv.org/abs/1506.02660}{{\tt 1506.02660}}.

\bibitem{Cruz-Martinez:2018rod}
J.~Cruz-Martinez, T.~Gehrmann, E.~W.~N. Glover and A.~Huss,
\newblock \emph{{Second-order QCD effects in Higgs boson production through
  vector boson fusion}},
\newblock Phys. Lett. B \textbf{781}, 672 (2018),
\newblock \doi{10.1016/j.physletb.2018.04.046},
\newblock \href{http://arxiv.org/abs/1802.02445}{{\tt 1802.02445}}.

\bibitem{Asteriadis:2021gpd}
K.~Asteriadis, F.~Caola, K.~Melnikov and R.~R\"ontsch,
\newblock \emph{{NNLO QCD corrections to weak boson fusion Higgs boson
  production in the H \textrightarrow{} b$ \overline{b} $ and H
  \textrightarrow{} WW$^{*}$ \textrightarrow{} 4l decay channels}},
\newblock JHEP \textbf{02}, 046 (2022),
\newblock \doi{10.1007/JHEP02(2022)046},
\newblock \href{http://arxiv.org/abs/2110.02818}{{\tt 2110.02818}}.

\bibitem{Han:1992hr}
T.~Han, G.~Valencia and S.~Willenbrock,
\newblock \emph{{Structure function approach to vector boson scattering in p p
  collisions}},
\newblock Phys. Rev. Lett. \textbf{69}, 3274 (1992),
\newblock \doi{10.1103/PhysRevLett.69.3274},
\newblock \href{http://arxiv.org/abs/hep-ph/9206246}{{\tt hep-ph/9206246}}.

\bibitem{Bolzoni:2010xr}
P.~Bolzoni, F.~Maltoni, S.-O. Moch and M.~Zaro,
\newblock \emph{{Higgs production via vector-boson fusion at NNLO in QCD}},
\newblock Phys. Rev. Lett. \textbf{105}, 011801 (2010),
\newblock \doi{10.1103/PhysRevLett.105.011801},
\newblock \href{http://arxiv.org/abs/1003.4451}{{\tt 1003.4451}}.

\bibitem{Bronnum-Hansen:2023vzh}
C.~Br\o{}nnum-Hansen, M.-M. Long and K.~Melnikov,
\newblock \emph{{Scale dependence of non-factorizable virtual corrections to
  Higgs boson production in weak boson fusion}},
\newblock JHEP \textbf{11}, 130 (2023),
\newblock \doi{10.1007/JHEP11(2023)130},
\newblock \href{http://arxiv.org/abs/2309.06292}{{\tt 2309.06292}}.

\bibitem{Long:2023mvc}
M.-M. Long, K.~Melnikov and J.~Quarroz,
\newblock \emph{{Non-factorizable virtual corrections to Higgs boson production
  in weak boson fusion beyond the eikonal approximation}},
\newblock JHEP \textbf{07}, 035 (2023),
\newblock \doi{10.1007/JHEP07(2023)035},
\newblock \href{http://arxiv.org/abs/2305.12937}{{\tt 2305.12937}}.

\bibitem{Gates:2023iiv}
L.~Gates,
\newblock \emph{{On Evaluation of Nonfactorizable Corrections to Higgs Boson
  Production via Vector Boson Fusion}},
\newblock Phys. Lett. B \textbf{846}, 138191 (2023),
\newblock \doi{10.1016/j.physletb.2023.138191},
\newblock \href{http://arxiv.org/abs/2305.04407}{{\tt 2305.04407}}.

\bibitem{Andersen:2006ag}
J.~R. Andersen and J.~M. Smillie,
\newblock \emph{{QCD and electroweak interference in Higgs production by gauge
  boson fusion}},
\newblock Phys. Rev. D \textbf{75}, 037301 (2007),
\newblock \doi{10.1103/PhysRevD.75.037301},
\newblock \href{http://arxiv.org/abs/hep-ph/0611281}{{\tt hep-ph/0611281}}.

\bibitem{Andersen:2007mp}
J.~R. Andersen, T.~Binoth, G.~Heinrich and J.~M. Smillie,
\newblock \emph{{Loop induced interference effects in Higgs Boson plus two jet
  production at the LHC}},
\newblock JHEP \textbf{02}, 057 (2008),
\newblock \doi{10.1088/1126-6708/2008/02/057},
\newblock \href{http://arxiv.org/abs/0709.3513}{{\tt 0709.3513}}.

\bibitem{Harlander:2008xn}
R.~V. Harlander, J.~Vollinga and M.~M. Weber,
\newblock \emph{{Gluon-Induced Weak Boson Fusion}},
\newblock Phys. Rev. D \textbf{77}, 053010 (2008),
\newblock \doi{10.1103/PhysRevD.77.053010},
\newblock \href{http://arxiv.org/abs/0801.3355}{{\tt 0801.3355}}.

\bibitem{Greiner:2016awe}
N.~Greiner, S.~H\"oche, G.~Luisoni, M.~Sch\"onherr and J.-C. Winter,
\newblock \emph{{Full mass dependence in Higgs boson production in association
  with jets at the LHC and FCC}},
\newblock JHEP \textbf{01}, 091 (2017),
\newblock \doi{10.1007/JHEP01(2017)091},
\newblock \href{http://arxiv.org/abs/1608.01195}{{\tt 1608.01195}}.

\bibitem{Greiner:2015jha}
N.~Greiner, S.~H\"oche, G.~Luisoni, M.~Sch\"onherr, J.-C. Winter and V.~Yundin,
\newblock \emph{{Phenomenological analysis of Higgs boson production through
  gluon fusion in association with jets}},
\newblock JHEP \textbf{01}, 169 (2016),
\newblock \doi{10.1007/JHEP01(2016)169},
\newblock \href{http://arxiv.org/abs/1506.01016}{{\tt 1506.01016}}.

\bibitem{Andersen:2017kfc}
J.~R. Andersen, T.~Hapola, A.~Maier and J.~M. Smillie,
\newblock \emph{{Higgs Boson Plus Dijets: Higher Order Corrections}},
\newblock JHEP \textbf{09}, 065 (2017),
\newblock \doi{10.1007/JHEP09(2017)065},
\newblock \href{http://arxiv.org/abs/1706.01002}{{\tt 1706.01002}}.

\bibitem{Andersen:2018tnm}
J.~R. Andersen, T.~Hapola, M.~Heil, A.~Maier and J.~M. Smillie,
\newblock \emph{{Higgs-boson plus Dijets: Higher-Order Matching for High-Energy
  Predictions}},
\newblock JHEP \textbf{08}, 090 (2018),
\newblock \doi{10.1007/JHEP08(2018)090},
\newblock \href{http://arxiv.org/abs/1805.04446}{{\tt 1805.04446}}.

\bibitem{Andersen:2018kjg}
J.~R. Andersen, J.~D. Cockburn, M.~Heil, A.~Maier and J.~M. Smillie,
\newblock \emph{{Finite Quark-Mass Effects in Higgs Boson Production with
  Dijets at Large Energies}},
\newblock JHEP \textbf{04}, 127 (2019),
\newblock \doi{10.1007/JHEP04(2019)127},
\newblock \href{http://arxiv.org/abs/1812.08072}{{\tt 1812.08072}}.

\bibitem{Chen:2021azt}
X.~Chen, A.~Huss, S.~P. Jones, M.~Kerner, J.~N. Lang, J.~M. Lindert and
  H.~Zhang,
\newblock \emph{{Top-quark mass effects in H+jet and H+2 jets production}},
\newblock JHEP \textbf{03}, 096 (2022),
\newblock \doi{10.1007/JHEP03(2022)096},
\newblock \href{http://arxiv.org/abs/2110.06953}{{\tt 2110.06953}}.

\bibitem{Andersen:2022zte}
J.~R. Andersen, H.~Hassan, A.~Maier, J.~Paltrinieri, A.~Papaefstathiou and
  J.~M. Smillie,
\newblock \emph{{High energy resummed predictions for the production of a Higgs
  boson with at least one jet}},
\newblock JHEP \textbf{03}, 001 (2023),
\newblock \doi{10.1007/JHEP03(2023)001},
\newblock \href{http://arxiv.org/abs/2210.10671}{{\tt 2210.10671}}.

\bibitem{Becker:2020rjp}
K.~Becker \emph{et~al.},
\newblock \emph{{Precise predictions for boosted Higgs production}},
\newblock SciPost Phys. Core \textbf{7}, 001 (2024),
\newblock \doi{10.21468/SciPostPhysCore.7.1.001},
\newblock \href{http://arxiv.org/abs/2005.07762}{{\tt 2005.07762}}.

\bibitem{Buckley:2021gfw}
A.~Buckley \emph{et~al.},
\newblock \emph{{A comparative study of Higgs boson production from
  vector-boson fusion}},
\newblock JHEP \textbf{11}, 108 (2021),
\newblock \doi{10.1007/JHEP11(2021)108},
\newblock \href{http://arxiv.org/abs/2105.11399}{{\tt 2105.11399}}.

\bibitem{Asteriadis:2024nbg}
K.~Asteriadis, A.~Behring, K.~Melnikov, I.~Novikov and R.~R\"ontsch,
\newblock \emph{{QCD corrections to Higgs boson production and
  H\textrightarrow{}bb\textasciimacron{} decay in weak boson fusion}},
\newblock Phys. Rev. D \textbf{110}(5), 054017 (2024),
\newblock \doi{10.1103/PhysRevD.110.054017},
\newblock \href{http://arxiv.org/abs/2407.09363}{{\tt 2407.09363}}.

\bibitem{Nason:2009ai}
P.~Nason and C.~Oleari,
\newblock \emph{{NLO Higgs boson production via vector-boson fusion matched
  with shower in POWHEG}},
\newblock JHEP \textbf{02}, 037 (2010),
\newblock \doi{10.1007/JHEP02(2010)037},
\newblock \href{http://arxiv.org/abs/0911.5299}{{\tt 0911.5299}}.

\bibitem{Frixione:2013mta}
S.~Frixione, P.~Torrielli and M.~Zaro,
\newblock \emph{{Higgs production through vector-boson fusion at the NLO
  matched with parton showers}},
\newblock Phys. Lett. B \textbf{726}, 273 (2013),
\newblock \doi{10.1016/j.physletb.2013.08.030},
\newblock \href{http://arxiv.org/abs/1304.7927}{{\tt 1304.7927}}.

\bibitem{Jager:2022acp}
B.~J\"ager and J.~Scheller,
\newblock \emph{{Electroweak corrections and shower effects to Higgs production
  in association with two jets at the LHC}},
\newblock JHEP \textbf{09}, 191 (2022),
\newblock \doi{10.1007/JHEP09(2022)191},
\newblock \href{http://arxiv.org/abs/2208.00013}{{\tt 2208.00013}}.

\bibitem{Hoche:2021mkv}
S.~H\"oche, S.~Mrenna, S.~Payne, C.~T. Preuss and P.~Skands,
\newblock \emph{{A Study of QCD Radiation in VBF Higgs Production with Vincia
  and Pythia}},
\newblock SciPost Phys. \textbf{12}(1), 010 (2022),
\newblock \doi{10.21468/SciPostPhys.12.1.010},
\newblock \href{http://arxiv.org/abs/2106.10987}{{\tt 2106.10987}}.

\bibitem{Chen:2021phj}
T.~Chen, T.~M. Figy and S.~Pl\"atzer,
\newblock \emph{{NLO multijet merging for Higgs production beyond the VBF
  approximation}},
\newblock Eur. Phys. J. C \textbf{82}(8), 704 (2022),
\newblock \doi{10.1140/epjc/s10052-022-10671-9},
\newblock \href{http://arxiv.org/abs/2109.03730}{{\tt 2109.03730}}.

\bibitem{Jager:2020hkz}
B.~J\"ager, A.~Karlberg, S.~Pl\"atzer, J.~Scheller and M.~Zaro,
\newblock \emph{{Parton-shower effects in Higgs production via Vector-Boson
  Fusion}},
\newblock Eur. Phys. J. C \textbf{80}(8), 756 (2020),
\newblock \doi{10.1140/epjc/s10052-020-8326-7},
\newblock \href{http://arxiv.org/abs/2003.12435}{{\tt 2003.12435}}.

\bibitem{Bittrich:2021ztq}
C.~Bittrich, P.~Kirchgae\ss{}er, A.~Papaefstathiou, S.~Pl\"atzer and S.~Todt,
\newblock \emph{{Soft QCD effects in VBS/VBF topologies}},
\newblock Eur. Phys. J. C \textbf{82}(9), 783 (2022),
\newblock \doi{10.1140/epjc/s10052-022-10741-y},
\newblock \href{http://arxiv.org/abs/2110.01623}{{\tt 2110.01623}}.

\bibitem{vanBeekveld:2023chs}
M.~van Beekveld and S.~Ferrario~Ravasio,
\newblock \emph{{Next-to-leading-logarithmic PanScales showers for Deep
  Inelastic Scattering and Vector Boson Fusion}},
\newblock JHEP \textbf{02}, 001 (2024),
\newblock \doi{10.1007/JHEP02(2024)001},
\newblock \href{http://arxiv.org/abs/2305.08645}{{\tt 2305.08645}}.

\bibitem{Denner:2014cla}
A.~Denner, S.~Dittmaier, S.~Kallweit and A.~M\"uck,
\newblock \emph{{HAWK 2.0: A Monte Carlo program for Higgs production in
  vector-boson fusion and Higgs strahlung at hadron colliders}},
\newblock Comput. Phys. Commun. \textbf{195}, 161 (2015),
\newblock \doi{10.1016/j.cpc.2015.04.021},
\newblock \href{http://arxiv.org/abs/1412.5390}{{\tt 1412.5390}}.

\bibitem{Glashow:1978ab}
S.~L. Glashow, D.~V. Nanopoulos and A.~Yildiz,
\newblock \emph{{Associated Production of Higgs Bosons and Z Particles}},
\newblock Phys. Rev. D \textbf{18}, 1724 (1978),
\newblock \doi{10.1103/PhysRevD.18.1724}.

\bibitem{Eichten:1984eu}
E.~Eichten, I.~Hinchliffe, K.~D. Lane and C.~Quigg,
\newblock \emph{{Super Collider Physics}},
\newblock Rev. Mod. Phys. \textbf{56}, 579 (1984),
\newblock \doi{10.1103/RevModPhys.56.579},
\newblock [Addendum: Rev.Mod.Phys. 58, 1065--1073 (1986)].

\bibitem{HAN1991167}
T.~Han and S.~Willenbrock,
\newblock \emph{Qcd correction to the $pp\to wh$ and $zh$ total cross
  sections},
\newblock Physics Letters B \textbf{273}(1), 167 (1991),
\newblock \doi{https://doi.org/10.1016/0370-2693(91)90572-8}.

\bibitem{Brein:2003wg}
O.~Brein, A.~Djouadi and R.~Harlander,
\newblock \emph{{NNLO QCD corrections to the Higgs-strahlung processes at
  hadron colliders}},
\newblock Phys. Lett. B \textbf{579}, 149 (2004),
\newblock \doi{10.1016/j.physletb.2003.10.112},
\newblock \href{http://arxiv.org/abs/hep-ph/0307206}{{\tt hep-ph/0307206}}.

\bibitem{Ciccolini:2003jy}
M.~L. Ciccolini, S.~Dittmaier and M.~Kramer,
\newblock \emph{{Electroweak radiative corrections to associated WH and ZH
  production at hadron colliders}},
\newblock Phys. Rev. D \textbf{68}, 073003 (2003),
\newblock \doi{10.1103/PhysRevD.68.073003},
\newblock \href{http://arxiv.org/abs/hep-ph/0306234}{{\tt hep-ph/0306234}}.

\bibitem{Denner:2011id}
A.~Denner, S.~Dittmaier, S.~Kallweit and A.~Muck,
\newblock \emph{{Electroweak corrections to Higgs-strahlung off W/Z bosons at
  the Tevatron and the LHC with HAWK}},
\newblock JHEP \textbf{03}, 075 (2012),
\newblock \doi{10.1007/JHEP03(2012)075},
\newblock \href{http://arxiv.org/abs/1112.5142}{{\tt 1112.5142}}.

\bibitem{Ferrera:2011bk}
G.~Ferrera, M.~Grazzini and F.~Tramontano,
\newblock \emph{{Associated WH production at hadron colliders: a fully
  exclusive QCD calculation at NNLO}},
\newblock Phys. Rev. Lett. \textbf{107}, 152003 (2011),
\newblock \doi{10.1103/PhysRevLett.107.152003},
\newblock \href{http://arxiv.org/abs/1107.1164}{{\tt 1107.1164}}.

\bibitem{Ferrera:2013yga}
G.~Ferrera, M.~Grazzini and F.~Tramontano,
\newblock \emph{{Higher-order QCD effects for associated WH production and
  decay at the LHC}},
\newblock JHEP \textbf{04}, 039 (2014),
\newblock \doi{10.1007/JHEP04(2014)039},
\newblock \href{http://arxiv.org/abs/1312.1669}{{\tt 1312.1669}}.

\bibitem{Ferrera:2014lca}
G.~Ferrera, M.~Grazzini and F.~Tramontano,
\newblock \emph{{Associated ZH production at hadron colliders: the fully
  differential NNLO QCD calculation}},
\newblock Phys. Lett. B \textbf{740}, 51 (2015),
\newblock \doi{10.1016/j.physletb.2014.11.040},
\newblock \href{http://arxiv.org/abs/1407.4747}{{\tt 1407.4747}}.

\bibitem{Campbell:2016jau}
J.~M. Campbell, R.~K. Ellis and C.~Williams,
\newblock \emph{{Associated production of a Higgs boson at NNLO}},
\newblock JHEP \textbf{06}, 179 (2016),
\newblock \doi{10.1007/JHEP06(2016)179},
\newblock \href{http://arxiv.org/abs/1601.00658}{{\tt 1601.00658}}.

\bibitem{Gauld:2021ule}
R.~Gauld, A.~Gehrmann-De~Ridder, E.~W.~N. Glover, A.~Huss and I.~Majer,
\newblock \emph{{VH + jet production in hadron-hadron collisions up to order $
  {\alpha}_{\mathrm{s}}^3 $ in perturbative QCD}},
\newblock JHEP \textbf{03}, 008 (2022),
\newblock \doi{10.1007/JHEP03(2022)008},
\newblock \href{http://arxiv.org/abs/2110.12992}{{\tt 2110.12992}}.

\bibitem{Dicus:1988yh}
D.~A. Dicus and C.~Kao,
\newblock \emph{{Higgs Boson - $Z^0$ Production From Gluon Fusion}},
\newblock Phys. Rev. D \textbf{38}, 1008 (1988),
\newblock \doi{10.1103/PhysRevD.38.1008},
\newblock [Erratum: Phys.Rev.D 42, 2412 (1990)].

\bibitem{Kniehl:1990iva}
B.~A. Kniehl,
\newblock \emph{{Associated Production of Higgs and Z Bosons From Gluon Fusion
  in Hadron Collisions}},
\newblock Phys. Rev. D \textbf{42}, 2253 (1990),
\newblock \doi{10.1103/PhysRevD.42.2253}.

\bibitem{Altenkamp:2012sx}
L.~Altenkamp, S.~Dittmaier, R.~V. Harlander, H.~Rzehak and T.~J.~E. Zirke,
\newblock \emph{{Gluon-induced Higgs-strahlung at next-to-leading order QCD}},
\newblock JHEP \textbf{02}, 078 (2013),
\newblock \doi{10.1007/JHEP02(2013)078},
\newblock \href{http://arxiv.org/abs/1211.5015}{{\tt 1211.5015}}.

\bibitem{Hasselhuhn:2016rqt}
A.~Hasselhuhn, T.~Luthe and M.~Steinhauser,
\newblock \emph{{On top quark mass effects to $gg\to ZH$ at NLO}},
\newblock JHEP \textbf{01}, 073 (2017),
\newblock \doi{10.1007/JHEP01(2017)073},
\newblock \href{http://arxiv.org/abs/1611.05881}{{\tt 1611.05881}}.

\bibitem{Davies:2020drs}
J.~Davies, G.~Mishima and M.~Steinhauser,
\newblock \emph{{Virtual corrections to $gg\to ZH$ in the high-energy and
  large-$m_t$ limits}},
\newblock JHEP \textbf{03}, 034 (2021),
\newblock \doi{10.1007/JHEP03(2021)034},
\newblock \href{http://arxiv.org/abs/2011.12314}{{\tt 2011.12314}}.

\bibitem{Alasfar:2021ppe}
L.~Alasfar, G.~Degrassi, P.~P. Giardino, R.~Gr\"ober and M.~Vitti,
\newblock \emph{{Virtual corrections to $gg\to ZH$ via a transverse momentum
  expansion}},
\newblock JHEP \textbf{05}, 168 (2021),
\newblock \doi{10.1007/JHEP05(2021)168},
\newblock \href{http://arxiv.org/abs/2103.06225}{{\tt 2103.06225}}.

\bibitem{Davies:2026uxl}
J.~Davies, K.~Sch{\"o}nwald, M.~Steinhauser and D.~Stremmer,
\newblock \emph{{gg {\textrightarrow} ZH at NLO matched to parton showers with
  ggxy and POWHEG}},
\newblock JHEP \textbf{07}, 146 (2026),
\newblock \doi{10.1007/JHEP07(2026)146},
\newblock \href{http://arxiv.org/abs/2603.15762}{{\tt 2603.15762}}.

\bibitem{Caola:2017xuq}
F.~Caola, G.~Luisoni, K.~Melnikov and R.~R{\"o}ntsch,
\newblock \emph{{NNLO QCD corrections to associated $WH$ production and $H \to
  b \bar b$ decay}},
\newblock Phys. Rev. \textbf{D97}(7), 074022 (2018),
\newblock \doi{10.1103/PhysRevD.97.074022},
\newblock \href{http://arxiv.org/abs/1712.06954}{{\tt 1712.06954}}.

\bibitem{Gauld:2019yng}
R.~Gauld, A.~Gehrmann-De~Ridder, E.~W.~N. Glover, A.~Huss and I.~Majer,
\newblock \emph{{Associated production of a Higgs boson decaying into bottom
  quarks and a weak vector boson decaying leptonically at NNLO in QCD}},
\newblock JHEP \textbf{10}, 002 (2019),
\newblock \doi{10.1007/JHEP10(2019)002},
\newblock \href{http://arxiv.org/abs/1907.05836}{{\tt 1907.05836}}.

\bibitem{Behring:2020uzq}
A.~Behring, W.~Bizo{\'n}, F.~Caola, K.~Melnikov and R.~R{\"o}ntsch,
\newblock \emph{{Bottom quark mass effects in associated $WH$ production with
  the $H \to b\bar{b}$ decay through NNLO QCD}},
\newblock Phys. Rev. D \textbf{101}(11), 114012 (2020),
\newblock \doi{10.1103/PhysRevD.101.114012},
\newblock \href{http://arxiv.org/abs/2003.08321}{{\tt 2003.08321}}.

\bibitem{Kumar:2014uwa}
M.~C. Kumar, M.~K. Mandal and V.~Ravindran,
\newblock \emph{{Associated production of Higgs boson with vector boson at
  threshold N$^{3}$LO in QCD}},
\newblock JHEP \textbf{03}, 037 (2015),
\newblock \doi{10.1007/JHEP03(2015)037},
\newblock \href{http://arxiv.org/abs/1412.3357}{{\tt 1412.3357}}.

\bibitem{Das:2022zie}
G.~Das, C.~Dey, M.~C. Kumar and K.~Samanta,
\newblock \emph{{Threshold enhanced cross sections for colorless productions}},
\newblock Phys. Rev. D \textbf{107}(3), 034038 (2023),
\newblock \doi{10.1103/PhysRevD.107.034038},
\newblock \href{http://arxiv.org/abs/2210.17534}{{\tt 2210.17534}}.

\bibitem{Harlander:2014wda}
R.~V. Harlander, A.~Kulesza, V.~Theeuwes and T.~Zirke,
\newblock \emph{{Soft gluon resummation for gluon-induced Higgs Strahlung}},
\newblock JHEP \textbf{11}, 082 (2014),
\newblock \doi{10.1007/JHEP11(2014)082},
\newblock \href{http://arxiv.org/abs/1410.0217}{{\tt 1410.0217}}.

\bibitem{Das:2025wbj}
G.~Das, C.~Dey, M.~C. Kumar and K.~Samanta,
\newblock \emph{{Soft gluon resummation for gluon fusion $ZH$ production}},
\newblock Phys. Rev. D \textbf{113}(1), 014024 (2026),
\newblock \doi{10.1103/mxxw-bvm3},
\newblock \href{http://arxiv.org/abs/2501.10330}{{\tt 2501.10330}}.

\bibitem{Das:2025qym}
G.~Das, C.~Dey, M.~C. Kumar and K.~Samanta,
\newblock \emph{{Threshold improved $Z H$ production at the LHC}}  (2025),
\newblock \doi{10.21468/SciPostPhysCommRep.14},
\newblock \href{http://arxiv.org/abs/2508.20676}{{\tt 2508.20676}}.

\bibitem{Banerjee:2026hzp}
P.~Banerjee, C.~Dey, N.~Koirala, M.~C. Kumar and V.~Pandey,
\newblock \emph{{Associated $ZH$ production in gluon fusion process at
  NLO+NLL}}  (2026),
\newblock \href{http://arxiv.org/abs/2605.24131}{{\tt 2605.24131}}.

\bibitem{Luisoni:2013cuh}
G.~Luisoni, P.~Nason, C.~Oleari and F.~Tramontano,
\newblock \emph{{$HW^{\pm}$/HZ + 0 and 1 jet at NLO with the POWHEG BOX
  interfaced to GoSam and their merging within MiNLO}},
\newblock JHEP \textbf{10}, 083 (2013),
\newblock \doi{10.1007/JHEP10(2013)083},
\newblock \href{http://arxiv.org/abs/1306.2542}{{\tt 1306.2542}}.

\bibitem{Granata:2017iod}
F.~Granata, J.~M. Lindert, C.~Oleari and S.~Pozzorini,
\newblock \emph{{NLO QCD+EW predictions for HV and HV +jet production including
  parton-shower effects}},
\newblock JHEP \textbf{09}, 012 (2017),
\newblock \doi{10.1007/JHEP09(2017)012},
\newblock \href{http://arxiv.org/abs/1706.03522}{{\tt 1706.03522}}.

\bibitem{Hespel:2015zea}
B.~Hespel, F.~Maltoni and E.~Vryonidou,
\newblock \emph{{Higgs and Z boson associated production via gluon fusion in
  the SM and the 2HDM}},
\newblock JHEP \textbf{06}, 065 (2015),
\newblock \doi{10.1007/JHEP06(2015)065},
\newblock \href{http://arxiv.org/abs/1503.01656}{{\tt 1503.01656}}.

\bibitem{Goncalves:2015mfa}
D.~Goncalves, F.~Krauss, S.~Kuttimalai and P.~Maierh{\"o}fer,
\newblock \emph{{Higgs-Strahlung: Merging the NLO Drell-Yan and Loop-Induced
  0+1 jet Multiplicities}},
\newblock Phys. Rev. D \textbf{92}(7), 073006 (2015),
\newblock \doi{10.1103/PhysRevD.92.073006},
\newblock \href{http://arxiv.org/abs/1509.01597}{{\tt 1509.01597}}.

\bibitem{Astill:2016hpa}
W.~Astill, W.~Bizon, E.~Re and G.~Zanderighi,
\newblock \emph{{NNLOPS accurate associated HW production}},
\newblock JHEP \textbf{06}, 154 (2016),
\newblock \doi{10.1007/JHEP06(2016)154},
\newblock \href{http://arxiv.org/abs/1603.01620}{{\tt 1603.01620}}.

\bibitem{Astill:2018ivh}
W.~Astill, W.~Bizo{\'n}, E.~Re and G.~Zanderighi,
\newblock \emph{{NNLOPS accurate associated HZ production with $ H\to
  b\overline{b} $ decay at NLO}},
\newblock JHEP \textbf{11}, 157 (2018),
\newblock \doi{10.1007/JHEP11(2018)157},
\newblock \href{http://arxiv.org/abs/1804.08141}{{\tt 1804.08141}}.

\bibitem{Alioli:2019qzz}
S.~Alioli, A.~Broggio, S.~Kallweit, M.~A. Lim and L.~Rottoli,
\newblock \emph{{Higgsstrahlung at NNLL'$+$NNLO matched to parton showers in
  GENEVA}},
\newblock Phys. Rev. D \textbf{100}(9), 096016 (2019),
\newblock \doi{10.1103/PhysRevD.100.096016},
\newblock \href{http://arxiv.org/abs/1909.02026}{{\tt 1909.02026}}.

\bibitem{Zanoli:2021iyp}
S.~Zanoli, M.~Chiesa, E.~Re, M.~Wiesemann and G.~Zanderighi,
\newblock \emph{{Next-to-next-to-leading order event generation for VH
  production with H {\textrightarrow}$ b\overline{b} $ decay}},
\newblock JHEP \textbf{07}, 008 (2022),
\newblock \doi{10.1007/JHEP07(2022)008},
\newblock \href{http://arxiv.org/abs/2112.04168}{{\tt 2112.04168}}.

\bibitem{Brein:2012ne}
O.~Brein, R.~V. Harlander and T.~J.~E. Zirke,
\newblock \emph{{vh@nnlo - Higgs Strahlung at hadron colliders}},
\newblock Comput. Phys. Commun. \textbf{184}, 998 (2013),
\newblock \doi{10.1016/j.cpc.2012.11.002},
\newblock \href{http://arxiv.org/abs/1210.5347}{{\tt 1210.5347}}.

\bibitem{Harlander:2018yio}
R.~V. Harlander, J.~Klappert, S.~Liebler and L.~Simon,
\newblock \emph{{vh@nnlo-v2: New physics in Higgs Strahlung}},
\newblock JHEP \textbf{05}, 089 (2018),
\newblock \doi{10.1007/JHEP05(2018)089},
\newblock \href{http://arxiv.org/abs/1802.04817}{{\tt 1802.04817}}.

\bibitem{Braun:2025afl}
J.~Braun, B.~Campillo~Aveleira, G.~Heinrich, M.~H{\"o}fer, S.~P. Jones,
  M.~Kerner, J.~Lang and V.~Magerya,
\newblock \emph{{One-Loop Calculations in Effective Field Theories with
  GoSam-3.0}},
\newblock SciPost Phys. Codeb. \textbf{62}, 1 (2026),
\newblock \doi{10.21468/SciPostPhysCodeb.62},
\newblock \href{http://arxiv.org/abs/2507.23549}{{\tt 2507.23549}}.

\bibitem{Kulesza:2015vda}
A.~Kulesza, L.~Motyka, T.~Stebel and V.~Theeuwes,
\newblock \emph{{Soft gluon resummation for associated $t \bar{t} H$ production
  at the LHC}},
\newblock JHEP \textbf{03}, 065 (2016),
\newblock \doi{10.1007/JHEP03(2016)065},
\newblock \href{http://arxiv.org/abs/1509.02780}{{\tt 1509.02780}}.

\bibitem{Broggio:2015lya}
A.~Broggio, A.~Ferroglia, B.~D. Pecjak, A.~Signer and L.~L. Yang,
\newblock \emph{{Associated production of a top pair and a Higgs boson beyond
  NLO}},
\newblock JHEP \textbf{03}, 124 (2016),
\newblock \doi{10.1007/JHEP03(2016)124},
\newblock \href{http://arxiv.org/abs/1510.01914}{{\tt 1510.01914}}.

\bibitem{Broggio:2016lfj}
A.~Broggio, A.~Ferroglia, B.~D. Pecjak and L.~L. Yang,
\newblock \emph{{NNLL resummation for the associated production of a top pair
  and a Higgs boson at the LHC}},
\newblock JHEP \textbf{02}, 126 (2017),
\newblock \doi{10.1007/JHEP02(2017)126},
\newblock \href{http://arxiv.org/abs/1611.00049}{{\tt 1611.00049}}.

\bibitem{Kulesza:2017ukk}
A.~Kulesza, L.~Motyka, T.~Stebel and V.~Theeuwes,
\newblock \emph{{Associated $t \bar{t} H$ production at the LHC: Theoretical
  predictions at NLO+NNLL accuracy}},
\newblock Phys. Rev. D \textbf{97}(11), 114007 (2018),
\newblock \doi{10.1103/PhysRevD.97.114007},
\newblock \href{http://arxiv.org/abs/1704.03363}{{\tt 1704.03363}}.

\bibitem{Kulesza:2018tqz}
A.~Kulesza, L.~Motyka, D.~Schwartl\"ander, T.~Stebel and V.~Theeuwes,
\newblock \emph{{Associated production of a top quark pair with a heavy
  electroweak gauge boson at NLO$+$NNLL accuracy}},
\newblock Eur. Phys. J. C \textbf{79}(3), 249 (2019),
\newblock \doi{10.1140/epjc/s10052-019-6746-z},
\newblock \href{http://arxiv.org/abs/1812.08622}{{\tt 1812.08622}}.

\bibitem{Broggio:2019ewu}
A.~Broggio, A.~Ferroglia, R.~Frederix, D.~Pagani, B.~D. Pecjak and I.~Tsinikos,
\newblock \emph{{Top-quark pair hadroproduction in association with a heavy
  boson at NLO+NNLL including EW corrections}},
\newblock JHEP \textbf{08}, 039 (2019),
\newblock \doi{10.1007/JHEP08(2019)039},
\newblock \href{http://arxiv.org/abs/1907.04343}{{\tt 1907.04343}}.

\bibitem{Kulesza:2020nfh}
A.~Kulesza, L.~Motyka, D.~Schwartl\"ander, T.~Stebel and V.~Theeuwes,
\newblock \emph{{Associated top quark pair production with a heavy boson:
  differential cross sections at NLO+NNLL accuracy}},
\newblock Eur. Phys. J. C \textbf{80}(5), 428 (2020),
\newblock \doi{10.1140/epjc/s10052-020-7987-6},
\newblock \href{http://arxiv.org/abs/2001.03031}{{\tt 2001.03031}}.

\bibitem{Frixione:2014qaa}
S.~Frixione, V.~Hirschi, D.~Pagani, H.~S. Shao and M.~Zaro,
\newblock \emph{{Weak corrections to Higgs hadroproduction in association with
  a top-quark pair}},
\newblock JHEP \textbf{09}, 065 (2014),
\newblock \doi{10.1007/JHEP09(2014)065},
\newblock \href{http://arxiv.org/abs/1407.0823}{{\tt 1407.0823}}.

\bibitem{Frixione:2015zaa}
S.~Frixione, V.~Hirschi, D.~Pagani, H.~S. Shao and M.~Zaro,
\newblock \emph{{Electroweak and QCD corrections to top-pair hadroproduction in
  association with heavy bosons}},
\newblock JHEP \textbf{06}, 184 (2015),
\newblock \doi{10.1007/JHEP06(2015)184},
\newblock \href{http://arxiv.org/abs/1504.03446}{{\tt 1504.03446}}.

\bibitem{Frederix:2018nkq}
R.~Frederix, S.~Frixione, V.~Hirschi, D.~Pagani, H.~S. Shao and M.~Zaro,
\newblock \emph{{The automation of next-to-leading order electroweak
  calculations}},
\newblock JHEP \textbf{07}, 185 (2018),
\newblock \doi{10.1007/JHEP11(2021)085},
\newblock [Erratum: JHEP 11, 085 (2021)],
\newblock \href{http://arxiv.org/abs/1804.10017}{{\tt 1804.10017}}.

\bibitem{Alwall:2014hca}
J.~Alwall, R.~Frederix, S.~Frixione, V.~Hirschi, F.~Maltoni, O.~Mattelaer,
  H.~S. Shao, T.~Stelzer, P.~Torrielli and M.~Zaro,
\newblock \emph{{The automated computation of tree-level and next-to-leading
  order differential cross sections, and their matching to parton shower
  simulations}},
\newblock JHEP \textbf{07}, 079 (2014),
\newblock \doi{10.1007/JHEP07(2014)079},
\newblock \href{http://arxiv.org/abs/1405.0301}{{\tt 1405.0301}}.

\bibitem{Frixione:2019fxg}
S.~Frixione, B.~Fuks, V.~Hirschi, K.~Mawatari, H.-S. Shao, P.~A. Sunder and
  M.~Zaro,
\newblock \emph{{Automated simulations beyond the Standard Model:
  supersymmetry}},
\newblock JHEP \textbf{12}, 008 (2019),
\newblock \doi{10.1007/JHEP12(2019)008},
\newblock \href{http://arxiv.org/abs/1907.04898}{{\tt 1907.04898}}.

\bibitem{Barnett:1987jw}
R.~M. Barnett, H.~E. Haber and D.~E. Soper,
\newblock \emph{{Ultraheavy Particle Production from Heavy Partons at Hadron
  Colliders}},
\newblock Nucl. Phys. B \textbf{306}, 697 (1988),
\newblock \doi{10.1016/0550-3213(88)90440-3}.

\bibitem{Dicus:1988cx}
D.~A. Dicus and S.~Willenbrock,
\newblock \emph{{Higgs Boson Production from Heavy Quark Fusion}},
\newblock Phys. Rev. D \textbf{39}, 751 (1989),
\newblock \doi{10.1103/PhysRevD.39.751}.

\bibitem{Dicus:1998hs}
D.~Dicus, T.~Stelzer, Z.~Sullivan and S.~Willenbrock,
\newblock \emph{{Higgs boson production in association with bottom quarks at
  next-to-leading order}},
\newblock Phys. Rev. \textbf{D59}, 094016 (1999),
\newblock \doi{10.1103/PhysRevD.59.094016},
\newblock \href{http://arxiv.org/abs/hep-ph/9811492}{{\tt hep-ph/9811492}}.

\bibitem{Balazs:1998sb}
C.~Balazs, H.-J. He and C.~P. Yuan,
\newblock \emph{{QCD corrections to scalar production via heavy quark fusion at
  hadron colliders}},
\newblock Phys. Rev. \textbf{D60}, 114001 (1999),
\newblock \doi{10.1103/PhysRevD.60.114001},
\newblock \href{http://arxiv.org/abs/hep-ph/9812263}{{\tt hep-ph/9812263}}.

\bibitem{Cal:2023mib}
P.~Cal, R.~von Kuk, M.~A. Lim and F.~J. Tackmann,
\newblock \emph{{qT spectrum for Higgs boson production via heavy quark
  annihilation at N3LL'+aN3LO}},
\newblock Phys. Rev. D \textbf{110}(7), 076005 (2024),
\newblock \doi{10.1103/PhysRevD.110.076005},
\newblock \href{http://arxiv.org/abs/2306.16458}{{\tt 2306.16458}}.

\bibitem{Das:2024pac}
G.~Das and A.~Sankar,
\newblock \emph{{Next-to-soft threshold effects on Higgs boson production via
  bottom quark annihilation}},
\newblock Phys. Rev. D \textbf{111}(7), 076008 (2025),
\newblock \doi{10.1103/PhysRevD.111.076008},
\newblock \href{http://arxiv.org/abs/2409.01553}{{\tt 2409.01553}}.

\bibitem{Harlander:2003ai}
R.~V. Harlander and W.~B. Kilgore,
\newblock \emph{{Higgs boson production in bottom quark fusion at
  next-to-next-to leading order}},
\newblock Phys. Rev. \textbf{D68}, 013001 (2003),
\newblock \doi{10.1103/PhysRevD.68.013001},
\newblock \href{http://arxiv.org/abs/hep-ph/0304035}{{\tt hep-ph/0304035}}.

\bibitem{Biello:2024vdh}
C.~Biello, A.~Sankar, M.~Wiesemann and G.~Zanderighi,
\newblock \emph{{NNLO+PS predictions for Higgs production through bottom-quark
  annihilation with MINNLOPS}},
\newblock Eur. Phys. J. C \textbf{84}(5), 479 (2024),
\newblock \doi{10.1140/epjc/s10052-024-12845-z},
\newblock \href{http://arxiv.org/abs/2402.04025}{{\tt 2402.04025}}.

\bibitem{Gavardi:2025zpf}
A.~Gavardi, R.~von Kuk and M.~A. Lim,
\newblock \emph{{Resumming transverse observables for NNLO+PS matching in
  GENEVA}},
\newblock Phys. Rev. D \textbf{114}(5), 054022 (2026),
\newblock \doi{10.1103/nlfd-2y6w},
\newblock \href{http://arxiv.org/abs/2505.14773}{{\tt 2505.14773}}.

\bibitem{Raitio:1978pt}
R.~Raitio and W.~W. Wada,
\newblock \emph{{Higgs Boson Production at Large Transverse Momentum in
  {QCD}}},
\newblock Phys. Rev. D \textbf{19}, 941 (1979),
\newblock \doi{10.1103/PhysRevD.19.941}.

\bibitem{Ng:1983jm}
J.~N. Ng and P.~Zakarauskas,
\newblock \emph{{A {QCD} Parton Calculation of Conjoined Production of Higgs
  Bosons and Heavy Flavors in $p \bar{p}$ Collision}},
\newblock Phys. Rev. D \textbf{29}, 876 (1984),
\newblock \doi{10.1103/PhysRevD.29.876}.

\bibitem{Kunszt:1984ri}
Z.~Kunszt,
\newblock \emph{{Associated Production of Heavy Higgs Boson with Top Quarks}},
\newblock Nucl. Phys. B \textbf{247}, 339 (1984),
\newblock \doi{10.1016/0550-3213(84)90553-4}.

\bibitem{Dittmaier:2003ej}
S.~Dittmaier, M.~Kr{\"a}mer and M.~Spira,
\newblock \emph{{Higgs radiation off bottom quarks at the Tevatron and the CERN
  LHC}},
\newblock Phys. Rev. \textbf{D70}, 074010 (2004),
\newblock \doi{10.1103/PhysRevD.70.074010},
\newblock \href{http://arxiv.org/abs/hep-ph/0309204}{{\tt hep-ph/0309204}}.

\bibitem{Dawson:2003kb}
S.~Dawson, C.~B. Jackson, L.~Reina and D.~Wackeroth,
\newblock \emph{{Exclusive Higgs boson production with bottom quarks at hadron
  colliders}},
\newblock Phys. Rev. \textbf{D69}, 074027 (2004),
\newblock \doi{10.1103/PhysRevD.69.074027},
\newblock \href{http://arxiv.org/abs/hep-ph/0311067}{{\tt hep-ph/0311067}}.

\bibitem{Wiesemann:2014ioa}
M.~Wiesemann, R.~Frederix, S.~Frixione, V.~Hirschi, F.~Maltoni and
  P.~Torrielli,
\newblock \emph{{Higgs production in association with bottom quarks}},
\newblock JHEP \textbf{02}, 132 (2015),
\newblock \doi{10.1007/JHEP02(2015)132},
\newblock \href{http://arxiv.org/abs/1409.5301}{{\tt 1409.5301}}.

\bibitem{Biello:2024pgo}
C.~Biello, J.~Mazzitelli, A.~Sankar, M.~Wiesemann and G.~Zanderighi,
\newblock \emph{{Higgs boson production in association with massive bottom
  quarks at NNLO+PS}},
\newblock JHEP \textbf{04}, 088 (2025),
\newblock \doi{10.1007/JHEP04(2025)088},
\newblock \href{http://arxiv.org/abs/2412.09510}{{\tt 2412.09510}}.

\bibitem{Badger:2021ega}
S.~Badger, H.~B. Hartanto, J.~Kry\'s and S.~Zoia,
\newblock \emph{{Two-loop leading-colour QCD helicity amplitudes for Higgs
  boson production in association with a bottom-quark pair at the LHC}},
\newblock JHEP \textbf{11}, 012 (2021),
\newblock \doi{10.1007/JHEP11(2021)012},
\newblock \href{http://arxiv.org/abs/2107.14733}{{\tt 2107.14733}}.

\bibitem{Badger:2024mir}
S.~Badger, H.~B. Hartanto, R.~Poncelet, Z.~Wu, Y.~Zhang and S.~Zoia,
\newblock \emph{{Full-colour double-virtual amplitudes for associated
  production of a Higgs boson with a bottom-quark pair at the LHC}},
\newblock JHEP \textbf{03}, 066 (2025),
\newblock \doi{10.1007/JHEP03(2025)066},
\newblock \href{http://arxiv.org/abs/2412.06519}{{\tt 2412.06519}}.

\bibitem{Deutschmann:2018avk}
N.~Deutschmann, F.~Maltoni, M.~Wiesemann and M.~Zaro,
\newblock \emph{{Top-Yukawa contributions to bbH production at the LHC}},
\newblock JHEP \textbf{07}, 054 (2019),
\newblock \doi{10.1007/JHEP07(2019)054},
\newblock \href{http://arxiv.org/abs/1808.01660}{{\tt 1808.01660}}.

\bibitem{Manzoni:2023qaf}
S.~Manzoni, E.~Mazzeo, J.~Mazzitelli, M.~Wiesemann and M.~Zaro,
\newblock \emph{{Taming a leading theoretical uncertainty in HH measurements
  via accurate simulations for $ \textrm{b}\overline{\textrm{b}}\textrm{H} $
  production}},
\newblock JHEP \textbf{09}, 179 (2023),
\newblock \doi{10.1007/JHEP09(2023)179},
\newblock \href{http://arxiv.org/abs/2307.09992}{{\tt 2307.09992}}.

\bibitem{Hamilton:2013fea}
K.~Hamilton, P.~Nason, E.~Re and G.~Zanderighi,
\newblock \emph{{NNLOPS simulation of Higgs boson production}},
\newblock JHEP \textbf{10}, 222 (2013),
\newblock \doi{10.1007/JHEP10(2013)222},
\newblock \href{http://arxiv.org/abs/1309.0017}{{\tt 1309.0017}}.

\bibitem{ATLAS:2025dix}
{ATLAS Collaboration},
\newblock \emph{{Modelling of top-quark-mediated Higgs boson production in
  association with bottom quarks}},
\newblock Tech. Rep. ATL-PHYS-PUB-2025-033, CERN, Geneva (2025).

\bibitem{Pagani:2020rsg}
D.~Pagani, H.-S. Shao and M.~Zaro,
\newblock \emph{{RIP $ Hb\overline{b} $: how other Higgs production modes
  conspire to kill a rare signal at the LHC}},
\newblock JHEP \textbf{11}, 036 (2020),
\newblock \doi{10.1007/JHEP11(2020)036},
\newblock \href{http://arxiv.org/abs/2005.10277}{{\tt 2005.10277}}.

\bibitem{Harlander:2011aa}
R.~Harlander, M.~Kramer and M.~Schumacher,
\newblock \emph{{Bottom-quark associated Higgs-boson production: reconciling
  the four- and five-flavour scheme approach}}  (2011),
\newblock \href{http://arxiv.org/abs/1112.3478}{{\tt 1112.3478}}.

\bibitem{Forte:2015hba}
S.~Forte, D.~Napoletano and M.~Ubiali,
\newblock \emph{{Higgs production in bottom-quark fusion in a matched scheme}},
\newblock Phys. Lett. \textbf{B751}, 331 (2015),
\newblock \doi{10.1016/j.physletb.2015.10.051},
\newblock \href{http://arxiv.org/abs/1508.01529}{{\tt 1508.01529}}.

\bibitem{Forte:2016sja}
S.~Forte, D.~Napoletano and M.~Ubiali,
\newblock \emph{{Higgs production in bottom-quark fusion: matching beyond
  leading order}},
\newblock Phys. Lett. \textbf{B763}, 190 (2016),
\newblock \doi{10.1016/j.physletb.2016.10.040},
\newblock \href{http://arxiv.org/abs/1607.00389}{{\tt 1607.00389}}.

\bibitem{Gauld:2021zmq}
R.~Gauld,
\newblock \emph{{A massive variable flavour number scheme for the Drell-Yan
  process}},
\newblock SciPost Phys. \textbf{12}(1), 024 (2022),
\newblock \doi{10.21468/SciPostPhys.12.1.024},
\newblock \href{http://arxiv.org/abs/2107.01226}{{\tt 2107.01226}}.

\end{thebibliography}


\end{document}